\documentstyle[12pt,epsf]{article}
\def\photonatomright{\begin{picture}(3,1.5)(0,0)
                                \put(0,-0.75){\tencircw \symbol{2}}
                                \put(1.5,-0.75){\tencircw \symbol{1}}
                                \put(1.5,0.75){\tencircw \symbol{3}}
                                \put(3,0.75){\tencircw \symbol{0}}
                      \end{picture}
                     }

\def\photonatomup{\begin{picture}(1.5,3)(0,0)
                             \put(-0.75,3){\tencircw \symbol{3}}
                             \put(-0.75,1.5){\tencircw \symbol{2}}
                             \put(0.75,1.5){\tencircw \symbol{0}}
                             \put(0.75,0){\tencircw \symbol{1}}
                   \end{picture}
                  }

\def\photonright{\begin{picture}(30,1.5)(0,0)
                     \multiput(0,0)(3,0){10}{\photonatomright}
                  \end{picture}
                 }

\def\photonrighthalf{\begin{picture}(30,1.5)(0,0)
                     \multiput(0,0)(3,0){5}{\photonatomright}
                  \end{picture}
                 }

\def\photonup{\begin{picture}(1.5,30)(0,0)
                  \multiput(0,0)(0,3){10}{\photonatomup}
               \end{picture}
              }

\def\fermionurr{\begin{picture}(30,15)(0,0)
                        \put(-30,-15){\vector(2,1){15}}
                        \put(-15,-7.5){\line(2,1){15}}
                  \end{picture}
                 }

\def\fermiondrr{\begin{picture}(30,15)(0,0)
                        \put(0,0){\vector(2,-1){15}}
                        \put(15,-7.5){\line(2,-1){15}}
                  \end{picture}
                 }

\newenvironment{Feynman}[3]{\begin{center}
                            \setlength{\unitlength}{#3 mm}
                            \begin{picture}(#1)(#2)
                            \thicklines
                           }{\end{picture} \end{center}}

\def\theequation{\arabic{section}.\arabic{equation}}
\newcommand{\ezero}{\setcounter{equation}{0}}
\newcommand{\vph}{ \vphantom{\int\limits_t^t} }
\newcommand{\vsph}{ \vphantom{\int_e^e} }
\newcommand{\Vph}{ \vphantom{\int\limits_{T_A}^{T^A} }}
\newcommand{\mr}{\mathrm}

\newcommand {\ltau} {\mbox{${\mr L}_{ \tau} $}}

\newcommand{\nll}{\nonumber \\}

\newcommand {\Ql} {\mbox{$Q^2_{l}  $}}
\newcommand {\ql} {\mbox{$Q^2_{l}  $}}
\newcommand {\yl} {\mbox{$y  _{l}  $}}
\newcommand {\xl} {\mbox{$x   _{l}  $}}

\newcommand {\qm} {\mbox{$Q^2_{m}  $}}
\newcommand {\ym} {\mbox{$y  _{m}  $}}
\newcommand {\xm} {\mbox{$x   _{m}  $}}
\newcommand {\xh} {\mbox{$x  _{h}  $}}
\newcommand {\yh} {\mbox{$y  _{h}  $}}
\newcommand {\Qh} {\mbox{$Q^2_{h}  $}}
\newcommand {\qh} {\mbox{$Q^2_{h}  $}}
\newcommand {\Qt} {\mbox{$Q^2_{\tau}$}}
\newcommand {\xjb} {\mbox{$x_  {\mathrm{JB}}$}}
\newcommand {\yjb} {\mbox{$y_  {\mathrm{JB}}$}}
\newcommand {\qjb} {\mbox{$Q^2_{\mathrm{JB}}$}}

\newcommand {\LQZ} {\mbox{${\mr L}_{ \tau} $}}

\newcommand {\litwo} {\mbox{${\mr{ {Li}}}_{2} $}}
\newcommand {\lh} {\mbox{${\mr L}_{\mathrm h }$}}
\newcommand {\ljb} {\mbox{${\mr L}_{\mathrm{JB}}$}}
\newcommand {\lm} {\mbox{${\mr L}_{\mathrm m }$}}
\newcommand {\lhone}{\mbox{${\mr L}_{\mathrm{h_1}}$}}
\newcommand {\Lh}{\mbox{${\mr L}_{\mathrm h}$}}
\newcommand {\Lb}{\mbox{${\mr L}_{\mathrm {\beta} }$}}

\newcommand {\oa}  {${\cal O}({\alpha}  )$}
\newcommand {\oaa} {${\cal O}({\alpha}^2)$}
\newcommand{\nn}{\noindent}

\newcommand{\bq}{\begin{equation}}
\newcommand{\eq}{\end{equation}}
\newcommand{\ba}{\begin{eqnarray}}
\newcommand{\ea}{\end{eqnarray}}

\newcommand {\TT}   {\mbox{$y_h S$}}

\begin{document}
\thispagestyle{empty}
\noindent
DESY 94--115
 
\noindent
CERN--TH. 7339/94
 
\noindent
IC/94/154
 
\noindent
updated March 1996
\vfill
{ \huge
\bf \begin{center}
 
Model independent QED corrections
to the process $ep \longrightarrow eX \, ^{\displaystyle \dag}$
 
\vspace*{1.4cm}
\end{center}
     } 
\nn
{\large
Arif Akhundov$\;^{1,2}$,
$\;$
Dima Bardin$\;^{3,4}$,
$\;$
Lida Kalinovskaya$\;^{4}$,
$\;$
Tord~Riemann$\,^3$}
\\
\vspace*{0.5cm}
 
\begin{normalsize}
\begin{tabbing}
$^1$  \=
International Centre for Theoretical Physics,
Strada Costiera 11, I--34014 Trieste, Italy
\end{tabbing}
\begin{tabbing}
$^2$  \=
Institute of Physics, Academy of Sciences of Azerbaijan,
pr. Azizbekova 33,
\\   \>
AZ--370143 Baku, Azerbaijan
\end{tabbing}
\begin{tabbing}
$^3$  \=
DESY -- Institut f\"ur Hochenergiephysik,
  Platanenallee 6, D--15738 Zeuthen, Germany
\end{tabbing}
\begin{tabbing}
$^4$ \=
Laboratory for Theoretical Physics, JINR,
ul. Joliot-Curie 6, RU--141980 Dubna, Russia
\end{tabbing}
\end{normalsize}
 
\vspace*{1.5cm}
\vfill
\thispagestyle{empty}
{\large
\centerline{\bf ABSTRACT}
}
\vspace*{.2cm}
 
\small
\nn
We give an exhaustive presentation of the semi-analytical approach
to the model independent leptonic QED corrections to deep inelastic
neutral current lepton-nucleon scattering.
These corrections include photonic bremsstrahlung from and
vertex corrections to the lepton current of the order \oa\ with soft
photon exponentiation.
A common treatment of these radiative corrections in several variables
-- leptonic, hadronic, mixed, Jaquet-Blondel variables -- has been
developed
and double differential cross sections are calculated.
In all sets of variables  we use some structure functions, which depend
on the
hadronic variables  and which do not have to be
 defined in the quark parton model.
The remaining numerical integrations are twofold (for leptonic
variables) or
onefold (for all other variables).
For the case of hadronic variables, all phase space integrals have been
performed analytically.
Numerical results are presented for a large kinematical range,
covering fixed target as well as collider experiments
at HERA or LEP$\otimes$LHC, with a special emphasis on HERA physics.
\normalsize
\vfill
\vspace*{.5cm}
 
\bigskip
\vfill
\footnoterule
\nn
$^{\dag}$~Supported by the Heisenberg-Landau fund.
{\small
\begin{tabbing}
Emails: \= {\tt akhundov@ictp.trieste.it, bardin@ifh.de,}
\\      \> {\tt lika@thsun1.jinr.dubna.su, riemann@ifh.de}
\end{tabbing}
}
\newpage
\thispagestyle{empty}
$\;$
\vspace{2cm}
\begin{Large}
 \begin{center}
{
\sf
 To the memory of our colleague and friend \\
 \bigskip
 
  {Oleg Fedorenko}  \\
 \bigskip
 
   1951--1994
}
\end{center}
\end{Large}
\newpage
\newpage
\tableofcontents
\newpage
\section
{Introduction
\label{intr}}
\begin{figure}[t]
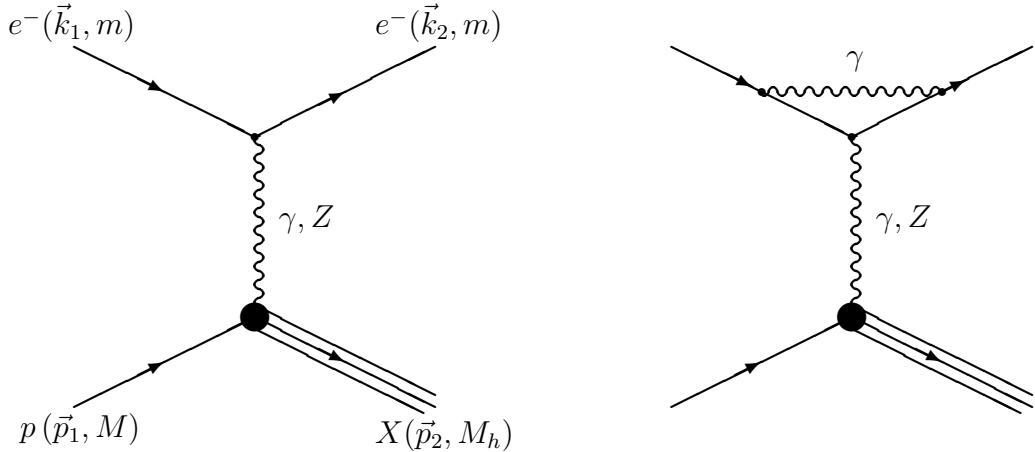

\vspace*{1.cm}
\begin{minipage}[tbhp]{7.8cm}{
\begin{center}
\begin{Feynman}{60,60}{0,0}{0.8}
%
\put(00,60){\fermiondrr}
\put(-11,62){$e^-(\vec k_1,m)$}  
\put( 50,62){$e^-(\vec k_2,m)$}
\put( 50,-06){$X(\vec p_2,M_h)$}
\put(-09,-05){$p \, (\vec p_1,M)$}
\put(34,30){$\gamma, Z$}
\put(60,60){\fermionurr}
\put(30.5,15){\photonup}
\put(30,15){\circle*{5}}
\put(30,17){\line(2,-1){30}}
\put(30,13){\line(2,-1){28}}
\put(30,45){\circle*{1.5}}
\put(30,15){\fermionurr}
\put(30,15){\fermiondrr}
\end{Feynman}
\end{center}
}\end{minipage}
\begin{minipage}[tbhp]{7.8cm} {
\begin{center}
\begin{Feynman}{60,60}{0,0}{0.8}
%
\put(34,30){$\gamma, Z$}
\put(29,57){$\gamma$}
\put(00,60){\vector(2,-1){13.5}}
\put(11,54.5){\line(2,-1){19.}}
%
\put(45,52.5){\line(2,1){15}}
\put(30,45){\vector(2,1){19.}}
\put(14.5,52.4){\photonright}
\put(45.0,52.4){\circle*{1.5}}
\put(15.0,52.4){\circle*{1.5}}
\put(30,45){\circle*{1.5}}
\put(30.5,15){\photonup}
\put(30,15){\circle*{5}}
\put(30,17){\line(2,-1){30}}
\put(30,13){\line(2,-1){28}}
\put(30,15){\fermionurr}
\put(30,15){\fermiondrr}
\end{Feynman}
\end{center}
}\end{minipage}
\vspace*{1.cm}
%
\caption{\it
Deep inelastic scattering of electrons off protons:
(a) Born diagram, (b) leptonic vertex correction.
\label{fig1}
}
\end{figure}

\begin{figure}[thbp]
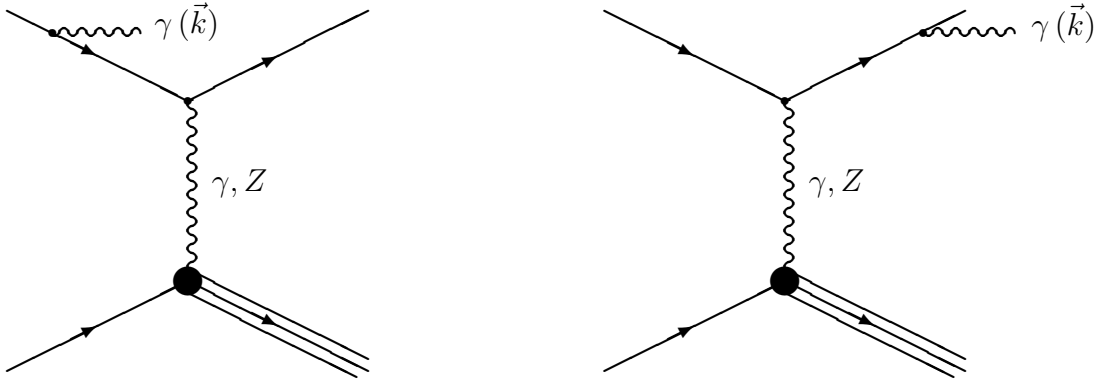

\vspace*{1.cm}
\begin{minipage}[tbh]{7.8cm}{
\begin{center}
\begin{Feynman}{60,60}{0,0}{0.8}
%
\put(00,60){\fermiondrr}
\put(60,60){\fermionurr}
\put(30.5,15){\photonup}
\put(30,15){\circle*{5}}
\put(24.5,56.25){$\gamma \, (\vec k)$}
\put(34,30){$\gamma, Z$}
\put(30,17){\line(2,-1){30}}
\put(30,13){\line(2,-1){28}}
\put(30,45){\circle*{1.5}}
\put(30,15){\fermionurr}
\put(30,15){\fermiondrr}
\put(7.5,56.25){\circle*{1.5}}
\put(7.0,56.25){\photonrighthalf}
\end{Feynman}
\end{center}
}\end{minipage}
\begin{minipage}[tbh]{7.8cm} {
\begin{center}
\begin{Feynman}{60,60}{0,0}{0.8}
%
\put(00,60){\fermiondrr}
\put(60,60){\fermionurr}
\put(30.5,15){\photonup}
\put(71,56){$\gamma \, (\vec k)$}
\put(34,30){$\gamma, Z $}
\put(30,15){\circle*{5}}
\put(30,17){\line(2,-1){30}}
\put(30,13){\line(2,-1){28}}
\put(30,45){\circle*{1.5}}
\put(30,15){\fermionurr}
\put(30,15){\fermiondrr}
\put(53.0,56.25){\photonrighthalf}
\put(53.0,56.25){\circle*{1.5}}
\end{Feynman}
\end{center}
}\end{minipage}
\vspace*{.5cm}
%
\caption{\it
The two leptonic bremsstrahlung diagrams.
\label{fig4}
}
\end{figure}

\ezero
The first deep inelastic scattering experiments have been performed
at SLAC in
1968 and lead to the discovery of the parton structure of
nucleons~\cite{taylor}.
The cross section of the reaction of figure~\ref{fig1},
\bq
e(k_1) + p(p_1) \rightarrow e(k_2) + X(p_2),
\label{eqborn}
\eq
was determined at a transferred momentum of $\sqrt{Q^2}\sim 3$~GeV.
The measurements
were based on  the registration of the scattered electron's
energy and angle.
The physical interpretation of the data took into account contributions
from
two important competing processes:
the bremsstrahlung contribution
to~(\ref{eqborn})
of figure~\ref{fig4}
with
non-observed photon(s),
\bq
e(k_1) + p(p_1) \rightarrow e(k_2) + X(p_2) + n\gamma(k),
\label{eqdeep}
\eq
and also the elastic radiative tail, i.e. photonic bremsstrahlung in the
elastic channel:
\bq
e(k_1) + p(p_1) \rightarrow e(k_2) + p(p_2) + n\gamma(k).
\label{eqERT}
\eq
 
Semi-analytical expressions for the cross sections of these radiative corrections
to~(\ref{eqborn}) had been calculated in the classical work of
Mo and Tsai~\cite{motsai}.
They were of considerable importance for the interpretation of the data.
In the long period between
    the SLAC experiment and the start-up              of HERA,
the experimental devices for the study of
deep inelastic  lepton-nucleon
scattering had a relatively simple structure.
Therefore,
there was little need to
radically improve the treatment of the radiative corrections as developed
in~\cite{motsai}.
On the other hand, rising experimental accuracy and considerably
higher beam energies had to be met with corresponding improvements
in the treatment of radiative corrections.
 
\bigskip
 
Originally motivated by the needs of the BCDMS experiment~\cite{BCDMS},
a series of papers~\cite{AB}--\cite{ABS5}
was published by the Dubna group in the seventies and  eighties.
They all are devoted to the  model independent semi-analytical calculation of
QED corrections arising from the reactions~(\ref{eqdeep})
and~(\ref{eqERT}).
The QED corrections connected with the electrons are treated for processes
with virtual exchange of a photon between the lepton and the hadronic
system.
Applications to high energy muon scattering were first published in
~\cite{II}--\cite{IV}.
In~\cite{BS},
a technique was developed
for a covariant treatment
of the
corrections, thus preventing the introduction of an unphysical parameter
 $\bar{\omega}$; this parameter was used in~\cite{motsai} to
divide the contribution from~(\ref{eqdeep})
into two parts  which describe soft and hard photons separately.
A covariant form of the peaking approximation of~\cite{motsai} and of
 soft photon exponentiation  have been introduced in~\cite{Shum}.
With the momentum transfer of $\sqrt{Q^2} \sim 10$ GeV in
    the BCDMS experiment,
the $Z$-boson exchange had to be taken into account in the Born
cross section.
Therefore, the structure functions became modified, and a new one
describing
parity violation had to be introduced~\cite{Derman}.
All these developments, together with
the contributions from QED lepton pair production~\cite{ABS4}
and the \oaa\ corrections of the process~(\ref{eqERT})
                            ~\cite{ABS5}
    have been included
in the semi-analytical program {\tt TERAD86}~\cite{TERAD86}.
A detailed comparison of the Fortran program {\tt FERRAD}, which is
based on the approach of Mo and Tsai with {\tt TERAD86}
may be found in~\cite{badelek}.
 
So far, the theoretical description of deep inelastic scattering
relied on a picture, where leptons with some couplings (electrical
charge, weak neutral couplings) interact via a neutral current with
a hadronic system, being described by structure functions.
These contain the quark couplings and distributions.
In this sense, a {\em factorization} of the matrix
element into a leptonic and a hadronic part occured.

 
\bigskip
 
In a next step,
the complete treatment of electroweak Standard Model corrections
to deep inelastic
lepton-nucleon scattering was performed
                    in~\cite{BFS},
thereby
covering both the complete \oa\ photonic
corrections with soft photon exponentiation and the
full set of weak one loop insertions.
Such a complete treatment of the weak corrections unavoidably spoils
the aforementioned factorization between the electronic and the
hadronic part of the cross
section formula, which is the spiritual basis of the introduction of
structure functions. See~\cite{zfpc42} for details. Therefore,
the model independent approach to the radiative corrections has to be
given up
in favor of the quark parton model.
The calculations in~\cite{BFS} go beyond the model independent
approach in a second respect:
A complete treatment of the photonic corrections
includes
photonic radiation from the hadronic state, containing as a part the
electron-quark
interference.
The numerical calculations in the quark parton model approach
were realized in the Fortran program {\tt ASYMETR}~\cite{BFS}.
At that time, the sophisticated treatment of the tiny
effects which are covered in {\tt ASYMETR} but not in {\tt TERAD86}
was not justified by experimental needs so that the program
{\tt ASYMETR} did not
find a broader attention.
The same happened with the early
study published in~\cite{consoli}, where
a leading logarithmic calculation of leptonic QED corrections
for $ep$ scattering at HERA energies
was performed.
Recently, initiated by the HERA Physics Workshops,
the Dubna-Zeuthen group
updated the
treatment of the weak one loop corrections and the renormalization
scheme
used in
{\tt ASYMETR}.
Further, the QED part was recalculated, carefully checked, and
considerably compactified~\cite{zfpc42},~\cite{zfpc44};
the related programs are {\tt DISEPNC} and {\tt DISEPCC}. These
two programs use the weak library {\tt DIZET}~\cite{dizet}.

\bigskip
 
Recently,
the experimental techniques have been improved considerably and
much higher beam energies are reached.
In deep inelastic fixed target experiments, transferred momenta
of
$\sqrt{Q^2}\sim 10$ GeV and
$\sqrt{Q^2}\sim 17$ GeV
were obtained at CERN~\cite{SPS} and Fermilab~\cite{FERMILAB}.
At HERA~\cite{WS79,WS87,WS91},
the mass scale of the weak gauge bosons is in reach,
$\sqrt{Q^2}\sim 100$ GeV.
As did the SLAC experiment, the recent fixed target experiments rely
on
the
observation of the scattered electrons (muons) (with the notable
exclusion of
the neutral current neutrino scattering experiments~\cite{CHARM}).
By comparison, the HERA detectors H1~\cite{H1} and ZEUS~\cite{ZEUS}
represent a new generation.
They allow to detect
the scattered electrons, the hadronic final state, and also photons
with high precision.
 Thus,
for the physical analysis of the $ep$ collisions at HERA one may use
not only the familiar electron variables,
\ba
Q_l^2 = (k_1 - k_2)^2, \hspace{1.cm}
y_l = \frac{p_1 (k_1 - k_2)}{ p_1 k_1} ,
\hspace{1.cm} x_l =  \frac{Q_l^2}{y_l S},
\label{qxyl}
\ea
where
\ba
     s =-(k_1 + p_1)^2 = S+m^2+M^2 \approx 4E_eE_p,
\label{s}
\ea
but also the kinematical variables from the hadron measurement,
\ba
Q_h^2 = (p_2 - p_1)^2
, \hspace{1.cm}
y_h = \frac{p_1 (p_2 - p_1)}{ p_1 k_1} ,
\hspace{1.cm} x_h =  \frac{Q_h^2}{y_h S},
\label{qxyh}
\ea
or some composition of both, the so-called mixed
variables~\cite{Max,variables}:
\ba
Q_m^2 = \ql, \hspace{1.cm}
\ym = y_h,
\hspace{1.cm} x_m =  \frac{Q_l^2}{y_h S}.
\label{qxy2}
\ea
Here $E_e,m$ and $E_p,M$ are the energies and masses of incident
electron and proton (see figure~\ref{fig1}).
 
\begin{figure}[bhtp]
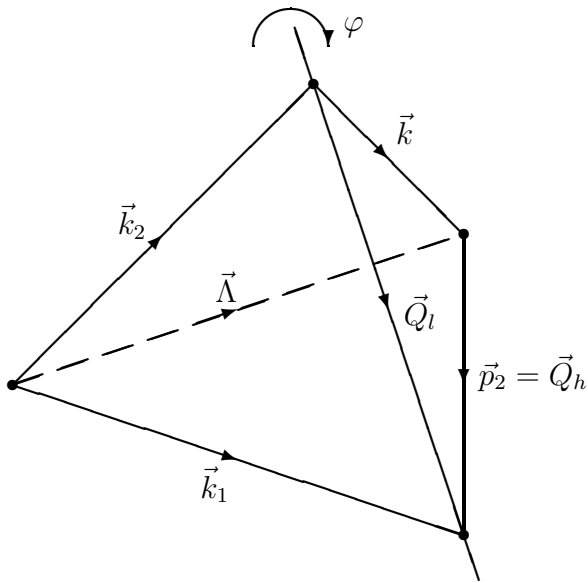

\begin{minipage}[bht]{12.cm}{
\begin{center}
\begin{Feynman}{70,75}{0,0}{1.}
%
\thicklines 
\put(00,20){\line(1, 1){40}} 
\put(00,20){\line(3,-1){60}} 
\put(60,00){\line(0, 1){40}} 
\put(40,60){\line(1,-1){20}} 
\put(40,60){\line(1,-3){20}} 
%
\multiput(00,20)(6,2){10}{\line(3,1){4}}  
\put(60,00){\line(1,-3){2.0}}  
\put(40,60){\line(-1,3){2.5}}  
\put(37,65){\oval(10,10)[t]}
\put(42,65){\vector(0,-1){0}}
\put(44,67){$\varphi$}
%
\put(20,40){\vector(1, 1){0}}
\put(14,40){${\vec k}_2$}
\put(30,10){\vector(3,-1){0}}
\put(25,05){${\vec k}_1$}
\put(60,20){\vector(0,-1){0}}
\put(62,20){${\vec p}_2={\vec Q}_h$}
\put(50,50){\vector(1,-1){0}}
\put(51,52){${\vec k}$}
\put(50,30){\vector(1,-3){0}}
\put(52,28){${\vec Q}_l$}
\put(30,30){\vector(3, 1){0}}
\put(27,31){${\vec \Lambda}$}
\put(00,20){\circle*{1.5}}
\put(60,00){\circle*{1.5}}
\put(60,40){\circle*{1.5}}
\put(40,60){\circle*{1.5}}
%
\end{Feynman}
\end{center}
}\end{minipage}
\vspace*{1.cm}
\caption{\it
Spatial configuration of the momenta in reaction~(1.2)
in the proton rest system.
}
\label{tetra}
\end{figure}
%

Another useful
set of hadronic variables has been introduced by Jaquet and
Blondel~\cite{JB}:
\ba
\qjb =
\frac {({\vec p}_{2}^{\perp})^2} {1-\yh},
\hspace{1.cm}
\yjb = \yh,
\hspace{1.cm}
\xjb &=& \frac{\qjb}{ \yh S}.
\label{xjb}
\ea
 
Different choices of variables make no difference for the determination
of
the cross section of reaction~(\ref{eqborn}) in the Born approximation.
Although, there are huge differences in the predictions  for the
radiative corrections,
because the kinematics becomes quite different. This may be seen from
the tetrahedron of momenta  which is shown in figure~\ref{tetra}.
For vanishing photon momentum $k$, the simple Born kinematics is
recovered.
The differences concern the {\em calculation} of the
corrections, but also, and maybe more important,
their {\em numerical values.}
In addition, the advanced HERA detectors allow the application of
dedicated experimental cuts in the event selection  which represent
further
potential problems for the calculation of radiative corrections.
A satisfactory treatment relies almost exclusively on the use of
the Monte-Carlo technique.
Nevertheless,
from the data with sometimes complicated cuts,
it is usually possible to reconstruct
some sufficiently inclusive and smooth cross sections.
These cross sections
then may be the subject of further
study by a semi-analytical approach as is advocated here.
 
The HERA physics workshops in 1987~\cite{WS87} and 1991~\cite{WS91}
lead to an
 enhanced activity both in the calculation of radiative corrections and in the
comparison between the results of different authors~\cite{Hubert}.
Several authors obtained new results for
the
radiative corrections using quite different techniques, which we quote
here
for completeness:
leading logarithmic calculations~\cite{Berends}--\cite{Kripfganz},
    Monte-Carlo approaches
~\cite{HERACLES}--\cite{KRONOS}, and
semi-analytical approaches
~\cite{zfpc42,zfpc44,TERAD91,ABBCK,BOSP,EPRC91}
for both     the neutral current and the charged current reactions at
HERA.
Eventually, all     groups were able to reproduce the numerical results
of the
above mentioned series of papers
of the Dubna-Zeuthen group
with reasonable precision~\cite{brighton} and, of course,
to go beyond in many other aspects~\cite{Hubert}.
Considerable parts of the present article have been worked out during
and after the 1991 workshop on HERA physics.

\bigskip
 
In the following, we will restrict ourselves to neutral current
scattering.
The                 Born cross section is:
\ba
\frac {d^2 \sigma_{\mr B}} {dy dQ^2} =
        \frac{2 \pi \alpha^{2}}{{S}}
        \sum_{i=1}^3 {\cal A}_i(x,Q^2) \frac{1}{Q^4}\;
        {\cal S}_{i}^{\mr B} (y,Q^2),
\label{eqBor0}
\ea
with the kinematical factors
\ba
{\cal S}_{1}^{\mr B}(y,Q^2) &=& Q^2,
\label{eqS3c1}
\nll
{\cal S}_{2}^{\mr B}(y,Q^2) &=& 2(1-y)S^2,
\label{eqS3c2}
\nll
{\cal S}_{3}^{\mr B}(y,Q^2) &=& 2Q^{2} (2-y)S.
\nonumber
\label{eqScB}
\ea
%
The generalized structure functions
${\cal A}_i(x,Q^2)$ describe the
electroweak interactions
of the electrons with the protons
via the exchange of a photon or $Z$ boson and will be defined later.

\bigskip
 
It is well-known from the above mentioned earlier
studies, e.g. from~\cite{zfpc42},
that a treatment of the leptonic photonic corrections
covers the bulk of the complete corrections to this cross section.
Fortunately, both types of corrections -- weak loop insertions and
QED corrections related to the hadronic current --
are relatively small. For a large part of the kinematical region they are
below the experimental accuracy.
If the experimental intention is a study of the hadronic current, one
may
be interested in a model independent description of the cross section
which not necessarily uses
the quark parton model.
With the last remarks, the line of thought which we will
follow from now on in this article is indicated:
\\
{\em
This article is
devoted to complete, model independent, semi-analytical calculations
of leptonic corrections to neutral current deep inelastic
lepton-nucleon
scattering in different kinematical variables.}
By {\em complete}  we mean the full \oa\ corrections
with soft photon
exponentiation  which are not
restricted to the leading logarithmic approximation.
By {\em semi-analytical}  we understand that the
                 Monte-Carlo technique is not used.
We perform as many analytical
integrations as is possible for a given set of kinematical variables.
In order to get
a double differential cross section, one has to perform three
phase space integrations.
In principle,
 one is interested to describe the cross sections with structure
functions  which may have an arbitrary dependence on the
variables $\xh, \qh$.
Then, for the case of {\em leptonic}
variables, only one analytical integration, for the
case of {\em Jaquet-Blondel} variables
 or for {\em mixed} variables -- two integrations,
 and for {\em hadronic} variables
all three phase space integrations may be performed analytically.
 
\bigskip
 
In the above discussion, the characteristic elements of the calculation of real
bremsstrahlung corrections have been introduced:
 
\begin{itemize}
\item
Choice of a reasonable phase space parameterization with a practical set of
internal variables  which are to be integrated over;
\item
Choice of the order of integration  and complete understanding of the
corresponding kinematical boundaries, by necessity without
neglect of masses;
\item
Separation of the infrared singular part of the bremsstrahlung integral
with use of a special rest system which has to be chosen appropriately;
\item
Dedicated performance of the various hard bremsstrahlung integrations
which are regulated or finite in the soft photon part of the phase space;
\item
Calculation of the infrared divergent correction with a reasonable
regularization procedure.
Consecutively, compensation of the infrared
singularity with that of the virtual corrections and elimination of the
soft and hard photon separation with establishment
of the lorentz invariance of
the net correction.
\end{itemize}
 
\bigskip
 
In order to give an impression of the spirit of the approach
we quote here an expression which contains the complete leptonic QED
corrections
to order \oa\ with soft-photon exponentiation
to the Born cross section~(\ref{eqBor0}):
\ba
\frac{d^2 \sigma}{d \yh d \qh}
=
\frac{d^2 \sigma_{\mr B}}{d \yh d \qh }
\exp \left[ \frac{\alpha}{\pi} \delta_{\mr {inf}} (\yh,\qh) \right]
+ \frac{2 \alpha^3}{S} \sum_{i=1}^3 \frac{1}{Q_h^4}
{\cal A}_i(\xh,\qh) {\cal S}_i(\yh,\qh).
\nonumber
\label{d2wh}
\ea
The soft photon corrections factorize with the Born cross section.
They are exponentiated in order to
take into account the multiple soft photon emission:
\ba
\delta_{\mr {inf}} (\yh,\qh)
=
\left( \ln\frac{\qh}{m^2}   - 1 \right) \ln(1-\yh).
\label{dinwh}
\ea
The hard bremsstrahlung corrections are taken into account to order
\oa. They do not factorize but lead to modifications of the functions
${\cal S}_i^{\mr B}$ while the generalized structure functions
${\cal A}_i(x,Q^2)$ remain unchanged:
\ba
{\cal S}_1(\yh,\qh)
&=&
\qh
\Biggl[
\frac{1}{4}\ln^2\yh -\frac{1}{2}\ln^2(1-\yh) -\ln\yh\ln(1-\yh)
-\frac{3}{2}\litwo(\yh) +\frac{1}{2}\litwo(1)
\nll & &
- \frac{1}{2} \ln \yh \lh + \frac{1}{4} \left( 1 + \frac{2}{\yh} \right)
\lhone
+\left(1- \frac{1}{\yh} \right) \ln \yh
   - \left(1+\frac{1}{4\yh}\right)
\Biggr],
\nll
\label{w1h}
\nll
{\cal S}_2(\yh,\qh)
&=&
S^2
\Biggl[
-\frac{1}{2}\yh\ln^2\yh -(1-\yh)\ln^2(1-\yh) -2(1-\yh)\ln\yh\ln(1-\yh)
-\yh\litwo(1)
\nll & &
-~(2-3\yh)\litwo(\yh)
+ \yh  \ln \yh \lh
+ \frac{\yh}{2} (1-\yh) \lhone
-\frac{\yh}{2} (2-\yh) \ln \yh
\Biggr],
\nll
\label{w2h}
\nll
{\cal S}_3(\yh,\qh)
&=&
S \qh
\Biggl\{
-(2-\yh)\Biggl[
-\frac{1}{2}\ln^2\yh +\ln^2(1-\yh) +2\ln\yh\ln(1-\yh)
+3\litwo(\yh)
\nll & &
-~\litwo(1) +\ln\yh  +\ln\yh \lh
\Biggr]
+ \frac{3}{2} \yh \lhone
   + \yh - \frac{7}{2}
+ 2~(1-2\yh) \ln \yh
\Biggr\},
\nll
\label{w3h}
\ea
with $\lh = \ln (\qh/m^2)$ and $\lhone = \lh + \ln[\yh/(1-\yh)]$.
 
Since the structure functions depend on the hadronic variables and
are hardly to be integrated over together with radiative corrections,
such a compact result may be obtained only in
hadronic                      variables.
 
\bigskip
 
The purpose of the present article is fourfold:
 
\begin{itemize}
\item
We give a systematic presentation of the model independent
approach to the leptonic QED
corrections in terms of leptonic variables, thereby
compactifying known results and transforming them into a modern
terminology.
\item
The complicated kinematics of deep inelastic scattering is
explained in great detail for leptonic, mixed, hadronic,
and Jaquet-Blondel variables.
We systematically retain both the electron and the proton masses if needed.
 This
is necessary for many intermediate steps of the calculation of
QED corrections.
\item
The complete leptonic QED corrections to order \oa\ with soft photon
exponentiation in terms of mixed, hadronic, and Jaquet-Blondel
variables are reviewed here for the first time\footnote{
With the exception of references~\cite{jbpl} and~\cite{teup}, which
contain short collections of the main formulae.
}.
\item
The contents of this article is
the theoretical framework for the Fortran program {\tt TERAD91}
\cite{TERAD91}.
This program is used here
for a systematic numerical study of the QED corrections for a wide
kinematical range: fixed target deep inelastic $\mu p$~scattering, $ep$~
scattering
at HERA, and $ep$ scattering at LEP$\otimes$LHC~\cite{leplhc}.
Naturally, the
main emphasis will be on applications to HERA physics.
\end{itemize}
 
\bigskip
 
The article is organized as follows.
After an introduction of notations and of the Born cross section
the radiative process~(\ref{eqdeep}) is described in section~\ref{sigm}.
In section~\ref{kineI}, the phase space is parameterized such that
the analytical integrations we aim at may be performed. A first
integration will be carried out there.
After this, the further treatment of the phase space
will depend on the choice of kinematical variables.
In section~\ref{irreI}, we isolate and handle the infrared singularity,
thereby defining soft and hard cross section parts in a covariant way.
The soft photon exponentiation is performed.
All this will be done separately for leptonic, mixed, and hadronic
variables.
The integrated cross section in the three above-mentioned variables
is derived in section~\ref{sec5}.
There, for the case of mixed and hadronic variables, a second analytical
integration is performed.
Numerical results for the QED corrections at three different
accelerator scenarios will be compared.
In section~\ref{photo}, we introduce the necessary
modifications for an accurate treatment of the photoproduction region.
Section~\ref{kineII} contains a parameterization of the phase space,
which deviates from that introduced in section~\ref{kineI}. This
parameterization will allow us to derive compact formulae for the
QED corrections in terms of Jaquet-Blondel and
hadronic variables. For the case of hadronic variables  we
            perform all three phase space integrations
analytically,
thus obtaining a double differential
cross section formula for the QED corrections without any
integration being left.
Section~\ref{disc} contains a discussion. It  includes a comparison
with estimates which are based on the leading logarithmic approximation,
a study of the influence of different choices of structure functions
on the relative size of the radiative corrections, an example of
cuts on the photon kinematics, and an outlook.
In appendix~\ref{appa},
we derive several  phase space parameterizations, which are used in
the calculations.
       In
 appendix~\ref{appb}, the
kinematical boundaries for the different variable sets are studied.
The
photons are treated totally inclusively in the main body of this
article.
Appendix~\ref{appd} is an exclusion to this.
It contains
some formulae with cuts on the photonic variables.
Several tables of integrals which have to be used in order to perform
the many                           integrations are listed in
appendices~\ref{appe} and~\ref{appg} for the different phase
space parameterizations.
For completeness, we collect
the cross sections
in leading logarithmic approximation which have been used for comparisons
in appendix~\ref{lla}.
\section
{Matrix elements and differential cross sections
\label{sigm}}
\setcounter{equation}{0}
\subsection
{The Born process $ep \rightarrow e X$
\label{kineborn}
}
 
In order to define the notations,
   we collect some basic formulae for the Born
process~(\ref{eqborn}).
The matrix element which corresponds to the diagram shown in
figure~1a is
\ba
{\cal M}^{\mr B}
&=&
 \frac{s_e e^2}{(2\pi)^{2}}
(2\pi)^3 \langle p_2| {\cal J}_{\mu} |p_1 \rangle \frac{1}{Q^2}
{\bar{u}}(k_2) \left\{ |Q_e|  \gamma_{\mu} + \chi \gamma_{\mu}
\left(v_e+a_e \gamma_5 \right) \right\} u(k_1),
\label{matrixborn}
\ea
where $v_{e}$ and $a_{e}$ are the vector and axial-vector
couplings of the electron to the $Z$~boson:
\ba
v_{e}=1-4 |Q_{e}| \sin^{2}\theta_{W}, \hspace{1.cm} a_{e}=1.
\label{veae}
\ea
Here, $\theta_{W}$ is the weak mixing angle
$\sin^2 \theta_W = 1 - M_W^2/M_Z^2$ and
$s_e |Q_e| = Q_{e} = -1$ is the electron charge.
Further,
\ba
\chi =  \chi (Q^2) =
{G_\mu \over\sqrt{2}}{M_{Z}^{2} \over{8\pi\alpha}}{Q^2 \over
{Q^2+M_{Z}^{2}}},
\label{chiq}
\ea
%
with the finestructure constant $\alpha=1/137.06$,
the Fermi constant $G_{\mu}= 1.16637\cdot 10^{-5}$ GeV$^{-2}$, and the
$Z$ boson mass $M_Z$.
In the numerical examples, we will use the following values:
$\sin^2 \theta_W = 0.228$, $M_Z=91.173$ GeV.
 
The cross section of the reaction~(\ref{eqborn}),
\ba
d\sigma_{\mr B}
= 2 \frac {(2\pi)^2  p_1^0}  {\sqrt{\lambda_S}}
 \sum_{\mr {spins}} \left| {\cal M}^{\mr B} \right|^2
\frac{d{\vec k}_2}{2k_2^0} \prod_i \frac{d{\vec p}_i}{2p_i^0}
\delta^4 (k_1+p_1-k_2-\sum_i p_i),
\ea
\ba
\lambda_S = S^2 - 4m^2M^2,
\label{lambdas}
\ea
depends on three variables, e.g. on $S$, the momentum transfer $Q$, and
the energy transfer $\nu$:
\ba
S = -2 p_1 k_1, \hspace{1.cm} Q^2 = (p_2 - p_1)^2, \hspace{1.cm}
\nu = - 2 p_1 Q.
\label{sqh}
\ea
%
In the following, we will use also the Bjorken scaling variables $y$
and $x$, the parton momentum fraction,
\ba
y= \frac{\nu}{S}, \hspace{1.cm} x=\frac{Q^2}{\nu},
\label{xy}
\ea
with
\ba
Q^2 = x y S.
\label{xyqs}
\ea
 
The phase space element of the Born process is:
\ba
d\Gamma_{\mr B} &=& \frac{d\vec{k}_2}{2k_2^0} \prod_i \frac{d\vec{p}_i}
{2p_i^0}
\delta^{4}(k_1+p_1-k_2- \sum_i p_i)
\nll
&=&
\frac{d\vec{k}_2}{2k_2^0} \frac{d\vec{p}_2}{2p_2^0} dM_h^2
\delta^{4}(k_1+p_1-k_2-  p_2) d\Gamma_h
\nll
&=&
\frac{\pi S}{2 \sqrt{\lambda_S}} dy dQ^2 d\Gamma_h.
\label{gamma}
\ea
In the last step, the integral over $p_2$ has been performed with the aid of
the $\delta$-function. The integral over $k_2$ is performed in
appendix~\ref{ph1}. Further, we introduced
the phase space element of the hadron system in the final state
$d\Gamma_h$ and its invariant mass $M_h$:
\bq
d\Gamma_h = \prod_i \frac{d\vec{p}_i}{2p_i^0}
            \delta^{4}(p_2-\sum_ip_i),
\hspace{1.cm}
            M_h^2 =- p_2^2.
\label{dgmh}
\eq
In the following, the $d\Gamma_h$
                   will become part of the definition of the
hadronic structure functions.
 
With our definitions we follow~\cite{bilenk},
\ba
\frac{p_1^0}{M} \sum_{\mr {spins}} \left| {\cal M}^{\mr B} \right|^2
=
\frac{e^4}{(2\pi)^4 Q^4} \left(
{\cal S}_{\mu \nu}^{\mr B \gamma} W_{\mu \nu}^{\gamma}
+ 2 \chi {\cal S}_{\mu \nu}^{{\mr{BI}}} W_{\mu \nu}^{I}
+ \chi^2 {\cal S}_{\mu \nu}^{{\mr{BZ}}} W_{\mu \nu}^{Z} \right),
\label{swb}
\ea
where ${\cal S}_{\mu \nu}^{\mr {B A}}$ and $W_{\mu \nu}^{A}$,
$A=\gamma,I,Z$,
are the corresponding components of the leptonic and hadronic tensors.
 
The phenomenological hadronic tensors $W_{\mu \nu}^{A}$~\cite{Derman},
\ba
{W}_{\mu\nu}^{A}
&=& (2\pi)^6 \frac{p_1^0}{M}
    \sum \int \langle p_1| {\cal J}_{\mu}^{\gamma,Z}|p_2 \rangle
    \langle p_2| {\cal J}_{\nu}^{\gamma,Z}|p_1 \rangle d\Gamma_h
\nll
&\equiv&
    {W}_{1}^{A}(\delta_{\mu \nu}-{Q_{\mu}Q_{\nu} \over Q^{2}})+
    {{W}_{2}^{A} \over M^{2}}(p_{1\mu}-{p_{1}Q \over Q^{2} }Q_{\mu})
    (p_{1\nu}-{p_{1}Q \over Q^{2}}Q_{\nu})
+ {{W}_{3}^{A} \over 2M^{2}}
e_{{\mu}{\nu}{\rho}{\sigma}}p_{1{\rho}}Q_{\sigma},
\nll
\ea
describe the deep inelastic interactions of unpolarized nucleons with
a photon and a $Z$ boson.
The real scalar functions $W_a^A$ depend on
                           the invariants $Q^{2}$ and $\nu $.
 
After neglecting in the contractions in~(\ref{swb}) those terms from
the
squared
$Z$ exchange, which are proportional to $a_e^2 m^2$,
one observes a factorization of the leptonic and hadronic
parts of the Born cross section:
\ba
\frac {d^2 \sigma_{\mr B}} {dy dQ^2} =
        \frac{2 \pi \alpha^{2}}{{\lambda_S}} S
        \sum_{i=1}^3 {\cal A}_i(x,Q^2) \frac{1}{Q^4}\;
        {\cal S}_{i}^{\mr B} (y,Q^2).
\label{eqBorn}
\ea
The factorized functions ${\cal S}_i^{\mr B}$ are:
\ba
{\cal S}_{1}^{\mr B}(y,Q^2) &=& Q^2-2m^2,
\label{eqS3B1}   \\
{\cal S}_{2}^{\mr B}(y,Q^2) &=& 2[(1-y)S^2-M^{2}Q^{2}],
\label{eqS3B2}                        \\
{\cal S}_{3}^{\mr B}(y,Q^2) &=& 2Q^{2} (2-y)S.
\label{eqS3B}
\ea
 
The factorized hadronic functions ${\cal A}_i(x,Q^2)$ describe the
electroweak interactions
of leptons with beam charge $Q_e$
and a lepton beam polarization $\xi$
via the exchange of a photon or $Z$ boson with unpolarized nucleons:
%
\begin{eqnarray}
\begin{array}{rclcl}
{\cal A}_{1}(x,Q^2)
&\equiv&
{(2MW_{1})}&=&2  {\cal F}_{1}^{\mathrm{NC}}(x,Q^2),
  \\
{\cal A}_{2}(x,Q^2)
&\equiv& \frac{\displaystyle 1}{\displaystyle yS}
  (\nu W_{2}) &=&\frac{\displaystyle 1}{
\displaystyle
yS}
{\cal F}_{2}^{\mathrm{NC}}(x,Q^2),
\\
{\cal A}_{3}(x,Q^2)
&\equiv& \frac{
\displaystyle
1}{
\displaystyle
2 Q^2}{({\nu}W_{3})}&=&\frac{
\displaystyle
1}{
\displaystyle
2 Q^2}
{\cal F}_{3}^{\mathrm{NC}}(x,Q^2).
\end{array}
\label{deffi}
\end{eqnarray}
 
The generalized structure functions
${\cal F}_{i}^{\mathrm{NC}}(x,Q^2)$
are:
\begin{eqnarray}
{\cal F}_{1,2}^{\mathrm{NC}}(x,Q^2)
&=& F_{1,2}(x,Q^2) + 2 |Q_{e}| \left( v_{e} + \lambda a_e \right)
\chi(Q^2) G_{1,2}(x,Q^2)
\nll & &+~4 \left( v_{e}^{2} + a_{e}^{2} + 2 \lambda v_e a_e \right)
\chi^2(Q^2) H_{1,2}(x,Q^2),
\label{f112}
\\
{\cal F}_3^{\mathrm{NC}}(x,Q^2)
&=& -2~\mbox{\rm sign}(Q_l) \Biggl\{
 |Q_{e}|
\left( a_{e} + \lambda v_e \right) \chi(Q^2) xG_{3}(x,Q^2)
\nll
& &
+2
\left[2v_{e} a_e + \lambda \left(v_e^2 + a_e^2 \right) \right]
\chi^2(Q^2) xH_{3}(x,Q^2)\Biggr\},
\label{f123}
\end{eqnarray}
with $\lambda=\xi \,
\mbox{\rm sign}(Q_l) $.
In the above formulae, the $Q_l$ is the charge of the lepton beam.
 
The contribution of the weak loop corrections to~(\ref{eqborn}) is
not within the
scope of the present article.
We only mention that these corrections may be covered by an inclusion of
real-valued weak form factors
into the definitions of the weak neutral couplings~(\ref{veae}).
Such a program has been carried through first in~\cite{zfpc42}, and was
described and updated also in~\cite{EWRC}.
 
The generalized structure functions
${\cal F}_{i}^{\mathrm{NC}}(x,Q^2)$ may also be used to describe
some new phenomena, e.g. the virtual exchange
of an additional heavy gauge boson $Z'$~\cite{zprime}.
 
The running of the QED coupling~$\alpha$
may be taken into account by a real form
factor:
\ba
\alpha(Q^2)
&=& 
\frac{\alpha}
{1- \sum_f Q_f^2 N_f \Delta F_f(Q^2)},
\label{vac0}
\\
\Delta F_{f}(Q^2)
&=&
\frac{\alpha}{\pi}
\left[-\frac{5}{9}+\frac{4}{3}\frac{m_f^2}{Q^2}+\frac{1}{3}
\beta_f
\left(1-\frac{2m_f^2}{Q^2}\right)
\ln \frac{\beta_f+1}{\beta_f-1} \right],
\label{vac1}
\\
\beta_f
&=&
\sqrt{1+\frac{4m_f^2}{Q^2}} .
\label{vac2}
\ea
The sum in~(\ref{vac0}) extends over all charged fermions $f$,
$Q_f$ is the
corresponding electric charge,
and $N_f$ the color factor: $N_f=3,1$ for quarks and
 lep\-tons, respectively.
The heavy fermions practically decouple in~(\ref{vac0}).
For light fermions,
 with $m_f^2 << Q^2$, the following approximate formula is valid:
\ba
\Delta F_f(Q^2)
=
\frac{\alpha}{3\pi}
\left( \ln\frac{Q^2}{m_f^2}-\frac{5}{3}\right).
\label{runalf}
\ea
 
The numerical factors
in~(\ref{f112})--(\ref{f123})
are chosen such that some often used definitions
of the structure functions in the quark parton model are matched.
Assuming here the Callan-Gross relation, it is
\ba
\begin{array}{rcccl}
\vphantom{\int\limits_t^t}
2x F_1(x,Q^2) &=&
F_2(x,Q^2) &=&
x \sum\limits_q |Q_q|^2 \, [ q(x,Q^2) + {\bar q}(x,Q^2) ],
\\
\vphantom{\int\limits_t^t}
2x G_1(x,Q^2) &=&
G_2(x,Q^2) &=&
x \sum\limits_q |Q_q| \, v_q \, [ q(x,Q^2) + {\bar q}(x,Q^2) ],
\\
\vphantom{\int\limits_t^t}
2x H_1(x,Q^2) &=&
H_2(x,Q^2) &=&
x \sum\limits_q \frac{1}{4} \left(v_q^2 + a_q^2 \right)
[ q(x,Q^2) + {\bar q}(x,Q^2) ],
\\
&&
\vphantom{\int\limits_t^t}
x G_3(x,Q^2) &=& x \sum\limits_q |Q_q| \, a_q \,
[ q(x,Q^2) - {\bar q}(x,Q^2) ],
\\
&&
\vphantom{\int\limits_t^t}
x H_3(x,Q^2) &=& x \sum\limits_q \frac{1}{2} \, v_q a_q \,
[ q(x,Q^2) - {\bar q}(x,Q^2) ].
\end{array}
\label{qpdis}
\ea

These definitions should  help the reader to find a
link to other articles on the subject, which often prefer the use
of a slightly different notation
(remind that we use the definition $a_f=1$ for all fermions).
A comprehensive presentation of the basic formulae for the description
of deep inelastic scattering may be found in~\cite{jblec}.

The allowed region of the variables $y$ and $Q^2$ is derived in
appendix~\ref{b13} and shown in figure~\ref{yq}.
In the following, sometimes the exact expressions in the
electron and the proton masses are needed.
So, in the figures with kinematics we will not apply the
ultra-relativistic approximation.
 
\begin{figure}[tbhp]
\begin{center}
\mbox{
\epsfysize=9.cm
\epsffile[0 0 530 530]{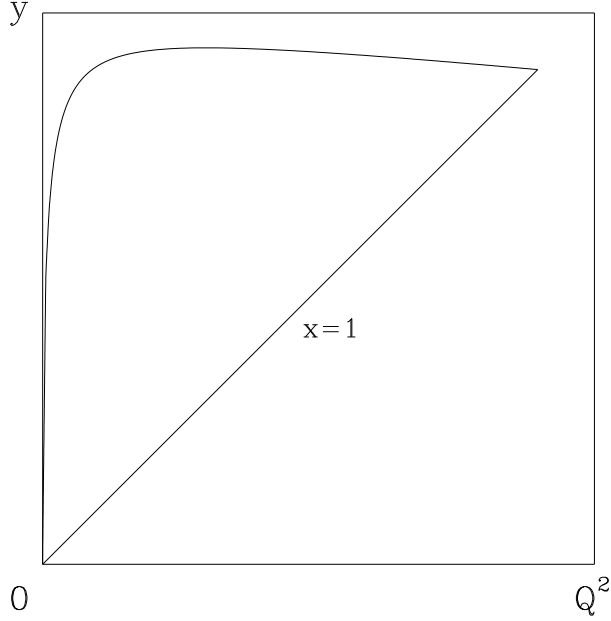}
}
\end{center}
\caption{\it
Physical region of the variables $y$ and $Q^2$.
\label{yq}
}
\end{figure}
 
\begin{figure}[tbhp]
\begin{center}
\mbox{
\epsfysize=9.cm
\epsffile[0 0 530 530]{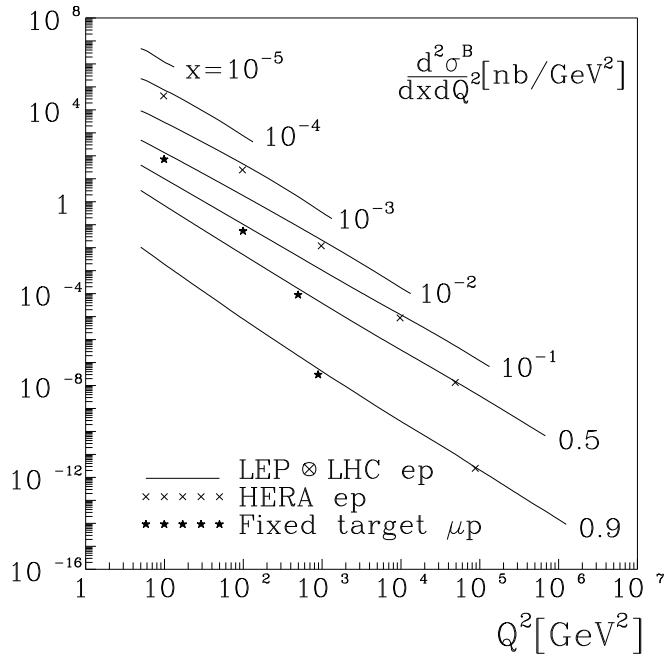}
}
\end{center}
\caption{\it
Born cross section $d^2\sigma^{\rm B}/dxdQ^2$
for deep inelastic neutral current scattering
as function of \,  $\ln \, Q^2$ with $x$ as a parameter.
}
\label{bcs3}
\end{figure}

In figure~\ref{bcs3},
we show the Born cross section~(\ref{eqBorn}), rewritten as a function
of $x$ and $Q^2$:
\ba
\frac {d^2 \sigma_{\mr B}} {dx dQ^2}
&=&
\frac{2 \pi \alpha^2}{x Q^4}
\Biggl\{
2\left( 1 - y - xy\frac{M^2}{S}\right)
{\cal F}_{2}^{\mathrm{NC}}(x,Q^2)
+ 2xy^2\left(1-2\frac{m^2}{Q^2}\right)
{\cal F}_{1}^{\mathrm{NC}}(x,Q^2)
\nll
& &~+~
 y(2-y){\cal F}_{3}^{\mathrm{NC}}(x,Q^2) \Biggr\}.
\label{bf12}
\ea
At small $Q^2$ the neglect of the $Z$ exchange is a good approximation:
\ba
\frac {d^2 \sigma_{\mr B}^{\gamma}} {dx dQ^2}
&=&
\frac{2 \pi \alpha^2}{x Q^4}
\left[
2\left(1-y\right)-2xy\frac{M^2}{S}
+ \left(1-2\frac{m^2}{Q^2}\right)
\left(1+4x^2\frac{M^2}{Q^2} \right)
\frac{y^2}{1+R} \right]
{\cal F}_{2}^{\gamma}(x,Q^2),
\nll
\label{bf1a}
\ea
where $R$ is the ratio of the cross sections with virtual exchange of
longitudinal and transverse photons, respectively:
\ba
R(x,Q^2) = \frac{\sigma_L}{\sigma_T}
=
\left(
1+4x^2\frac{M^2}{Q^2}
\right) \frac{
{\cal F}_{2}^{\gamma}(x,Q^2)
}{
2x{\cal F}_{1}^{\gamma}(x,Q^2)
} - 1.
\label{rqcd}
\ea
This cross section formula becomes
in the ultra-relativistic approximation:
\ba
\frac {d^2 \sigma_{\mr B}^{\gamma}} {dx dQ^2}
=
\frac{2 \pi \alpha^2}{x Q^4}
\left[ Y_+
{\cal F}_{2}^{\gamma}(x,Q^2)
-y^2{\cal F}_{L}^{\gamma}(x,Q^2) \right] ,
\label{xgBorn}
\ea
where the notations
\ba
{\cal F}_{L}^{\gamma}(x,Q^2) &=&
{\cal F}_{2}^{\gamma}(x,Q^2) -2x{\cal F}_{1}^{\gamma}(x,Q^2)
\nll &=&
2x{\cal F}_{1}^{\gamma}(x,Q^2) R(x,Q^2)
\label{fl}
\ea

\noindent
and
\ba
Y_{\pm} = 1 \pm (1-y)^2
\label{ypm}
\ea
are   introduced.
The cross section~(\ref{bf12}) becomes in the ultra-relativistic
approximation:
\ba
\frac {d^2 \sigma_{\mr B}} {dx dQ^2}
=
\frac{2 \pi \alpha^2}{x Q^4}
\left[ Y_+
{\cal F}_{2}^{\mathrm{NC}}(x,Q^2)
+ Y_- {\cal F}_{3}^{\mathrm{NC}}(x,Q^2) \right] .
\label{xqBorn}
\ea
Here, we additionally assume the validity of
the Callan-Gross relation~\cite{callan}:
\ba
{\cal F}_{2}^{\mathrm{NC}}(x,Q^2)
= 2 x {\cal F}_{1}^{\mathrm{NC}}(x,Q^2).
\label{bjli}
\ea
In the numerical examples,
we will cover
   three typical kinematical regions: that of fixed target muon
   scattering
($S=1000$~GeV$^2$), of the $ep$ collider HERA
($S=4\cdot 30 \cdot 820$~GeV$^2$),
and of the $ep$ collider project LEP$\otimes$LHC
($S=4\cdot 50 \cdot 7000$~GeV$^2$).
The solid curves in figure ~\ref{bcs3} are calculated for
LEP$\otimes$LHC. They
start at $Q^2=5$ GeV$^2$, the approximate lower
limit of applicability of the structure functions which are derived from
the available data.
For a given value of $x$, the highest value of $Q^2$ depends on
$S$. This boundary derives  from~(\ref{xyqs}), taken at $y=1$, and
   is seen in the figure.
Further, the actual value of $S$ has an influence on the cross sections
~(\ref{xqBorn})
via the functions $Y_{\pm}$.
This leads to different predictions for the fixed target and HERA
kinematics.
The stars (for fixed targets) and crosses (for HERA)
in the figure indicate the corresponding cross section
values at the maximal allowed $Q^2$ for given $x$.
The first curve which shows a prediction for the fixed target case, e.g.,
is that for $x=10^{-2}$. Smaller values of $x$ are outside the
kinematical range for $Q^2 > 5$ GeV$^2$.
The influence of $S$ on the predictions is minor
(in the variables and scales, which are  chosen here).
In the double logarithmic representation, the cross sections are
nearly linear functions over the complete kinematical region.
The cross section behaviour is determined by the characteristic
$1/(x Q^4)$ dependence.
The $1/Q^4$ arises from the
photon propagator (and, at very high energies, also from the $Z$
propagator).
The event rates are highest for the smallest values of $x$ and
$Q^2$.
With the lower bound on $Q^2$ being fixed, the accessible kinematical
range in $x$ and $Q^2$ rises considerably with the available
beam energies, although the event rates at the highest values of $x$
and $Q^2$ are substantially suppressed.
 
In sum, the $ep$ collider with the highest accessible beam energies
 will open the field not only for a study of very deep inelastic
 scattering, but also, at high rates, the study of the region of very
small $x$.
 
The cross section predictions may depend crucially
on the choice of            structure functions. For the
illustrational
purposes, we decided to use the MRS ${\mr D}_-^{\prime}$
parameterization in the DIS scheme~\cite{lib1}
as it is compiled in~\cite{pdflib1}.
The influence of the structure functions on
the theoretical predictions, including the numerical
values of the QED corrections, will be a subject of the discussion.
\subsection
{The radiative process $ep \rightarrow e X \gamma$
\label{kinerad}
}
The differential cross section for the scattering of electrons
off protons~(\ref{eqdeep}), originating from the Feynman diagrams
of~figure~\ref{fig4},
may be derived from the following matrix element:
\ba
{\cal M}^{\mr R}
&=&
\frac{s_e Q_e e^3}{(2\pi)^{7/2}}
\epsilon_{\alpha}(k)(2\pi)^3 \langle p_2| {\cal J}_{\mu} |p_1 \rangle
\frac{1}{\qh}
\nll & &
\times~
{\bar{u}}(k_2)
\Biggl\{ |Q_e|
\left[
\frac{1}{z_1} \gamma_{\mu} \left( 2 k_{1\alpha}
-{\hat{k}} \gamma_{\alpha} \right)
-
\frac{1}{z_2}  \left( 2 k_{2\alpha} + \gamma_{\alpha}{\hat{k}} \right)
\gamma_{\mu} \right]
\nll
& & +~
\chi
\Biggl[
\frac{1}{z_1} \gamma_{\mu}
\left(v_e+a_e \gamma_5 \right)
\left( 2 k_{1\alpha}
-{\hat{k}} \gamma_{\alpha} \right)
-
\frac{1}{z_2}  \left( 2 k_{2\alpha} + \gamma_{\alpha}{\hat{k}} \right)
\gamma_{\mu}
\left(v_e+a_e \gamma_5 \right)
\Biggr]
\Biggr\}
u(k_1),
\nll
\label{matrixcorr}
\ea
where the electron propagators have the denominators $z_1$ or $z_2$:
\ba
z_1 = -2k_1 k, \hspace{1.cm} z_2 = -2k_2 k.
\label{y1y2}
\ea
\vspace*{1.cm}
The corresponding cross section\footnote
{
The squaring of the Feynman diagrams has been performed with the
program for algebraic manipulations
{\tt SCHOONSCHIP}~\cite{SCHOONSCHIP}.
},
\ba
d\sigma_{\mr R}
=
2 \frac {(2\pi)^2 p_1^0}
{\sqrt{\lambda_S}}
 \sum_{\mr {spins}} \left| {\cal M}^{\mr R} \right|^2
\delta^4 (k_1+p_1-k_2-\sum_i p_i-k) \frac{d{\vec k}_2}{2k_2^0}
\frac{ d{\vec k}}{2k^0}\prod_i \frac{d{\vec p}_i}{2p_i^0},
\ea
may be rewritten as follows:
\ba
 d\sigma_{\mr R}
   &=&   {8 \alpha^{3}\over { \pi^2 \sqrt{\lambda_S}} }\frac{1}{Q_h^4}
   {\cal S} d \Gamma,
\label{eq3.3}
\ea
where
\ba
{\cal S} =
\frac{1}{2} M \left(
{\cal S}_{\mu \nu}^{\gamma} W_{\mu \nu}^{\gamma}
+ 2 \chi {\cal S}_{\mu \nu}^{\rm I} W_{\mu \nu}^{I}
+ \chi^2 {\cal S}_{\mu \nu}^{\rm Z} W_{\mu \nu}^{Z} \right),
\label{sw}
\ea
and
\ba
d\Gamma &=&
\frac{d\vec{k}_2}{2k_2^0} \frac{d\vec{p}_2}{2p_2^0} \frac{d\vec{k}}{2k^0}
 dM_h^2
\delta^{4}(k_2+p_2+k-k_1-p_1).
\label{gammar}
\ea
The
sum~(\ref{sw}) may be treated exactly in the masses of electron $m$
and proton $M$.
%
As mentioned above, the Born cross section~(\ref{eqBorn}) exhibits a nice
factorization property.
We have shown explicitly,
that the same is true for the radiative cross section
after a neglection of the tiny terms proportional
to $\chi^2 a_e^2 m^2/z_{1(2)}$ and $\chi^2 a_e^2 m^4/z_{1(2)}^2$:
\ba
{\cal S} \equiv
{\cal S}({\cal E}, {\cal I},
z_1,z_2)
 =
\sum_{i=1}^3 {\cal A}_i(\xh,\qh)
{\cal S}_{i}({\cal E}, {\cal I},
z_1,z_2).
\label{eqS}
\ea
The neglected terms are due to the $Z$ exchange.
They are vanishing at small $Q^2$ since the $Z$ propagator is
suppressed there compared to the photon propagator.
So,
the corresponding approximation is sufficiently well even in the
photoproduction region.
In that region one has to retain, however,
 all particle masses in the cross section contributions from
photon exchange.
We should mention in addition that certain bremsstrahlung
corrections, which are
proportional to $m^2/z_{1(2)}^2$
yield finite contributions to the cross section even at
larger $Q^2$; this may be seen from an inspection of the tables
of integrals in appendix~\ref{appe}.
 
Besides on $z_1$ and $z_2$,
the radiator functions ${\cal S}_{i}({\cal E}, {\cal I},z_1,z_2)$
depend on two additional two-dimensional sets of variables
${\cal E}$ and ${\cal I}$:
\ba
{\cal S}_{1}({\cal{E,I}
},z_{1},z_{2})
&=&~1+\frac{1}{z_{1}z_{2}}(Q^4_h-4m^{4})
    +\left( \frac{1}{z_{1}}-\frac{1}{z_{2}} \right)
     \left[ \frac{1}{2}(\Qh + \Ql)-2m^{2} \right]  \nll
& &-~  m^{2}(\Qh-2m^{2}) \left( \frac{1}{z_{1}^{2}}
    +\frac{1}{z_{2}^{2}} \right),
\label{eqs1z}
\\
{\cal S}_{2}({\cal{E,I}},z_{1},z_{2})
& = &-~2M^{2}+\frac{1}{z_{1}z_{2}}
\biggl\{ \Qh \bigl[ (1-\yh)S^2 + (1-\yl)(1-\yl+\yh) S^2  \nll
& &-~  2M^{2}( \Qh+2m^{2})\bigr]
  + 2m^{2}\Bigl[ 2(1-\yl) + \yh(\yl-\yh)\Bigr] S^2 \biggr\}  \nll
& &-~ \frac{1}{z_{1}}\Bigl[\yh S^2 +M^{2}(\Qh+\Ql)\Bigr]
-\frac{1}{z_{2}}\Bigl[\yh(1-\yl)S^2-M^{2}(\Qh+\Ql) \Bigr]      \nll
& &-~ 2m^{2}\frac{1}{z_{1}^{2}}\Bigl[ (1-\yl)(1-\yl+\yh)S^2-M^{2}
        \Qh \Bigr]
\nll & &
-~2m^{2}\frac{1}{z_{2}^{2}}
 \Bigl[(1-\yh)S^2-M^{2} \Qh \Bigr]  ,
\label{eqs2z}
\\
{\cal S}_{3}({\cal{E,I}},z_{1},z_{2})
&=&  S\Biggl\{
2\Qh \frac{1}{z_{1}z_{2}}(\Qh + 2m^{2})(2-\yl)  \nll
& &+~   \frac{1}{z_{1}}\Bigl[  2 \Qh - \yh (\Qh + \Ql)\Bigr]
     -\frac{1}{z_{2}}\Bigl[  2(1-\yl)\Qh + \yh (\Qh + \Ql)\Bigr]  \nll
& &-~ 2 m^{2} \Qh \left[\frac{1}{z_{1}^2}(2-2\yl+\yh   )
     +\frac{1}{z_{2}^2}(2 -\yh)\right] \Biggr\}.
\label{eqs3z}
\ea
The
first two of the functions
${\cal S}_i$ are the result of an exact
calculation in both the masses $m$ and $M$ for the case of photon
exchange.
 For $Z$ exchange,
 they contain certain harmless approximations in the electron mass, which
 were mentioned before~(\ref{eqS}).
 The same is with the ${\cal S}_3$.
 
 These radiators
 ${\cal S}_i$
 will be the central objects of the
derivations in the following sections.
In their arguments,
the symbol $\cal E$ stands for
 two {\it external} variables, which will not be
integrated over, while $\cal I$ denotes two {\it internal} variables.
These latter two
become,
together with $z_2$ (or $ z_1$), part of the integration measure.
In the first part of the paper, the variable sets
$\cal E$  and
$\cal I$
 will be selected out of the following
set of variables:
\bq
\begin{array}{rclcrclcrcl}
\vphantom{\int\limits_t^t}
Q_l^2
&=&
(k_1 - k_2)^2,
&\hspace{1.cm}&
\yl
&=&
\frac{
\displaystyle
-2p_1 Q_l}{
\displaystyle
S},
&\hspace{1.cm}&
\xl
&=&
\frac{\ql}{
\displaystyle
\yl S},
\\
\vphantom{\int\limits_t^t}
Q_h^2
&=&
(p_2 - p_1)^2,
&\hspace{1.cm}&
\yh
&=&
\frac{
\displaystyle
-2p_1 Q_h}{
\displaystyle
S},
&\hspace{1.cm}&
\xh
&=&
\frac{\qh}{
\displaystyle
\yh S}.
\end{array}
\label{qxy}
\eq
 
The
radiators ${\cal S}_i$ are the main ingredient of the subsequent
integration over the radiative phase space.
The functions ${\cal S}_1$ and ${\cal S}_2$ agree analytically with
expressions obtained earlier~\cite{III}).
Further, the Monte Carlo program {\tt HERACLES}~\cite{HERACLES}
is based on formulae, which are equivalent to
our functions ${\cal S}_i$ ($i$=1,2,3)~\cite{privat}.
 
The following sections will be devoted to
a number of analytical integrations,
aiming at semi-analytical, compact expressions for the QED corrections.
 
\subsection
{The elastic radiative tail $ep \rightarrow e p \gamma$
\label{tail}
}
A possible background process for deep inelastic scattering is
the {\em elastic} scattering~(\ref{eqERT}) from which
the so-called elastic radiative tail originates.
The corresponding formulae have been derived in~\cite{II,ABBCK}.
They may be obtained from the formulae for reaction~(\ref{eqdeep})
applying
the following relations between the generalized structure functions
${\cal A}_{i}(\xh,\Qh)$ ($i$=1,2,3)
and the generalized elastic form factors $ {\cal A}_{i}(\Qh)$:
\ba
{\cal A}_{1}(\xh,\Qh) & = &
\xh \, {\cal A}_{1}(\Qh) \, {\delta}(1-\xh) ,\\
{\cal A}_{2,3}(\xh,\Qh) & = &
\frac{\xh}{\qh} \, {\cal A}_{2,3}(\Qh) \, {\delta}(1-\xh) ,
\ea
where ${\cal A}_{i}(\Qh)$ are defined by formulae
(20)--(22) of~\cite{ABBCK}.
 
Further, from~(\ref{eq313})--(\ref{eq316}) of the subsequent section,
one may get formulae (26)--(28) of~\cite{ABBCK}, the expressions for
 ${\cal S}_{1,2,3}^{\mr{el}}$ in the ultra-relativistic approximation.
The exact in $m$ and $M$ expressions for ${\cal S}_{1,2}^{\mr{el}}$
may be found
in~\cite{II}, equation~(37).
\section
{A covariant treatment of the phase space
\label{kineI}
}
\setcounter{equation}{0}
\subsection
{Phase space parameterization
\label{phI}
}
 
 
The cross section of reaction~(\ref{eqdeep}) is characterized by six
independent
invariants which may be taken as follows:
\ba
S, \, \yl, \, \ql, \, \yh, \, \qh, \, z_2.
\label{6inv}
\ea
In addition, we use
\ba
z_{1}     &=& z_{2} + \Qh - \Ql.
\label{eq0304}
\ea
 
We will now rewrite the phase space of process~(\ref{eqdeep}) such
that it
is
expressed by these variables.
The phase space is defined as follows:
\ba
d\Gamma &=& \frac{d\vec{k}_2}{2k_2^0} \frac{d\vec{k}}{2k^0}
    \frac{d\vec{p}_2}{2p_2^0} dM_h^2 \,
    \delta^{4}(k_1+p_1-k_2-k-p_2),
\label{gamma1}
\ea
where the $d\Gamma_h$ is already split away.
After rewriting $d\vec{p}_2/ 2p_2^0 = d^4p_2 \,
\delta(p_2^2 + M_h^2)$
and taking the integral over $d^4p_2$ using the $\delta$ function,
we arrive at
\ba
d\Gamma &=&
\frac{d\vec{k}_2}{2k_2^0} \frac{d\vec{k}}{2k^0}
\, \delta [(p_1+Q_h)^2+M_h^2]
   \,        dM_h^2.
\label{gamma2}
\ea
With a trivial calculation, which is explained in appendix~\ref{ph1}, one gets
\ba
\frac{d\vec{k}_2}{2k_2^0}
&=&
\frac{\pi S}{2 \sqrt{\lambda_S}} d\yl d\ql.
\label{gamma3}
\ea
Introducing the notation
\ba
d\Gamma_k =
\frac{d\vec{k}}{2k^0} \delta [(p_1+Q_h)^2+M_h^2] \delta[\qh -
(p_2-p_1)^2)],
\label{dgk}
\ea
we arrive at:
\ba
d\Gamma =
\frac{\pi S}{2 \sqrt{\lambda_S}} \, d\yl \, d\ql \,
dM_h^2 \, d\qh \, d\Gamma_k.
\label{gam1}
\ea
With $M_h^2 = -p_2^2 = M^2 +\yh S-\qh$, the differential $dM_h^2 d\qh$
may be replaced by $Sd\yh d\qh$.
In appendix~\ref{ph2}, we derive the following parameterization,
which will
be used for the subsequent integrations:
\ba
d\Gamma_k
&=&
\frac{dz_2}{2 \sqrt{R_z}},
\label{gam2}
\ea
where $R_z$ is a quadratic polynomial in $z_2$.
 
In sum, the phase space
in terms of the invariants $\yl, \ql, \yh, \qh$, together with
one additional variable $z_2$ (or $z_1$), is~\cite{III}:
\ba
\Gamma = \frac{\pi S^2}{4 \sqrt{\lambda_S}} \int
 d\yl \, d\ql \, d\yh \, d\qh \frac{dz_{1(2)}}{\sqrt{R_z}}.
\label{gamie}
\ea
\subsection
{An analytical integration
\label{phb}
}
 
 
Taking into account~(\ref{eqS}) and~(\ref{gamie}),
the differential cross section~(\ref{eq3.3})
for the process~(\ref{eqdeep})
                            can be written as follows:
\ba
 d\sigma_{\mr R} = \frac{2 \alpha^{3} S^2}{\pi \lambda_S}
{\cal S}({\cal E},{\cal I},z_1,z_2) \, d{\cal E} \, d{\cal I}
\frac{dz_{1(2)}}{Q_h^4 \sqrt{R_z}}.
\label{eq321}
\ea
Here and henceforth we use the following variables:
\bq
\begin{array}{rrccclcrcccl}
{\mr{leptonic \,  variables:}}&  \,
 {\cal E}
&=&
 {\cal E}_l
 &=&
 (\yl,\ql),
&\hspace{.5cm}&
 {\cal I}
 &=&
 {\cal I}_l
 &=&
 (\yh,\qh);
\nll
{\mr{ mixed \, variables:}}& \,
 {\cal E}
 &=&
 {\cal E}_m
 &=&
 (\yh,\ql),
&\hspace{.5cm}&
{\cal I}
&=&
{\cal I}_m
&=&
(\yl,\qh);
\nll
{\mr{hadronic \, variables:}}& \,
{\cal E}
&=&
{\cal E}_h
&=&
(\yh,\qh),
&\hspace{.5cm}&
                {\cal I}
                 &=&{\cal I}_h
                 &=&
                 (\yl,\ql).
\end{array}
\eq
For the case of mixed variables, we define in addition:
\bq
\xm = \frac{\ql}{\yh S}.
\label{xm}
\eq
The physical regions of ${\cal E}_a$ and ${\cal I}_a$, $a=l,m,h$, are
determined in appendix~\ref{appb}.
Here, we would only like to mention that the variable
$\xm$ may not be restricted to the interval [0,1]; see the
discussion in appendix~\ref{appb2}.
 
For the calculation of the double differential cross
section~(\ref{eqdeep})
one has to perform a threefold integration over the squared matrix
element~(\ref{eq3.3}) with the following integration variables:
$z_2$ (or $z_1$), and the two invariants in ${\cal I}_a$, $a=l,m,h$,
respectively.

The first integration is that over $z_2$.
Since the generalized structure functions
${\cal A}_{i}(\xh,\qh)$ are independent of $z_{1(2)}$, it is sufficient
to integrate the kinematical functions
${\cal S}_{i}({\cal E}, {\cal I},z_1,z_2)$ of
(\ref{eqs1z})--(\ref{eqs3z}).
The integration limits may be found in~(\ref{ax}),
and
the necessary table of integrals in appendix~\ref{appe1}.
 
The following expressions are the result of the integration:
\ba
{\cal S} ({\cal E,\cal I})  &=&  {1 \over \pi}
{\int{dz_{1(2)} \over {\sqrt{R_{z}}}}
 {\cal S}({\cal E,\cal I},z_1,z_2)}
   \equiv      \sum_{i=1}^3
             {\cal A}_{i}(\xh,\Qh) {\cal S}_{i}({\cal E,\cal I})\;,
\label{eq313}
\ea
where the ${\cal S}_i({\cal E,\cal I})$ are
%
\begin{eqnarray}
{\cal S}_{1}( {\cal{E, I}})
&=& \Biggl\{ \frac{1}{\sqrt{C_{2}}}
\label{eq314}
\left[2 m^2 - \frac{1}{2}
\left( \Qh+\Ql \right) + \frac{Q_h^4-4m^{4}}{\Qh-\Ql}\right]
\nll
& &-~m^{2}(\Qh-2m^2)\frac{B_{2}}{C_{2}^{3/2}} \Biggr\}
 +\frac{1}{\sqrt{A_2}}
 - \Biggl\{  (1)  \leftrightarrow - (1-\yl)  \Biggr\},
\\
{\cal S}_{2}( {\cal{E, I}})
&=& \Biggl\{\frac{1}{\sqrt{C_{2}}} \Bigl[ M^{2}(\Qh + \Ql)-
  \yh (1-\yl) S^2 \Bigr]
  \nll
& & +~
\frac{1}{(\Qh-\Ql)\sqrt{C_{2}}}
 \Biggl[ \Qh \Bigl[ (1) \left[ (1) - \yh \right] S^2  \nll
& & +~ (1-\yl) \left[ (1-\yl) + \yh \right] S^2
 - 2 M^{2}(\Qh+2m^2) \Bigr]   \nll
& & +~2m^2 S^2 \Bigl[ \left[ (1)-\yh \right]
    \left[(1-\yl) + \yh \right]
 +   (1)(1-\yl) \Bigr]  \Biggr]
\nll
& &-~2m^{2} \frac{B_{2}}{C_{2}^{3/2}}
     \Bigl[    (1) \left[ (1) - \yh \right] S^2 - M^{2}\Qh \Bigr]
\Biggr\}
- \frac{2M^2}{\sqrt{A_2}}
\nll
& &
-~\Biggl\{ (1) \leftrightarrow - (1-\yl)  \Biggr\},
\\
{\cal S}_{3}({\cal{E, I}})
&=&  \Biggl\{\frac{S}{\sqrt{C_{2}}}
\Biggl[\frac{2  \Qh ( \Qh+2m^2)
\left[ (1) + (1-\yl)  \right] }
{\Qh-\Ql}
   - 2 (1-\yl) \Qh -  \yh(\Qh+\Ql)\Biggr] \nonumber \\
& &-~ 2m^2 S \Qh\frac{B_{2}}{C_{2}^{3/2}}
   \left[ 2 (1) - \yh \right] \Biggr\}
+ ~\Biggl\{  (1) \leftrightarrow - (1-\yl) \Biggr\},
\label{eq316}
\end{eqnarray}
where~\cite{III}
\begin{eqnarray}
A_2 &=& \lambda_q, \nonumber       \\
B_2 &=&\left\{ 2 M^2 \Ql ( \Ql - \Qh )
      +     (1-\yl) (\yl \Qh - \yh \Ql) S^2
 +S^2 (1) \Ql (\yl - \yh) \right\}
\nll
&\equiv&~
-~B_1
        \left\{ (1) \leftrightarrow - (1-\yl) \right\}, \nonumber  \\
C_2 &=&  \left\{  (1-\yl) \Qh - \Ql \left[ (1)- \yh \right] \Bigr]^2 S^2
     + 4m^2 \Bigl[ (\yl-\yh)(\yl \Qh-\yh \Ql) S^2
- M^2(Q^2_h - Q^2_l )^2  \right\}
\nll
   &\equiv&~
C_1 \left\{  (1) \leftrightarrow - (1-y_l) \right\}.
\nonumber
\end{eqnarray}
The kinematical function $\lambda_q$ and the coefficients
$A_i, B_i, C_i$, $i$=1,2, are derived in appendix~\ref{appa}.
Again, for the photon exchange and the $\gamma Z$ interference
cross sections, the radiators are exact in the masses $m,M$.
 
Formulae~(\ref{eq313})--(\ref{eq316}) describe the contribution of
the radiative process to neutral current deep inelastic scattering
without applying a cut on the photon kinematics. In~(\ref{eq313}),
an integration over the azimuthal angle $\varphi$ of the emitted
photon [as introduced in~(\ref{dgak})] is already performed.
  In principle, one could proceed here in a different way
and apply some simple photonic cuts. This is done in section~\ref{pc} and
appendix~\ref{appd}.
 
Further, we should indicate that the resulting expressions are
singular at $k \rightarrow 0$.
There, $z_1$ and $z_2$ vanish.
It may be seen from~(\ref{eq0304}) which relates $z_1$ and $z_2$
that \qh\ and \ql\ become equal in this limit.
The same happens with \yh\ and \yl\ [see~(\ref{qxy})].
This reflects the infrared singularity of the radiative cross section
 which deserves
a special treatment before we can perform additional numerical or
analytical integrations.
\section
{Removal of the infrared divergence
\label{irreI}
}
\setcounter{equation}{0}
\subsection
{The infrared divergence
\label{irres}}
In the last section, it has been shown that
the double differential cross section of the process~(1.2)
can be written in the following form:
\ba
\frac {d\sigma_{\mr R}} {d{\cal E}} =
        \frac{2 \alpha^{3}S^2}{{\lambda_S}} \int d{\cal I}
        \sum_{i=1}^3 {\cal A}_i(\xh,\qh) \frac{1}{Q_h^4}\;
        {\cal S}_{i}({\cal E, \cal I}).
\label{eqinf}
\ea
As was mentioned at the end of the foregoing section,
the integrand of~(\ref{eqinf}) diverges if
the kinematics is such that the photon momentum vanishes.
Then the kinematical functions
${\cal S}_{i}({\cal E,\cal I})$ become infrared singular as may be seen from
formulae~(\ref{eq314})--(\ref{eq316}).
Usually, in complete \oa\ QED calculations
the infrared singularity is treated as follows.
One introduces an (artificial) infrared cut-off parameter
$\epsilon$, which divides the photon phase space into two parts:
In the soft photon part, the photon momentum may become infinitesimal,
while in the other, the hard photon part, it is enforced to
remain finite.
The soft photon contribution to the cross section is treated
analytically; it depends on the cut-off which is introduced in a
specially chosen lorentz frame.
Thereby the lorentz invariance is spoiled intermediary in the calculation.
The soft photon contribution
contains the infrared singularity, which is compensated
by the vertex correction of figure~\ref{fig1}b.
The hard photon contribution
is calculated by some dedicated integration method
and depends also on the cut-off $\epsilon$.
Finally, the compensation of the various occurrences
of the infinitesimal parameter $\epsilon$
is performed numerically, often with the
need of an adjustment of it to the kinematical situation.
Examples of such an approach are~\cite{motsai} and~\cite{BOSP}.
 
For the calculations, which are aiming at the
leading logarithmic approximation, the soft photon problem
may be solved completely by an analytical calculation of all the
contributions.
This
       implies both an explicit analytical compensation
of the infrared cut-off
       parameter $\epsilon$
and the semi-analytical integration of the hard photon bremsstrahlung.
 
Here, the infrared problem will be treated with a {\em covariant}
method~\cite{BS} with similar features as that used in the leading
logarithmic calculations.
As mentioned above, a cut-off parameter $\epsilon$ will be invented.
Then, from the exact squared matrix element the terms,
which contain the infrared singularity are extracted. These are
 {\em of much simpler structure than the exact expression}.
The resulting simplified
     cross section part will be integrated analytically
over the {\em full} phase space, thereby
compensating the infrared singularity with the vertex correction
and, more complicated, eliminating the cut-off parameter
$\epsilon$. By construction, the infrared divergent bremsstrahlung
cross section
part contains hard and soft photon contributions. This is quite similar
to the leading logarithmic approximation.
 
The rest of the bremsstrahlung contribution is free of the
soft photon problem.
The integrand is quite complicated.
Nevertheless, one may  perform analytical integrations over all those
internal
variables, which are not an argument of the structure functions.
 
One may represent~(\ref{eqinf}) in the following form:
\ba
\frac {d^2 \sigma_{\mr R}} {d{\cal E}} =
\left(
\frac {d^2 \sigma_{\mr R}}          {d{\cal E}} -
\frac {d^2 \sigma_{\mr R}^{\mr {IR}}} {d{\cal E}}
\right)
+
\frac {d^2 \sigma_{\mr R}^{\mr {IR}}} {d{\cal E}}
\equiv
\frac {d^2 \sigma_{\mr R}^{\mr F }} {d{\cal E}} +
\frac {d^2 \sigma_{\mr R}^{\mr {IR}}} {d{\cal E}},
\label{4.2}
\ea
where
$ {d \sigma_{\mr R}^{\mr {IR}}} / {d{\cal E}} $
is the infrared divergent part to be defined below and
$ {d \sigma_{\mr R}^{\mr F}}  / {d{\cal E}} $
is finite at $ k\rightarrow 0 $.
Such a separation
is not unique.
The essentials of the approach will be the same for all the
different ways  of integrations over sets ${\cal I}$ of invariants,
although certain details of the specific calculations will differ.
 
Going back to~(\ref{eq3.3}) and~(\ref{gam1}), the fully differential
cross section may be written as
\ba
{d\sigma_{\mr R}}
 =
  \frac{4\alpha^3 S^2}{\pi \lambda_S}
  \frac{1} {Q_h^4} {\cal S}({\cal E,\cal I},z_1,z_2) \,
   d{\cal E} \, d{\cal I} \, d\Gamma_k.
\label{eq4.1}
\ea
From~(\ref{eqs1z})--(\ref{eqs3z}), one may get the limit of
${\cal S}({\cal E,\cal I},z_1,z_2)/Q_h^4$ at $k \rightarrow 0$:
\ba
\lim_{k \rightarrow 0} \frac{1}{Q_h^4}{\cal S}({\cal E,\cal I},z_1,z_2)
=  \sum_{i=1}^3 {\cal A}_i( x,Q^2 )
\frac{1}{Q^4}{\cal S}_i^{\mr B}(y, Q^2 ) {\cal F}^{\mr{ IR}}
(Q^2,z_1,z_2),
\label{sko}
\ea
where
\bq
 {\cal F}^{\mr{ IR}}(Q^2,z_1,z_2) = \frac{Q^2+2m^2}{z_1z_2} - m^2\left(
\frac{1}{z_1^2} + \frac{1}{z_2^2}\right) ,
\label{ir3}
\eq
and  $x$, $y$ and $Q^2$ are variables as defined
in the Born kinematics.
The factorized universal
 function ${\cal F}^{\mr{ IR}}$ is the well-known Low
factor~\cite{Low},
\bq
{\cal F}^{\mr{ IR}}
=
\left( \frac{k_1}{2k_1 k} -  \frac{k_2}{2k_2 k} \right)^2,
\label{low}
\eq
and contains the infrared
singularities of the photonic bremsstrahlung from the electron line.
 
Now, we define the infrared part of the cross section  as follows:
\ba
\frac{d^2\sigma_{\mr R}^{\mr{ IR}}} {d {\cal E}}
&=&
\frac{4 \alpha^3 S^2}{\pi \lambda_S} \int d{\cal I} \, d \Gamma_k
\sum_{i=1}^3 {\cal A}_i( x ,Q^2)
\frac{1}{Q^4}{{\cal S}_i^{\mr B}}(y,Q^2 ) {\cal F}^{\mr{ IR}}(Q^2,z_1,z_2)
\nll
&=&
\frac{d^2\sigma_{\mr B}} {d {\cal E}} \frac{2 \alpha S}{\pi^2}
 \int d{\cal I} \, d \Gamma_k {\cal F}^{\mr{ IR}}(Q^2,z_1,z_2)
\nll
&\equiv&
\frac{d^2\sigma_{\mr B}} {d\cal E} \frac{\alpha}{\pi}
\delta_{\mr R}^{\mr{ IR}}({\cal E}).
\label{irbe}
\ea
The integration range in~(\ref{irbe}) will be split into two parts:
\ba
\delta_{\mr R}^{\mr{ IR}}({\cal E})
&=&
\frac{2S}{\pi}
 \int d{\cal I} \, d \Gamma_k {\cal F}^{\mr{ IR}}(Q^2,z_1,z_2)
\theta (\epsilon-k_0)
\nll
& &+~\frac{2S}{\pi} \int d{\cal I} \, d \Gamma_k {\cal F}^{\mr{ IR}}
(Q^2,z_1,z_2) \theta (k_0-\epsilon)
\nll
&\equiv& \delta_{\mr{ soft}}^{\mr{ IR}}({\cal E},\epsilon)
+\delta_{\mr{ {hard}}}^{\mr{ IR}}({\cal E},\epsilon).
\label{irb2}
\ea
Here, the $k_0$ is chosen to be the energy of the emitted photon in the
proton rest frame~(\ref{eq06}),
\bq
\nonumber
k^0 = \frac{S}{2M} (\yl-\yh),
\label{k0}
\eq
and $\epsilon$ is an infinitesimal parameter, $\epsilon > 0, \epsilon
\rightarrow 0$.
The first term will constitute the soft photon contribution to
$\delta_{\mr{ soft}}^{\mr{ IR}}({\cal E})$:
\ba
\delta_{\mr{ soft}}^{\mr{ IR}}({\cal E},\epsilon)
&=&
\frac{2S}{\pi} \int d{\cal I}
\frac{d\vec k}{2k_0} \delta\left[\left(p_1+Q_h\right)^2+M_h^2\right]
\, \delta\left[\qh-\left(p_2-p_1\right)^2\right]
{\cal F}^{\mr{ IR}}(Q^2,z_1,z_2)
\theta (\epsilon-k_0)
\nll
&=&
\frac{2}{\pi} \int \frac{d\vec k}{2k_0}{\cal F}^{\mr {IR}}(Q^2,z_1,z_2)
\theta (\epsilon-k_0).
\label{irsoft}
\ea
The second part contains an integration over nearly the complete phase
space and implies also contributions from hard photons,
but with a considerably simplified integrand compared to the
complete bremsstrahlung integral.
It will also be dependent on the infrared cut-off $\epsilon$,
but is infrared finite for finite $\epsilon$:
\ba
\delta_{\mr {hard}}^{\mr {IR}}({\cal E},\epsilon)
&=&
\frac{S}{\pi} \int d{\cal I}
\frac{dz_{1(2)}}
{\sqrt{R_z}}{\cal F}^{\mr {IR}}(Q^2,z_1,z_2)\theta(k_0-\epsilon).
\ea
With the additional notation
\ba
{\cal F}^{\mr {IR}}({\cal E,\cal I})
&=&
\frac{S}{\pi} \int
\frac{dz_{1(2)}}
{\sqrt{R_z}}{\cal F}^{\mr {IR}}(Q^2,z_1,z_2)
\nll
&=&
\frac{Q^2+2m^2}{\Ql-\Qh} \Bigl(\ \frac{1}{\sqrt{C_1}}
   -\frac{1}{\sqrt{C_2}} \Bigr)\
   -m^{2}\Bigl(\ \frac{B_{1}}{C_{1}^{3/2}}
                +\frac{B_{2}}{C_{2}^{3/2}} \Bigr),
\label{point}
\ea
the hard part of the infrared divergent correction becomes:
\ba
\delta_{\mr {hard}}^{\mr {IR}}({\cal E},\epsilon)
&=&
 \int d{\cal I} \, {\cal F}^{\mr {IR}}({\cal E,\cal I})\theta
(k_0-\epsilon).
\label{irhard}
\ea
The integration of~(\ref{ir3}) over $z_2$ has been performed with the
table of integrals~\ref{appe1}.
The coefficients $B_i, C_i$, $i$=1,2 are introduced in
appendix~\ref{appa}.
 
For later use, the infrared divergent part of the corrections is quoted
also in a slightly more general notation:
\ba
\delta_{\mr R}^{\mr {IR}}({\cal E})
&=&
\frac{4}{\pi^2}   \frac{\sqrt{\lambda_S}}{S}  \int \frac{d\Gamma}{d{\cal E}}
{\cal F}^{\mr {IR}}(Q^2,z_1,z_2),
\label{iral1}
\ea
where the $Q^2$ on the right hand side is, by definition,
one of the external variables and the
${\cal F}^{\mr {IR}}(Q^2,z_1,z_2)$ is defined in~(\ref{ir3}).
\subsection
{The soft and hard parts of $\delta_{\mr R}^{\mr{{IR}}}$
\label{irre3}}
The soft photon contribution
$\delta^{\mr {IR}}_{\mr {soft}}({\cal E},\epsilon)$ will be considered
first.
The integration region is limited
to photons with an energy, which is smaller than $\epsilon$.
After inserting ${\cal F}^{\mr {IR}}$ as defined in~(\ref{ir3})
into~(\ref{irsoft}), one gets
\ba
 \delta^{\mr {IR}}_{\mr {soft}}({\cal E},\epsilon)
& = &
2 (2\pi)^2
\int \frac{d^3k}{(2\pi)^3k^0}
\theta(\epsilon-k^0)
\left[
 \frac{Q^2+2m^2}{(-2k_1k)(-2k_2k)}
- \frac{m^2}{(-2k_1k)^2}
       - \frac{m^2}{(-2k_2k)^2}
\right].
\nll
\label{d3k}
\ea
The expression~(\ref{d3k}) will be regularized now by
        changing into  the ($n-1$)-dimensional space~\cite{marciano}.
After an integration over ($n-3$) azimuthal angles,
two integrations remain to be performed:
one over the polar angle $\vartheta$ of the photon and the other over
the photon energy $k^0$:
\ba
 \delta^{\mr {IR}}_{\mr {soft}}({\cal E},\epsilon)
&\rightarrow&
\frac{2 (2\pi)^2}{(2\sqrt{\pi})^n \Gamma\left(n/2-1\right)}
\int_0^1d \alpha \frac{1}{\mu^{n-4}}
\int_0^{\epsilon} (k^0)^{n-5} dk^0 \int_0^{\pi}
(\sin \vartheta)^{n-3} d \vartheta
\nll
& &
\left[
 \frac{Q^2+2m^2}{
(k_{\alpha}^0)^2(1-\beta_{\alpha}\cos\vartheta_{\alpha})^2}
- \frac{m^2}{
(k_1^0)^2(1-\beta_1\cos\vartheta_1)^2}
       - \frac{m^2}{
(k_2^0)^2(1-\beta_2\cos\vartheta_2)^2}
                    \right].
\nll
\label{dir}
\ea
In order to simplify the angular integration, in the first term a
Feynman parameter integration is inserted,
which linearizes the angular dependence of the denominator:
\ba
\frac{1}{(-2k_1k)(-2k_2k)} =
\int_0^1 \frac{d\alpha}{[(-2k_1k)\alpha+(-2k_2k)(1-\alpha)]^2},
\label{kal27}
\ea
\ba
k_{\alpha} = k_1 \alpha + k_2 (1- \alpha).
\label{kal7}
\ea
Further,
the relations
\ba
-k_i k &=& k_i^0 k^0 \left(1 - \beta_i \cos \vartheta_i \right),
          \\
\beta_i &=& \frac {\left|{\vec k}_i\right|}{k_i^0}, \hspace{1cm}
i=1,2,\alpha,
\label{kalp2}
\ea
are used
where the $\beta_i$ are ($n-1$) dimensional velocities, and the
$\vartheta_i$ the corresponding spatial angles between the photon
three-momentum and ${\bar k}_i$.
We now go into the rest system of the proton,
where it is ${\vec p}_1 = 0$.
In this system,
the $\beta_i$ are expressed in terms of the invariants~(\ref{eq06}).
 
We get:
\ba
\frac{1}{\mu^{n-4}}
\int_0^{\epsilon}dk^0 (k^0)^{n-5} &=& \frac {(\epsilon/\mu)^{n-4}}{n-4}
\nll
&=& \frac{1}{n-4} \left[ 1 + (n-4) \ln\frac{\epsilon}{\mu} + \cdots
\right].
\label{n4}
\ea
Soft photon emission is isotropic.
For any of the three terms under the integral,
one has the freedom to choose a coordinate frame
               in the proton rest system with the
$(n-1)^{st}$ axis being parallel to ${\vec k}_i$,
i.e. $\vartheta_i=\vartheta$.
After a trivial change of variable, $\cos \vartheta=\xi$,
\ba
\int_0^{\pi} \frac{(\sin \vartheta)^{n-3} d\vartheta}
{(1-\beta_i \cos \vartheta)^2}
&=&
\int_{-1}^{1} \frac{(1-\xi^2)^{n/2-2}d\xi} {(1-\beta_i\xi)^2}
\nll
&=&
\int_{-1}^{1} \frac{d\xi} {(1-\beta_i\xi)^2} \left[1+\frac{1}{2}(n-4)
\ln(1-\xi^2)+\cdots\right].
\label{dxi}
\ea
From the above substitutions, one obtains the following expression
for $\delta^{\mr {IR}}_{\mr {soft}}({\cal E},\epsilon)$:
\ba
\delta^{\mr {IR}}_{\mr {soft}}({\cal E},\epsilon)
&=&
  \left[{\cal P}^{\mr{IR}}+\ln\frac{2\epsilon}{\mu} \right]
  \frac{1}{2} \int_{0}^{1} d\alpha
  \int_{-1}^{1} d\xi \; {\cal F}(\alpha,\xi)
+ \frac{1}{4} \int_{0}^{1} d\alpha
  \int_{-1}^{1} d\xi \; \mbox{ln}(1-\xi^2) \;
   {\cal F}(\alpha,\xi) .
\nll
\label{*}
\ea
Here, it is
\ba
{\cal F}(\alpha,\xi)  =
 \frac {Q^2+2m^2} {(k_{\alpha}^{0})^2(1-\beta_{\alpha}\xi)^2}
 - \frac {m^2} {(k_{1}^{0})^2(1-\beta_{1}\xi)^2}
 - \frac {m^2} {(k_{2}^{0})^2(1-\beta_{2}\xi)^2},
\label{*1}
\ea
and
\bq
{\cal P}^{\mr {IR}}=\frac{1}{n-4}+\frac{1}{2}\gamma_E
+\mbox{ln} \frac{1}{2\sqrt{\pi}}
\equiv -\frac{1}{2{\bar \epsilon}},
\label{eqn414}
\eq
where the ${\bar \epsilon}$ is a parameter, which is often used in the
${\overline{\mbox{MS}}}$ subtraction scheme.
The ${\cal P}^{\mr {IR}}$ contains
the pole term, which  corresponds
 to the infrared divergence, as well as the
  Euler constant $\gamma_E$. The arbitrary
parameter $\mu$ has the dimension of a mass.
 
After the integration over $\xi$ and $\alpha$
with the aid of the integrals of appendix~\ref{c2} one gets:
\ba
\delta^{\mr {IR}}_{\mr {soft}}({\cal E},\epsilon)
= 2\Bigl[ {\cal P}^{\mr {IR}}+\ln \frac{2\epsilon}{\mu} \Bigr]
\left[ (Q^2+2m^{2}) \lm        -1 \right]
+~
 {\cal S}_{\Phi}
 + \frac{1}{2 \beta_{1}} \;
  \mbox{ln}\frac{1+\beta_1}{1-\beta_1}
 +\frac{1}{2 \beta_{2}} \;
  \mbox{ln}\frac{1+\beta_2}{1-\beta_2}
,
\label{eqn415}
\ea
with
${\cal S}_{\Phi}$ defined in~(\ref{soft3}) and~(\ref{soft4}).
Further,
\ba
\beta_1 &=& \frac{\sqrt{\lambda_S}}{S} = \sqrt{1-\frac{4m^2M^2}{S^2}},
\\
\beta_2 &=& \frac{\sqrt{\lambda_l}}{(1-\yl)S}
= \sqrt{1-\frac{4m^2M^2}{(1-\yl)^2S^2}},
\label{bett2}
\\
\lm &=& \frac{1}{\sqrt{\lambda_m}} \ln \frac{\sqrt{\lambda_m} + Q^2}
{\sqrt{\lambda_m} - Q^2},
\label{lmlamm}
\\
{\lambda}_{m}
&=&
Q^2 (Q^2+4m^2).
\ea
 
The equation~(\ref{eqn415}) with ${\cal S}_{\Phi}$ defined in ~(\ref{soft3}) is exact.
In the ultra-relativistic limit 
it is
\ba
{\cal S}_{\Phi}
=
-\frac{1}{2}\mbox{ln}^{2}{\left(\frac{Q^2}{m^2} \frac{1}{1-y} \right)}
- \ln \frac{Q^2}{m^2} \ln \frac{S^2(1-y)^2}{M^2Q^2} - \litwo(1),
\label{soft4}
\ea
and one gets the following short expression:
\ba
\delta^{\mr {IR}}_{\mr {soft}}({\cal E},\epsilon)
&=& \left[ \ 2{\cal P}^{\mr {IR}}+2\mbox{ln}{\frac{2\epsilon}{\mu}}
  - \mbox{ln}{\frac{(1-y)S^2}{m^2M^2}}\right] \
    \left( \ \mbox{ln}{ \frac{Q^2}{m^2} } - 1 \right)
\nll
& &+~\frac{1}{2} \mbox{ln}^{2}{ \frac{Q^2}{m^2}}
   -~\frac{1}{2} \mbox{ln}^{2}{\left( 1-y \right)}
-\litwo(1),
\label{*8}
\ea
where
\bq
{\mr {Li}}_{2}(x)=-\int_{0}^{1} \frac{ \mbox{ln} (1-xy) }{y} dy.\nonumber
\label{eqn417a}
\eq
 
The infrared pole ${\cal P}^{\mr {IR}}$ in
$\delta^{\mr {IR}}_{\mr {soft}} ( {\cal E},\epsilon )$
has to cancel against a corresponding contribution from the QED vertex
correction:
\ba
\delta_{\mr {vert}}({\cal E})
 = -2 \left({\cal P}^{\mr {IR}}+\ln\frac{m}{\mu}\right)
 \left( \mbox{ln}{\frac{Q^2}{m^2}}-1 \right)
 -  \frac{1}{2} \mbox{ln}^{2}{\frac{Q^2}{m^2}}
 +  \frac{3}{2} \mbox{ln} {\frac{Q^2}{m^2}}
 + {\mr {Li}}_2(1)-2.
\label{*20}
\ea
This cancellation may be seen explicitly from the above expressions.
 
The parameter $\epsilon$ in~(\ref{*8})
has to be compensated by corresponding terms from
$\delta^{\mr {IR}}_{\mr {hard}}({\cal E},\epsilon)$.
 
In order to obtain the
$\delta^{\mr {IR}}_{\mr {hard}} ({\cal E} ,\epsilon) $,
one has to perform the integration of
$ {\cal F}^{\mr {IR}}({\cal E,\cal I})$ over the
 physical regions of the variable sets
$ {\cal I}_a (a=l,m,h)$ as derived in appendix~\ref{appb}, but with the
exclusion of a small region around the infrared point $\cal F$; see also
the figures in that appendix.
These integrations are tedious and will be commented on in
appendix~\ref{appe3}.
 
In the ultra-relativistic approximation,
one         gets the following expression for
         $\delta^{\mr {IR}}_{\mr {hard}}$ in
{\em leptonic} variables:
\ba
\delta^{\mr {IR}}_{\mr {hard}}({\cal E}_l ,\epsilon)
&=&  \left[ -2\mbox{ln}{ \frac{2\epsilon}{m}}
  +  \ln  \frac{ (1-\yl)y_l^2(1-\xl)^2 S^2}
     {(1- \yl\xl)(1-\yl(1-\xl)) m^2 M^2} \right]
     \left( \mbox{ln}\frac{\Ql}{m^2} - 1 \right)
\nll & &
+~ \frac{1}{2} \mbox{ln}^{2}{( 1 -\yl)}
-   \frac{1}{2} \mbox{ln}^{2}{\left[ \frac { 1-\yl(1 - \xl) } {1
 -\yl\xl}\right]}
\nll & &
 +~ {\mr {Li}}_{2}
 {\left[ \frac {(1-\yl)}{(1-\yl\xl)(1-\yl(1-\xl))}  \right]}
  -  {\mr {Li}}_2(1).
\label{eqn421}
\ea
In {\em mixed} variables, it is:
\ba
\delta^{\mr {IR}}_{\mr {hard}}({\cal E}_m,\epsilon) &=&
     \left[ -2\mbox{ln} { \frac{2\epsilon}{m}}
  +  \mbox{ln}   \frac{  (1-\yh)^2 ( 1 -\xm )S^2 }{m^2 M^2} \right]
     \left( \mbox{ln}\frac{\Ql}{m^2} - 1 \right)
\nll & &
-~\frac{1}{2} \mbox{ln}^{2}{( 1  - \xm )}
+ \mbox{ln}{( 1 - \yh )} \;
  \mbox{ln}{ ( 1 - \xm) }
  - {\mr {Li}}_{2} {\left[ \frac { \xm (1-\yh) }{\xm-1} \right]}.
\label{eqn422}
\ea
Finally, in {\em hadronic} variables:
\ba
\delta^{\mr {IR}}_{\mr {hard}}({\cal E}_h,\epsilon)
&=&  \left[ -2 \mbox{ln}{ \frac {2\epsilon}{m}}
  +  \mbox{ln}{\frac{(1-\yh)^2S^2}{m^2 M^2}} \right]
     \left( \mbox{ln}\frac{\Qh}{m^2} - 1 \right)
  +   \mbox{ ln} \, {\yh}
  +  {\mr {Li}}_2 {(1-\yh)}
  -  {\mr {Li}}_2 (1) + 1 .
\nll
\label{eqn423}
\ea
 
At this intermediate state of the calculation, we would only like
to remark that the dependencies of $\delta_{\mr{hard}}^{\mr{IR}}$
on the
$\ln(2\epsilon/m)$,
$\ln(S/m^2)$,
and $\ln(S/M^2)$ are fictitious.
They are compensated by corresponding dependencies of
$\delta_{\mr{soft}}^{\mr{IR}}$ completely; see~(\ref{*8}).
At the other hand, the $\ln(Q^2/m^2)$ becomes part of the final results.
%
%
\subsection{The net correction $\delta_{\mr R}^{\mr {IR}}$ and the
soft photon exponentiation
\label{softex}
}
%
For applications,
it is convenient to define a dimensionless radiative correction factor,
\ba
\delta_a        \equiv
\delta({\cal E}_a)=
 \frac{d^{2}{\sigma}_{\mr{theor}}/d{\cal E}_a}
      {d^{2}{\sigma}_{\mr B}    /d{\cal E}_a}-1 ,
\hspace{1.cm} a=l,m,h,
\label{61}
\ea
where
$ d^{2}{\sigma}_{\mr B}/d{\cal E} $
is the Born cross section of the process
(\ref{eqborn})  and
$ d^{2}{\sigma}_{\mr{theor}}/d{\cal E} $ is the
theoretical prediction for the measured cross section,
\ba
\frac{d^{2} {\sigma}_{\mr{theor}}}{d{\cal E}}
&=&
\left( 1+\frac{\alpha}{\pi}\;\delta_{\mr {vert}}\right)
\frac{d^{2} {\sigma}_{\mr B}}    {d{\cal E}}
+\frac{d^{2}{\sigma}_{\mr R}}    {d{\cal E}}.
\label{62a}
\ea
One may
collect within one expression all the terms, which are
explicitly proportional to the Born cross section;
we call it the
    {\em factorized part} $\delta^{\mr {VR}}$ of the total radiative
corrections and introduce it in the following way:
\ba
\frac{d^{2}{\sigma}_{\mr{theor}}}    {d{\cal E}}
&=&
\left[ 1+\frac{\alpha}{\pi}\;\delta_{\mr {vert}}({\cal E})\right]
\frac {d^2 \sigma_{\mr B }} {d{\cal E}} +
\frac {d^2 \sigma_{\mr{soft}}^{\mr{IR}}} {d{\cal E}} +
\frac {d^2 \sigma_{\mr{hard}}^{\mr{IR}}} {d{\cal E}} +
\frac {d^2 \sigma_{\mr{R}}^{\mr{F}}} {d{\cal E}}
\nll
&\equiv&
\left[1+\frac{\alpha}{\pi}\;\delta^{\mr {VR}}({\cal E})\right]
\frac{d^{2} {\sigma}_{\mr B}}    {d{\cal E}}
+\frac{d^{2}{\sigma}_{\mr R}^{\mr F}}    {d{\cal E}},
\label{62b}
\ea
with
\ba
 \delta_{\mr {VR}}           ({\cal E})
&=&
 \delta_         {\mr {vert}}({\cal E})
+\delta^{\mr {IR}}_{\mr {soft}}({\cal E},\epsilon)
+\delta^{\mr {IR}}_{\mr {hard}}({\cal E},\epsilon)
\nll
&\equiv&
 \delta_         {\mr {vert}}({\cal E})
+\delta^{\mr {IR}}_{\mr {R}}({\cal E})
.
\label{64c}
\ea
The net radiative correction factor~(\ref{61}) becomes to order \oa:
\ba
\delta_a
&=&
  \frac{\alpha}{\pi}\;\delta_{\mr {VR}}({\cal E}_a)
+ \frac{d^{2}{\sigma}_{\mr R}^{\mr F}}{d{\cal E}_a}
/ \frac{d^{2}{\sigma}_{\mr B}}{d{\cal E}_a}
\nll
&\equiv&
  \frac{\alpha}{\pi} \left[
\delta_{\mr {VR}}({\cal E}_a) +
\delta^{\mr {F}}_{\mr R}({\cal E}_a) \right]
.
\label{63b}
\ea
In
$\delta_{\mr {VR}}           ({\cal E}_a)$,
a dilogarithmic term $\delta_{\mr {inf}}({\cal E}_a)$ may be found,
which controls the multiple soft photon emission in the
bremsstrahlung process~(\ref{eqdeep}).
An exponentiation of this term leads to a significant numerical
improvement of
the QED predictions~\cite{Shum}:
\ba
\frac{\alpha}{\pi} \; \delta_{\mr {VR}}({\cal E})
\rightarrow
\label{dvr}
\frac{\alpha}{\pi} \; \delta_{\mr {VR}}^{\mr {exp}}({\cal E})
 =
{\mr {exp}} \left[\;\frac{\alpha}{\pi}
 \delta_{\mr {inf}}({\cal E}) \right]-1+\frac{\alpha}{\pi}\Bigl[
 \delta_{\mr {VR}} ({\cal E})
-\delta_{\mr {inf}}({\cal E})\Bigr] \; .
\ea
Finally, the radiative correction factor~(\ref{61}) with soft photon
exponentiation becomes:
\ba
\delta_a^{\mr{exp}}
&=&
  \frac{\alpha}{\pi} \left[
\delta^{\mr{exp}}_{\mr {VR}}({\cal E}_a) +
\delta^{\mr {F}}_{\mr R}({\cal E}_a) \right]
.
\label{444a}
\ea
 
The factorized correction~(\ref{64c}) follows from~(\ref{*8}),
 (\ref{*20}), and~(\ref{eqn421})--(\ref{eqn423})
and is different for the different sets of variables.
 
In {\em leptonic} variables, it is:
\ba
\delta_{\mr {VR}} ({\cal E}_l)
 &=& \delta_{\mr {inf}}({\cal E}_l)
 -\frac{1}{2}\mbox{ln}^{2} {\left[ \frac{1-\yl(1-\xl)}{1-\yl\xl}\right]}
 + {\mr {Li}}_{2}{\left[ \frac{1-\yl}{(1-\yl\xl)[1-\yl(1-\xl)]}\right]}
\nll & &
 +~\frac{3}{2} \mbox{ln}\frac{\Ql}{m^2} - {\mr {Li}}_2 (1)-2 \; ,
\label{65}
\ea
where
\ba
\delta_{\mr {inf}} ({\cal E}_l)=
\left( \mbox{ln}\frac{\Ql}{m^2}-1 \right)
     \mbox{ln} {\left[ \frac{y_l^2(1-\xl)^2}
     {(1- \yl\xl)[1-\yl(1-\xl)]}\right]}.
\label{66}
\ea
In {\em mixed} variables,
we can write~(\ref{dvr}) as follows:
\ba
\delta_{\mr {VR}}({\cal E}_m)
&=& \delta_{\mr {inf}}({\cal E}_m)
- \frac{1}{2}
 \mbox{ln}^{2}\left(\frac{1-\yh}{1-\xm}\right)
-~{\mr {Li}}_{2} {\left[ \frac {\xm (1-\yh)}{\xm - 1} \right]}
+\frac{3}{2} \mbox{ln}\frac{\Ql}{m^2} -2 \;,
\label{67}
\ea
where
\ba
\delta_{\mr {inf}}({\cal E}_m)=
\left( \mbox{ln}\frac{\Ql}{m^2} - 1 \right)
 \mbox{ln}{\left[ ( 1 - \yh ) ( 1 - \xm )\right]}.
\label{68}
\ea
In {\em hadronic} variables, finally:
\ba
\delta_{\mr {VR}}({\cal E}_h)& = & \delta^{\mr {inf}}({\cal E}_h)
 -  \frac{1}{2} \mbox{ln}^{2}{(1-\yh)}
   +  {\mr {Li}}_{2} {(1-\yh)}
+~ \mbox{ln}{\yh}
+ \frac{3}{2} \mbox{ln}\frac{\Qh}{m^2} - {\mr {Li}}_2 (1)-1 \; ,
\label{69}
\ea
where
\ba
 \delta_{\mr {inf}}({\cal E}_h)=
 \left( \mbox{ln}\frac{\Qh}{m^2} -1 \right)
       \mbox{ln}{(1-\yh)} .
\label{610}
\ea
\section
{
The net radiative correction $
\delta_{\mr R} = \delta^{\mr{exp}}_{\mr {VR}} + \delta_{\mr R}^{\mr {F}}$
\label{sec5}}
\setcounter{equation}{0}
In this section, the net correction will be discussed:
\ba
\delta_a^{\mr{exp}}
&=&
  \frac{\alpha}{\pi} \left[
\delta^{\mr{exp}}_{\mr {VR}}({\cal E}_a) +
\delta^{\mr {F}}_{\mr R}({\cal E}_a) \right]
.
\label{444b}
\ea
We will use the following abbreviations:
$\delta_{\mr{lep}}=\delta_l^{\mr{exp}}$,
$\delta_{\mr{mix}}=\delta_m^{\mr{exp}}$,
$\delta_{\mr{had}}=\delta_h^{\mr{exp}}$.
 
Before the discussion of the numerical results, some
preparations will be performed
with the infrared finite, hard part of the corrections.
This hard part of the cross section has been defined in~(\ref{4.2}):
\ba
\frac {d^2\sigma_{\mr R}^{\mr F}}{d{\cal E}}
&\equiv&
    \frac {d^2\sigma_{\mr R}    }{d{\cal E}}
  - \frac {d^2{\sigma^{\mr {IR}}_{\mr R}}}{d{\cal E}}
\label{*051}  \\
&=& \frac {2\alpha^3 S^2}{\lambda_S}
    \int d{\cal I} \sum_{i=1}^{3} \Bigl[ \;
    {\cal A}_{i}(\xh,\Qh)\frac{1}{Q^4_h}{\cal S}_{i}({\cal E,\cal I})
  - {\cal A}_{i}( x,Q^2 )\frac{1}{Q^4}\; {\cal S}_{i}^{\mr B}(y,Q^2 )\;
    {\cal F}^{\mr {IR}}({\cal E,\cal I}) \Bigr]
\nll
&\equiv&
\frac{\alpha}{\pi}
\delta_{\mr R}^{\mr F} \frac {d^2\sigma_{\mr B}}{d{\cal E}}.
\nonumber
\label{5.1}
\ea
From the explicit expressions~(\ref{eq314})--(\ref{eq316})
and~(\ref{point})
 one may see that both terms under the
integral contain a remnant of the soft photon singularity at
$\ql = \qh$, while their difference is finite by construction.
 
The subsequent calculations in this section
 will be organized in such a way that the
integrations over $\ql$ or $\qh$ remain the last ones, respectively.
Thus, the compensation of the harmless rests of the infrared
singularity appears to be quite transparent.
In the case of non-leptonic variables,
in the analytical integrations of the
hard bremsstrahlung no special care has to be devoted to
the soft photon problems.
 
In the following
subsections, formulae will be presented for the infrared free
part of the cross section in the different sets of variables.
\subsection
{Leptonic variables \label{flep}}
For the case of leptonic variables,
the generalized structure functions ${\cal A}_i(\xh,\qh)$ ($i$=1,2,3)
depend on the integration variables $ \yh $ and $ \qh $ and
the structure functions become part of the integrand
for the remaining twofold integral, which has to be performed
numerically.
For the variables $ {\cal E}_l=(\yl,\ql) $ and
$ {\cal I}_l=(\yh,\qh) $, the expression~(\ref{*051}) is the final
result:
\ba
\frac {d^2 \sigma_{\mr R}^{\mr F}}{d \yl d \ql}
&=& \frac {2\alpha^3 S^2}{\lambda_S}
     \int d\yh d\qh  \sum_{i=1}^{3} \Biggl[ {\cal A}_{i}(\xh,\qh)
     \frac{1}{Q^4_h} {\cal S}_{i}(\yl,\ql,\yh,\qh)
\nll
& &-~{\cal A}_{i}(\xl,\ql)\frac{1}{Q_{l}^4} \;
     {\cal S}_{i}^{B}(\yl,\ql)\;
     {\cal L}^{\mr {IR}}(\yl,\ql,\yh,\Qh) \Biggr] ,
\label{*052}
\ea
where the radiators
${\cal S}_{i}(\yl,\ql,$ $\yh,\Qh)$ are given
by~(\ref{eq314}) --(\ref{eq316}).
 
The expression for
${\cal L}^{\mr{IR}}(\yl,\ql,\yh,\Qh) $ is defined by~(\ref{point})
with $ Q^2=\Ql $; it is exact in both masses $m$ and $M$:
\ba
{\cal L}^{\mr {IR}}(\yl,\ql,\yh,\Qh)
&\equiv&
{\cal F}^{\mr {IR}}(\yl,\ql,\yh,\qh)|_{Q^2=Q_l^2}
\nll
&=&
\frac{\Ql+2m^2}{\Ql-\Qh} \Bigl(\ \frac{1}{\sqrt{C_1}}
   -\frac{1}{\sqrt{C_2}} \Bigr)\
   -m^{2}\Bigl(\ \frac{B_{1}}{C_{1}^{3/2}}
                +\frac{B_{2}}{C_{2}^{3/2}} \Bigr).
\label{eqn052a}
\ea
%
The two-dimensional numerical integration in~(\ref{*052}) has to be
 performed
over
the physical region~(\ref{ea42}); see figure~\ref{ilep}.
In the ultra-relativistic approximation, the boundaries are:
\ba
\begin{array}{rcccl}
0 &\leq& \yh &\leq& \yl,
\\
\frac{
\displaystyle
\yh}{
\displaystyle
\yl} \ql &\leq& \qh &\leq& \min
\left[ \yh S, \ql\left( 1 + \frac{
\displaystyle
\yl - \yh}{
\displaystyle
\xl} \frac{
\displaystyle
S}{
\displaystyle
M^2}
\right) \right].
\label{limel}
\end{array}
\ea
In the numerical integrations,
one has to leave out a small region around the phase space points
with $\ql=\qh$. There, the integrand is finite but occurs as the difference
of two divergent terms.
\subsubsection
{Discussion
\label{disl}
}
\begin{figure}[thbp]
\begin{center}
\mbox{
\epsfysize=9.cm
\epsffile[0 0 530 530]{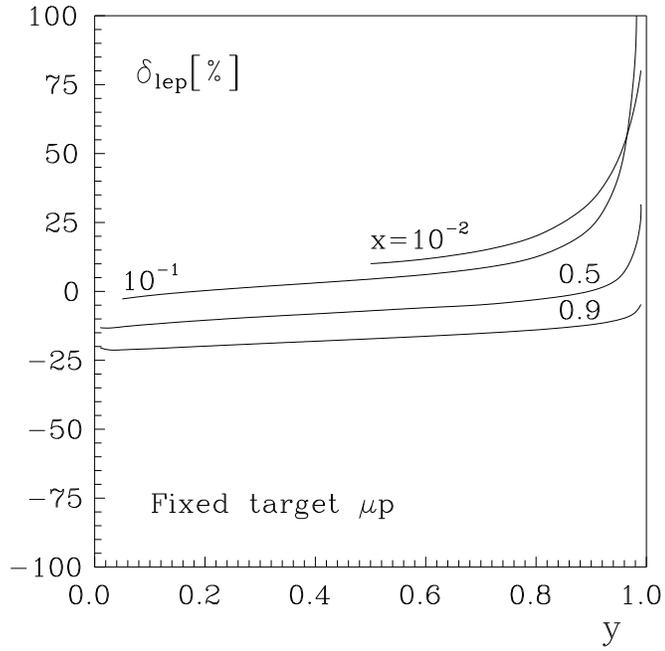}
}
\end{center}
\caption{\it
Radiative correction $\delta_{\mr{lep}}$ in \% for a
cross section measurement in terms of leptonic variables
at a fixed target $\mu p$ experiment.
}
\label{lcs1}
\end{figure}

\begin{figure}[hbtp]
\begin{center}
\mbox{
\epsfysize=9.cm
\epsffile[0 0 530 530]{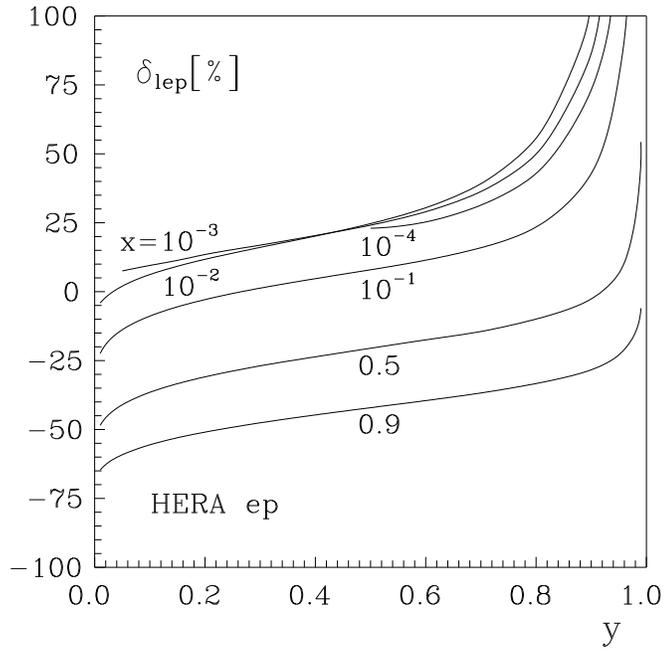}
}
\end{center}
\caption{\it
Radiative correction $\delta_{\mr{lep}}$ in \% for a
cross section measurement in terms of leptonic variables
at HERA.
}
\label{lcs2}
\end{figure}

\begin{figure}[thbp]
\begin{center}
\mbox{
\epsfysize=9.cm
\epsffile[0 0 530 530]{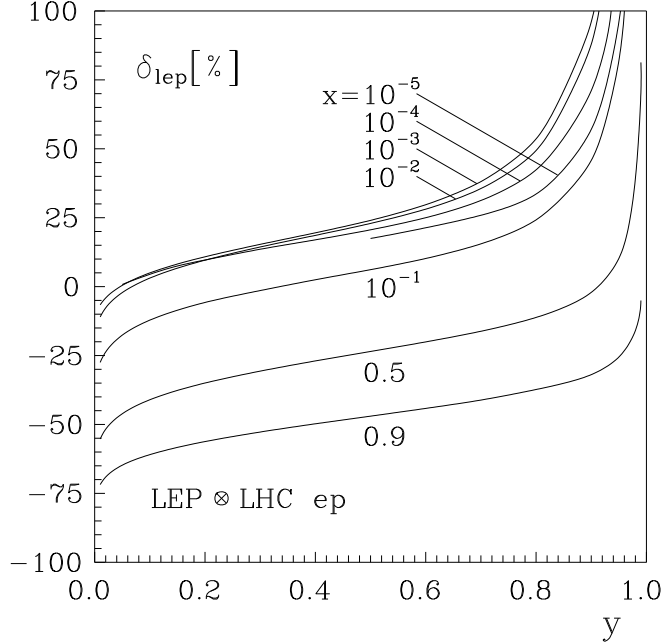}
}
\end{center}
\caption{\it
Radiative correction $\delta_{\mr{lep}}$ in \% for a
cross section measurement in terms of leptonic variables
at
LEP$\otimes$LHC.
}
\label{lcs3}
\end{figure}

The radiative corrections in leptonic variables are shown in
figures~\ref{lcs1}--\ref{lcs3} for a fixed target experiment, HERA, and
LEP$\otimes$LHC.
At HERA and LEP$\otimes$LHC, the corrections are very similar.
For small $x$,
the range of $y$ is reduced
due to the condition $Q^{2\min}= 5$ GeV$^2$ at a given value of
$x$. This tendency is more pronounced for the fixed target experiments.
The net corrections are smaller there because
for the fixed target experiments the $S$ is quite different. Further,
the muon mass sets the
scale in the leading logarithm $\ln(S/m^2)$, while at a collider it is
the electron mass.
 
From figure~\ref{ilep} and also
from~(\ref{k0}) it may be seen that for small
\yl\ the photon energy is strongly bound and only soft photons occur.
These corrections are negative.
Without the soft photon exponentiation, they
    would even diverge when \yl\ vanishes
and at the same time \xl\ approaches 1. This may be seen from
~(\ref{66}) as well.
At large \yl\ and small \xl\, the factorized correction seems to diverge also;
see the squared logarithm in~(\ref{65}).
This behaviour, however, is fictitious. It is compensated by
corresponding terms from the hard non-factorizing corrections.
The steep rise of the corrections at large \yl\ and small \xl\ is
completely due to the
Compton peak, which arises from the small \qh\ in the denominator of the
photon propagator. This is explained in detail in section~\ref{compt}.
From the $Z$-exchange diagrams, there is no
contribution to the Compton peak and no steep rise of the corrections
at large \ql. This may be explicitly seen in figure~3 of~\cite{zfpc42}.
 
%
 
At small values of \xl, the corrections are no longer monotonously
dependent on \xl.
In this region, the hard photon corrections are dominant.
They are not proportional to the Born cross section and an influence of
the properties of the structure functions at smaller values of $x$
may show up.
We will come back to this point in section~\ref{disc}.

The complete QED corrections in leptonic variables have been
calculated in the quark parton model in~\cite{zfpc42}.

\subsection
{Mixed variables. A second analytical integration
\label{fmix}
}
%
%
For the case of mixed variables, it is  ${\cal E}_{m}  = (\yh,\Ql)$
and the integrations have to be performed over the variables
${\cal I}_{m} = (\yl,\Qh) $.
The cross section~(\ref{*051}) becomes:
\ba
\frac {d^2 \sigma_{\mr R}^{\mr F}}{d\yh d\Ql}
&=& \frac {2\alpha^3 S^2}{\lambda_S}
    \int d\qh \, d\yl \sum_{i=1}^{3} \Biggl[
     {\cal A}_{i}(\xh,\Qh) \;\frac{1}{Q^4_h} \;
     {\cal S}_{i}(\yl,\ql,\yh,\qh)
\label{eqn053}                                 \nll
& &-~{\cal A}_{i}(x_m,\Ql) \frac{1}{Q_{l}^4} \;
     {\cal S}_{i}^{\mr B}(\yh,\Ql)\;
     {\cal L}^{\mr {IR}}(\yl,\ql,\yh,\qh) \Biggr].
\ea
Here, the definition~(\ref{xm}) of the variable $\xm$ is used,
$\xm=\ql/(\yh S)$.
 
In contrast to the case of leptonic variables,
in~(\ref{eqn053}) one may
        perform an analytical integration over the
variable $\yl$.
As it is discussed in appendix~\ref{b22} and may be seen in
figure~\ref{ylqha},
the integration region in~(\ref{eqn053}) must be split into
two regions~I and~II as long as the condition $\ql < \yh S$ is fulfilled,
which corresponds to $\xm \leq 1$.
If instead $\xm > 1$, there is only the region~I, see figure~\ref{ylqhb}.
Some technical details about the analytical integration
over $y_l$ may be found in appendix~\ref{appe4}.
 
In the ultra-relativistic approximation, one gets the
following expressions for $\xm<1$:
\ba
{\cal S}_{1}^{\mr I}(\Ql,\yh,\Qh)
&\equiv&
    \int_{ {y_l^{\min}}}^{y_l^{\max}}
  d \yl \, {\cal S}_{1}(\yl,\ql,\yh,\qh)
\nll
&=&\frac{1}{S}\Biggl[\frac{1}{2}\frac {\Ql}{\Ql-\Qh}
\left( 1+\frac{ Q_h^4}{ Q_l^4} \right)
       \left( {\mr L}  +{\mr L}_1-{\mr L}_2-1 \right)
\nll
 & &-~\frac{1}{2}\frac{\Ql}{\Qh} \left( 1+ \frac{ Q_h^4}{ Q_l^4} \right)
       \left( {\mr L}_t+{\mr L}_2 \right) +{\mr L}_{T}-{\mr L}_t
     +\frac{1}{2} \left( 1-\frac{\Qh}{\Ql} \right) \Biggr] ,
\label{m57}
\\
{\cal S}_{2}^{\mr I}(\Ql,\yh,\Qh)
&\equiv&
    \int_{ {y_l^{\min}}}^{y_l^{\max}}
    d\yl \, {\cal S}_{2}(\yl,\ql,\yh,\qh)
\nll
&=&
S \Biggl\{
\frac{1}{\Ql}\frac{\Qh-\yh\Ql}{\Ql-\Qh}
     \left( 1+\frac{ Q_h^4}{ Q_l^4} \right)
    \left( {\mr L}  +{\mr L}_1-{\mr L}_2-1 \right)
\nll
& &+~\frac{1}{\Qh}
\left( 1+\frac{ Q_h^4}{ Q_l^4} \right) \left[\yh(1+\frac{\Ql}{\Qh})
   -
\left( 1+\frac{\Ql}{\Qh} + \frac{\Qh}{\Ql} \right) \right]
     \left( {\mr L}_t+{\mr L}_2 \right)
\nll
& &+~\frac {1}{\Qh} \left(1+\frac{3}{2}\frac{\Qh}{\Ql}+3\frac{Q_h^4}{Q_l^4}
   +  2\frac{Q_h^6}{Q_l^6} \right)
\nll
& &-~ \frac{\yh}{\Qh}\left(
   \frac{\Ql}{\Qh }+2+4\frac{\Qh}{\Ql}+3\frac{Q_h^4}{Q_l^4}
     \right)
+~\frac{1}{2} \frac{ y_h^2}{\Qh} \left(
   2+\frac{\Ql}{\Qh}+2\frac{\Qh}{\Ql} \right)
\Biggr\},
\\
{\cal S}_{3}^{\mr I}(\Ql,\yh,\Qh)
& \equiv &
    \int_{ {y_l^{\min}}}^{y_l^{\max}}
    d\yl \, {\cal S}_{3}(\yl,\ql,\yh,\qh)
\nll
&=& \frac {\Ql}{\Ql-\Qh}\left( 1+\frac{ Q_h^4}{ Q_l^4}\right)
     \left( 2\frac{\Qh}{\Ql}-\yh \right)
   \left( {\mr L}   +{\mr L}_1 - {\mr L}_2-1 \right)
\nll
& &-~ \frac{\Ql}{\Qh} \left( 1+\frac{ Q_h^4}{ Q_l^4} \right)
\left( 2\frac{\Qh}{\Ql} -\yh +2   \right)
     \left( {\mr L}_t+{\mr L}_2 \right)
\nll
& &+~\frac{\ql}{\Qh}{\left(1+\frac{\Qh}{\Ql}\right)}^2 \left(
    2\frac{\Qh}{\Ql}-\yh \right)
   -\yh\left( \frac{\Ql}{\Qh}+1+2\frac{\Qh}{\Ql} \right)  ,
\label{eqn056}
\ea
\ba
{\cal S}_{1}^{\mr{II}}(\Ql,\yh,\Qh)
& \equiv &
    \int_{ {y_l^{\min}} }^{y_l^{\max}}
    d\yl \,  {\cal S}_{1}(\yl,\ql,\yh,\qh)
\nll
&=&\frac{1}{S}\Biggl[-\frac{1}{2}\frac {Q_l^4}{\Qh(\Ql-\Qh)}
\left(1+\frac{ Q_h^4}{ Q_l^4}\right)
\left( {\mr L}  -{\mr L}_1+{\mr L}_2-1 \right)
\nll
 & &-~\frac{1}{2}\frac{\Ql}{\Qh}\left(1+ \frac{ Q_h^4}{ Q_l^4}\right)
        {\mr L}_1 +   {\mr L}_{T}
     +\frac{1}{2}\left(1-\frac{\Ql}{\Qh}\right) \Biggr] ,
\\
{\cal S}_{2}^{\mr{II}}(\Ql,\yh,\Qh)
& \equiv &
    \int_{ {y_l^{\min}} }^{y_l^{\max}}
    d\yl \,  {\cal S}_{2}(\yl,\ql,\yh,\qh)
\nll
&=&
S\Biggl\{
-\frac{Q_l^4}{Q_h^4} \frac{(1-\yh)}{\Ql-\Qh}
     \left(1+\frac{ Q_h^4}{ Q_l^4}\right)
    \left({\mr L}  -{\mr L}_1+{\mr L}_2-1\right)
\nll
& &+~ \frac{1}{\Qh}\left(1+\frac{ Q_h^4}{ Q_l^4}\right)\left[\yh
\left(1+\frac{\Ql}{\Qh}\right)
   -\left(1+\frac{\Ql}{\Qh} + \frac{\Qh}{\Ql}\right) \right]
       {\mr L}_1
\nll 
& &+~ \frac {1}{\Ql} \left(\frac{3}{2}+\frac{\Qh}{\Ql}+3\frac{\Ql}{\Qh}
   +  2\frac{Q_l^4}{Q_h^4} \right)
\nll
& &-~  \frac{\yh}{\Qh}\left(
   \frac{3\Ql}{\Qh }+4+2\frac{\Qh}{\Ql}+\frac{Q_h^4}{Q_l^4}
     \right)
\nll
& &+~ \frac{1}{2} \frac{ y_h^2}{\Qh} \left(
   2+2\frac{\Ql}{\Qh}+\frac{\Qh}{\Ql} \right)
\Biggr\},
\\
{\cal S}_{3}^{\mr{II}}(\Ql,\yh,\Qh)
& \equiv &
    \int_{ {y_l^{\min}} }^{y_l^{\max}}
    d\yl \, {\cal S}_{3}(\yl,\ql,\yh,\qh)
\nll
&=& -~\frac {(2-\yh)}{(\Ql-\Qh)}\frac{Q_l^4}{\Qh}\left(1+\frac{ Q_h^4}{ Q_l^4}
\right)
    \left({\mr L}   -{\mr L}_1 + {\mr L}_2-1\right)
\nll
& &+~  \left(1+\frac{ Q_h^4}{ Q_l^4}\right)\left[
     \yh\frac{\Ql}{\Qh} -2\left(1+\frac{\Ql}{\Qh}\right) \right]
       {\mr L}_1
\nll
& & +~2(1-\yh)\frac{\Ql}{\Qh}{\left(1+\frac{\Qh}{\Ql}\right)}^2
     +\yh\left( 1-  \frac{\Ql}{\Qh} \right) .
\label{eqn511}
\ea
The subtracted part of the cross section also becomes
integrated
 over $\yl$ with the
aid of appendix~\ref{appe4}:
\ba
{\cal L}^{\mr {IR}}_{\mr I}(\Ql,\yh,\Qh)
&\equiv&
    \int_{ {y_l^{\min}} }^{y_l^{\max}}
    d \yl \, {\cal L}^{\mr {IR}}(\yl,\ql,\yh,\qh)
\nll
&=&\frac {1}{S(\Ql-\Qh)}
   \left[ -1 -\frac{\Ql}{\Qh}({\mr L}_t+{\mr L}_2)
   +({\mr L}  +{\mr L}_t+{\mr L}_1)\right] ,
\\
{\cal L}^{\mr {IR}}_{\mr{II}}(\Ql,\yh,\Qh)
&\equiv&
    \int_{ {y_l^{\min}} }^{y_l^{\max}}
    d \yl {\cal L}^{\mr {IR}}(\yl,\ql,\yh,\qh)
\nll
&=&\frac {1}{S(\Ql-\Qh)}
   \left[\ \frac{\Ql}{\Qh}+{\mr L}_1-\frac{\Ql}{\Qh}
   ({\mr L} +{\mr L}_2)\right]\ .
\label{eqn513}
\ea
 
The following abbreviations are used:
\ba
\nll
{\mr L}    =\mbox{ ln}\frac{\Ql}{m^2}, \hspace{.7cm}
{\mr L}_{t}=\mbox{ ln}\frac{\Ql}{\Qh},\hspace{.7cm}
{\mr L}_{T}=\mbox{ ln}\frac{1 }{\yh},
\label{eqnLOG}
\ea
\ba
\nll
{\mr L}_{1}=\mbox{ln}\frac{\Qh-\yh \ql}{ \mid \Ql-\Qh \mid} ,\hspace{.7cm}
{\mr L}_{2}=\mbox{ln}\frac{(1- \yh)\qh}{\mid \Ql-\Qh \mid} .
\label{eqn514}
\ea
 
The final formula for $ d^{2}\sigma^{\mr F}_{\mr R}/d\yh d\Ql $
can be written in the form:
\ba
\frac {d^2 \sigma_{\mr R}^{\mr F}} {d\yh d\ql}
&=&\frac{2 \alpha^{3} S^2}{\lambda_S}
   \Biggl\{\int_{Q_h^{2\min}}^{Q_l^2}
    d\Qh \sum_{i=1}^3 \Bigl[
   {\cal A}_i(\xh,\Qh)\frac{1}{Q^4_h}
    {\cal S}_{i}^{\mr{I}} (\yl,\Ql,\yh,\Qh)
\nll & &
-~{\cal A}_i(x_m,\ql)\frac{1}{Q^4_l}
    {\cal S}_{i}^{\mr B} (\yh,\Ql)
    {\cal L}_{\mr I}^{\mr {IR}}(\yl,\ql,\qh)\Bigr]
\nll
& &+~\int_{Q_l^2}^{y_h S}
d\Qh\sum_{i=1}^3 \Bigl[\
    {\cal A}_i(\xh,\Qh)\frac{1}{Q^4_h}
    {\cal S}_{i}^{\mr{II}}(\Ql,\yh,\Qh)
\nll & &
-~{\cal A}_i(x_m,\Ql) \frac{1}{Q^4_l}
    {\cal S}_{i}^{\mr B} (\Ql,\yh)
    {\cal L}_{\mr{II}}^{\mr {IR}}(\Ql,\yh,\Qh) \Bigr]\
   \Biggr\}.
\label{eqn515}
\ea
 
The above expressions again show remnants of the infrared
singularity at $\ql=\qh$. In contrast to the case of leptonic variables,
one may easily see that
the corresponding terms are multiplied by kinematical factors,
which in the limit $\ql=\qh$ become explicitly equal to the
kinematical Born functions ${\cal S}_i^{\mr B}$,
~(\ref{eqS3B1})--(\ref{eqS3B}).
This makes the finiteness of the
$d^2\sigma_{\mr R}^{\mr F}$ obvious.
%
\subsubsection
{Discussion
\label{dism}
}
\begin{figure}[bhtp]
\begin{center}
\mbox{
\epsfysize=9.cm
\epsffile[0 0 530 530]{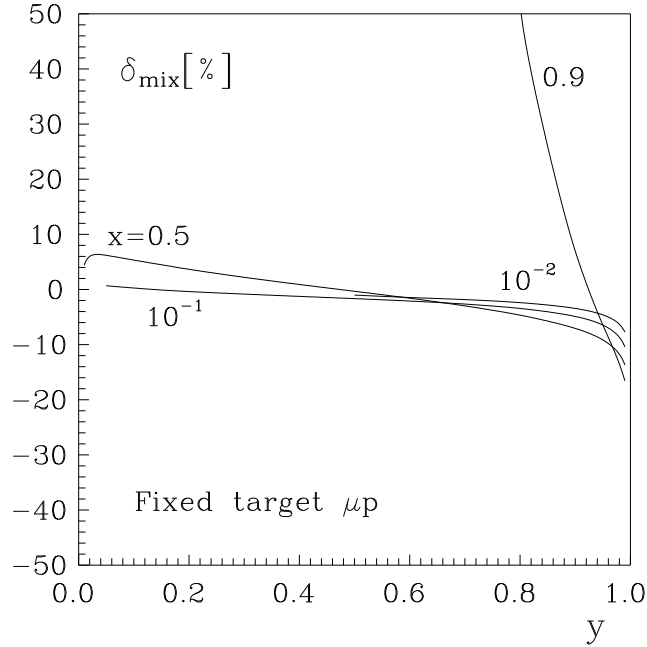}
}
\end{center}
\caption{\it
Radiative correction $\delta_{\mr{mix}}$ in \% for a
cross section measurement in terms of mixed variables
for fixed target $\mu p$ scattering.
}
\label{mcs1}
\end{figure}

\begin{figure}[thbp]
\begin{center}
\mbox{
\epsfysize=9.cm
\epsffile[0 0 530 530]{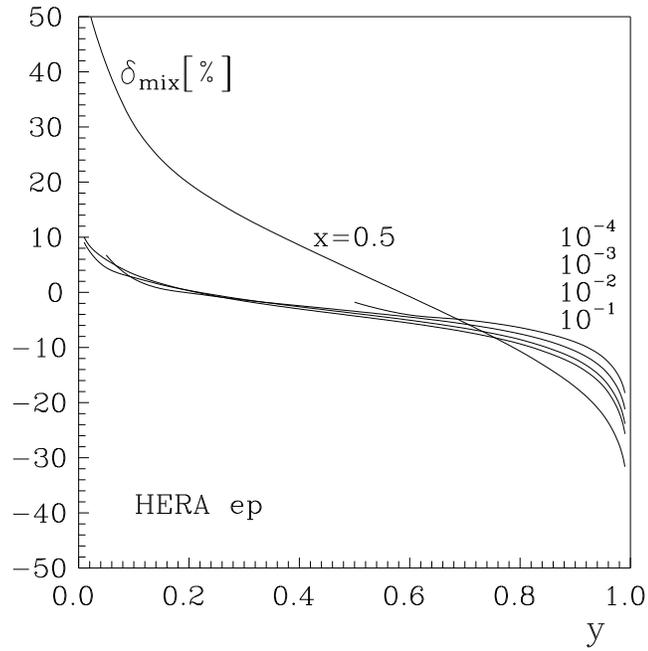}
}
\end{center}
\caption{\it
Radiative correction $\delta_{\mr{mix}}$ in \% for a
cross section measurement in terms of mixed variables
at HERA.
}
\label{mcs2}
\end{figure}

\begin{figure}[bhtp]
\begin{center}
\mbox{
\epsfysize=9.cm
\epsffile[0 0 530 530]{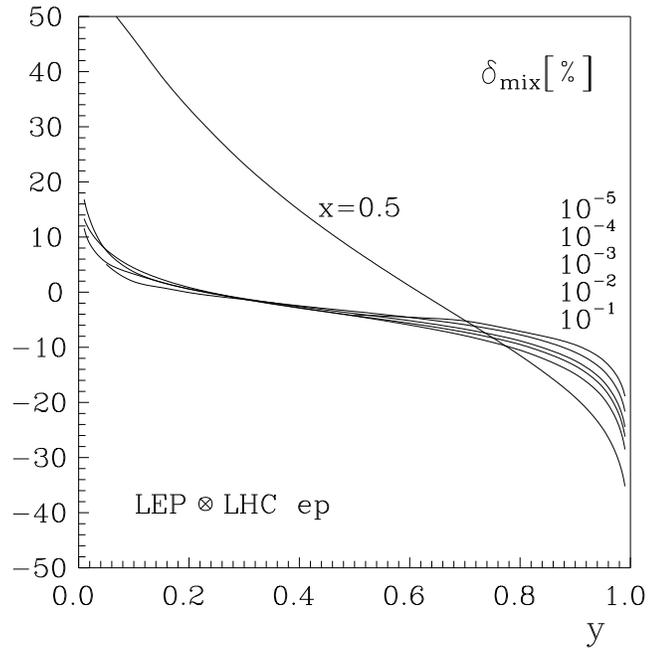}
}
\end{center}
\caption{\it
Radiative correction $\delta_{\mr{mix}}$ in \% for a
cross section measurement in terms of mixed variables
at LEP$\otimes$LHC.
}
\label{mcs3}
\end{figure}

The leptonic QED corrections in mixed variables are shown in
figures~\ref{mcs1}--\ref{mcs3} for three different kinematical situations.
As in leptonic variables,
at HERA and LEP$\otimes$LHC they are similar, while in the fixed
target case less pronounced.

The cross section depends on \ql\ and \yh.
From figure~\ref{ylqha}, and also
from~(\ref{k0}), it may be seen that for large
\yh\ the photon energy is strongly bound; we have only soft photon
corrections.
These corrections are negative there.
                               This is
contrary to the case of leptonic variables, where this happened at small
\yl.
Without the soft photon exponentiation, they
    would even diverge when \yh\ approaches 1.
This behaviour is enhanced, but less than for leptonic variables,
if $\xm$ approaches 1, too.
                                      This may be seen from
~(\ref{68}).
In principle, the Compton peak is present in mixed variables.
It is considerably suppressed
compared to the case of leptonic variables,
 see
section~\ref{compt}.
At small \yh, hard photon emission is possible and compensates for the
negative soft photon corrections.
The net correction is positive and remains moderate.
If in addition \xm\ approaches 1,
we observe that the corrections, if expressed in terms of the
$\delta$, rise steeply. This phenomenon is attributed to the following:
the Born cross section does not exist for $\xm>1$, while hard
photon
emission does (see also figure~\ref{emix}).
So, their ratio explodes and the $\delta$ becomes an
inappropriate variable.

In sum, the features of the radiative corrections in terms of mixed
variables are quite different from those in leptonic variables.
This is an instructive illustration of the general statement that
the radiative corrections depend substantial on the choice of variables,
in which they are determined.
In another context, namely for the neutrino-electron
charged current scattering, a similar observation was made
in~\cite{numue}.
 
The complete QED corrections in mixed variables have been
calculated in the quark parton model in~\cite{qpmmix}.

 
\subsection
{Hadronic variables. A second analytical integration
\label{fhad}
}
%
%
For the cross section in terms of
hadronic variables ${\cal E}_{h}=(\yh,\Qh)$,
the integrations have to be performed over
${\cal I}_{h}=(\yl,\Ql)$ and~(\ref{*051}) becomes:
\ba
\frac {d^2 \sigma_{\mr R}^{\mr F}} {d\yh d\Qh}
&=&\frac{2 \alpha^3 S^2 }{\lambda_S}
   \sum_{i=1}^3 \frac{ {\cal A}_i(\xh,\Qh)}{Q^4_h} \int d\ql \, d\yl
   \Bigl[ {\cal S}_{i}(\yl,\ql,\yh,\qh)
\nll
& &-~{\cal S}_i^{\mr B } (\yh,\Qh)
   {\cal H}^{\mr {IR}}(\yl,\ql,\yh,\qh) \Bigr],
\label{516}
\ea
where
\ba
{\cal H}^{\mr {IR}}(\yl,\ql,\yh,\qh)
&\equiv&
     {\cal F}^{\mr {IR}}(\yl,\ql,\yh,\qh)|_{Q^2=Q^2_h}
\nll
&=& \frac{\Qh+2m^2}{\Ql-\Qh}\Bigl( \frac{1}{\sqrt{C_{1}}}
                 -\frac{1}{ \sqrt{C_{2}}} \Bigr)
   -m^{2}\Bigl( \frac{B_{1}}{C_{1}^{3/2}}
               +\frac{B_{2}}{C_{2}^{3/2}}\Bigr).
\label{eqn517}
\ea
 
In~(\ref{516}) we again can perform the analytical integration over
$y_l$.
As in the case of mixed variables,
the region of integration in~(\ref{516}) must be split
into two parts (see appendix~\ref{b3g} and figure~\ref{yhqh}).
The limits for $\yl$ at given values of $\yh,\qh,\ql$ are naturally
the same as for the mixed variables.
The difference is with the $Q^2$, which interchange their roles.
 
After the integration over $\yl$, the following expression   for
$d^{2} \sigma_{\mr R}^{\mr F}/d\yh\Qh$ is obtained:
\ba
\frac {d^2 \sigma_{\mr R}^{\mr F}} {d\yh d\Qh}
&=& \frac{2 \alpha^{3} S^2}{\lambda_S}
    \sum_{i=1}^3 \frac{{\cal A}_i(\xh,\Qh)}{Q^4_h}
    \Biggl\{
    \int_{Q_l^{2\min}}^{Q_h^2} d\Ql
    \Bigl[ {\cal S}_{i}^{\mr I} (\ql,\yh,\Qh)
-~{\cal S}_{i}^{\mr B} (\yh,\Qh)
   {\cal H}_{\mr I}^{\mr {IR}}(\Ql,\yh,\Qh) \Bigr]
\nll
& &+ \int_{Q_h^2}^{Q_l^{2\max}} d\Ql
    \Bigl[ {\cal S}_i^{\mr{II}}(\Ql,\yh,\Qh)
   -{\cal S}_{i}^{\mr B}(\yh,\Qh)
    {\cal H}_{\mr{II}}^{\mr {IR}}(\Ql,\yh,\Qh) \Bigr]
   \Biggl\}.
\label{eqn518}
\ea
Here the functions
${\cal S}_{i}^{\mr {I,II}}(\Ql,\yh,\Qh) \; (i=1,\dots,3)$ are exactly
the same as those for mixed variables [see
~(\ref{m57})--(\ref{eqn511})].
The definition of the
${\cal H}^{\mr {IR}}_{\mr I}(\Ql,\yh,\Qh)$ differs slightly from that
of the $
{\cal L}^{\mr {IR}}_{\mr I}(\Ql,\yh,\Qh)$. The integrals
which are calculated with the aid of appendix~\ref{appe4} differ also
and the result in terms of hadronic variables is:
\ba
{\cal H}^{\mr {IR}}_{\mr I}(\Ql,\yh,\Qh)
&\equiv& \int_{ {y_l^{\min}}_{\mr I } }^{y_l^{\max}}
d\yl \,  {\cal H}^{\mr {IR}}(\yl,\ql,\yh,\qh)
\nll
&=&\frac{1}{S(\Ql-\Qh)}\left[ -1+\frac{\Qh}{\Ql}
({\mr L}  +{\mr L}_t+{\mr L}_1)
- {\mr L}_t-{\mr L}_2  \right],
\\
{\cal H}^{\mr {IR}}_{\mr{II}}(\Ql,\yh,\Qh)
&\equiv& \int_{ {y_l^{\min}} _ {\mr{II}} }^{y_l^{\max}}
d\yl \, {\cal H}^{\mr {IR}}(\yl,\ql,\yh,\qh)
\nll
&=&\frac{1}{S(\Ql-\Qh)}\left[ \frac{\Ql}{\Qh}
+\frac{\Qh}{\Ql}{\mr L}_{1}-  {\mr L}  -{\mr L}_2  \right] .
\label{eqn520}
\ea
The logarithms ${\mr L},{\mr L}_a, a=1,2,t$ are defined
in ~(\ref{eqnLOG})--(\ref{eqn514}).
 
The remarks about the remnants of                   the infrared
singularity and its numerical treatment may be simply taken over from
the case of the mixed variables.
 
The reader may realize that there is one interesting difference
between mixed and hadronic variables:
in the latter case, one may                             perform
also the last           integration analytically, thus    avoiding
any numerical part of the calculation.
In fact, we have done also this last step, but with
     a different parameterization of the phase space; see
                           section~\ref{irIIh}.
\subsubsection
{Discussion
\label{dish}
}
\begin{figure}[bhtp]
\begin{center}
\mbox{
\epsfysize=9.cm
\epsffile[0 0 530 530]{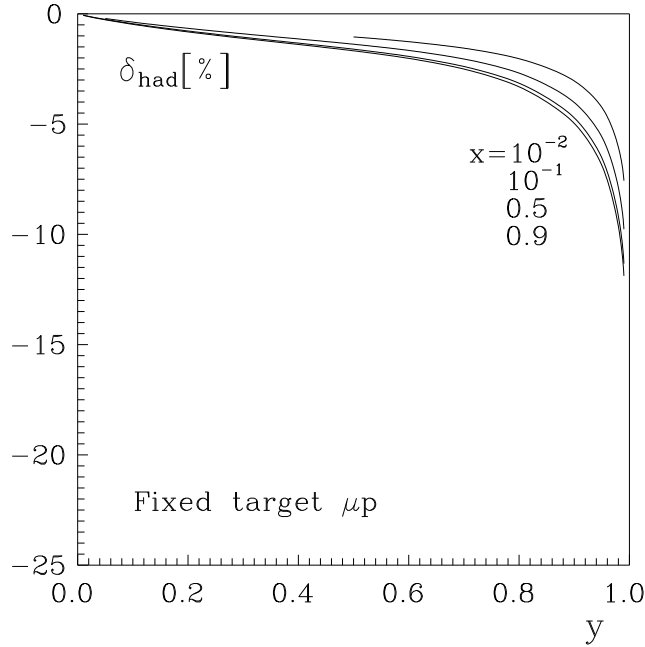}
}
\end{center}
\caption{\it
Radiative correction $\delta_{\mr{had}}$ in \% for a
cross section measurement in terms of hadronic variables
for fixed target $\mu p$ scattering.
}
\label{hcs1}
\end{figure}

\begin{figure}[tbhp]
\begin{center}
\mbox{
\epsfysize=9.cm
\epsffile[0 0 530 530]{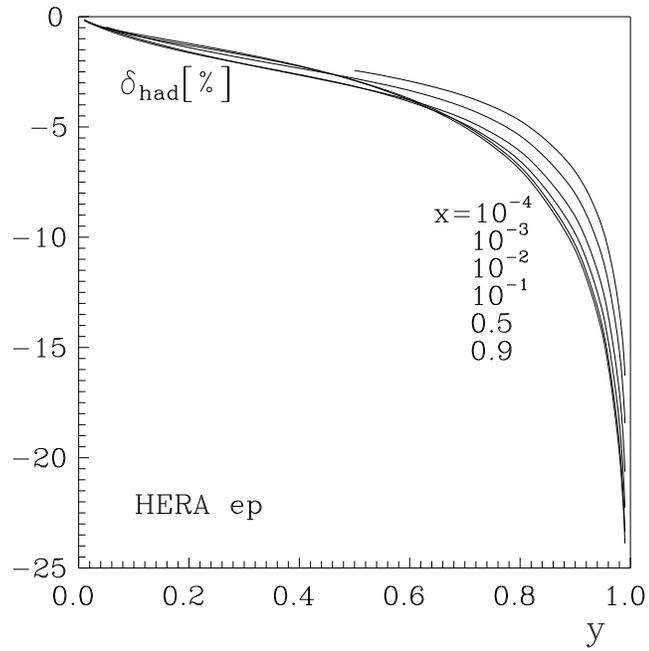}
}
\end{center}
\caption{\it
Radiative correction $\delta_{\mr{had}}$ in \% for a
cross section measurement in terms of hadronic variables
at HERA.
}
\label{hcs2}
\end{figure}

\begin{figure}[bhtp]
\begin{center}
\mbox{
\epsfysize=9.cm
\epsffile[0 0 530 530]{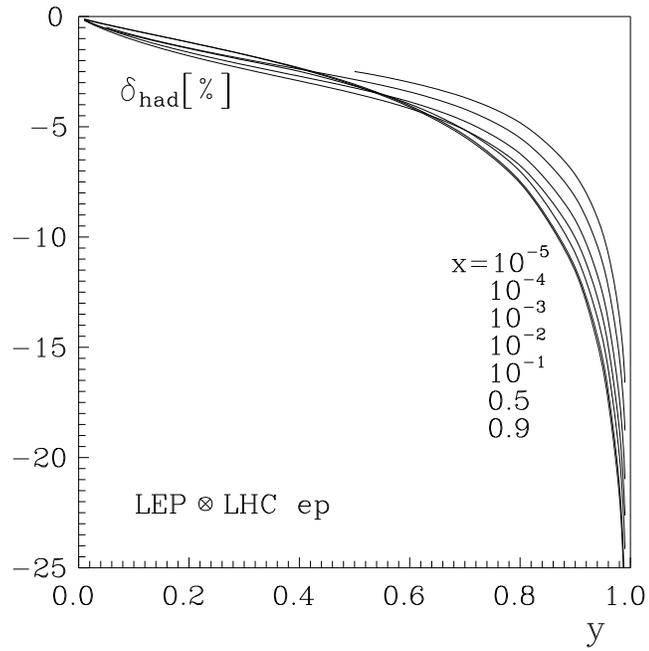}
}
\end{center}
\caption{\it
Radiative correction $\delta_{\mr{had}}$ in \% for a
cross section measurement in terms of hadronic variables
at LEP$\otimes$LHC.
}
\label{hcs3}
\end{figure}

The leptonic QED corrections in hadronic variables are shown in
figures ~\ref{hcs1}-\ref{hcs3}.
They are considerably smaller than those  which are determined in terms of
leptonic
variables.
From figure~\ref{yhqh} and also
from~(\ref{k0}), it may be seen that for small
\yh\ the photon energy is strongly bound; we have only soft photon
corrections there.
This is the origin of the similarity of the gross behaviour of the
corrections in hadronic and mixed variables with each other.
Without the soft photon exponentiation, the corrections
    would      diverge when \yh\ approaches 1.
This may be seen from~(\ref{610}). There is no explicit
dependence of the dominant terms on \xh.
Although, with the factor $\ln \qh/m^2$, one may explain that at higher
\xh\ the corrections are slightly more pronounced.
Another singularity of the factorizing part of the corrections
~(\ref{69})--(\ref{610}) is located at $\yh=0$.
As may be seen from the figures, it is compensated for in
 the net corrections.
Again, as was mentioned in the case of leptonic variables, this
singularity is cancelled by a corresponding behaviour of the hard
non-factorizing bremsstrahlung.
In fact, at $\yh=0$ the corrections vanish even.
 
This may be explicitly seen from the formulae of
 section~\ref{irIIh}, where a completely
integrated analytical expression for the corrections in
hadronic variables
is derived.
 
Finally, we should mention that a Compton peak may not be developed
since \qh\ is fixed.
\section
{
Photoproduction
\label{photo}}
\setcounter{equation}{0}
The $ep$ scattering process at vanishing photon momentum,
$Q^2 \approx 0$, proceeds via the exchange of nearly real photons and
is called photoproduction.
 
The process kinematics usually is described with the $Q^2$ as defined
from the leptonic variables and the invariant mass $W$ of the
compound system consisting of the photon and the hadrons,
\ba
Q^2 = \ql &=& (k_1 - k_2)^2,
\\
W^2 &=& -(Q_l + p_1)^2
\nll
    &=& M^2 +(1-\xl)\yl S = M^2 + \yl S - \ql.
\label{w2}
\ea
For a fixed value of $W^2 << S$, the minimal value of $Q^2$~(\ref{q2le})
may become extremely small. With~(\ref{eq21}),
$W^2 \geq (M+m_{\pi})^2$, it remains non-vanishing:
\ba
\ql^{\min}(W^2) \approx m^2 \, \frac{[W^2-M^2]^2}{S^2}
> 0.
\label{q2min}
\ea
The integrated deep inelastic cross section is strictly finite.
 
For example, in the reaction
$ep \rightarrow ep\rho$ at HERA energies
the possible values of $Q^2$
       extend  from 10$^{-15}$ to 10$^5$ GeV$^2$~\cite{schuler}.
In such a kinematical region,
the description of the radiative corrections must be exact in both
the proton mass $M$ and the electron mass $m$.
For this reason, we derive here
formulae for the QED corrections in leptonic
 variables
which remain valid at extremely small $Q^2$.
The chosen integration
variables differ slightly from those which we used so far:
\ba
  {\cal E}={\cal E}_l^{\prime}=(W^2,\Ql), \hspace{1cm}
                                   {\cal I}={\cal I'}_l=(M_h^2,\Qh).
\label{eqphot1}
\ea
The invariant mass $M_h^2$ of the hadronic system is
\bq
M_h^2 = -(p_1 + Q_h)^2 = M^2 + \yh S - \Qh.
\label{eq28}
\eq
It will be convenient to
treat the infrared singularity in a lorentz system
 where the above mentioned
compound system is at rest:
${\vec p}_2+ {\vec k}=0$.
The necessary kinematical relations are derived in the appendices
~\ref{app11} and~\ref{kineme}.
In the following it will be described
how the three different contributions to the
corrections
are treated: the infrared divergent correction
$\delta_{\mr{soft}}^{\mr{IR}}$,
the factorized hard part
$\delta_{\mr{hard}}^{\mr{IR}}$,
and the finite rest of the hard bremsstrahlung contribution.
 
The removal of the infrared divergence proceeds technically as it is
described in section~\ref{irre3}  but will be performed in the rest
system introduced here.
For the soft part of $\delta^{\mr {IR}}$ one gets again the expression
~(\ref{eqn415}). The velocities
$\beta_i$, which are introduced in~(\ref{kalp2})
are now to be expressed by~(\ref{fo6}) with
the invariants~(\ref{lambpr}).
The result is:
\ba
\beta_1=\sqrt{1-\frac{4m^2M^2}{(S-\Ql)^2}},\hspace{1cm}
\beta_2=\sqrt{1-\frac{4m^2M^2}{[S(1-\yl)+\Ql]^2}}.
\label{67-2}
\ea
Further, for ${\cal S}_{\Phi}$ the exact definition~(\ref{soft3})
has to be used
in terms of the momenta~(\ref{fo6}).
 
Integrating~(\ref{irhard}) threefold
as it is explained in appendix~\ref{appfo}
one obtains the following expression:
\ba
\delta^{\mr {IR}}_{\mr {hard}}(\yl,\Ql,\epsilon)
&=& 2\left[ -\mbox{ln}{ \frac{2\epsilon}{m}}
  +  \mbox{ln} { \frac{ W^2-(M+m_{\pi})^2}{m\sqrt{W^2}}}\right]
    \left(\frac{1+{\beta}^2}{2\beta}\Lb-1 \right) ,
\label{eqhard}
\ea
where
\ba
   \Lb = \mbox{ ln } \frac{\beta +1}{\beta -1},\hspace{1cm}
   \beta=\sqrt{1+\frac{4m^2}{\Ql}}.
\label{k120v2}
\ea
 
Further, we need the exact
QED vertex correction, which contains  $\delta_{\mr {vert}}$~\cite{III}:
\ba
\delta_{\mr {vert}}({\beta})
&=&
 -2\left({\cal P}^{\mr {IR}}+\ln\frac{m}{\mu}\right)
    \left(\frac{1+{\beta}^2}{2\beta}\Lb-1 \right)
  + \frac{3}{2} {\beta}\Lb  -2
\nll
 & &  -~
  \frac{1+{\beta}^2}{2\beta}\left[\Lb
 \mbox{ln}{\frac{4{\beta}^2}{{\beta}^2-1}}
 + {\mr {Li}}_{2}\left(\frac{1+\beta}{1-\beta}\right)
 - {\mr {Li}}_{2}\left(\frac{1-\beta}{1+\beta}\right)\right].
\label{vertex}
\ea
Finally
one has to take into account also the
contribution from the anomalous magnetic moment of the
electron~\cite{zerwas}.
This vertex correction is needed only for the virtual photon exchange,
i.e. one may neglect the axial vector part in the matrix
element~(\ref{matrixborn}):
\ba
\gamma_{\mu}
&\rightarrow&
\delta_{\mr{vert}}(\beta) \gamma_{\mu}
+
i m {\cal V}_{\mr{anom}} (\beta) \frac{(k_1 + k_2)_{\mu}}{Q^2},
\\
{\cal V}_{\mr{anom}} (\beta)
&=&
- \frac{{\mr L}_{\beta}}{\beta}.
\label{anomm}
\ea
The cross section contribution from the anomalous magnetic moment
is\footnote{
The corresponding equation~(47) in~\cite{III} was not
correct.}:
\ba
 \frac {d^2 \sigma_{\mr {anom}}} {d\yl d\Ql} =
        \frac{2  \alpha^{3}S}{{\lambda_S Q_l^4}}
\frac{m^2}{\xl Q_l^2}
{\cal V}_{\mr{anom}} (\beta)
        \left[
        2\beta^2 y_l^2\xl F_1(\xl,\ql)
   -(2-\yl)^2
           F_2 (\xl,\ql)   \right].
\label{68-2}
\ea
 
Combining the three factorizing corrections, one gets
for the factorized part $\delta_{\mr VR}$ the following expression:
\ba
\delta_{\mr {VR}} (\yl,\Ql )
&=& \delta_{\mr{vert}} + \delta^{\mr {IR}}_{\mr {hard}} +
\delta^{\mr {IR}}_{\mr {soft}}
\nll
 &=& \delta_{\mr {inf}}(\yl,\Ql)
+~\frac{1}{2 \beta_{1}} \;
  \mbox{ln}\frac{1+\beta_1}{1-\beta_1}
 +\frac{1}{2 \beta_{2}} \;
  \mbox{ln}\frac{1+\beta_2}{1-\beta_2}
+ {\cal S}_{\Phi}
\nll & &
  +~\frac{3}{2} {\beta}\Lb  -2
  - \frac{1+{\beta}^2}{2\beta}\left[\Lb
 \mbox{ln}{\frac{4{\beta}^2}{{\beta}^2-1}}
 + {\mr {Li}}_{2}\left(\frac{1+\beta}{1-\beta}\right)
 - {\mr {Li}}_{2}\left(\frac{1-\beta}{1+\beta}\right)\right] ,
\nll
\label{65-2}
\ea
where
\ba
\delta_{\mr {inf}} (\yl,\Ql)=
     2 \ln { \frac{ W^2-(M+m_{\pi})^2}{m\sqrt{W^2}}}
    \left(\frac{1+{\beta}^2}{2\beta}\Lb-1 \right).
\label{66-2}
\ea
Finally, the infrared free part of the cross
section~(\ref{eqborn})
can be written as follows:
\ba
\frac {d^2 \sigma_{\mr R}^{\mr F}}{d \yl d \ql}
&=& \frac {2\alpha^3 S }{\lambda_S}
     \int dM_h^2 d\qh  \sum_{i=1}^{3} \Biggl[ {\cal A}_{i}(\xh,\qh)
     \frac{1}{Q^4_h} {\cal S}_{i}(\yl,\ql,\yh,\qh)
\nll
& &-~{\cal A}_{i}(\xl,\ql)\frac{1}{Q_{l}^4} \;
     {\cal S}_{i}^{B}(\yl,\ql)\;
     {\cal L}^{\mr {IR}}(\yl,\ql,\yh,\Qh) \Biggr] ,
\label{*052p}
\ea
where $ {\cal S}_{i}(\yl,\ql,\yh,\Qh)\;(i=1,2,3)$ are given
by~(\ref{eq314})--(\ref{eq316}), and
$ {\cal L}^{\mr{IR}}(\yl,\ql,\yh,\Qh) $ is defined by~(\ref{eqn052a}).
We would like to remind that these expressions are
exact in both masses $m$ and $M$
for the photon
exchange contribution.
 
In sum, the cross section is:
\ba
\frac {d^2 \sigma_{\mr R}}{d \yl d \ql}
=
\frac {d^2 \sigma_{\mr {anom}}} {d\yl d\Ql} +
\frac{d^{2}{\sigma}_{\mr R}^{\mr F}}{d\yl d\ql}
+
\frac{d^{2} {\sigma}_{\mr B}}    {d\yl d\ql}
\left\{
{\mr {exp}} \left[\frac{\alpha}{\pi}
 \delta_{\mr {inf}}({\cal E}) \right]
 -1+\frac{\alpha}{\pi}\Bigl[
 \delta_{\mr {VR}} ({\cal E})
-\delta_{\mr {inf}}({\cal E})\Bigr]
\right\} .
\ea
Here, we should also mention that the structure functions
must have a certain behaviour in the limit of small $Q^2$.
From the relation between the total cross section of
photon-proton scattering
and the structure function $F_2$ at small $Q^2$~\cite{schuler},
\ba
\sigma_{\gamma p}^{T} (W) = \frac{4 \pi \alpha}{Q^2}
F_2(x,Q^2) |_{Q^2=0},
\label{sppt}
\ea
one concludes that the structure functions should vanish with vanishing
$Q^2$.
One possibility is to
multiply them by a global suppression factor~\cite{Prokh}:
\ba
{\cal F}_{1,2}^{\mathrm{NC}}(x,Q^2)
\rightarrow
   [1 - \exp(-aQ^2)]
{\cal F}_{1,2}^{\mathrm{NC}}(x,Q^2), \hspace{1cm}
a=3.37 \,  {\mr GeV}^{-2} .
\label{prokh}
\ea
In our numerical results, we follow this prescription.
\subsection
{Discussion
\label{disqh}
}
\begin{figure}[bhtp]
\begin{center}
\mbox{
\epsfysize=9.cm
\epsffile[0 0 530 530]{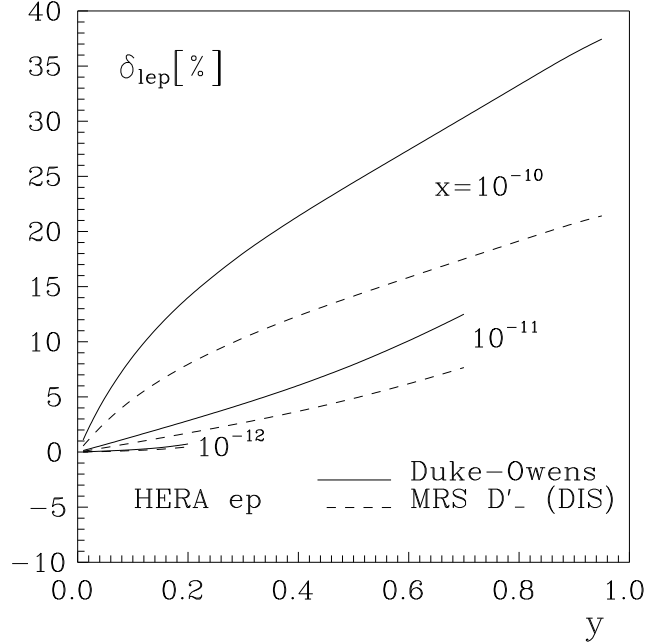}
}
\end{center}
\caption{\it
Radiative correction for photoproduction at HERA with two different
sets of structure functions.
\label{pcs}
}
\end{figure}

The QED corrections in the case of photoproduction are shown
 in figure~\ref{pcs} for HERA energies.
The structure functions~\cite{lib1,lib2}
are taken from~\cite{pdflib1} and~\cite{pdflib2}.
 
As was already
discussed in the section on leptonic variables,
soft photon emission is located near $\yl=0$.
With rising \yl\ more and more hard photons may be emitted  and the
influence of the structure functions via the non-factorizing hard
corrections becomes more pronounced.
It is further interesting to observe that at the extreme small values of
\xl\ as discussed here  the radiative corrections vanish at $\yl=0$.
For the anomalous vertex correction~(\ref{68-2}) and the exponentiated
soft photon correction~(\ref{66-2})  this follows immediately from the
limits in~(\ref{k120v2}):
$\beta \rightarrow \infty$ and ${\mr L}_{\beta} \rightarrow
2/\beta$.
 Here, it is important to notice that the \ql\ becomes small even
compared to the electron mass.
In fact, with the aid of~(\ref{w2}),
(\ref{q2min}) may be well
approximated by $\ql^{\min} \sim m^2 y_l^2 << m^2$ for small \yl.
Further, the $\delta_{\mr{hard}}^{\mr{IR}}$ of~(\ref{eqhard})
vanishes explicitly.
This reflects the general statement that for sufficiently small
\yl\ energetic photons cannot be emitted.
For the rest of the corrections,
we see no trivial argument why they should vanish or
compensate each other; but they do.
A further interesting feature is of kinematical origin.
At very small \xl, the \ql\ is also bound strongly.
As may be seen in figure~\ref{ilep}, then the allowed values of \yh\ are
bound from above. In the ultra-relativistic approximation  this property
gets lost.
 
\section
{An alternative treatment of the phase space
\label{kineII}}
\setcounter{equation}{0}
So far, one unique phase space parameterization has been used
for all three different
sets of external variables.
The infrared problem was treated in
the proton rest system.
 Compared to this,
for leptonic variables the rest system which was introduced for the treatment
of the photoproduction process lead  to considerable simplifications.
 
Here, we change to a completely different phase space parametrization, which
is specially adapted to the case of hadronic and Jaquet-Blondel variables.
The leptonic degrees of freedom will be integrated over at the very
beginning of the calculation.
\subsection
{Hadronic variables. The phase space
\label{hb2}
}
Wherever possible, the ultra-relativistic kinematics will be used
in this section.
We start from the slightly rewritten expression~(\ref{gammar}):
\ba
\Gamma =  \int \frac{d\vec{k_2}}{2k_2^0} \frac{d\vec{k  }}{2k  ^0}
       \delta^4(k_1+p_1-k_2-p_2-k) d^4 p_2 \delta(p_2^2+M_h^2) dM_h^2.
\label{gmtot}
\ea
Now, the integration variables we are interested in will be introduced.
For that purpose, we consider the photon and the final state electron
to be a compound system.
The corresponding rest frame $R$ is defined by the three-momentum
relation
$\vec{k}_2^R + \vec{k}^R = 0$.
The integration variables ${\vec k}_2, {\vec k}$ may be replaced by the following
variables:
\begin{itemize}
\item
the invariant mass of the ($\gamma e$) compound system $\tau=-(k_2+k)^2$;
\item
the photonic angles $\vartheta_R$, $\varphi_R$ in the
rest frame $R$ of this compound system.
\end{itemize}
The phase space of the variables $k_2,k$ may be parameterized as follows:
\ba
\int \frac{d\vec{k_2}}{2k_2^0} \frac{d\vec{k  }}{2k  ^0}
\delta^4(k_1+p_1-k_2-p_2-k)
&=&
\int d\tau \delta(\tau+\Lambda^2) d\Gamma_{\gamma e}.
\label{spac1}
\ea
Here, the phase space element of the $(\gamma e)$-compound
system is introduced:
\ba
d\Gamma_{\gamma e}= \frac{d\vec{k_2}}{2k_2^0} \frac{d\vec{k}}{2k^0}
       \delta^4(\Lambda-k_2-k).
\label{dgeg}
\ea
In the rest frame $R$ of the compound system, (\ref{dgeg}) may be rewritten
as it is explained in appendix~\ref{ph3}:
\ba
\int
d\Gamma_{\gamma e}
&=&
\frac{\sqrt{\lambda(\tau,m^2,0)}}{8\tau}
\int_{-1}^{1  } d \cos \vartheta_R \int_{0}^{2\pi}d
\varphi_R,
\nll
\sqrt{\lambda(\tau,m^2,0)}
&=&
\tau-m^2.
\label{dgr}
\ea
In a next step, the hadronic momentum transferred
$\qh$ is introduced
into~(\ref{gmtot}), and the corresponding  $\delta$-function under
the $p_2$ integral is exploited.
In appendix~\ref{ph4}, it is shown:
\ba
\int d^4 p_2 \delta(p_2^2+M_h^2)
\delta\left[(k_1+p_1-p_2)^2 + \tau \right] \delta \left[ \qh-(p_2-p_1)^2
\right]
=
 \frac{\pi}{2\sqrt{\lambda_S}}.
\label{pi2s}
\ea
Then, with the relation $y_h S = \qh + M_h^2 - M^2$ from~(\ref{eq28})
one gets the identity
$d \qh d M_h^2=S d \qh d\yh$.
This introduces
 the last     of the hadronic variables into the integration measure:
\ba
\Gamma = \frac{\pi^2 S} {4 \sqrt{\lambda_S}}
\int d \yh d \qh \int d\tau \frac{1}{4 \pi} \frac{\tau - m^2}{\tau}
\int {d\cos{\vartheta_R} d{\varphi_R}}.
\label{gmto1}
\ea
The physical region of $(\yh,\qh,\tau)$ is derived in
appendix~\ref{b33}.
\subsection
{A twofold angular integration
\label{hjb3}
}
In hadronic variables, the double differential cross section of the
process~(\ref{eqdeep}) becomes:
\ba
\frac {d^2 \sigma_{\mr R}} { d\yh d\qh }
= \frac{2 \alpha^{3}}{S Q^4_h} \int\! d\tau  \sum_{i=1}^3 A_i(x_h,Q^2_h)
{\cal S}_{i}(\yh,\qh,\tau),
\label{dsigR}
\ea
 where
\ba
 {\cal S }_i  (\yh,\qh,\tau)
 =\frac{1}{4 \pi} \frac{z_2 }{\tau}
 \int {d\cos{\vartheta_R}}    {d{\varphi_R}} \;
 {\cal S}_i(\yh,\qh,\tau,\cos{\vartheta_R},\varphi_R ).
\label{eq313a}
\ea
The functions $ {\cal S}_i(\yh,\qh,\tau,\cos{\vartheta_R},\varphi_R )$
arise from the functions $ {\cal S}_i({\cal E,I},z_1,z_2)$ after the following
substitutions for $\yl,\ql$, and $z_1$ in~(\ref{eqs1z})--(\ref{eqs3z})
in accordance with appendix~\ref{a12}
and with the definitions~(\ref{y1y2}), (\ref{qxy}):
\ba
\ql &=& \qh +z_2 - z_1,
\label{z12}
\\
z_1 &=& \frac{z_2}{2 \tau} \left( \qh + \tau+m^2 - \sqrt{\lambda_1}
\cos \vartheta_R \right),
\label{z1}
\\
1 - \yl &=& \frac{1}{2}(1-\yh) \left(1+\frac{m^2}{\tau}\right)
+ \frac{z_2}{2\tau S} \sqrt{\lambda_{\tau}}
\left( \cos \vartheta_p \cos \vartheta_R
+ \sin \vartheta_p \sin \vartheta_R \sin \varphi_R
\right) ,
\label{yl}
\ea
where
\ba
\cos \vartheta_p = \frac{S(\qh-z_2) - \yh S (\qh+\tau+ m^2)}
{\sqrt{\lambda_{\tau}} \sqrt{\lambda_1}}.
\label{tep}
\ea
After performing the above insertions, the ${\cal S}_{i}$ may be
integrated
over the two photon angles
with the aid of the tables of integrals of appendix~\ref{d1}.
In the ultra-relativistic approximation, the result is:
\ba
{\cal S}_1(\yh,\qh,\tau)
& = & \qh \left[ \frac{1}{z_2} \left(\LQZ -2\right)
 +  \frac{1}{4} \frac{\tau - m^2 }{\tau^2} \right]  +  \frac{1}{4}
+ \frac{1}{2} \left(1-\frac{\qh}{Q^2_\tau}\right) \LQZ,
\label{s123a}
\\
\label{s123b}
{\cal S}_2(\yh,\qh,\tau)
& = & S^2 \Biggl\{ 2(1-\yh) \left[ \frac{1}{z_2} \left( \LQZ - 2 \right)
  + \frac{1}{4}\frac{\tau - m^2}{\tau^2} \right] - \frac{\mbox{$Q^4_{\mr h}$}}
   {\mbox{$Q^6_{\tau}$}} (\LQZ-3)
\\
& &+~\frac{\qh} {\mbox{$Q^4_{\tau}$}} \left[ 1-(1-\yh)(\LQZ -3) \right]
-  \frac{1}{Q^2_\tau}\left[ (2-\yh) \LQZ -\frac{1}{2}
(1-\yh)\left(7-\yh \right)
\right] \Biggr\},
\nll
{\cal S}_3(\yh,\qh,\tau)
& = & S \Biggl\{ 2 \qh  (2-\yh) \left[ \frac{1}{z_2} (\LQZ - 2)
    + \frac{1}{4} \frac{\tau - m^2}{\tau^2} \right]
- \yh \left( \LQZ - \frac{1}{2} \right)
\nll
& & +~\frac{\qh}{Q^2_\tau}  \left[ 2  - \left( \LQZ - 2 \right)
\left( 2-\yh+\frac{2 \qh } { Q^2_\tau }\right)  \right] \Biggr\},
\label{s123x}
\ea
where
\ba
\vph
\LQZ &=& \mbox{ln} {\frac{Q^4_\tau} {m^2 \tau}},
\label{ltau}
\\  \vph
Q^2_\tau &=& \qh + \tau -m^2.
\label{qtau2}
\ea
The infrared singularity is located at $z_2=0$ and has to be treated
appropriately.
\subsection
{The infrared divergence
\label{hjbx}
}
As was done in section~\ref{irres}, the
 following ansatz will be used in order to separate
the infrared divergent parts in~(\ref{dsigR}):
\ba
\frac {d^2 \sigma_{\mr R}} { d\yh d\qh }
&=&
 \frac{2 \alpha^{3}}{S Q_h^4}   \sum_{i=1}^3 A_i(x_h,Q^2_h)
\Biggl\{ {\cal S}_{i}^{\mr B}(\yh,\qh)
\delta_{\mr R}^{\mr {IR}}(\yh,\qh)
\nll & &
+~\int\! d\tau \left[{\cal S}_{i}(\yh,\qh,\tau)-
{\cal S}_{i}^{\mr B}(\yh,\qh){\cal F}^{\mr {IR}}(\yh,\qh,\tau)\right]\Biggr\}
.
\label{had2}
\ea
Here, we define the $\delta_{\mr R}^{\mr {IR}}(\yh,\qh)$
in complete analogy with~(\ref{iral1}) as
the sum of an infrared divergent  soft  and a finite
hard contribution:
\ba
\delta_{\mr {soft}}^{\mr {IR}}(\yh,\qh,\epsilon)
&=&
\frac{2}{\pi} \int \frac{d^3k}{2k^0} {\cal F}^{\mr {IR}}(\qh,z_1,z_2)
\theta(\epsilon-k^0),
\label{irhs}
\\
\delta_{\mr{hard}}^{\mr {IR}}(\yh,\qh,\epsilon)
&=&
\frac{4}{\pi^2}   \frac{\sqrt{\lambda_S}}{S}  \int \frac{d\Gamma}{d{\cal E}}
{\cal F}^{\mr {IR}}(Q^2,z_1,z_2)
\theta(k^0-\epsilon)
\nll
&=&
\int d\tau
\frac{1}{4\pi}\frac{\tau-m^2}{\tau}\int d\cos \vartheta_R d\varphi_R
{\cal F}^{\mr {IR}}(Q^2,z_1,z_2)
\theta(k^0-\epsilon).
\label{iralh}
\ea
The ${\cal F}^{\mr {IR}}(\yh,\qh,\tau)$ will be determined below.
The expression for $\delta_{\mr {soft}}^{\mr {IR}}(\yh,\qh,\epsilon)$
is the equivalent of~(\ref{irsoft})  and
$\delta_{\mr {hard}}^{\mr {IR}}(\yh,\qh,\epsilon)$ follows immediately
from~(\ref{iral1}).
Further,
\ba
{\cal F}^{\mr {IR}}(\qh,z_1,z_2)
&=&
 \frac{\qh}{z_1z_2} - m^2\left(\frac{1}{z_1^2} +
 \frac{1}{z_2^2}\right).
\label{irh3}
\ea
The hard part of the infrared divergent correction will be integrated
first.
The integrals over the photon angles in
$\delta_{\mr {hard}}^{\mr {IR}}(\yh,\qh,\epsilon)$ may be performed with the
aid of~(\ref{z1}) and of the table of integrals in appendix~\ref{d1}:
\ba
\delta_{\mr {hard}}^{\mr {IR}}(\yh,\qh,\epsilon)
&=&
\int d\tau {\cal F}^{\mr {IR}}(\yh,\qh,\tau)\theta(k^0-\epsilon),
\label{irhh}
\\
{\cal F}^{\mr {IR}}(\yh,\qh,\tau)
&=&
\frac{1}{4\pi}\frac{\tau-m^2}{\tau}\int d\cos \vartheta_R d\varphi_R
{\cal F}^{\mr {IR}}(\qh,z_1,z_2)
\nll
&=&
\frac{1}{z_2} \left(\ltau-2\right)-\frac{1}{Q_{\tau}^2}\ltau +
\frac{1}{\tau},
\label{ftauh}
\ea
and $\ltau$
and $Q_{\tau}^2$ are
defined in~(\ref{ltau}) and~(\ref{qtau2}).
 
The boundaries for the integration over $\tau$ may be found in~(\ref{tau2}).
The potentially infrared divergent term in
$\delta_{\mr{hard}}^{\mr {IR}}(\yh,\qh,\epsilon)$
contains a dependence on $1/z_2$ and is
integrated with the aid of the table of integrals~\ref{d2}:
\ba
\int_{{\bar z}_2}^{\tau^m} \frac{d z_2}{z_2}\left(\ltau-2\right)
&=&
 2~(\Lh-1 )\ln\left[\frac{\qh}{2m\epsilon}\left(\frac{1}{\yh}-1\right)
\right]
-\frac{1}{2}\ln^2\left[\frac{\Qh}{m^2}\left(\frac{1}{\yh}-1\right)\right]
\nll
& &
-~2\litwo
\left(1-\frac{1}{\yh}\right)-\litwo(1).
\label{had7}
\ea
With the aid of appendix~\ref{d3}, the explicitly finite part of
$\delta_{\mr{hard}}^{\mr {IR}}(\yh,\qh,\epsilon)$
may finally also be integrated.
We will not quote the result separately, but rather treat this contribution
together with the finite hard cross section in the next section.
 
Now, the infrared divergence in
$\delta_{\mr {soft}}^{\mr {IR}}(\yh,\qh,\epsilon)$
will be treated.
For this purpose, the $R$ system of the foregoing section is used.
As derived in appendix~\ref{a12},
the photon energy becomes:
\bq
k^{0,{R}}= |{\vec k^R}| =
\frac{\tau-m^2}{2\sqrt{\tau}}
=\frac{z_2}{2\sqrt{\tau}} < \epsilon.
\label{k0t}
\eq
Besides $k^R$, we will need the four momenta $k_1^R, k_2^R$ in the
limit $z_2 \to 0$, see ~\ref{3m2}:
\ba
\begin{array}{rclcrclclcl}
k_2^{0,{R}} &\rightarrow&
m, &\hspace{.7cm}&
|{\vec k}_2^{R}|=|{\vec k^R}| \sim
z_2 &\rightarrow& 0, &\hspace{.7cm}&
\beta_2 \rightarrow 0,&&
\\
k_1^{0,{R}} &\rightarrow&
\frac{\displaystyle \qh}{\displaystyle 2m}, &\hspace{.7cm}&
|{\vec k}_1^{R}| &\rightarrow&
\frac{\qh}{\displaystyle 2m}, &\hspace{.7cm}&
1-\beta_1^2 \rightarrow \frac{\displaystyle 4m^4}
{\displaystyle (Q_h^2+2m^2)^2}.&&
\label{isovec}
\end{array}
\ea
Since the final state lepton is practically at rest there,
the following isotropic
relation holds:
\ba
-2kk_2 \rightarrow 2mk^{0,R}.
\label{isotr}
\ea
It is this
property, which simplifies the angular integration.
In~(\ref{d3k}), it is not necessary to introduce the Feynman parameter
which finally lead to the complicated function $S_{\Phi}$ of~(\ref{soft3}).
Instead of~(\ref{dir}), one gets:
\ba
 \delta^{\mr {IR}}_{\mr {soft}}({\cal E},\epsilon)
&\rightarrow&
\frac{2 (2\pi)^2}{(2\sqrt{\pi})^n \Gamma\left(n/2-1\right)}
\frac{1}{\mu^{n-4}}
\int_0^{\epsilon} (k^{0,R})^{n-5} dk^{0,R} \int_0^{\pi}
(\sin \vartheta)^{n-3} d \vartheta_1
\nll
& &
\times~\left[
 \frac{Q^2+2m^2}{
 m k_1^{0,R}(1-\beta_1\cos\vartheta_1)}
- \frac{m^2}{
(k_1^{0,R})^2(1-\beta_1\cos\vartheta_1)^2
} - 1
                    \right],
\nll
\label{dir7}
\ea
 where one has to use $ Q^2 \rightarrow \qh$.
Instead of~(\ref{*}) and~(\ref{*1}), it is now
\ba
\delta^{\mr {IR}}_{\mr {soft}}({\cal E},\epsilon)
&=&
  \left[{\cal P}^{\mr {IR}} +\ln \frac{2\epsilon}{\mu} \right]
  \frac{1}{2}
  \int_{-1}^{1} d\xi \;
{\cal F}(\xi)
+ \frac{1}{4}
  \int_{-1}^{1} d\xi \; \mbox{ln}(1-\xi^2) \;
{\cal F}(\xi),
\label{*7}
\ea
with
\ba
{\cal F}(\xi)  =
 \frac {2(Q^2_h+2m^2)}{Q^2_h} \frac{1}
{1-\beta_1 \xi}
 - \frac {4m^4}{Q^4_h} \frac{1}{(1-\beta_{1}\xi)^2}
 - 1.
\label{*17}
\ea
 
The integrations in~(\ref{*7}) may be performed with the aid of
appendix~\ref{c2}:
\ba
\delta_{\mr {soft}}^{\mr {IR}}(\yh,\qh,\epsilon)
=
2\left[{\cal P}^{\mr {IR}}+\ln\frac{2\epsilon}{\mu}\right]\left(\Lh-1\right)
-
{{\mr L}^2_{\mathrm h}}
+\Lh-\litwo(1)+1,
\label{had5}
\ea
where ${\cal P}^{\mr {IR}}$ is introduced in~(\ref{eqn414}), and
\ba
 {\mr L}_{\mr h}=\mbox{ ln}\frac{\Qh}{m^2}.
\label{hadl}
\ea
The cut-off parameter $\epsilon$
may be seen to be cancelled against the corresponding terms in~(\ref{had7}).
 
In order to get an infrared finite and well-defined cross section
the photonic vertex correction  $\delta_{\mr {vert}}$ has to be added
to the radiative cross section~(\ref{had2}).
It is given by~(\ref{*20}), again with the replacement $Q^2
\rightarrow \qh$.
In the sum, the infrared divergence cancels:
\bq
\delta_{\mr {soft}}^{\mr {IR}}(\yh,\qh,\epsilon)
+
\delta_{\mr {vert}}(\qh)=
 2\left(\Lh-1\right)\ln\frac{2\epsilon}{m}-\frac{3}{2}
{{\mr L}^2_{\mathrm h}}
+\frac{5}{2}\Lh-1.
\label{had6}
\eq
\subsection
{A third analytical integration
\label{irIIh}}
In~(\ref{had2}), the last remaining integral is that over $\tau$ or,
equivalently, over $z_2$.
Since the structure functions depend on hadronic variables  which are here at
the same time the external variables  the last integral may be performed
analytically.
For this purpose one has to integrate the functions
${\cal S}_{i}(\yh,\qh,\tau)$ as given by~(\ref{s123a})--(\ref{s123x}).
The cross section may be reordered such that the infrared singular terms
in the finite hard cross section explicitly compensate each other:
\ba
\frac {d^2 \sigma_{\mr R}} { d\yh d\qh }
&=&
\frac {d^2 \sigma_{\mr B}} { d\yh d\qh }
\frac{\alpha}{\pi} \left[ \delta_{\mr {soft}}^{\mr {IR}}(\yh,\qh,\epsilon)
+ \int_{{\bar z}_2}^{\tau^m} \frac{d z_2}{z_2}\left(\ltau-2\right)\right]
\nll
& &~+
\frac{2 \alpha^{3}}{S Q_h^4}\sum_{i=1}^3 A_i(x_h,Q^2_h)\int_{m^2}^{\tau^m}
d\tau\left[{\cal S}_{i}(\yh,\qh,\tau)-
{\cal S}_{i}^{\mr B}(\yh,\qh)\frac{1}{z_2}\left(\ltau-2\right)\right].
\nll
\label{had4}
\ea
The resulting cross section is rather compact  and the last integrations over
$\tau$ may be easily performed with the aid of appendix~\ref{d3}.
 
The net cross section is:
\ba
\frac{d^2 \sigma}{d \yh d \qh}
=
\frac{d^2 \sigma_{\mr B}}{d \yh d \qh }
\exp \left[ \frac{\alpha}{\pi} \delta_{\mr {inf}} (\yh,\qh) \right]
+ \frac{2 \alpha^3}{S} \sum_{i=1}^3 \frac{1}{Q_h^4}
{\cal A}_i(\xh,\qh) {\cal S}_i(\yh,\qh).
\label{d2sh}
\ea
The $\delta_{\mr {inf}}(\yh,\qh)$ and the ${\cal S}_i(\yh,\qh)$
are defined in~(\ref{dinwh}) and~(\ref{w3h}).
 
We would like to remark that these purely analytical
formulae for the leptonic QED corrections in hadronic variables
to neutral current deep inelastic scattering are an extremely
esthetic result.
They are published here for the first time.
Numerically, the corrections agree completely with those derived
within the approach of section ~\ref{fhad}.
 
As was mentioned above, the corrections vanish at $y \rightarrow 0$.
That this is really the case may be seen by inspecting the analytical
expressions.
%
%
\subsection
{Jaquet-Blondel variables
\label{jb2}
}
From hadronic variables, one may easily change
to Jaquet-Blondel variables~(\ref{xjb})
in the phase space~(\ref{gmto1}).
The expression for \qjb\ in terms of hadronic variables is derived in
appendix~\ref{appjb}:
\ba
\yjb &=& \yh,
\label{jbin1}
\\
\qjb &=& \qh - \frac{\yh}{1-\yh}(\tau-m^2).
\label{jbinh}
\ea
The physical boundaries in the phase space integral~(\ref{gmto1}),
\ba
\Gamma = \frac{\pi^2} {4} \int d\yjb \, d\qjb  \int d\tau
\frac{1}{4 \pi}
\frac{\tau - m^2}{\tau} \int {d\cos{\vartheta_R} \, d{\varphi_R}},
\label{gmto3}
\ea
are also derived there.
From the above relations, it may be seen that the Jaquet-Blondel variables
and the hadronic variables are related in a one-to-one correspondence,
with $\tau$ being a parameter.
 
In Jaquet-Blondel variables, the double differential cross section of
process~(\ref{eqdeep}) reads:
\ba
\frac{ d^2 \sigma_{\mr R}} {d \yjb d \qjb}
= \frac{ 2 \alpha^3}{S} \int d \tau \sum_{i=1}^3 {\cal A}_i(\xh,\qh)
  \frac{1}{Q_h^4} {\cal S}_i(\yjb,\qjb,\tau).
\label{eq4.x}
\ea
The ${\cal S}_i(\yjb,\qjb,\tau)$
may be trivially obtained from~(\ref{s123a})--(\ref{s123x}) with the relations
\ba
\yh &=& \yjb,
\label{jbhhh}
\\
\qh &=& \qjb + \frac{\yjb}{1-\yjb} (\tau-m^2).
\label{jbhh}
\ea
The explicit expressions are~\cite{jbpl}\footnote
{
The equation~(\ref{S3}) corrects the equation (17) in~\cite{jbpl}.
The error was
created in the manuscript. Fortran program and numerical
results were not influenced by this.
}:   
\ba
{\cal S}_1(\qjb,\yjb,\tau) & = &
    \qjb
    \left[ \frac{1}{z_2} \left(\LQZ -2\right)
 +  \frac{z_2 }{4 \tau^2} \right]
 +  \frac{ 1 - 8 \yjb }{4(1-\yjb)}  
 +  \LQZ \left( \frac{z_2}{2 \Qt }+\frac{\yjb}{1-\yjb}\right),
\label{S1}
\\
{\cal S}_2(\qjb,\yjb,\tau) & = &
      S^2 \Biggl\{  2(1-\yjb)
    \left[ \frac{1}{z_2} \left( \LQZ - 2 \right)
  + \frac{z_2}{4\tau^2} \right]   
  - \frac{1}{\Qt} \left[ 2( \LQZ -2)
  +  \frac{1}{2}(1-{\yjb}^2) \right] \nll
 & &+~\frac{\qjb} {\mbox{$Q^4_{\mr  \tau} $}} (1-\yjb)
    \left[ 1 -  (1+\yjb)(\LQZ -3) \right]
  - \frac{\mbox{$Q^4_{\mr {JB}}$}}
         {\mbox{$Q^6_{\mr \tau}$}} (1-\yjb)^2 (\LQZ-3)
                     \Biggr\},
\nll
\label{S2}
\\
{\cal S}_3(\qjb,\yjb,\tau) & = &
      S \Biggl\{ 2 \qjb (2-\yjb)
      \left[ \frac{1}{z_2} (\LQZ - 2)
    + \frac{z_2}{4\tau^2} \right]             
 +  \frac{\yjb( 1 + \yjb^2 )} { 1-\yjb } \LQZ +5   \nll
& &-~ \frac{ 7 \yjb }{2(1-\yjb)}-(1-\yjb)(5+2\yjb)  \nll
& &+~ \frac{\qjb}{\Qt} (1-\yjb) \left[(1-\yjb)
                              \left( 3-\frac{2 \qjb }{ \Qt }\right)
  (\LQZ - 2)
  + 12-5 \LQZ \right]
                     \Biggr\}.
\label{S3}
\ea
 
The integral over
$\tau$ has to be performed numerically as long as one cannot neglect
the difference between $\qjb$ and $Q_h^2$ --    the structure functions
are depending on $Q_h^2$ and thus on $\tau$.
 
The integrand in~(\ref{eq4.x}) is infrared singular at $\tau=m^2$, which
corresponds to $z_2=0$.
The treatment of the infrared singularity may be performed in full analogy with
the foregoing section.
In the infrared singular point, $z_2 \rightarrow 0$, the hadronic and the
Jaquet-Blondel variables agree and the
$\delta_{\mr {soft}}^{\mr {IR}}$ in these variables do so:
\ba
\delta_{\mr {soft}}^{\mr {IR}}(\yjb,\qjb,\epsilon)
&=&
\frac{1}{\pi} \int \frac{d^3k}{k^0} {\cal F}^{\mr {IR}}(\qjb,z_1,z_2)
\theta(\epsilon-k^0)
\nll
&=&
2\left[{\cal P}^{\mr {IR}}+\ln\frac{2\epsilon}{\mu}\right]
\left(\ljb-1\right) -
{{\mr L}^2_{\mathrm{JB}}}
+\ljb-\litwo(1)+1.
\label{softj}
\ea
 
The
$\delta_{\mr {hard}}^{\mr {IR}}(\yjb,\qjb,\epsilon)$ differs slightly from the
hadronic correction:
\ba
\delta_{\mr{hard}}^{\mr {IR}}(\yjb,\qjb,\epsilon)
&=&
\frac{4}{\pi^2}   \frac{\sqrt{\lambda_S}}{S}  \int \frac{d\Gamma}{d{\cal E}}
{\cal F}^{\mr {IR}}(\qjb,z_1,z_2)
\theta(k^0-\epsilon)
\nll
&=&
\int d\tau
\frac{1}{4\pi}\frac{\tau-m^2}{\tau}\int d\cos \vartheta_R d\varphi_R
{\cal F}^{\mr {IR}}(\qjb,z_1,z_2)
\theta(k^0-\epsilon).
\label{iralha}
\ea
In fact, it is
\ba
{\cal F}^{\mr {IR}}(\qjb,z_1,z_2)
&=&
 \frac{\qjb}{z_1z_2} - m^2\left(\frac{1}{z_1^2} +
 \frac{1}{z_2^2}\right)
\nll
&=&
{\cal F}^{\mr {IR}}(\qh,z_1,z_2) - \frac{\yjb}{1-\yjb} \frac{1}{z_1}.
\label{irh3a}
\ea
Using the results of the calculation of the hadronic correction,  it is
sufficient to explicitly integrate
the difference term in~(\ref{irh3a})
 over $\varphi_R$ and $\cos\vartheta_R$
with
the aid of appendix~\ref{d1}. Then,
\ba
\delta_{\mr {hard}}^{\mr {IR}}(\yjb,\qjb,\epsilon)
&=&
\int d\tau {\cal F}^{\mr {IR}}(\yjb,\qjb,\tau)\theta(k^0-\epsilon),
\label{irjbh}
\ea
where
\ba
{\cal F}^{\mr {IR}}(\yjb,\qjb,\tau)
&=&
\frac{1}{4\pi}\frac{\tau-m^2}{\tau}\int d\cos \vartheta_R d\varphi_R
{\cal F}^{\mr {IR}}(\qjb,z_1,z_2)
\nll
&=&
\frac{1}{z_2} \left(\ltau-2\right)-\frac{1}{(1-\yjb)}
\frac{\ltau}{Q_{\tau}^2} + \frac{1}{\tau}.
\label{ftau}
\ea
The $\ltau$ and $Q_{\tau}^2$ are defined in~(\ref{ltau}) and~(\ref{qtau2}),
respectively;
furthemore the substitutions  ~(\ref{jbhhh}) and ~(\ref{jbhh}) have to be used.
The integral over $\tau$ may be performed with appendix~\ref{d2}
for the potentially infrared singular terms and
with appendix~\ref{d3} for the others; but now within the
boundaries~(\ref{boutau}).
 
The result is:
\ba
\delta_{\mr{ {hard}}}^{\mr {IR}}(\yjb,\qjb,\epsilon)
&=&
2\left(\ljb-1\right)\ln\frac{\tau^{\max}}{2m\epsilon}-\frac{1}{2}\ln^2
 \frac{\tau^{\max}}{m^2}+\ln\frac{\tau^{\max}}{m^2}
\nll & &+~\ln X_{\mr {JB}}\left[
\ln(1-\yjb)-\ln X_{\mr {JB}}-\ljb\right]-\litwo(X_{\mr {JB}})-
2\litwo(1-X_{\mr {JB}}),
\nll
\label{hardj}
\ea
where
\ba
\ljb &=& \ln\frac{\qjb}{m^2},
\\ X_{\mr {JB}} &=& \frac{1}{\xjb\yjb}\left[1-\xjb(1-\yjb)\right],
\\ \tau^{\max} &=& (1-\xjb)(1-\yjb) S.
\label{lxt}
\ea
 
The photonic vertex correction  $\delta_{\mr {vert}}$ is given
by~(\ref{*20}), but with the replacement $Q^2 \rightarrow \qjb$.
 
The net factorized part is
\ba
\delta_{\mr{VR}}(\yjb,\qjb)
&=&
\delta_{\mr {vert}}
+
\delta_{\mr{ {soft}}}^{\mr {IR}}(\yjb,\qjb,\epsilon)
+
\delta_{\mr{ {hard}}}^{\mr {IR}}(\yjb,\qjb,\epsilon)
\nll \vspace{0.3cm}
&=&  \delta^{\mr{inf}}(\qjb,\yjb)
-\mbox{ln}^2X_{\mr{JB}}
+\mbox{ln}X_{\mr{JB}} \left[\mbox{ln}(1-\yjb)-1\right]
\nll  \vspace{0.3cm}
& &-~\frac{1}{2}
 \mbox{ln}^2 \left[ \frac{(1 - \xjb)(1 - \yjb)}{\xjb\yjb} \right]
\nll
& &+~ \frac{3}{2}\ljb
-   {\mr{ Li}}_2 (X_{\mr{JB}})
- 2 {\mr{ Li}}_2 (1-{X_{\mr{JB}}}) - 1 ,
\label{dvrj}
\ea
where
\bq
\delta_{\mr {inf}}(\yjb,\qjb)=\ln\frac{(1-\xjb)(1-\yjb)}{1-\xjb(1-\yjb)}
\left(\ln\frac{\qjb}{m^2}-1\right).
\label{dinfj}
\eq
 
The finite part of the radiative cross section in Jaquet-Blondel variables is
\ba
\frac {d^2 \sigma_{\mr R}^{\mr F}}{d\yjb d\qjb}
&=& \frac {2\alpha^3}{S}\int d\tau\sum_{i=1}^{3} \Biggl[ {\cal A}_{i}(\xh,\qh)
\frac{1}{Q^4_h} {\cal S}_{i}(\yjb,\qjb,\tau)
\nll
& &-~ {\cal A}_{i}(\xjb,\qjb)\frac{1}{Q_{\mr {JB}}^4} \;
     {\cal S}_{i}^{B}(\yjb,\qjb)\;
     {\cal F}^{\mr {IR}}(\yjb,\qjb,\tau) \Biggr].
\label{fjb}
\ea
 
The net cross section is:
\ba
\frac {d^2 \sigma_{\mr R}}{d\yjb d\qjb}
&=&
\frac{d^{2} {\sigma}_{\mr B}}    {d\yjb d\qjb}
\left\{
{\mr {exp}} \left[\frac{\alpha}{\pi}
 \delta_{\mr {inf}}(\yjb,\qjb) \right]
 -1+\frac{\alpha}{\pi}\Bigl[
 \delta_{\mr {VR}} (\yjb,\qjb)
-\delta_{\mr {inf}}(\yjb,\qjb)\Bigr]
\right\}
\nll
& &+~
\frac{d^{2}{\sigma}_{\mr R}^{\mr F}}{d\yjb d\qjb}.
\ea
\subsubsection
{Discussion
\label{disj}
}
\begin{figure}[bhtp]
\begin{center}
\mbox{
\epsfysize=9.cm
\epsffile[0 0 530 530]{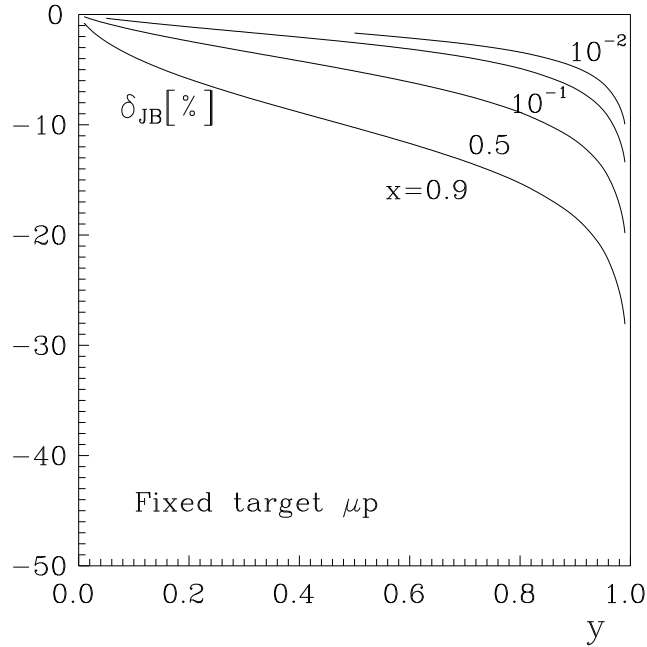}
}
\end{center}
\caption{\it
Radiative cross section for fixed target deep inelastic neutral current
scattering in terms of Jaquet-Blondel variables.
}
\label{jbcs1}
\end{figure}

\begin{figure}[tbhp]
\begin{center}
\mbox{
\epsfysize=9.cm
\epsffile[0 0 530 530]{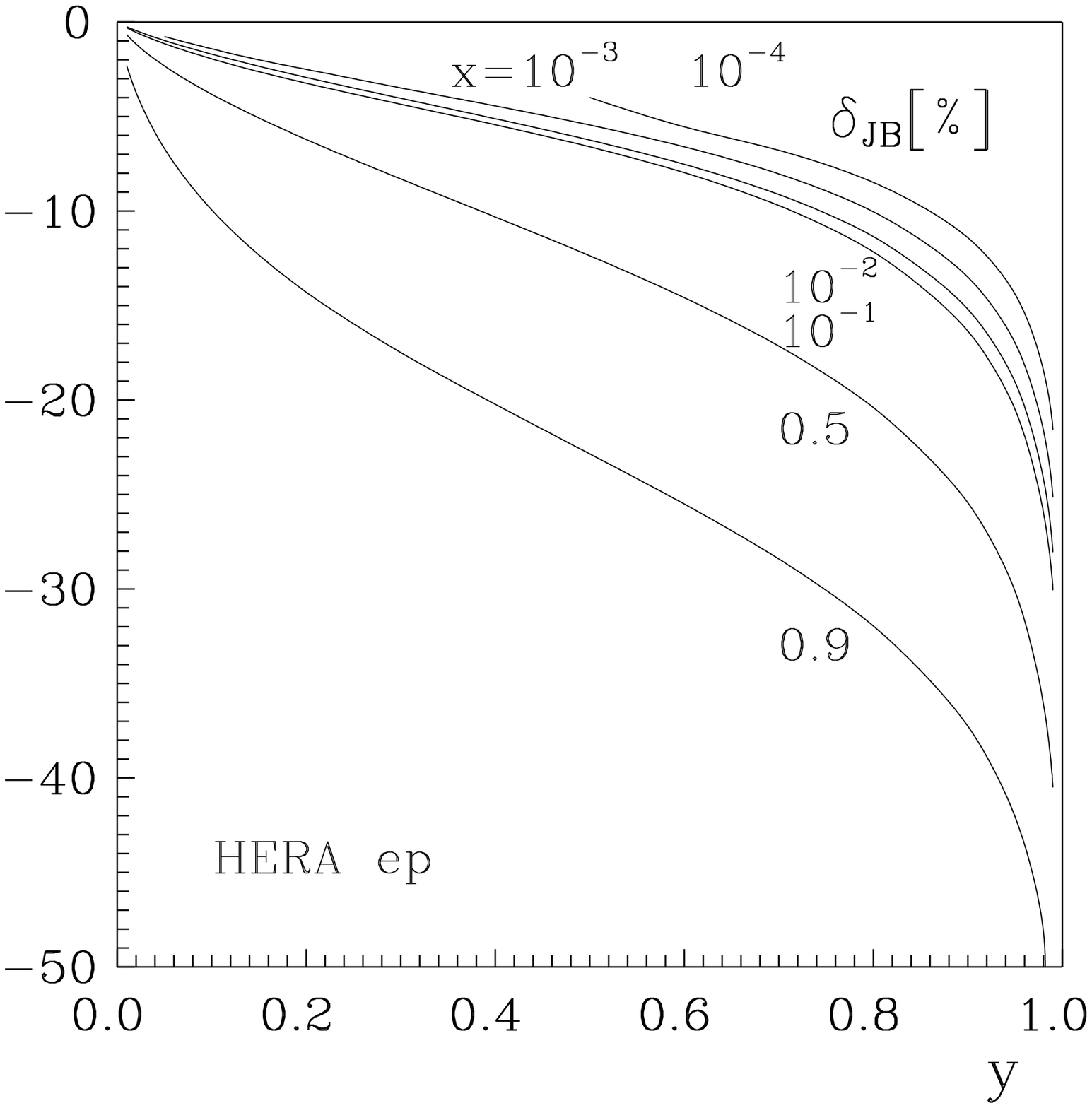}
}
\end{center}
\caption{\it
Radiative cross section for deep inelastic neutral current scattering
at HERA
in terms of Jaquet-Blondel variables.
}
\label{jbcs2}
\end{figure}

\begin{figure}[tbhp]
\begin{center}
\mbox{
\epsfysize=9.cm
\epsffile[0 0 530 530]{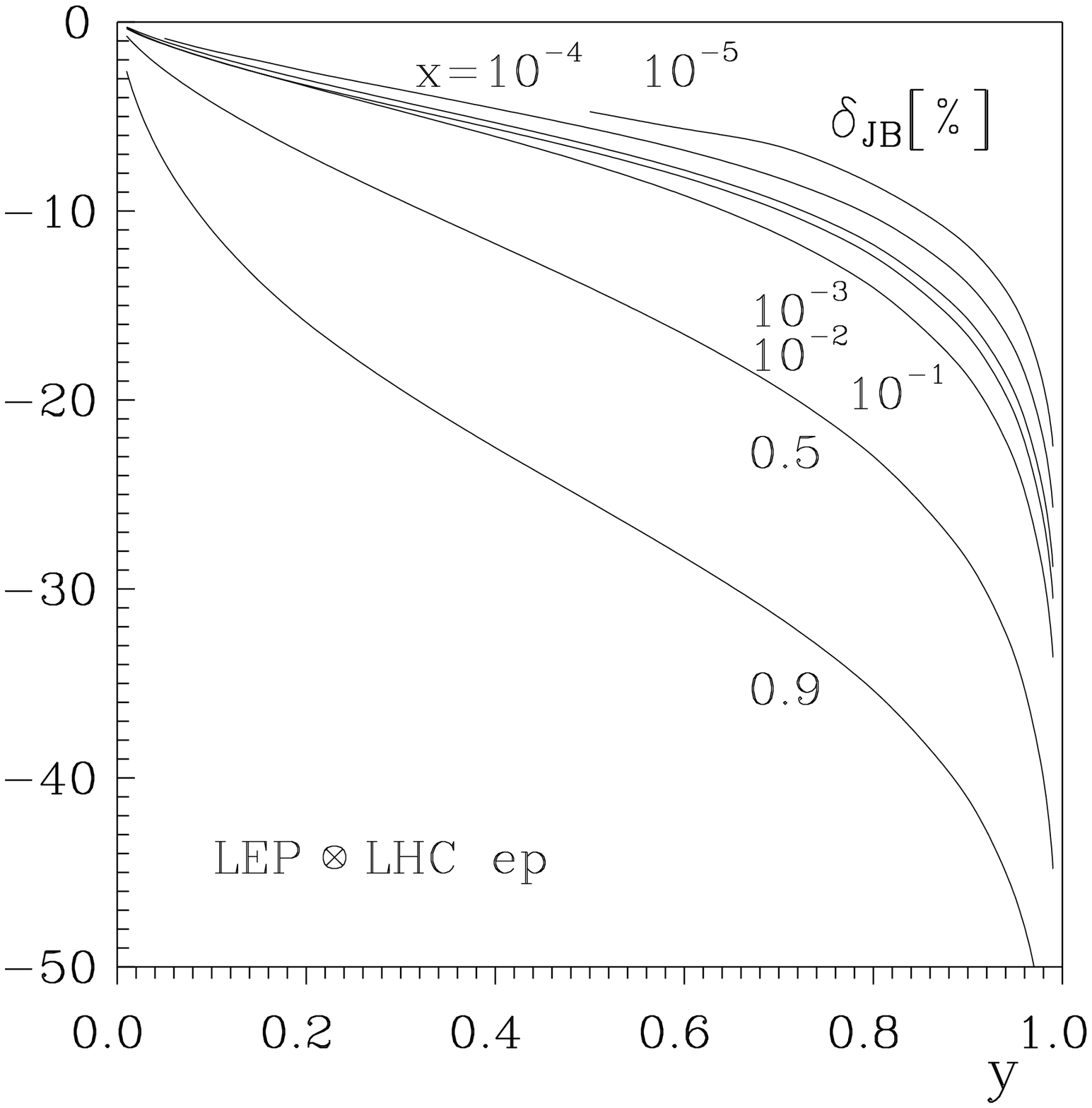}
}
\end{center}
\caption{\it
Radiative cross section for deep inelastic neutral current scattering
at LEP$\otimes$LHC
in terms of Jaquet-Blondel variables.
}
\label{jbcs3}
\end{figure}

The leptonic QED corrections in Jaquet-Blondel variables are shown in figures~
\ref{jbcs1}--\ref{jbcs3}.
They behave quite similar to the corrections in hadronic variables.
At small $y$, this follows from the fact that the variables become equal
there. Thus, the vanishing of the corrections at $\yh \rightarrow 0$
is understood from the hadronic case.
As was discussed with the mixed and hadronic variables, the soft photon
corner is at \yjb\  approaching 1.
From~(\ref{dinfj}) one may see that the dominant soft photon corrections
are negative and the more pronounced, the larger the \xjb. This makes some
difference to the hadronic variables.
Again as in all the previous cases, the corrections for fixed target
scattering
 are different
from the collider cases due to the smaller value of $S$ and to the
difference of the muon and     electron
masses.
The collider results are quite similar to each other.
\section
{Discussion
 \label{disc}}
\ezero
In the foregoing sections, we discussed some of the basic features of
the leptonic QED corrections to the neutral current deep inelastic
$ep$ scattering in terms of different sets of kinematical variables.
 
Basic features of
the magnitudes of the corrections and their behaviour as functions of the
kinematics have been discussed there.
 
\vfill
 
\newpage
 
In this section, we discuss some additional features of the corrections:
\begin{itemize}
\item
the dependence of the corrections on the structure functions;
\item
comparison with the leading logarithmic approximation;
\item
photonic cuts in leptonic variables;
\item
the Compton peak;
\item
other radiative corrections.
\end{itemize}
 
\subsection{The influence of structure functions on the QED corrections
\label{si}
}
\begin{figure}[tbhp]
\begin{center}
\mbox{
\epsfysize=9.cm
\epsffile[0 0 530 530]{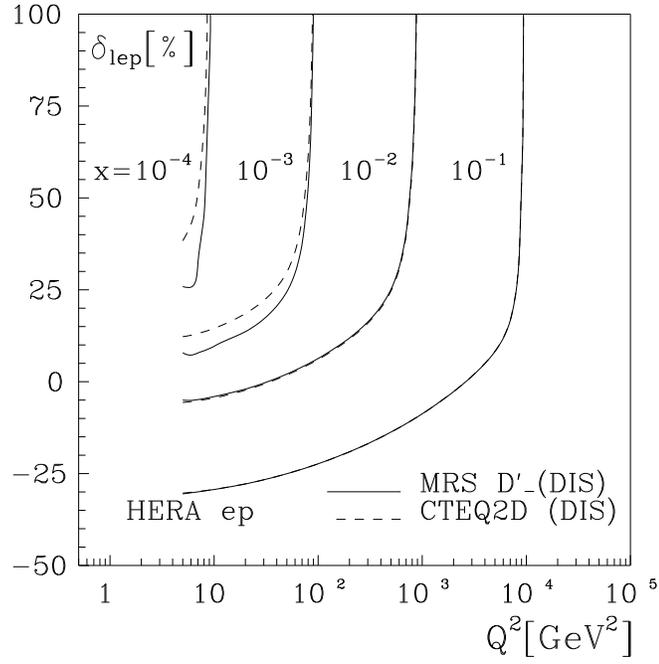}  
}
\end{center}
\caption{\it
Comparison of the radiative corrections $\delta_{\mr{lep}}(x,Q^2)$
at HERA
in terms of two different sets of structure functions.
\label{fdis1}
}
\end{figure}

\begin{figure}[tbhp]
\begin{center}
\mbox{
\epsfysize=9.cm
\epsffile[0 0 530 530]{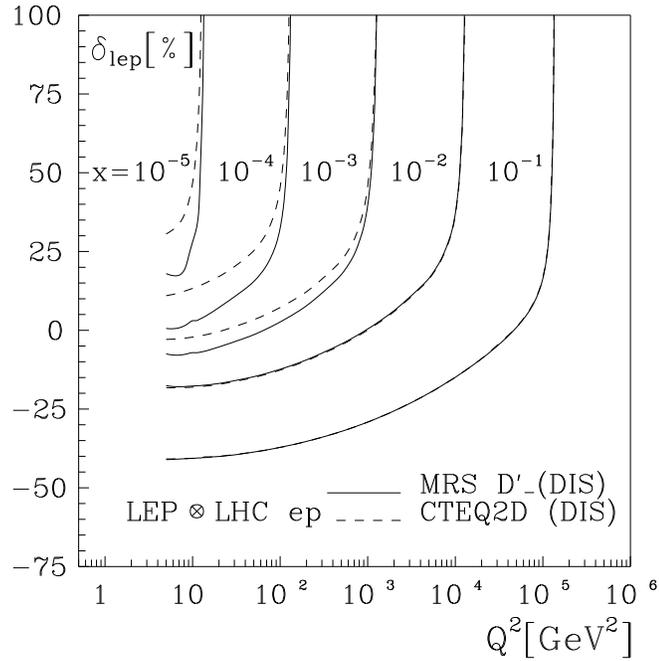}    
}
\end{center}
\caption{\it
Comparison of the radiative corrections $\delta_{\mr{lep}}(x,Q^2)$
at LEP$\otimes$LHC
 in terms of two different sets of structure functions.
\label{fdis1a}
}
\end{figure}

In figures~\ref{fdis1} and~\ref{fdis1a}, the radiative corrections
in leptonic variables at HERA and LEP$\otimes$LHC are parameterized
with two different sets
of structure functions~\cite{lib1,pdflib1,lib3,pdflib3}.
Both sets were derived before HERA started operation.
If there are deviations between predictions with different structure
functions, then they are due to the contribution from hard photon
emission.
For soft photons, the correction factorizes and the structure
function dependence drops out in the corrections, when expressed in form
of the ratios $\delta$.
In experimentally unexplored regions  the predicted structure functions
rely heavily on the predictive power of perturbative QCD.
Especially for small values of $x$ problems may arise.
These are reflected in rising deviations of corrections with different
structure functions when $x$ becomes smaller. While, at intermediate and
larger values of $x$, the dependence of the corrections on the choice of
structure functions becomes minor.
For this reason we changed the presentation of the corrections and chose
the set of variables $Q^2, x$, thereby losing sensitivity to $y$.

\begin{figure}[tbhp]
\begin{center}
\mbox{
\epsfysize=9.cm
\epsffile[0 0 530 530]{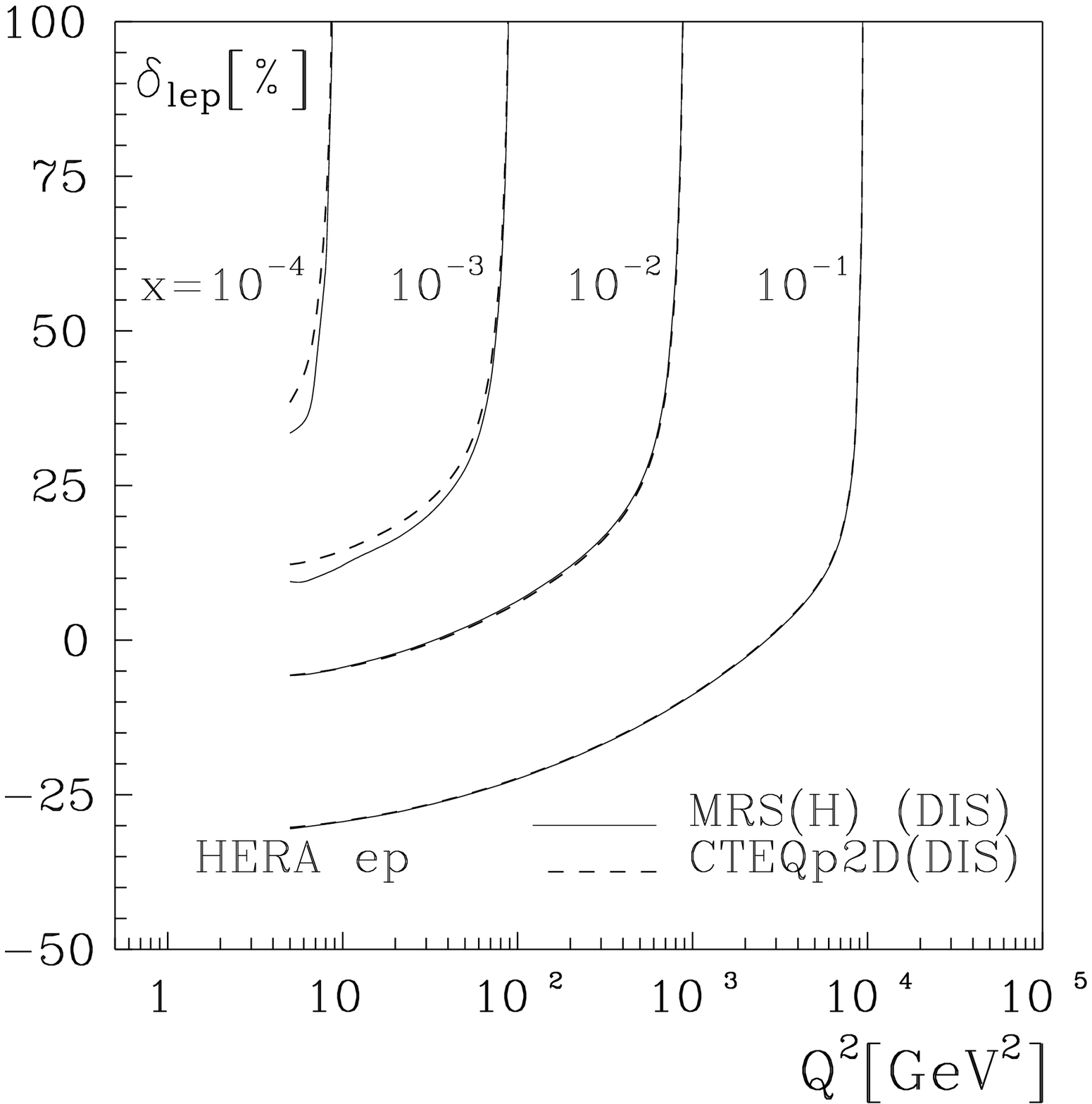}   
}
\end{center}
\caption{\it
Comparison of the radiative corrections $\delta_{\mr{lep}}(x,Q^2)$
at HERA in terms of two different sets of HERA-improved structure
functions.
\label{fdis11}
}
\end{figure}

\begin{figure}[tbhp]
\begin{center}
\mbox{
\epsfysize=9.cm
\epsffile[0 0 530 530]{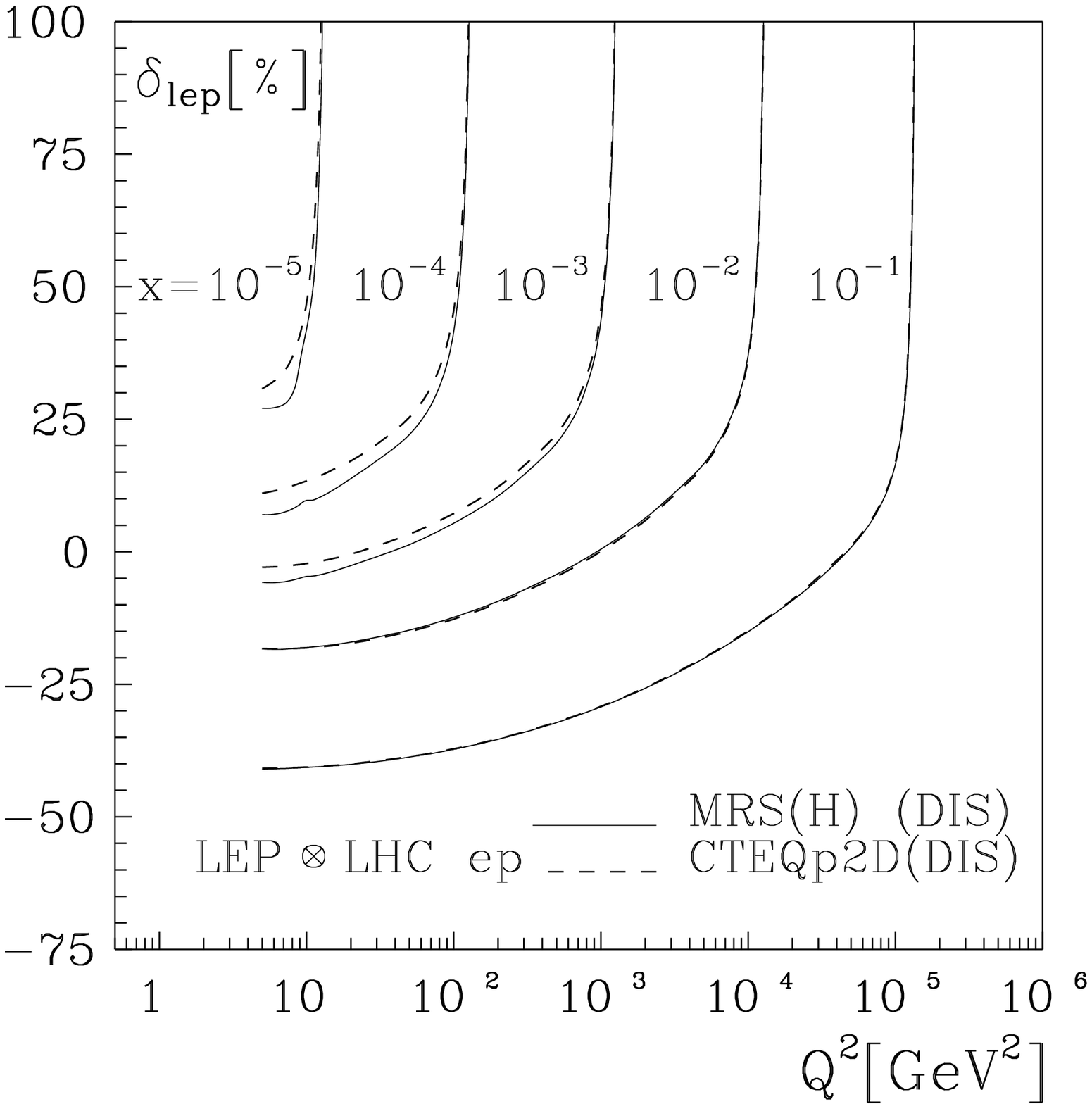}    
}
\end{center}
\caption{\it
Comparison of the radiative corrections $\delta_{\mr{lep}}(x,Q^2)$
at LEP$\otimes$LHC in terms of two different sets of HERA-improved
structure functions.
\label{fdis11a}
}
\end{figure}

Figures~\ref{fdis11} and~\ref{fdis11a} show the same corrections as
figures~\ref{fdis1} and~\ref{fdis1a}, respectively. But now with
structure functions
~\cite{lib4,pdflib4,lib3,pdflib5}
whose derivations took into account HERA data at
low $x$. The disagreement of the predictions at small $x$ remains
but    is much smaller.
 
\begin{figure}[tbhp]
\begin{center}
\mbox{
\epsfysize=9.cm
 \epsffile[0 0 530 530]{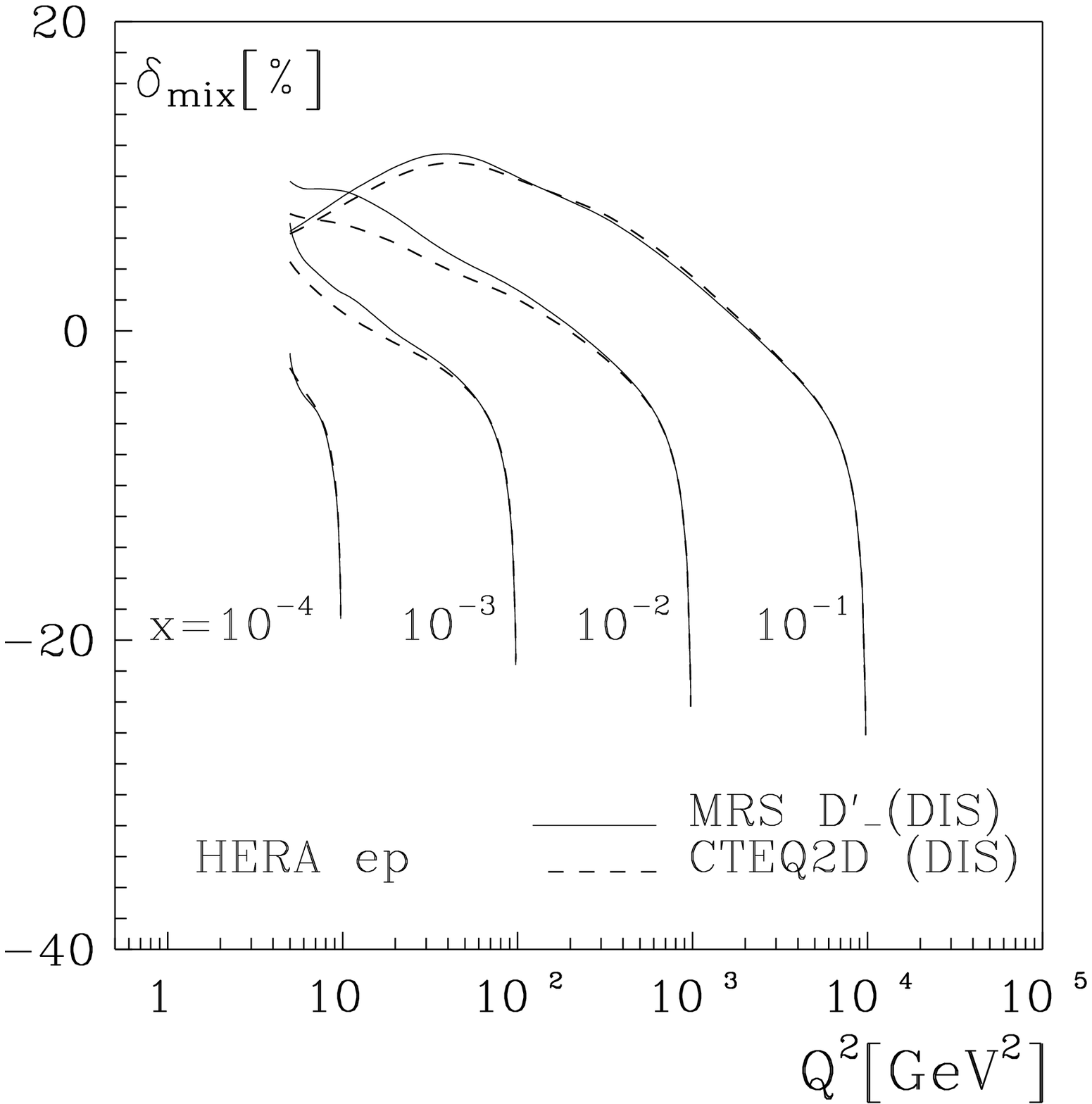}
}
\end{center}
\caption{\it
Comparison of the radiative corrections $\delta_{\mr{mix}}(x,Q^2)$
at HERA in terms of two different sets of structure functions.
\label{fdism}
}
\end{figure}

\begin{figure}[tbhp]
\begin{center}
\mbox{
\epsfysize=9.cm
\epsffile[0 0 530 530]{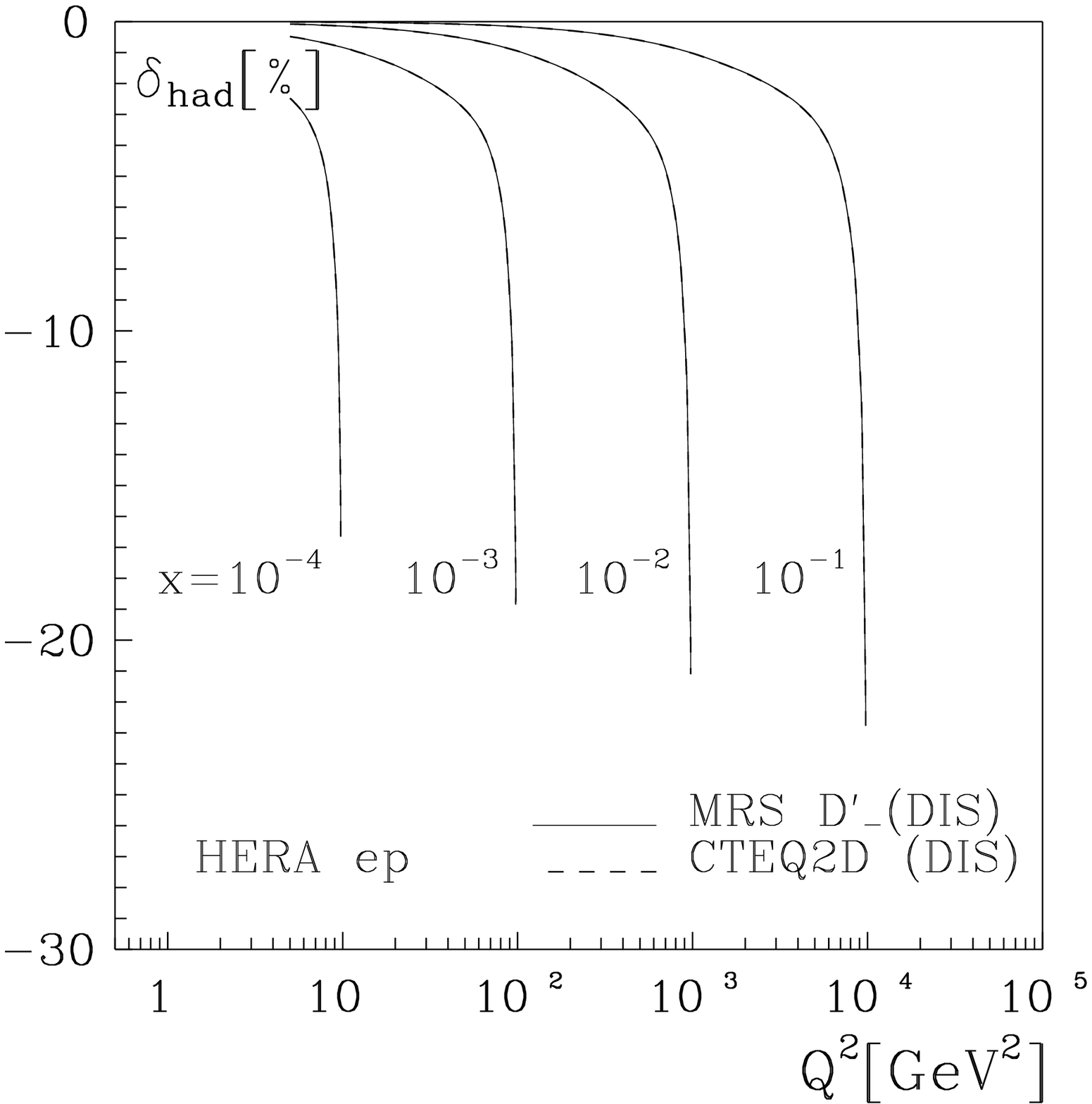}
}
\end{center}
\caption{\it
Radiative corrections $\delta_{\mr{had}}(x,Q^2)$
at HERA
in terms of two different sets of structure functions.
\label{fdis3}
}
\end{figure}

\begin{figure}[tbhp]
\begin{center}
\mbox{
\epsfysize=9.cm
\epsffile[0 0 530 530]{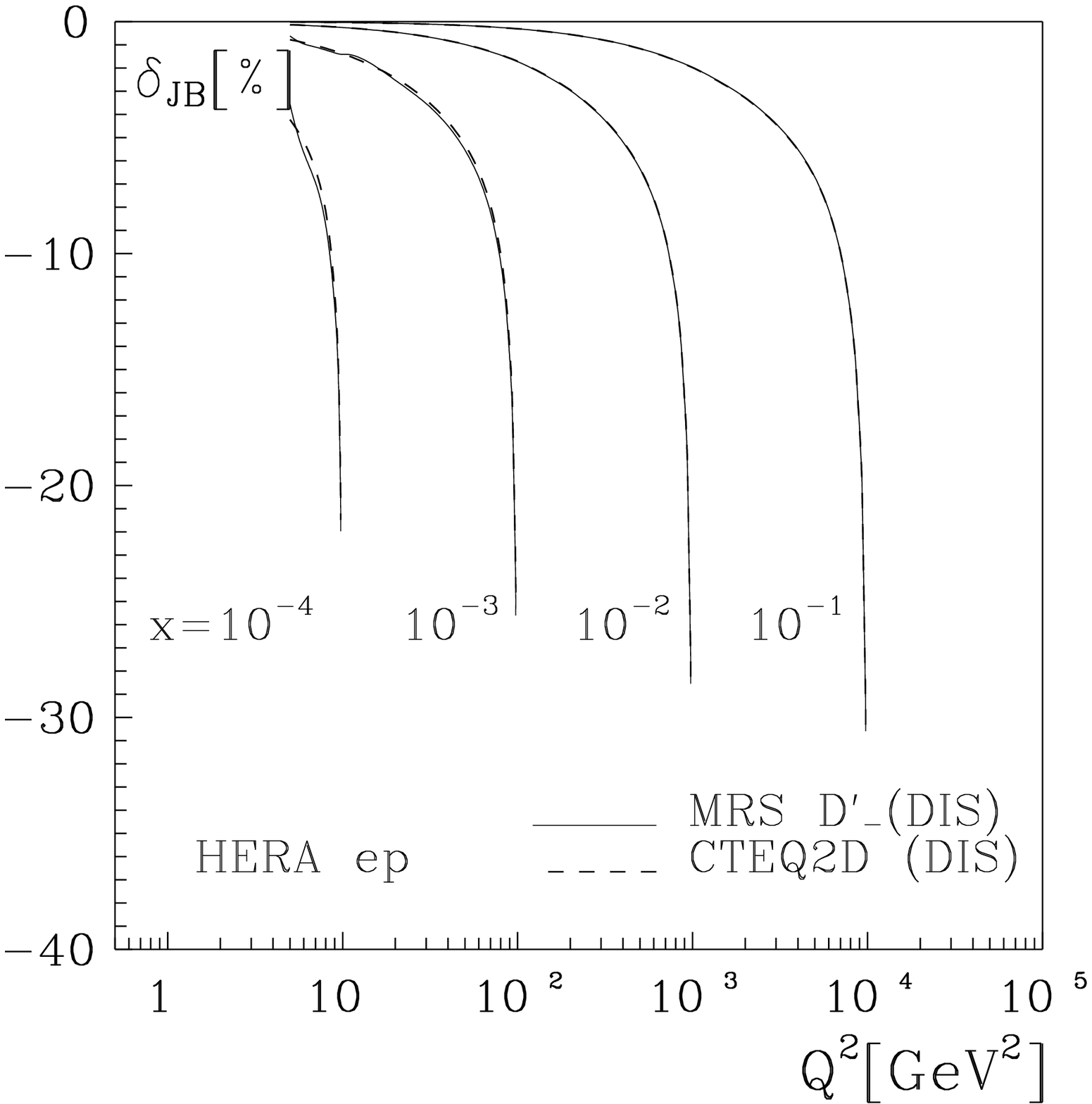}
}
\end{center}
\caption{\it
Radiative corrections $\delta_{\mr{JB}}(x,Q^2)$
at HERA
in terms of two different sets of structure functions.
\label{fdis4}
}
\end{figure}

A similar comparison for the other sets of variables is
shown in figures~\ref{fdism}--\ref{fdis4}.
The
structure functions~\cite{lib1,pdflib1,lib3,pdflib3} which were not
yet improved with the HERA data are used.
The dependence of the corrections in mixed variables
on the choice of structure functions
is  quite substantial
but less pronounced than in the corresponding
 case in leptonic variables.
For the hadronic and Jaquet-Blondel variables the influence
is reasonably small.
This follows from the smaller contribution
of hard photon emission.
The net corrections are negative.
This proves the dominance of the factorizing
soft and virtual photonic contributions. Their
dependence on the structure functions cancels when $\delta$ is
calculated.
 
The dependence of the QED corrections on the structure functions
is not only influenced by different estimates of the QCD predictions
for the parton distribution evolutions.
The corrections show additional dependences on
the low-$Q^2$ behaviour of the structure functions,
the choice of $Q^2$ in the argument of the parton distributions,
soft photon exponentiation, and
the running of the QED coupling constant.
 
We use the corrections in mixed variables for an illustration of this.
The Fortran program {\tt TERAD91}~\cite{TERAD91}
allows us to switch on and off the
above indicated dependencies
 of the cross sections with the following flags:
\begin{itemize}
\item[(i)]
For {\tt IVAR}=0, the low-$Q^2$ behaviour of the structure functions remains
untouched; for {\tt IVAR}=2, the modification~(\ref{prokh}) is applied.
\item[(ii)]
For {\tt ITERAD}=0, the structure functions are assumed to depend on \ql.
A correct choice is {\tt ITERAD}=1, which makes the structure functions
dependent on \qh.
\item[(iii)]
With the settings {\tt IEXP}=0,1 the soft photon exponentiation
as it is introduced in ~(\ref{dvr}) is switched off
or on.
\item[(iv)]
For {\tt IVPOL}=0,1 the running of the QED coupling constant with the $Q^2$
of the t-channel          momentum flow is switched off or on.
\end{itemize}
In the figures, the flag settings are quoted as follows:
({\tt IVAR,ITERAD,IEXP,IVPOL}).

\begin{figure}[tbhp]
\begin{center}
\mbox{
\epsfysize=9.cm
\epsffile[0 0 530 530]{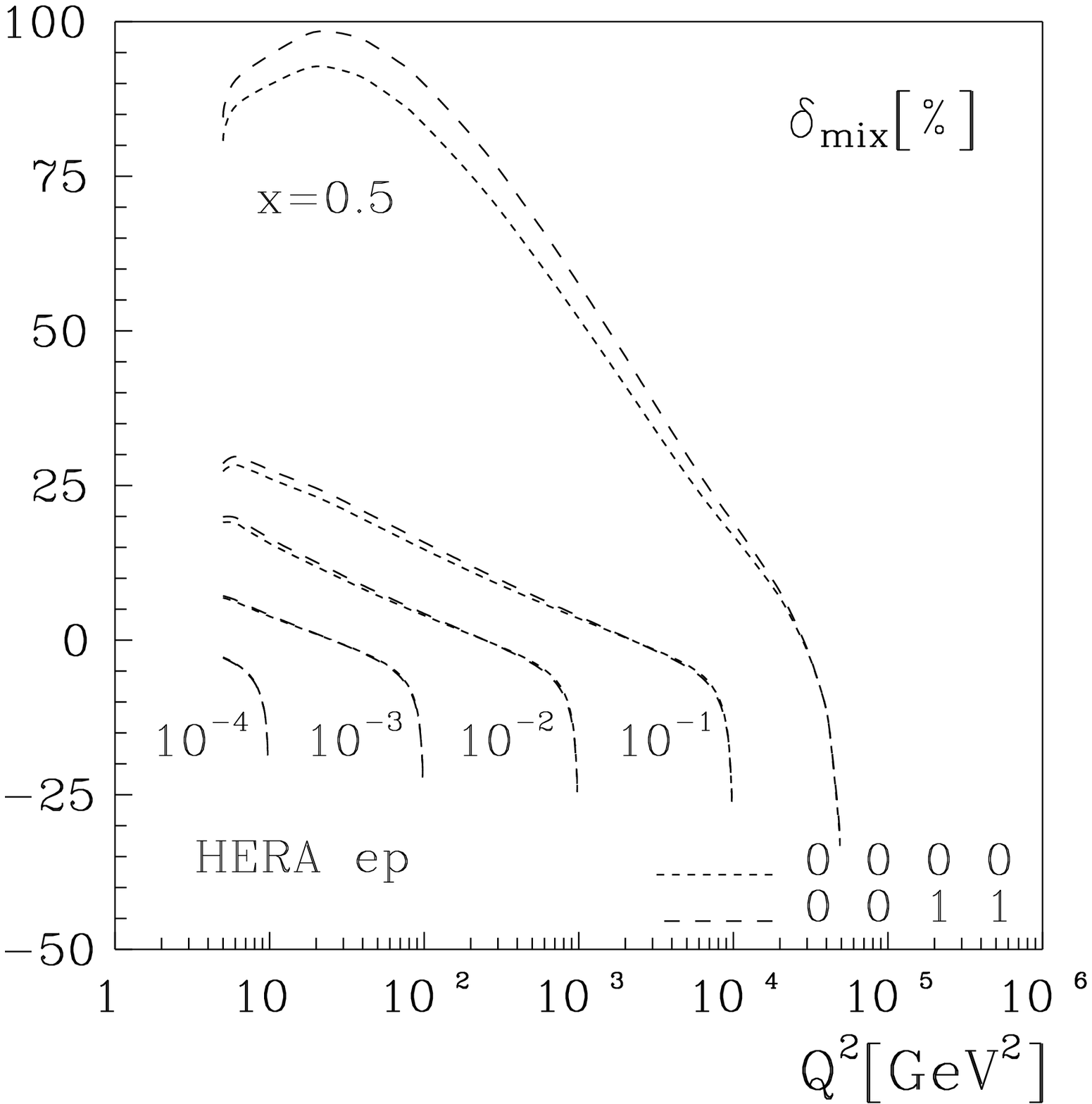}
}
\end{center}
\caption{\it
Comparison of the radiative corrections $\delta_{\mr{mix}}(x,Q^2)$
at HERA with two parameterizations of structure functions.
Flag settings ({\tt IVAR,ITERAD,IEXP,IVPOL}) are explained in the text.
\label{fdic1}
}
\end{figure}

Figure~\ref{fdic1} shows the dependence of the corrections on both the
running of the QED coupling constant
and the soft photon exponentiation.
Both corrections are small and tend to compensate each other
for large values of $y$.
For $\xm=0.5$ and the maximal value of $Q^2$, the values are:
-34.4\% for (0000), -29.8\% for (0010), and -32.7\% for (0011).
 
\begin{figure}[tbhp]
\begin{center}
\mbox{
\epsfysize=9.cm
 \epsffile[0 0 530 530]{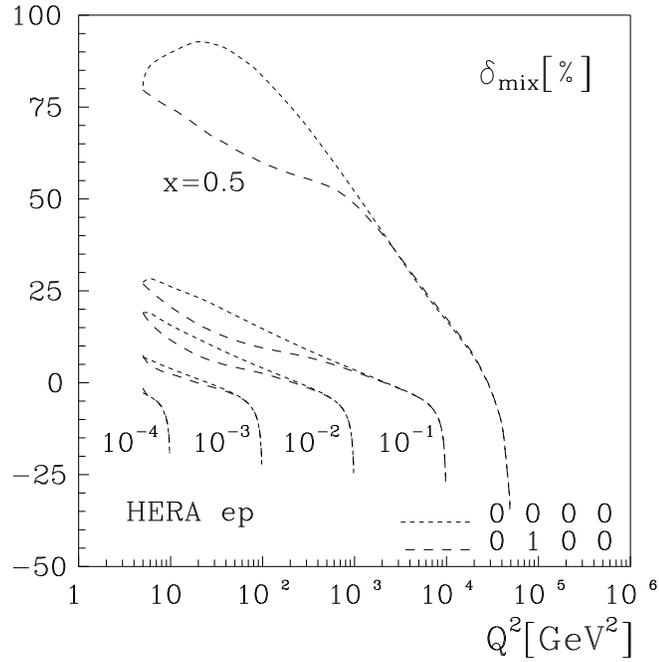}
}
\end{center}
\caption{\it
Comparison of the radiative corrections $\delta_{\mr{mix}}(x,Q^2)$
at HERA with two parameterizations of structure functions;
the flag settings are explained in the text.
\label{fdic2}
}
\end{figure}

Figure~\ref{fdic2} shows the dependence of the corrections on the
choice of the $Q^2$ in the structure functions under the integral for
the hard photon emission.
There is a strong dependence                when hard photon emission
contributes substantially to the corrections.
We see also a faking behaviour of the corrections at smaller values of
$Q^2$, which is more pronounced at larger values of $x$.
At $Q^2=5$ GeV$^2$, the parton distributions are artificially frozen
since below this value the evolution may not be controlled.
Thus, there is no room for a variation of $Q^2$ at all, and both
flag settings become equal.
 
\begin{figure}[tbhp]
\begin{center}
\mbox{
\epsfysize=9.cm
\epsffile[0 0 530 530]{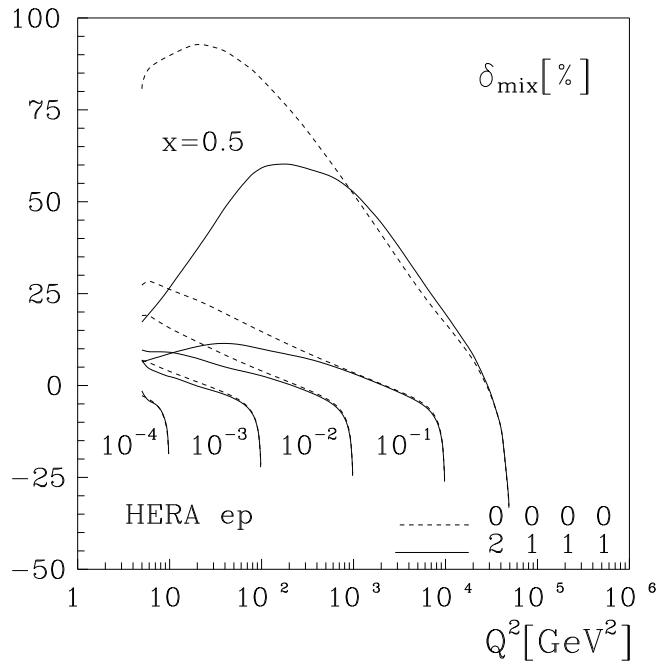}
}
\end{center}
\caption{\it
Comparison of the radiative corrections $\delta_{\mr{mix}}(x,Q^2)$
at HERA with two parameterizations of structure functions.
Flag settings are explained in the text.
\label{fdic3}
}
\end{figure}

Finally, figure~\ref{fdic3} shows the most realistic choice of the
flags mentioned. As is evident from the above discussion and a comparison
with the foregoing figures, the low-$Q^2$
modifications are extremely important for a correct description of the
corrections at small $Q^2$ and large $x$.
 
%
To summarize this part of the discussion, at least in certain regions
of the phase space it is extremely important to ensure an iterative
improvement \hfill of the various \hfill  components of the
%
%
description of
deep inelastic scattering, including the interplay of structure functions
and QED corrections.
\subsection{A comparison with leading logarithmic approximations
\label{ll}
}
An interesting and practically important question concerns the
accuracy of the leading logar\-ith\-mic approximations (LLA).
For a comparison, we use the Fortran program {\tt HELIOS}~\cite{HELIOS}
with the same structure functions        as in {\tt TERAD}.
For the special purpose here
soft photon exponentiation, other higher order corrections,
the low-$Q^2$ modifications etc.
are excluded.
 
\begin{figure}[tbhp]
\begin{center}
\mbox{
\epsfysize=9.cm
 \epsffile[0 0 530 530]{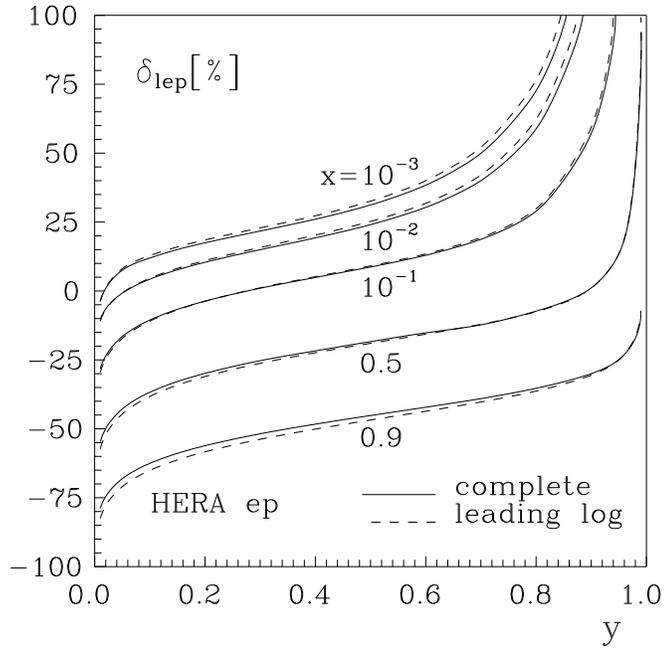} 
}
\end{center}
\caption{\it
Comparison of
radiative corrections $\delta_{\mr{lep}}(x,Q^2)$
at HERA from a complete \oa\ and a LLA calculation.
\label{fdis5}
}
\end{figure}

\begin{figure}[tbhp]
\begin{center}
\mbox{
\epsfysize=9.cm
 \epsffile[0 0 530 530]{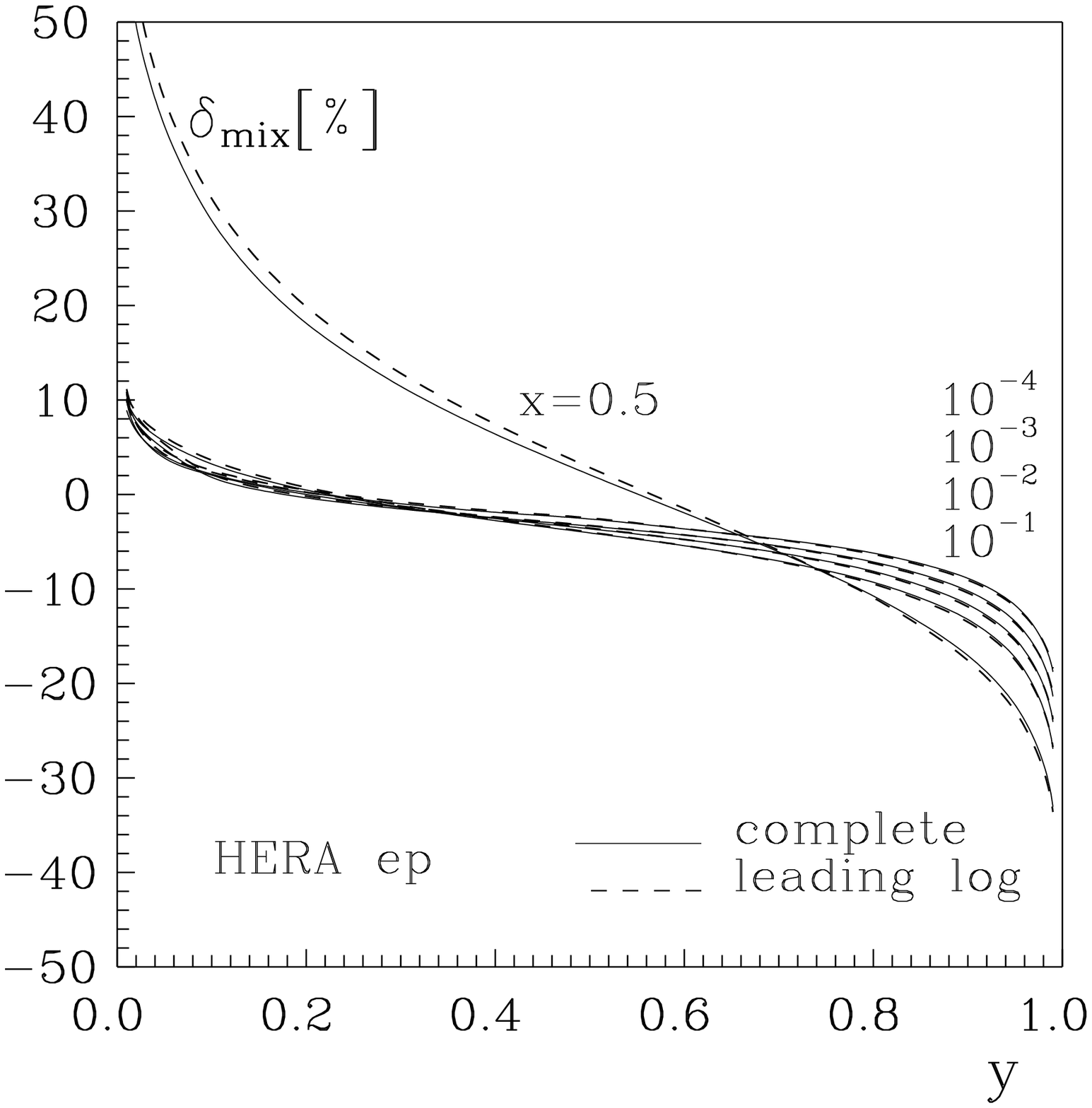}    
}
\end{center}
\caption{\it
Comparison of
radiative corrections $\delta_{\mr{mix}}(x,Q^2)$
at HERA from a complete \oa\ and a LLA calculation.
\label{fdis6}
}
\end{figure}

\begin{figure}[tbhp]
\begin{center}
\mbox{
\epsfysize=9.cm
 \epsffile[0 0 530 530]{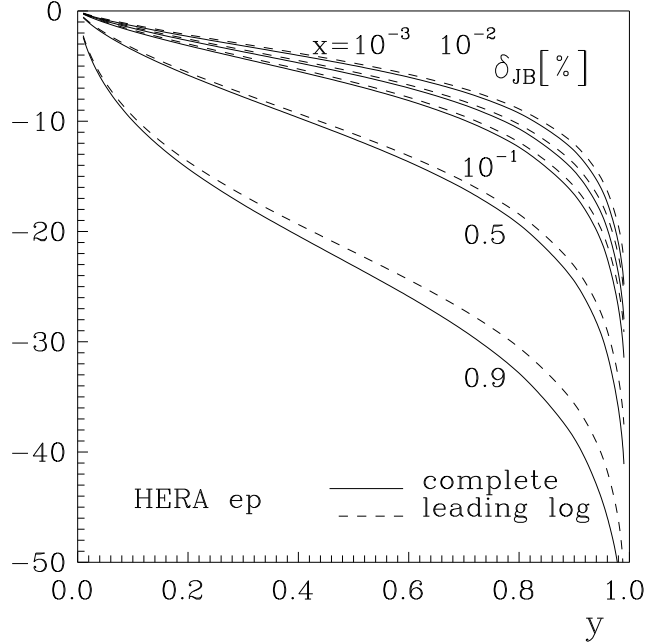} 
}
\end{center}
\caption{\it
Comparison of
radiative corrections $\delta_{\mr{JB}}(x,Q^2)$
at HERA from a complete \oa\ and a LLA calculation.
\label{fdis7}
}
\end{figure}

The comparison is performed for leptonic, mixed, and Jaquet-Blondel
variables and shown in figures~\ref{fdis5} --~\ref{fdis7}.
For the comparison, the structure functions~\cite{lib3,pdflib6} and the
following flags in {\tt TERAD} have been chosen:
({\tt IVAR,ITERAD,IEXP,IVPOL})=(0100).
 
Over a wide kinematical range, the leading logarithmic approximation
works quite well and is completely sufficient for a description
of the experimental data.
For intermediate $x$ in leptonic and small $x$ in mixed and hadronic
variables, the accuracy of the LLA is best.
 
One may also conclude
from the figures that a complete \oa\  calculation will become necessary
if the experimental accuracy reaches the level of one per cent or better.
 
%
 
\subsection{Photonic cuts
\label{pc}
}
\begin{figure}[thbp]
\begin{center}
\mbox{
\epsfysize=9.cm
\epsffile[0 0 530 530]{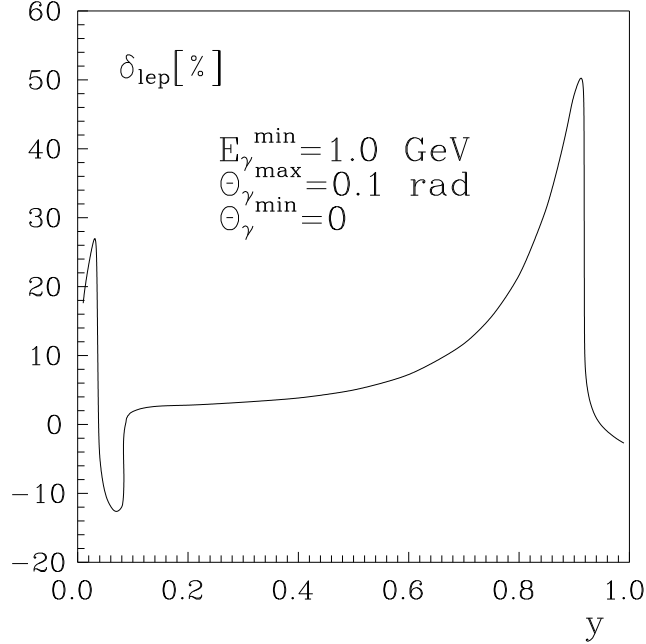}
}
\end{center}
\caption{\it
Radiative cross section for deep inelastic neutral current scattering
at HERA in leptonic variables with photonic cuts; $x$=0.001.
}
\label{cutcs}
\end{figure}

An instructive illustration of some basic features of the QED
corrections is figure~\ref{cutcs}\footnote
{
The figure was made with data from {\tt TERAD91}. Later, it was
reproduced by the Monte-Carlo program
{\tt HERACLES}~\cite{Hubert,HERACLES}.
We should like to thank H.~Spiesberger for his careful analysis of the
figure, which we used                when preparing this section.
}.
It shows the dependence of the
correction $\delta_{\mr{lep}}$ on \yl\ in presence of
cuts on the photon momentum  which are introduced in
appendix~\ref{appd}.
 
To be definite, we chose $x=0.001$ and HERA kinematics. The quark
distributions are~\cite{lib2,pdflib2}.
It is assumed that photons may be observed if they have
an energy larger than $E_m=1$ GeV
and are inside a
cone with an opening angle of $\theta_{\gamma}^{\max}=0.1$ rad.
The cut conditions are inactive as long as the photon energy is smaller
than $E_m$.
The left-hand side of the second of~(\ref{eqA.6}) corresponds to this.
As long as $z_E^{\min}$ is negative, the cut is fictitious.
It starts to be influential if, in the ultra-relativistic limit,
\ba
\yl = \yh + \frac{E_m}{E_e}.
\label{ymem}
\ea
Since \yh\ is integrated over, this condition weakens to the following one:
\ba
\yl = \frac{E_m}{E_e} = 0.033.
\label{cute}
\ea
At this value of \yl\ the rise of the cross section is sharply
interrupted.
The collinear initial and final state
photons \hfill may get an energy \hfill larger than $E_m$. \hfill
If so they become cut
when
being

\noindent
emitted inside the cut cone.
The collinear final state photons escape from the cut cone and
become non-observable if the electron scattering angle
$\theta_e^{\prime}$,
\ba
\cos
\theta_e^{\prime} &=& \frac{1}{E_e^{\prime}}
                      \left[(1-\yl)E_e - \xl\yl E_p\right],
\label{thee}
\\ 
E_e^{\prime} &=&
                      \left[(1-\yl)E_e + \xl\yl E_p\right],
\label{eepr}
\ea
becomes larger than
$\theta_{\gamma}^{\max}$.
The value of \yl\ above which this is the case and
the collinear final state photons become more and more
unconstrained by our cuts is:
\ba
\yl =
\frac{
\left(1-
\cos\theta_e^{\prime}\right)E_e
}{
\left(1-
\cos\theta_e^{\prime}\right)E_e
        +
\left(1+
\cos\theta_e^{\prime}\right)\xl E_p} = 0.084.
\label{ylfi}
\ea
Above this value, the contribution from
final state radiation to the cross section rises.
 
Further, the emission of photons due to the Compton mechanism grows.
This contribution peaks at large \yl\ and small \xl\
(see the next section). It is due to the t-channel exchange of photons
with practically vanishing $Q^2$.
Thus, it comes mainly from events with real photons whose transverse
momenta are balanced with those of the scattered electrons and
the longitudinal ones are opposite.
If now $\yl$ becomes larger than the value
\ba
\yl =  1 -
\frac{
\left(1-
\cos\theta_e^{\prime}\right)E_e
}{
\left(1-
\cos\theta_e^{\prime}\right)E_e
        +
\left(1+
\cos\theta_e^{\prime}\right)\xl E_p}
=1- 0.084=0.916,
\label{ylff}
\ea
these photons enter into the cut cone and will not contribute to the measured
cross section. The Compton peak starts to become cut.
\subsection{The Compton peak
\label{compt}
}
The Compton peak has been discussed first in~\cite{motsai}.
The corresponding cross section enhancement is due to energetic
final state radiation from the lepton.
It is related to large $\yl$ and small $x_l$.
As will be shown here, the origin of the Compton peak is deeply
connected with the lower bound on the \qh\, which is reached when
the integral over \qh\ is performed in order to get a cross section
in leptonic variables. The extremely small $Q^2$ of the virtual
photon makes
the kinematical situation
resembling the scattering of light at matter.
 
As may be also
seen from the figures, the Compton peak occurs in leptonic and
mixed variables, while it is absent in hadronic and Jaquet-Blondel
variables.
The reason is that
in the last two sets of variables, the \qh\ is an external variable
and thus not free to vary
down to the kinematical limit.
 
The Compton peak arises from the integration of the photon propagator
$d\qh / Q_h^4$ over $\qh$ in the contribution from the
$\gamma$ exchange in formulae~(\ref{*052}) and~(\ref{eqn053}).
We will indicate here the essential steps only.
For the structure functions,
Bjorken scaling will be assumed:
\ba
F_i(\xh,\qh) \approx F_i(\xh).
\label{bjc}
\ea
Further, for an integration over \qh\ at fixed \xh\
the following change of
integration variables is necessary:
\ba
d\yh d\qh = d\xh d\qh \frac{\qh}{x^2_h S}.
\label{q1}
\ea
The variables \xh\ and \qh\ vary within limits  which are derived in
appendix~\ref{b1x}.
 
Using~(\ref{bjc}) and~(\ref{q1}),
the $\gamma$ exchange part of the radiative cross section may be
rewritten as follows:
 
\ba
\frac{d^2 \sigma_{\mr{QED}}(\gamma)}{d\yl d\ql}
&=&
\frac{d^2 \sigma_{\mr{B}}(\gamma)}{d\yl d\ql}
\frac{\alpha}{\pi}
{\hat{\delta}}_{\mr{VR}}
+
\frac{d^2 \sigma_{\mr R}^{\mr F}(\gamma)}{d\yl d\ql},
\\
\vspace{0.3cm}
\frac{d^2 \sigma_{\mr R}^{\mr F}(\gamma)}{d\yl d\ql}
&=&
\frac{2 \alpha^3 S}{\lambda_S} \int d\xh d\qh  \nll
 &&\times~
\Biggl\{
\frac{2}{x_h^2 Q_h^4} F_1(\xh) S_1(\yl,\ql;\yh,\qh)
+ \frac{1}{\xh Q_h^4}F_2(\xh) S_2(\yl,\ql;\yh,\qh)
\nll
& & -~\left[
\frac{2}{x_l^2 Q_l^4} F_1(\xl)  S_1^B(\yl,\ql)
+
\frac{1}{\xl Q_l^4}F_2(\xl) S_2^B(\yl,\ql)
\right]
{\cal L}^{\mr{IR}}(\yl,\ql;\yh,\qh)
\Biggr\}. \nll
\label{e4r}
\ea
The correction ${\hat{\delta}}_{\mr{VR}}$ may be found in
~\cite{II,zfpc42}.
For a discussion of the Compton peak, only the first two terms under the
integral in~(\ref{e4r}) are relevant.
Examining their variations as functions of \qh\ one may conclude
that they are of         two types.
From the integrands $(1; 1/\sqrt{C_1}; 1/\sqrt{C_2}) d\qh/\qh$,
a $\ln(\qh^{\min}/\qh^{\max})$ emerges, while from the integrands
$(1; 1/\sqrt{C_1}$; $1/\sqrt{C_2}) d\qh/Q_h^4$ a $1/{Q_h^{2\min}}$.
The latter, however, are screened by factors $m^2$ or $M^2$.
 
In~\cite{II,zfpc42}, the following formula has been derived after
integration of~(\ref{e4r}) over \qh\ in the ultra-relativistic
approximation:
\ba
\frac{d^2 \sigma_{\mr{QED}}(\gamma)}{d\yl d\ql}
&=&
\frac{2\alpha^3}{\yl Q_l^4}
\Biggl\{
Y_+ \Biggl[
\ln\frac{(1-\xl)^2 y_l^2}{x_l^2(1-\yl)}
\left(
\ln\frac{\ql}{m_e^2}
-1\right)
+\frac{3}{2}
\ln\frac{\ql}{m_e^2} -\frac{1}{2} \ln^2 (1-\yl) -2
\Biggr]
F_2(\xl)
\nll & & +~2 Y_+
\left(
\ln\frac{\ql}{m_e^2}
-1\right)
\int_1^{1/x_l} dz
\frac{
F_2(z\xl)-F_2(\xl)}{z-1}
\nll & &+~\int_1^{1/x_l} dz \frac{dz}{z^2}
\left[ U_e(\gamma;1,1-\yl) +U_e(\gamma;\yl-1,-1)\right]  F_2(z\xl)
\Biggr\},
\label{qpms}
\ea
where
\ba
U_e(\gamma;a,b)
&=&
-y_l^2
\left(
\ln\frac{\ql}{m_e^2}
-1\right)
-\frac{y_l^2(a^2+b^2)(z-1)}{2ab}+\frac{y_l^2}{2z}\ln\frac{\yl(z-1)S}
{\xl M^2}
\nll & &
-~\frac{y_l^3}{2(1-z)}
\left(
\ln\frac{\ql}{m_e^2} +\ln\frac{(za-b)^2}{(z-1)y_l^2} \right)
-\frac{\yl}{b}\left[a\yl-\frac{y_l^2}{2z}-z(a^2+b^2)\right]
\ln\frac{b^2S^2}{m^2M^2},
\nll
\label{ue}
\ea
and
\ba
z &=& \frac{\xh}{\xl}.
\ea
From the exact boundaries of appendix~\ref{b1x}
one gets the following relations:
\ba
\qh^{\max} (\yl,\ql,\xh)
&\approx&
\frac{\xh}{\xl}\ql,
\label{qhma}
\\
\qh^{\min} (\yl,\ql,\xh)
&\approx&
\frac{
\xh\xl \ql}
     {\left(\frac{\xh}{\xl} -1\right)\ql + x_h^2  M^2}  M^2.
\label{qhmm}
\ea
Therefore, one of the logarithmically peaking integrands in~(\ref{e4r})
is:
\ba
\int_{Q_h^{2\min}}^{Q_h^{2\max}} \frac{d\qh}{\qh}
\approx \ln \frac{(\xh-\xl)\yl S+(\xh M)^2}
                                              {(\xl M)^2}
\approx
        \ln \frac{(z-1)\yl S}{\xl M^2}.
\label{e42}
\ea
This contribution is explicitly seen in~(\ref{ue}).
Two additional logarithmically enhanced contributions
may be found in the
last term in ~(\ref{ue}).
The above mentioned screening of the other potential sources of
kinematical peaks in the variable \qh\ may also be seen in~(\ref{qpms})
-- the inverse powers of $M^2$ are absent.
 
The argument of the logarithm at the right hand side of~(\ref{e42})
may become very
large at $\xl \approx 0, \yl \approx 1$ ($\xh > \xl$).
This is the origin of the Compton peak.
We would like to stress that the cross section depends on the proton
mass $M$ on purely kinematical grounds.
It has nothing to do with assumptions in the quark parton model.
In the above considerations we used one implicit assumption.
Namely, for the argument to hold, the parton distribution must be a
slowly varying, non-vanishing function at smallest $\qh$.
This is exactly, what they aren't.
We remember here about the discussion
of~(\ref{q2min}). Thus, the Compton peak is at least partly regularized
by this damping of the quark distributions.
In view of lack of a reliable model for the parton distributions
in the region of smallest $\qh$  it is not evident which theoretical
prediction of the radiative corrections is best.
Maybe one should look inversely onto the problem: a clean measurement
of the Coulomb peak could teach us something about the parton
distributions which give rise to it~\cite{BLH}.
We do not think that there are strong arguments to favor
$\Lambda_{\mr{ QCD}}$
as a scale of damping of the parton distributions at small \qh\
as was proposed in~\cite{johmix}.
 
So far, we discussed the case of leptonic variables.
In section~\ref{dism}, it was mentioned that the Compton peak in mixed
variables is much less pronounced but present.
The cross section in mixed variables has been integrated once, over
\yl, in~(\ref{m57})--(\ref{eqn515}). Only in one of the two regions, in region~I,
the \qh\ may become small. There, again the integration starts from
$\qh^{\min}$.
 The minimal value of \qh\ has been
determined in~(\ref{qhmin}). With the aid of~(\ref{eq25}), in the ultra-relativistic
limit in $m_e$  it is $y_l^{\max}(\ql)=1-M^2 \ql/S^2$ and
\ba
\qh^{\min} &=&\Ql \left(1-y_l^{\max}+\yh\right)
\nll &=& \xm y_h^2 (1+r\xm) S \approx \xm y_h^2 S.
\label{qmmin}
\ea
The corresponding kinematical singularity is located at $\xm \rightarrow 0,
\yh \rightarrow 0$.
The extreme value of \qh\ for mixed variables is scaled by a factor of the
order of $S/M^2$ compared to the case of leptonic variables and the
peak is much less pronounced in the  experimentally accessible kinematical
ranges.
\subsection{Summary and outlook
\label{lack}
}
In this article, we gave a comprehensive presentation of the \oa\
leptonic QED corrections to deep inelastic $ep$ scattering.
The detailed analysis of the kinematics of deep inelastic
scattering is of a general interest. It has been performed
  exactly in both the electron and proton masses
in order to scope both with extreme kinematical situations and the
characteristic singularities of the totally differential cross section.
Then, for several sets of kinematical variables
semi-analytical integrations have been performed.
The remaining two integrals for leptonic variables have to be taken
numerically  while for mixed and Jaquet-Blondel variables one
integral remains. The hadronic case has been solved completely
analytically.
Most of the results of this article have not been published before.
 
It was not the intention
to                            cover all     relevant radiative
corrections.
 
Here  we would like to mention which corrections have been left out of
the presentation:
\begin{itemize}
\item
Electroweak one loop corrections~\cite{zfpc42}.
These have to be treated in the
quark parton model.
\item
The QED corrections which are not related to the lepton legs --
the lepton-quark interference and the quarkonic radiation.
Their calculation rests also on the quark parton model.
The corrections are known in leptonic variables~\cite{zfpc42} and in
mixed variables~\cite{qpmmix}.
In other variables they are under study.
\item
Within the LLA approach, there exist predictions for QED corrections
also in other variables;
for an overview, see~\cite{HECTOR}. 
A semi-analytical,
complete \oa\ calculation in these variables seems to be nontrivial.
\item
A discussion of the Monte-Carlo approach has been left out completely.
For this topic, we refer to the overviews given in~\cite{WS87,WS91}. 
\item
Further, there is the wide field of higher order corrections.
Their importance depends on the variables chosen.
For leptonic variables, 
They have certainly to be included in certain regions of the phase
space if an accuracy of one per cent or better is aimed at.
The soft photon exponentiation is part of this  and the hard photon
contributions beyond \oa\ are certainly needed in LLA. Partial solutions
are obtained in
~\cite{Kripfganz,bluemho,kuraev,HECTOR}.
\item
Aiming at an accuracy of 1\% or better, one has to concern about
QCD corrections to the parton distributions.
This has been done in the {\tt HECTOR} project~\cite{HECTOR}.
\item
Finally, one should mention that for many physical applications
charged current
scattering  has to be described
with the same accuracy as the neutral current one.
A lot of theoretical work remains to be done here. So far we may
refer only to~\cite{zfpc44,Hubert}.
\end{itemize}
\bigskip
{\em To summarize},
in this article we treated the numerically largest QED radiative
corrections semi-analytically
in several of the most important kinematical variables.
This gives an important tool for a satisfactory description of
fixed target and
HERA data and the possibility to estimate the corrections at much
higher beam energies.
The most important next (or alternative) steps have been indicated.

\bigskip

Note added in March 1996.
\\
We corrected misprints in the following formulae: (2.25), (A.23),
(A.26).

\section*{Acknowledgements}
We would like to acknowledge the participation of N.M.~Shumeiko in the
long first period of work on the project, when many of the basic
techniques had been developed in the context of corrections in leptonic
variables.
We would like to thank
A.~Arbuzov,
B.~Badelek,
F.A.~Berends,
J.~Bl\"umlein,
P.~Christova,
J.~Feltesse,
W.~Hol\-lik,
M.~Klein,
G.~Kra\-mer,
W.~Kras\-ny,
E.~Kuraev,
K.~Kurek,
A.~Leike,
D.~Lehner,
L.~Lipatov,
L.~Mo,
H.~Spies\-ber\-ger
for numerous, valuable discussions on radiative corrections and $ep$
scattering.
A.A., D.B. and L.K. thank P.~S\"oding, DESY--IfH~Zeuthen
and DESY--Hamburg for the
opportunity to work at DESY on this project and for financial support.
A.A. would like to thank S.~Randjbar-Daemi, the International
Atomic Energy Agency and UNESCO for the hospitality
extended to him at the International Centre for Theoretical Physics, Trieste
and for financial support.
 
\newpage
 
\appendix
\def\theequation{\Alph{section}.\arabic{equation}}
 
\section
{Kinematics and phase space
\label{appa}
}
\ezero
 
\subsection{Kinematics and phase space for section~3
\label{appa1}
}
Throughout this section, all definitions and relations are understood
to be exact in the electron and proton masses.
The presentation follows~\cite{III}.
 
Let us introduce the following kinematical $\lambda$-functions:
\bq
\begin{array}{rclcl}
\vphantom{\int\limits_t^t}
\lambda_k &\equiv& \lambda[ -(p_1+k)^2, -p_1^2, -k^2 ]
        &=&
        (\yl-\yh)^2S^2,
\nll
\vphantom{\int\limits_t^t}
\lambda_S
&\equiv&
\lambda[ -(p_1+k_1)^2, -p_1^2, -k_1^2 ]
        &=&
        S^2 - 4 m^2 M^2,
\nll
\vphantom{\int\limits_t^t}
\lambda_l  
&\equiv& \lambda[ -(p_1+k_2)^2, -p_1^2, -k_2^2 ]
        &=&
        (1-\yl)^2S^2 - 4 m^2 M^2,
\nll
\vphantom{\int\limits_t^t}
\lambda_q  
&\equiv& \lambda[ -(p_1+Q_l)^2, -p_1^2, -Q^2_l ]
        &=&
        y_l^2S^2 + 4 M^2 Q^2_l,
\nll
\vphantom{\int\limits_t^t}
\lambda_h &\equiv& \lambda[ -(p_1+p_2)^2, -p_1^2, -p_2^2 ]
        &=&
        y_h^2 S^2 + 4 M^2 Q_h^2,
\nll
\vphantom{\int\limits_t^t}
\lambda_{\tau}  
&\equiv& \lambda[ -(p_1+\Lambda)^2,-p_1^2,-\Lambda^2]
        &=&
        (1-\yh)^2S^2-4M^2\tau  ,
\end{array}
\label{lamb1}
\eq
where
\bq
-\Lambda^2 = \tau = -(k_2+k)^2 = z_2+m^2
\label{dftau}
\eq
and
\ba
\lambda (x,y,z)
   &=& x^2 + y^2 + z^2 - 2xy - 2xz - 2yz
\nll
   &=& (x-y-z)^2 - 4yz
\nll
   &=& [x - (\sqrt y + \sqrt z )^2][x - (\sqrt y - \sqrt z )^2].
\label{eqlambda}
\ea
In the proton rest system, $\vec{p}_1=0$, the following relations hold:
\bq
\displaystyle
\begin{array}{rclcrcl}
\vphantom{\int\limits_t^t}
|\vec{k}|
&=&
\frac{\displaystyle 1}{\displaystyle 2M}\sqrt{\lambda_k},
&\hspace{1cm}&
k^0
&=&
 \frac{\displaystyle S}{\displaystyle 2M}(   \yl-\yh),
\nll
\vphantom{\int\limits_t^t}
{|\vec{k}_1|}
&=&
\frac{\displaystyle 1}{\displaystyle 2M} \sqrt{\lambda_S},
&\hspace{1cm}&
k_1^0
&=&
\frac{\displaystyle S}{\displaystyle 2M},
\nll
\vphantom{\int\limits_t^t}
|\vec{k}_2|
&=&
{\frac{\displaystyle 1}{\displaystyle 2M}}
\sqrt{\lambda_{l}},
&\hspace{1cm}&
k_2^0
&=&
\frac{\displaystyle S}{\displaystyle 2M}(1-\yl),
\nll
\vphantom{\int\limits_t^t}
|\vec{Q}_l|
&=&
\frac{\displaystyle 1}{\displaystyle 2M}\sqrt{\lambda_q},
&\hspace{1cm}&
Q_l^0
&=&
\frac{\displaystyle S}{\displaystyle 2M}\yl,
\nll
\vphantom{\int\limits_t^t}
|\vec p_2|
&=&
|\vec Q_h|
 =
\frac{\displaystyle 1}{\displaystyle 2M} \sqrt{\lambda_h},
&\hspace{1cm}&
p_2^0
&=&
M+\frac{\displaystyle S}{\displaystyle 2M}\yh,
\nll
\vphantom{\int\limits_t^t}
|\vec{\Lambda}|
&=&
\frac{\displaystyle 1}{\displaystyle 2M}\sqrt{\lambda_{\tau}},
&\hspace{1cm}&
\Lambda^0
&=&
\frac{\displaystyle S}{\displaystyle 2M}(1-\yh).
\end{array}
\label{eq06}
\eq
They may be derived with the aid of the relation~\cite{byckling}
\bq
4 M_A^2 |\vec{p}_{\mr B}|^2 =
  \lambda[-(p_{\mr A}+p_{\mr B})^2, -p_{\mr A}^2,
            -p_{\mr B}^2 ] |_{\vec{p}_{\mr A}=0}.
\label{momentum}
\eq
Thus, to any of the momenta in the rest system of the proton
corresponds one of the relativistic invariant
$\lambda$-functions.
From~(\ref{lamb1}) and~(\ref{eq06}) it
follows that the invariants~(\ref{6inv}) totally fix
the spatial configuration of momenta of reaction~(\ref{eqdeep})
which in the proton rest system can be drawn as a
momentum tetrahedron; see figure~\ref{tetra}.
\subsection{Kinematics for section~6
\label{app11}
}
For the calculation of QED corrections to the
photoproduction process sets of external and integration variables are
used
which differ from the set ${\yl,\ql,\yh,\qh}$.
Further,
the
infrared singularity is treated
in a different rest frame.
For these reasons,
some additional
kinematical relations are needed here.
The presentation follows~\cite{III}.
 
Let us introduce the following $\lambda$-functions:
\bq
\begin{array}{rclcl}
\vphantom{\int\limits_t^t}
\lambda_k'
&\equiv&
\lambda[ -p_2^2, -(p_2+k)^2, -k^2 ]
        &=&
        (W^2 - M_h^2)^2,
\nll
\vphantom{\int\limits_t^t}
\lambda_1'
&\equiv&
\lambda[ -(p_1-k_2)^2, -(p_2+k)^2, -k_1^2 ]
        &=&
        (S-\ql)^2 - 4 m^2 W^2,
\nll
\vphantom{\int\limits_t^t}
\lambda_2' 
&\equiv& \lambda[ -(p_1+k_1)^2, -(p_2+k)^2, -k_2^2 ]
        &=&
  \left[(1-\yl)S+\ql\right]^2 - 4 m^2 W^2
\nll &&&=& \left(S-W^2+M^2\right)^2-4m^2W^2.
\end{array}
\label{lambpr}
\eq
With the aid of them and of~(\ref{momentum}),
the following relations may be found in the rest frame of the compound
system, which is defined
by the condition ${\vec k}^R+{\vec p}_2^R=0$:
\ba
\begin{array}{rclcrcl}
\Vph
|{\vec k}^{R}|
&=&
\frac{\displaystyle \sqrt{\lambda_k'}}{\displaystyle 2\sqrt{W^2}},
&\hspace{.7cm}&
k^{0,{R}}
&=&
\frac {\displaystyle W^2-M_h^2}{\displaystyle 2\sqrt{W^2}},
\\
\Vph
|{\vec k}_1^{R}|
&=&
 \frac{
\displaystyle
\sqrt{\lambda_1'}
       }{
\displaystyle
2\sqrt{W^2}
              },
&\hspace{.7cm}&
k_1^{0,{R}}
&=&
\frac{\displaystyle S-\Ql}{\displaystyle 2\sqrt{W^2}},
\\
\Vph
|{\vec k}_2^{R}|
&=&
 \frac{
\displaystyle
\sqrt{\lambda_2'}
       }{
\displaystyle
2\sqrt{W^2}
              },
&\hspace{.7cm}&
k_2^{0,{R}}
&=&
\frac{
\displaystyle
(1-\yl)S+\Ql}{
\displaystyle
2\sqrt{W^2}}.
\end{array}
\label{fo6}
\ea
The invariant mass of the compound system at rest is $W^2=-(k+p_2)^2$.
%
\subsection{Kinematics for section~7
\label{a12}
}
In the approach of section~\ref{kineII} to the phase space, the
first two integrations will be performed over
the photonic angles in the rest frame ${R}$ of the
($\gamma e$)~compound system:
${\vec \Lambda}^{R} = {\vec k}^{R} + {\vec k}_2^{R} =0$.
For that purpose, the expressions for some
four momenta and invariants are needed
in that system  in order to finally
express $\yl, \ql, z_1$ in~(\ref{eqs1z})--(\ref{eqs3z}).
The velocities  which are used in the
infrared divergent part of the cross section are also calculated
in the $R$ frame.
 
The necessary $\lambda$-functions are:
\bq
\begin{array}{rclcl}
\vphantom{\int\limits_t^t}
\lambda_k''
&\equiv&
\lambda[ -(\Lambda-k)^2, -\Lambda^2, -k^2 ]
        &=&
(\tau - m^2)^2,
\nll
\vphantom{\int\limits_t^t}
\lambda_1
&\equiv&
\lambda[ -(\Lambda-k_1)^2, -\Lambda^2, -k_1^2 ]
        &=&
\left( \qh + z_2 \right)^2 + 4 m^2 \qh,
\nll
\vphantom{\int\limits_t^t}
\lambda_2  
&\equiv&
\lambda[ -(\Lambda-k_2)^2, -\Lambda^2, -k_2^2 ]
        &=&
(\tau - m^2)^2,
\end{array}
\label{lambp2}
\eq
and the $\lambda_{\tau}$ has been introduced in~(\ref{lamb1}).
 
The four-momenta are:
\ba
\begin{array}{rclcrcl}
\Vph
|{\vec k}^{R} |
&=&
\frac{
\displaystyle
\sqrt{\lambda_k''}}{
\displaystyle
2 \sqrt{\tau}},
&\hspace{.7cm}&
k^{0,{R}}&=&\frac{\displaystyle\tau-m^2}{\displaystyle 2 \sqrt{\tau}}
=\frac {\displaystyle z_2}{
\displaystyle
2 \sqrt{\tau}},
\\
\Vph
|{\vec k}_1^{R}|
&=&
\frac{
\displaystyle
\sqrt{\lambda_1}}{
\displaystyle
2 \sqrt{\tau}},
&\hspace{.7cm}&
k_1^{0,{R}}
&=&
\frac{
\displaystyle
\qh+\tau+m^2}{
\displaystyle
2 \sqrt{\tau}},
\\
\Vph
|{\vec k}_2^{R}|
&=&
\frac{
\displaystyle
\sqrt{\lambda_2}}{
\displaystyle
2 \sqrt{\tau}},
&\hspace{.7cm}&
k_2^{0,{R}}
&=&
\frac{
\displaystyle
\tau+m^2}{
\displaystyle
2 \sqrt{\tau}},
\\
\Vph
|{\vec p}_1^{R}|
&=&
\frac{
\displaystyle
\sqrt{\lambda_{\tau}}}
{
\displaystyle
2 \sqrt{\tau}},
&\hspace{.7cm}&
p_1^{0,{R}}
&=&
\frac{
\displaystyle
(1-\yh) S}{
\displaystyle
2\sqrt{\tau}}.
\end{array}
\label{3m2}
\ea
In the ${R}$  system the following cartesian coordinates are chosen:
the $z$-axis be parallel to ${\vec k}_1^{R}$  and the $yz$-plane
spanned by the vectors ${\vec k}_1^{R}, {\vec p}_1^{R}$;
the angle between the $z$-axis and ${\vec p}_1^{R}$ be $\vartheta_p$.
 
Then it is:
\ba
\vsph
k
&=& \left\{ k^{0,{R}} \sin \vartheta_{R} \cos \varphi_{R},\;
k^{0,{R}} \sin \vartheta_{R} \sin \varphi_{R},\; k^{0,{R}} \cos \vartheta_{R},\;
 k^{0,{R}}
\right\},
\\  \vsph
k_1
&=& \left\{ 0,\; 0,\; | {\vec k}_1^{R} |,\; k_1^{0,{R}} \right\},
\\      \vsph
k_2
&=& \left\{- k^{0,{R}} \sin \vartheta_{R} \cos \varphi_{R},\;
-k^{0,{R}} \sin \vartheta_{R} \sin \varphi_{R},
\; -k^{0,{R}} \cos \vartheta_{R},\; k_2^{0,{R}}
\right\},
\\
\vsph
p_1
&=& \left\{ 0,\;  | {\vec p}_1^{R} | \sin \vartheta_p,\;
| {\vec p}_1^{R} | \cos \vartheta_p,\; p_1^{0,{R}} \right\}.
\label{k1p1k}
\ea
From these expressions, one may easily derive~(\ref{z12})--(\ref{tep}).
 
\subsection
{Some phase space parameterizations
\label{phrm}
}
\subsubsection
{The integral $d^3k_2/2k_2^0$
\label{ph1}
}
The following phase space
integral has to be calculated in section~\ref{phI}:
\ba
I
&=& \int \frac{d^3k_2}{2k_2^0}
=
\int \frac{|{\vec k}_2|}{2} dk_2^0 d\cos \vartheta d\varphi.
\label{i1}
\ea
It is:
\ba
\ql = -2m^2+2 k_1^0 k_2^0 \left( 1 - \beta_1\beta_2 \cos\vartheta_{12}\right).
\ea
One has the freedom to identify $\vartheta_{12} = \vartheta$.
In the rest system of the proton the formulae of
appendix~\ref{appa1} may be used. It follows:
\ba
\yl = 1-\frac{2M}{S} k_2^0, \hspace{1.cm} S = 2 M k_1^0.
\ea
These two relations allow to replace the energies introduced above by the
invariants \yl\ and \ql.
The corresponding two three-momenta may also be expressed with the aid
of~\ref{eq06}:
\ba
|{\vec k}_1| = \frac{\sqrt{\lambda_S}}{2M},
\hspace{1.cm}
|{\vec k}_2| =  \frac{\sqrt{\lambda_l}}{2M}.
\ea
Inserting these relations into~(\ref{i1}), one gets:
\ba
I = \frac{\pi S}{2\sqrt{\lambda_S}} \int d\yl d\ql.
\label{i2}
\ea
\subsubsection
{The integral $d\Gamma_k$
\label{ph2}
}
 
 In section 3.1, the following expression  has to be simplified:
\ba
d\Gamma_k = \frac {d\vec{k}}{2k^0} \,
\delta \left[(p_1+Q_h)^2+M_h^2\right]
\delta \left[\qh-(p_2-p_1)^2\right].
\label{dgamk}
\ea
With
$ Q_l = Q_h + k $ it is
in the proton rest system:
\ba
d\Gamma_k  &=&
 \frac{1}{2} k_0 dk_0 \, d\cos \vartheta_\gamma  \, d\varphi
  \, \delta \left[(p_1+Q_l-k)^2+M_h^2\right]
   \delta \left[\qh-(Q_l-k)^2      \right]                \nll
&=&
 \frac{1}{2} k_0 dk_0 \, d\cos \vartheta_\gamma \, d\varphi
  \, \delta \left[(-M^2 + 2p_1(Q_l-k)
               +(Q_l-k)^2 + M_h^2 \right]
   \delta \left[\qh-(Q_l-k)^2      \right]                \nll
&=&
   \frac{1}{2} k_0 dk_0 \, d\cos \vartheta_\gamma \, d\varphi
  \, \delta \left[(-M^2 + 2p_1(Q_l-k) + \qh + M_h^2 \right]
   \delta \left[ Q_h^2 - Q_l^2 + 2 Q_l k \right]                              \nll
&=& \frac{1}{2} k_0 \frac{1}{2M}
    \frac{d\varphi}{2 |Q_l| k_0}
 =  \frac{d\varphi}{ 8 M |{\vec Q}_l| }.
\label{dg}
\ea
Here
$\varphi$ is the angle between the planes defined by the three-momenta
($\vec{k}_1,\vec{k}_2$) and ($\vec{k},\vec{p}_2$),
see figure~\ref{tetra}.
 
 Now the integration over $dz_2$ may be introduced
with the aid of an auxiliary
$\delta$-function $\delta(z_2+2k_2k)$:
\ba
d\Gamma_k
= \frac {dz_2} {8M|{\vec Q}_l|} d\varphi \,\delta(z_2+2k_2Q_l-2k_2Q_h).
\label{dgak}
\ea
The following coordinate system will be chosen.
The $z$ axis is parallel to ${\vec Q}_l$.
The vector ${\vec k}_2$ stays
 within the $xz$ planeand has an angle
$\vartheta_p$
with respect to the $z$ axis.
The projection of the vector ${\vec Q}_h={\vec p}_2$ into the $xy$ plane
has angle $\varphi$ to the $x$ axis, and the vector itself has the angle
$\vartheta$ with respect to the $z$ axis.
Then, it will be
\ba
{\vec k}_2 \cdot {\vec Q}_h
=
{\vec k}_2 \cdot {\vec p}_2
=
|{\vec k}_2| | {\vec p}_2| (\sin \vartheta_p \sin \vartheta \cos \varphi +
\cos \vartheta_p \cos \vartheta),
\label{veca}
\ea
and
\ba
d\Gamma_k
&=& \frac {dz_2}{16 M |\vec{Q}_l| |\vec{k}_2| |\vec{p}_2|
|\sin \vartheta_p \sin \vartheta \sin \varphi |}
\nll
&\equiv& \frac {dz_{2}}{16 M |\vec{Q}_l \cdot (\vec{k}_2 \times \vec{p}_2 )|}.
\label{gaka}
\ea
%
 The three-momenta used here and in the
following may be calculated with the aid of
the $\lambda$-functions~(\ref{lamb1}).
The
$d\Gamma_k$ is related to
the volume $V_T$ of the tetrahedron of momenta as shown
in figure~\ref{tetra}:
\bq
V_T \equiv
\frac{1}{6}
\left|
\vec{Q}_l \cdot (\vec{k}_2 \times \vec{p}_2 ) \right|
=
\frac{1}{24 M} \sqrt{R_z}.
\label{sqrz}
\eq
Explicitly,
\ba
V_T   =    \frac{1}{6}\Biggl\{ \vec{Q}_l^2
  \vec{k}_2^2\vec{p}_2^2 - \vec{Q}_l^2(\vec{k}_2\vec{p}_2)^2
- \vec{k}_2^2(\vec{p}_2\vec{Q}_l)^2
- \vec{p}_2^2(\vec{Q}_l \vec{k}_2)^2
    + 2(\vec{Q}_l\vec{k}_2)(\vec{k}_2\vec{p}_2)(\vec{p}_2\vec{Q}_l)
\Biggr\}^{1/2}.
\label{qkp}
\ea
 
The positive-definite function $R_z$ in~(\ref{sqrz})
is the Gram determinant of $4$-vectors $k_1$, $p_1$, $k_2$, $p_2$,
\bq
R_z \equiv -\Delta_4(k_1,p_1,k_2,p_2) = - 16
\left| \begin{array}{cccc}
          k_1^2 & (k_1p_1) & (k_1k_2) & (k_1p_2)  \\
       (p_1k_1) & p_1^2    & (p_1k_2) & (p_1p_2)  \\
       (k_2k_1) & (k_2p_1) & k_2^2    & (k_2p_2)  \\
       (p_2k_1) & (p_2p_1) & (p_2k_2) & p_2^2
       \end{array} \right|.
\label{determinant}
\eq
It may be expressed by the ${G}$-function introduced in
\cite{byckling}:
\bq
R_z = - \frac{1}{16M^4} {G}
(\lambda_{l}, \lambda_h, \lambda_{\tau},\lambda_q, \lambda_S, \lambda_k).
\label{rzg}
\eq
For further use, it is convenient to write $R_z$ in a
form which exhibits its explicit dependence on $z_1$ or $z_2$:
\ba
R_z &=& -
A_1 z_1^2 + 2 B_1 z_1 - C_1 \equiv -
A_2 z_2^2 + 2 B_2 z_2 - C_2,
\label{rz}
\\
A_2 &=&  \lambda_q \equiv A_1,
\label{eq09}
 \\
B_2 &=&\Bigl\{ 2 M^2 \Ql ( \Ql - \Qh )
      +     (1-\yl) (\yl \Qh - \yh \Ql) S^2
 +S^2 (1) \Ql (\yl - \yh) \Bigr\}
\nll
&\equiv&~
-~B_1
        \Bigl\{ (1) \leftrightarrow - (1-\yl) \Bigr\},
\label{eq10}
\\
C_2 &=&  \Bigl\{  (1-\yl) \Qh - \Ql \left[ (1)- \yh \right] \Bigr]^2 S^2
     + 4m^2 \Bigl[ (\yl-\yh)(\yl \Qh-\yh \Ql) S^2
- M^2(Q^2_h - Q^2_l )^2  \Bigr\}
\nll
   &\equiv&~
C_1 \Bigl\{  (1) \leftrightarrow - (1-y_l) \Bigr\},
\label{eq11}
\ea
with the discriminant
\bq
D_z = B_{1,2}^2 - A_{1,2} C_{1,2} = \frac{1}{64M^4} \lambda(\lambda_S,
\lambda_{l},\lambda_q)
\lambda(\lambda_q,\lambda_h,\lambda_k).
\label{eq12}
\eq
With~(\ref{sqrz}),
the phase space
takes the form
\bq
d\Gamma_k =
 2 \times \frac{dz_{2}}{4\sqrt{R_z}}
  = \frac{dz_{2}}{2\sqrt{R_z}},
\label{newka}
\eq
where the overall factor of 2 at the right hand side corresponds to
$\varphi \in [0,\pi] \leftrightarrow z_{1(2)} \in [z_{1(2)}^{\min},
z_{1(2)}^{\max}]$.
The latter boundaries for the integration over $z_{1(2)}$ are derived in
appendix~\ref{c1}.
 
\subsubsection
{The compound phase space volume $d\Gamma_{\gamma e}$
\label{ph3}
}
The following phase space
integral has to be calculated in section~\ref{hb2}:
\ba
I &=& \int d\Gamma_{\gamma e}
=
\int \frac{d\vec k_2}{2k_2^0}\frac{d\vec k}{2k^0} \delta^4(\Lambda-k_2-k)
\nll
&=&~\int \frac{d\vec k_2}{2k_2^0} \delta[(\Lambda-k_2)^2]
=
\int \frac{|{\vec k}_2|}{2} dk_2^0 d\cos \vartheta d \varphi
\delta[(\Lambda-k_2)^2] .
\label{ii1}
\ea
In the rest system of the $(\gamma e)$ compound one may use
the expressions for the energies and momenta~(\ref{3m2}).
In this system, it is
\ba
\int dk_2^0 \delta[(\Lambda-k_2)^2]
&=&
\int dk_2^0 \delta(-\tau-m^2+2\sqrt{\tau}k_2^0)
  =~\frac{1}{2\sqrt{\tau}}.
\ea
With $|{\vec k}_2| = \sqrt{\lambda_2} / 2\sqrt{\tau}$ and
$\lambda_2 = \lambda(\tau,m^2,0) = (\tau-m^2)^2$, see appendix~\ref{a12},
one gets:
\ba
I &=& \frac{\tau - m^2}{8\tau} \int d\cos\vartheta d\varphi.
\label{ii2}
\ea
\subsubsection
{The integral $d^3p_2/2p_2^0$
\label{ph4}
}
In section~\ref{hb2}, the following integral has to be calculated:
\ba
I &=& \int d^4p_2 \delta\left(p_2^2 + M_h^2\right)
\delta\left[(p_2-p_1)^2-\qh\right]
\delta\left[(p_2-p_1-k_1)^2+\tau\right]
\nll
&=&~ 2 \pi \int \frac{|{\vec p}_2|}{2} dp_2^0 d\cos\vartheta
\delta\left[(p_2-p_1)^2-\qh\right]
\delta\left[(p_2-p_1-k_1)^2+\tau\right].
\ea
In the rest system of the proton it is
$\tau=-(p_2-p_1-k_1)^2 = M_h^2 + S - 2Mp_2^0 + 2p_2k_1$.
In appendix~\ref{appa1}, the expressions for the four-momenta $p_2$ and $k_1$ in
this frame in terms of invariants have been derived.
It is
$-2p_2k_1 =  -(\sqrt{\lambda_h\lambda_S}/2M^2)\cos\vartheta + \ldots$
This relation allows to use the second of the $\delta$-functions for the
calculation of the integral over $\cos\vartheta$:
\ba
\int d\cos\vartheta \delta\left[-(p_2-p_1-k_1)^2+\tau\right]
&=& \frac{2M^2}{\sqrt{\lambda_h\lambda_S}}.
\ea
With
$ |{\vec p}_2|=\sqrt{\lambda_h}/2M $ and
$-(p_2-p_1)^2+\qh = -2M p_2^0 +M^2 + M_h^2 +\qh$,
the last $\delta$-function allows to perform the
integration over $dp_2^0$ and one gets:
\ba
I &=& \frac{\pi}{2 \sqrt{\lambda_S}}.
\ea
\subsubsection
{Notations
\label{nota}
}
Here, we confront some old and new notations.
This may prove useful if someone tries to read the
preprints~\cite{AB},~\cite{II}:
\ba
\begin{array}{rclrclrcl}
S_l &=& \yl S
,\hspace{.7cm}
&
T &=& \yh S
,\hspace{.7cm}
&
t &=& \qh,
\\
Q^2 &=& \ql
,\hspace{.7cm}
&
X &=& (1-\yl) S
,\hspace{.7cm}
&
Y &=& \ql,
\\
S_X = S-X &=& \yl S.   &&&&&&
\end{array}
\label{notat}
\ea
\section
{Kinematic boundaries
\label{appb}
}
\ezero
We now come to the derivation of the many different kinematical boundaries
which
are connected with the different choices of the integration variables
and their
sequential order.
By $v^{\max(\min)}(V_1,V_2,\ldots)$,
the maximum (minimum) value of
$v$ will be denoted
which may be taken at fixed values of $V_1, V_2,\ldots$
Further, the extreme value of variable $v$ is
${\bar{v}}$; it depends only on $S$ and possibly on $M,m$.
 
The physical regions of invariants~(\ref{6inv}) may be found from
the conditions of existence
of the momentum tetrahedron depicted in~figure~\ref{tetra}.
This tetrahedron is defined by the triangles
$\Delta(\vec {k}_1,-\vec{k}_2,-\vec{Q}_l)$,
$\Delta(\vec {Q}_l,-\vec{p}_2,-\vec {k})$,
and
$\Delta(\vec{k}_1,-\vec{\Lambda},-\vec{p}_2)$
in the proton rest system.
The areas of these triangles are equal to
\begin{eqnarray*}
\frac{1}{4} \sqrt{-\lambda( \vec{k}_1^2, \vec{k}_2^2, \vec{Q}_l^2 )},
\hspace{1cm}
\frac{1}{4} \sqrt{-\lambda( \vec{Q}_l^2,{\vec{p_2}}^2, \vec{k}^2 )},
\hspace{1cm}
\frac{1}{4} \sqrt{-\lambda( \vec{k}_1^2, \vec{\Lambda}^2,{\vec{p_2}}^2)},
\label{triangle1}
\end{eqnarray*}
correspondingly; they exist if
\bq
\lambda(\vec{k}_1^2,\vec {k}_2^2,\vec {Q}_l^2) \leq 0,
\hspace{1cm}
\lambda(\vec{Q}_l^2,{\vec{p_2}}^2, \vec{k}^2) \leq 0,
\hspace{1cm}
\lambda(\vec{k}_1^2, \vec{\Lambda}^2,{\vec{p_2}}^2 ) \leq 0.
\label{eq14a}
\eq
With the expressions~(\ref{lamb1}) for the squared three-momenta in
terms of $\lambda$-functions, the following boundary conditions
are derived:
\bq
\lambda(\lambda_S,\lambda_{l},\lambda_q) \leq 0,
\hspace{1cm}
\lambda(\lambda_q,\lambda_h,\lambda_k) \leq 0,
\hspace{1cm}
\lambda ( \lambda_S,\lambda_{\tau}, \lambda_{h}) \leq 0.
\label{eq13}
\eq
The boundary equation for the first condition is:
\bq
M^2 Q_l^4 +  \Ql \yl S^2 + m^2 y_l^2 S^2 - \lambda_S \Ql = 0.
\label{eq22}
\eq
The boundary equation for the second condition is:
\bq
 y_l^2 S^2 \qh
 +  y_h^2 S^2\ql
 - M^2(\Ql-\Qh)^2- \yl \yh (\Ql+\Qh) S^2
 =0.
\label{eq39}
\eq
It is symmetric with respect to the interchanges $(\yl \leftrightarrow
\yh)$ and $(\ql \leftrightarrow \qh)$.
 
The third boundary condition may be rewritten:
\bq
(2M^2 z_2 + \yh S^2 + 2M^2 \Qh)^2 - \lambda_S\lambda_h \leq 0.
\label{eqc3}
\eq
The $z_2$ is known to be positive definite (see appendix~\ref{c11}).
The maximal value $z_2^{\max}(\yh,\qh)$
 of $z_2$ at given values of $(\yh,\qh)$ is
one of the roots $z_2^{\pm}(\yh,\qh)$ of~(\ref{eqc3}),
\bq
z_2^{\pm}(\yh,\qh)
 = \frac{1}{2M^2} \left[-(\yh S^2+2M^2 \Qh) \pm
            \sqrt{ \lambda_S \lambda_h }\right],
\label{eqc4}
\eq
namely $z_2^+$.
It is positive definite if
\bq
M^2 Q_h^4 +  \Qh \yh S^2 + m^2  y_h^2 S^2 - \lambda_S \Qh \leq 0.
\label{eqc5}
\eq
{}From the zero of~(\ref{eqc5}) the
third boundary equation results.
By the replacements
$\qh \leftrightarrow \ql$ and $\yl \leftrightarrow \yh$ it transforms
into the boundary equation of the first condition.
\subsection
{Boundaries for $z_1, z_2$
\label{c1}
}
\subsubsection
{Boundaries for $z_{1(2)}(\yl,\ql,\yh,\qh)$
\label{c11}
}
Here we closely follow~\cite{III}.
{}From~(\ref{eq12}) and~(\ref{eq13}) one may conclude
$D_z\geq 0$.
The invariants $z_1$ and $z_2$ are
positive definite. This may be understood easiest from the general
statement that an invariant of the form
$I=-2pk$ is positive definite when the momenta $p,k$ are on the mass
shell. One may also use an argument based on
the photon energy in~(\ref{3m2}), which is proportional to $z_2$.
For $z_1$  the same property may be derived by a replacement of $k_2$
in favour of $k_1$ in the arguments  which lead to the lower boundary
of $z_2$ (i.e. in section~\ref{a12}).
 
{}From~(\ref{lamb1}) and~(\ref{eq09})  it follows $A_{1,2}>0$.
The $R_z$  which is
introduced in~(\ref{sqrz}) and rewritten in~(\ref{rz})
is proportional to the volume of the tetrahedron in
figure~\ref{tetra}, i.e. it is also positive definite.
This volumee vanishes at the boundary values of the
physical regions of the invariants.
Rewriting $R_z$,
\ba
R_z = \lambda_q
\left[ z_{1(2)}^{\max} - z_{1(2)} \right]
\left[ z_{1(2)} - z_{1(2)}^{\min} \right],
\label{rzm}
\ea
one may derive an expression for the boundary values of $z_1$ and $z_2$
at fixed values of \yl,
\ql, \yh, \qh:
\bq
z_{1,2}^{\max,\min}(\yl,\ql;\yh,\qh)=\frac{B_{1,2} \pm \sqrt{D_z}}{A_z}
= \frac{C_{1,2}}{B_{1,2}\mp \sqrt{D_z}}.
\label{ax}
\eq
The $D_z$ in the above relation must be positive.
Since the $z_{1(2)}$ are positive
definite, the $B_{1(2)}$ must be positive, too.
Finally, from the second equality
in the above chain, the $C_{1(2)}$ are also seen to be positive.
In the ultra-relativistic approximation, the latter is also evident from
~(\ref{eq11}).
 
We mention here that the integration limits for $z_2$ at fixed values
of \yh, \qh\
will be derived in appendix~\ref{apph}
and at fixed \yjb, \qjb\ in appendix~\ref{appjb}. They are also
influenced by the photonic cuts, which are introduced for leptonic
variables in section
~\ref{photo}; see appendix~\ref{appd}.
 
\subsubsection
{Boundaries for $z_{1(2)}(\yl,\ql,\yh,\qh)$
with
photonic cuts
\label{appd}
}
In section~\ref{sec5} we argued that cuts may be applied
 on the
momentum of the bremsstrahlung
photon.
This is of some interest in the case of leptonic variables
            in the region of small $x$ and large $y$.
In this appendix, the necessary formulae will be derived.
 
We define the photonic cuts by the following conditions:
\begin{eqnarray}
\begin{array}{rcccl}
E_{\gamma}
&\geq& E_m
&=&    E_{\gamma}^{\;\min},
\\
\cos (\theta_{\gamma}^{\;\min})=  c_{\min}
\geq \cos \theta_{\gamma}
&\geq& c_{\max} &=&
\cos(\theta_{\gamma}^{\;\max}).
\label{cut2}
\end{array}
\end{eqnarray}
With these cuts,
photons with an energy larger than $
E_{\gamma}^{\;\min}$ and within a cone defined by
$\cos(\theta_{\gamma}^{\;\max,\min})$
are {\em excluded} from the observed cross section.
 
Both experiments at HERA have a $\gamma$ counter
which may register photons with an energy $ E_{\gamma}\geq 1$ GeV
 and an
emission angle $ 0\leq \theta_\gamma \leq 0.5$ mrad in the HERA
laboratory system.
 
In order to incorporate the cuts~(\ref{cut2}) into the cross section
calculations one needs two invariants  which depend exclusively
on $E_{\gamma}$ and $\theta_{\gamma}$ and, possibly, on
$E_{e}$ and $E_{p}$,
the energies of the electron and proton beams
in the laboratory system.
 
One candidate is $z_1$:
\ba
z_{1}=-2 k_{1} k
&=&
2E_{e}E_{\gamma}(1-\beta_{e}\cos\theta_{\gamma}) ,
\label{eqA.3}
\\
\beta_{e}
&=&
\sqrt{1-\frac{m^2}{E_e^2}}.
\label{eqA.4}
\ea
Another one may be found to be
\ba
I_{\gamma}=-2 p_{1} k = (\yl-\yh) S
&=&
2E_{p}E_{\gamma}
            (1+\beta_{p}
           \cos\theta_{\gamma}) ,
\label{eqA.2}
\\
\beta_{p}
&=&
\sqrt{1-\frac{M^2}{E_p^2}}.
\ea
Dividing~(\ref{eqA.3}) by~(\ref{eqA.2})  one gets a first
condition  which is formulated completely in terms of
integration variables and measurable parameters.
A second one may be obtained by eliminating the
$\cos\theta_{\gamma}$ from
~(\ref{eqA.3}) with the aid of~(\ref{eqA.2}):
\bq
\begin{array}{rcccl}
\Vph
z_c^{\min} \equiv \frac{\displaystyle 
E_{e}(1-\beta_{e}c_{\min})}
{\displaystyle
E_{p}(1+\beta_{p}c_{\min})}S(\yl -\yh)
           & \leq  &
z_{1}
           & \leq  &
\frac {
\displaystyle
E_{e}(1-\beta_{e}c_{\max})} {
\displaystyle
E_{p}(1+\beta_{p}c_{\max})} S(\yl-\yh)
\equiv z_c^{\max},
\label{eqA.5}
\\
\Vph 
z_E^{\min} \equiv 2E_{e}E_{m}(1+\frac{
\displaystyle
\beta_{e}}
{
\displaystyle
\beta_{p}})
-\frac{
\displaystyle
E_{e}\beta_{e}}
{
\displaystyle
E_{p}\beta_{p}}S(\yl -\yh)
           & \leq &
z_{1}.              &&
\end{array}
\label{eqA.6}
\eq
Taking into account
$z_2=z_1+\ql-\qh$  one may realize
that~(\ref{eqA.5})
represents cut  conditions
on the first integration variable.
 
The actual  modifications of the analytical integrations over $z_{1(2)}$
due to the cuts may be found in~\ref{appe1}.
After this first integration,
the {\em finite hard }cross section part undergoes
two numerical integrations, which are not essentially affected
by the cuts.
Concerning the treatment of the infrared singularity, nothing changes
compared to the presentation in section~\ref{irreI}.
The reason is simple. We subtracted from and added to the complete
bremsstrahlung correction a simplified expression, well adapted
to the infrared problem. Now, in the presence of cuts, we have the
freedom to subtract and add the {\em same} expressions.
Here, it is of importance that the applied cuts do not influence
the infrared behaviour since the contribution
of hard photons is manipulated.
 
Concluding,
\ba  \vph
{\hat z}_{1}^{\max} &=&
\min \left\{
z_1^{\max}(\yl,\ql,\yh,\qh),
z_{c}^{\max}(E_e,E_p,c_{\max},\yl,\yh)
\right\},
\label{z1ma}
\nll \vph
{\hat z}_{1}^{\min} &=&
\max \left\{
z_1^{\min}(\yl,\ql,\yh,\qh),
z_{c}^{\min}(E_e,E_p,c_{\min},\yl,\yh),
z_E^{\min}(E_e,E_p,\yl,\yh)
 \right\},
\label{z1mi}
\nll    \vph
{\hat z}_2^{\min(\max)} &=& z_1^{\min(\max)} +\ql-\qh.
\label{z2ai}
\ea
\subsubsection
{Boundaries for $z_{2}(\yh,\qh)$
\label{c14}
}
The following boundaries are used for the calculation of cross
sections in hadronic and Jaquet-Blondel variables  where the
integration over $\tau$ is the third one.
After               two integrations over photonic angles
the ultra-relativistic approximations are justified and will be
applied.
For given values of \yh\ and \qh, the upper limit of $z_2$
is given in~(\ref{eqc4})
while the lower one
is zero; compare the expression for $k^{0,R}$ in~(\ref{3m2}):
\ba
0\leq z_2\leq      \xh (1-\yh)S .
\label{tauz}
\ea
The boundary values of $z_2$
define also those for
the variable $\tau$  when the latter is used as
the argument of the
last integration  at given values  of either \yh, \qh\ or
\yjb, \qjb.
\subsection
{Boundaries for leptonic variables
\label{kineme}
}
{}From now on, we will use the following notations
for external variables:
if a variable is maximal or minimal as a function of the other one
(and of $S$ and the masses),
the superscripts min or max will be used to indicate this.
Absolute extremes  which depend only on $S$ and the masses
                                                get a bar.
 
The
maxima and minima of an internal variable at fixed values of the other
internal variable and of the external variables get the subscripts
I, III, and II, IV, respectively. In the figures  these functions are
boundaries.
If the integration order is changed       the corresponding curves
in the figures get a different analytical meaning.
This will become
      clear in the following discussions of integration regions.
%
\subsubsection
{External variables
${\cal E}_l = (\yl,\Ql)$
\label{b13}
}
We start with the derivation of absolute bounds for $W^2$.
 
For the invariant hadronic mass~$M_h^2$  the lower limit is defined
by the kinematical
threshold of hadron production, i.e. the mass of the lightest
strongly interacting particle, the pion.
The upper limit of $M_h^2$,
for a given $W^2$, may be derived from the expression
for $k_0$ in the              rest system~(\ref{fo6}).
It is reached for $k_0=0$:
\ba
\left(M+m_{\pi}\right)^2 \leq M_h^2 \leq W^2.
\label{lmh}
\ea
The above inequality may also be interpreted as the absolute
lower limit of $W^2$.

The upper
bound for $W^2=-(p_1+k_1-k_2)^2$
follows from a consideration of the defining four momentum squared
in the centre-of-mass system:
\bq
\left(M+m_{\pi}\right)^2  \leq W^2 \leq {\bar{W}^2} =  (\sqrt{s} - m)^2
\equiv (\sqrt{S+M^2+m^2}-m)^2.
\label{eq21}
\eq
 
One may eliminate $W^2$ from~(\ref{eq21}) with the aid
of~(\ref{w2})\footnote{
If not stated differently, the pion mass is neglected here and
in the following.
}:
\bq
0 \leq \yl S - \Ql \leq \left( \sqrt{S+M^2+m^2}-m \right)^2 - M^2.
\label{eq19}
\eq
The boundary equation~(\ref{eq22})
originally depends on the variables  we are interested in:
 \yl\ and \ql.
It may be pursued once more.
The boundaries for \yl\ at a given
value of \ql\ shall be determined.
The lower bound of \yl\ results from
~(\ref{eq19}); the upper limit is the solution of
~(\ref{eq22}):
\bq
y_l^{\min}(\ql) = \frac{Q_l^2}{S} \leq \yl \leq y_l^{\max}(\ql),
\label{eq24}
\eq
where
\bq
y_l^{\max}(\ql)
= \frac{1}{2m^2} \left[ \frac{1}{S}\sqrt{\lambda_S\lambda_m        } -
\Ql \right] \leq {\bar{\yl}}.
\label{eq25}
\eq
Here  the ${\bar{\yl}}$ is the maximum of $\yl$ and
may be determined by differentiation of~(\ref{eq25}) with
respect to $\ql$:
\ba
{\bar{\yl}}= \frac{S-2mM}{S}.
\label{baryl}
\ea
 
The absolute minimum of \ql\ is zero.
The condition $y_l^{\min}(\ql) = y_l^{\max}(\ql)$ is fulfilled at
$\ql = 0$ and also at
the absolute maximum of \ql. It may be used for a determination of
the limits of the latter variable:
\bq
0 \leq \Ql \leq \frac {\lambda_S} {S+M^2+m^2} \equiv
{\bar{\ql}}.
\label{eq23}
\eq
 
\begin{figure}[tbhp]
\begin{center}
\mbox{
\epsfysize=9.cm
\epsffile[0 0 530 530]{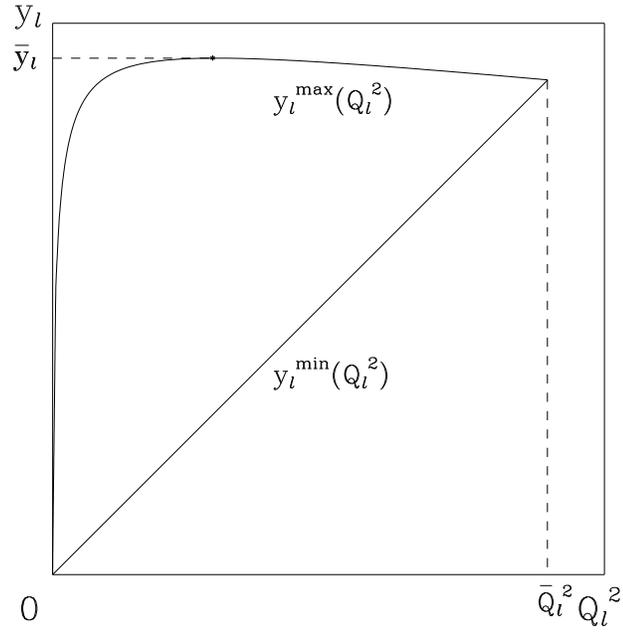}
}
\end{center}
\caption{\it
Physical region ($\ql,\yl$) for the cross section in
leptonic variables.
\label{xxxxx}
}
\end{figure}

The physical region
${\cal E}_l = (\ql,\yl)$
is shown in figure~\ref{xxxxx}.
%
%
\subsubsection
{External variables
${\cal{E}}_l^{'} = (W^2,\Ql)$
\label{b11}
}
The bounds for $W^2$, (\ref{eq21}),
 have been derived in the foregoing appendix.
 
Those for \ql\ at a given value of $W^2$ may be obtained from
~(\ref{eq22}) by replacing
the \yl\ by $W^2$ with the aid of
~(\ref{w2}):
\ba
\ql^{\max(\min)} (W^2)
&=&
\frac{
S\left(S-W^2+M^2\right) - 2 m^2 \left(W^2+M^2\right)
\pm \sqrt{\lambda_S\lambda_2'}}
{2\left( S+M^2+m^2\right)}.
\label{q2le}
\ea
{}From the lower limit in~(\ref{q2le})  the estimation of the lowest
accessible $Q^2$
~(\ref{q2min})
in the photoproduction region was derived.
 
\begin{figure}[tbhp]
\begin{center}
\mbox{
\epsfysize=9.cm
\epsffile[0 0 530 530]{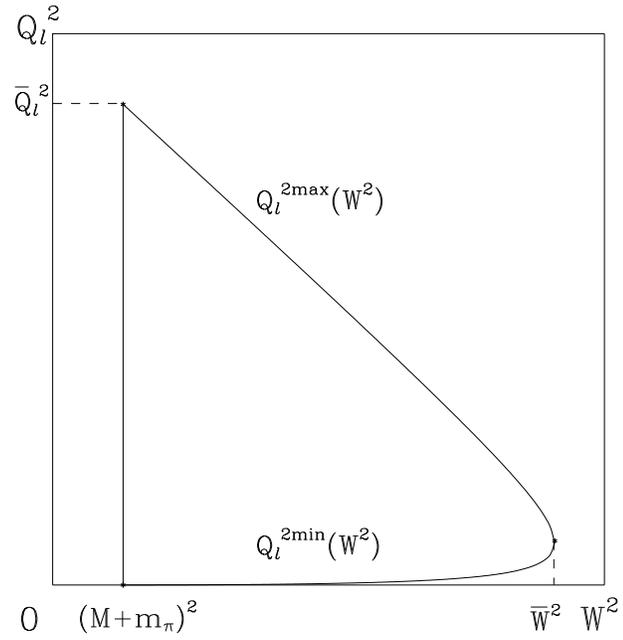}
}
\end{center}
\caption{\it
Physical region ($W^2,\ql$) for the photoproduction process.
\label{figfot1}
}
\end{figure}

The physical region ${\cal{E'}}_l = (W^2,\Ql)$ is shown in
figure~\ref{figfot1}.
 
\subsubsection
{Integration variables
${\cal I}_l = (\yh,\Qh)$
\label{b12}
}
For the hard photon corrections in leptonic variables, the last
two integrations are numerically performed over
${\cal I}_l = (\yh,\Qh)$.
The hard part of the infrared divergent contribution is integrated
analytically; thereby
the last integration
 is that over \yh.
 
Inserting~(\ref{eq28}) into~(\ref{lmh}), one gets a first condition
the left hand side of which is equivalent to $\xh=1$:
\bq
0 \leq \yh S - \Qh \leq W^2-M^2.
\label{eq27}
\eq
The~(\ref{eq39}) defines a pair of intersecting
straight lines :
\bq
\Qh_{\mr {I,II}} = \Ql + \frac{S}{2M^2}( \yl - \yh)
(\yl S \pm \sqrt{\lambda_q}).
\label{eq30}
\eq
{}From the second relation, the lower bound of \qh\ is obtained:
\ba
\qh^{\min} (\yh,\yl,\ql) = \Qh_{\mr{II}}.
\label{ea44}
\ea
The upper bound of \qh\ depends on the actual value of \yh.
The two curves $\qh=\yh S$ and $\qh=\Qh_{\mr{I}}$ meet at
the maximum value of \qh\, which is reached at the following value of \yh:
\ba
\vph
\yh_{d}(\yl,\ql)
&=& \frac{ \yl (\yl S+\sqrt \lambda_q) + 2 \Ql r }
               {\yl S+\sqrt \lambda_q + 2 M^2 },
\label{yhd}
\\  \vph
r&=& M^2/S.
\label{ryhd}
\ea
At $\yh > \yh_d$ it is
$\qh^{\max} (\yh,\yl,\ql)=\qh_{\mr I}$.
 
\begin{figure}[tbhp]
\begin{center}
\mbox{
\epsfysize=9.cm
\epsffile[0 0 530 530]{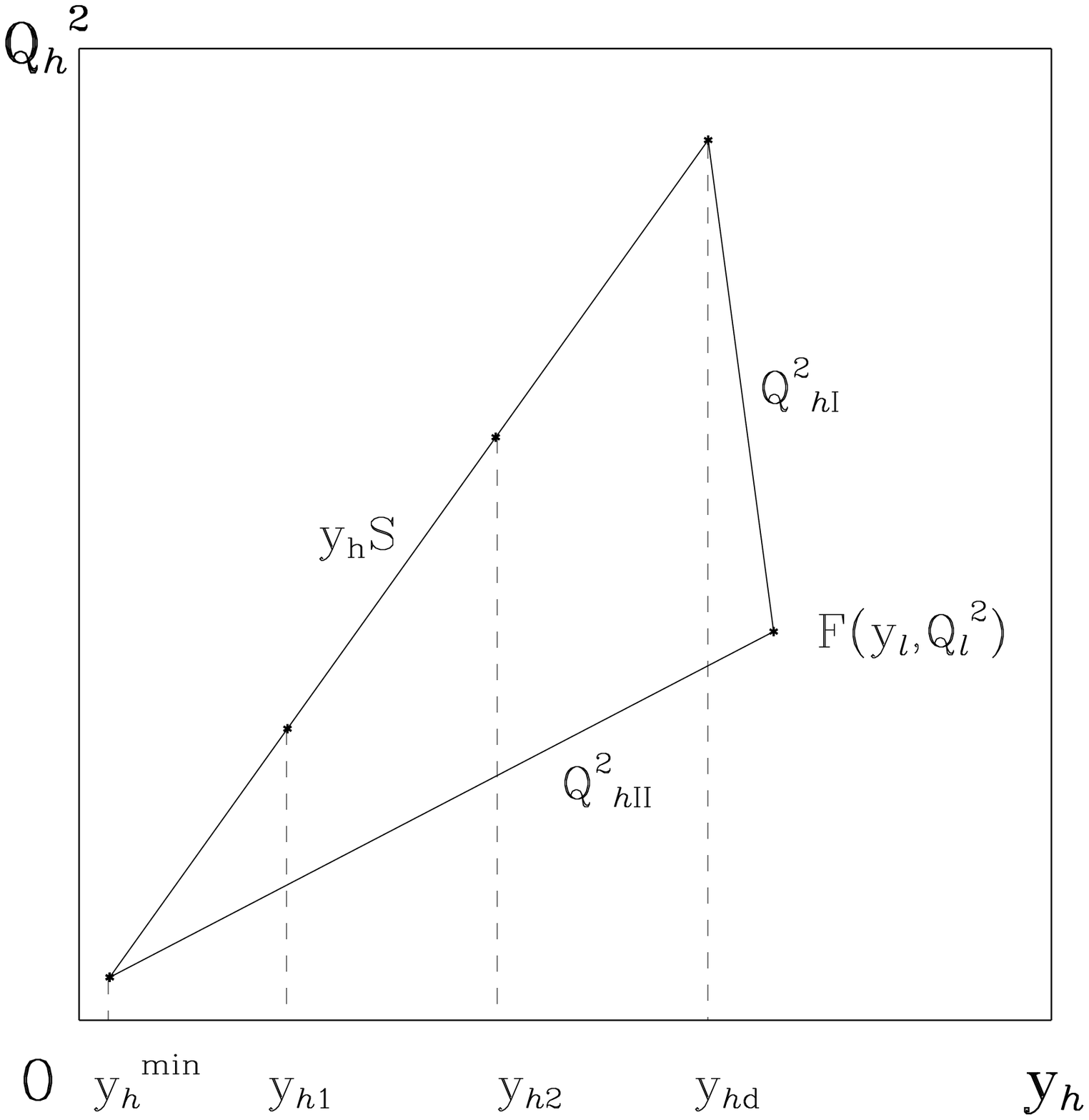}
}
\end{center}
\caption{\it
Integration region ($\yh,\qh$)
for the cross section in leptonic variables.
\label{ilep}
}
\end{figure}

The minimum of \yh\ may be obtained
{}from the equality  $\qh_{\mr{II}} = \yh S$; see figure~\ref{ilep}:
\ba
y_h^{\min}(\yl,\ql)=
\frac {
\displaystyle
\yl (\yl S-\sqrt{\lambda_q})+2\ql r}{
\displaystyle
\yl S- \sqrt{\lambda_q}+2 M^2}.
\label{ea43}
\ea
The maximum of \yh\ at a fixed value of \yl\ is
equal to \yl\ and independent of \ql; this may be seen from the
expression for $k^0$ in~(\ref{eq06}).
The infrared singularity is located at
${\cal F}(\yl,\ql).$
When calculating the hard part~(\ref{irhard}) of the infrared divergent
cross section contribution, one has to take into account an additional
condition:
\ba
k^0 \geq \epsilon > 0.
\ea
In the approach to the phase space where the boundaries of
this appendix are used
the infrared problem is treated in the
proton rest frame.
{}From the expression for $k^0$
in~(\ref{eq06})  the following
modification of the integration boundary results:
\ba
y_h^{\max} = \yl   \rightarrow
y_h^{\max} (\epsilon) =\yl - \frac{2 \epsilon M}{S}.
\label{epsl}
\ea
 
The net physical region for ${\cal I}_l=( \yh, \Qh )$ is:
\ba
\begin{array}{rcccl}
\vph
\Qh_{\mr{II}}&\leq&\Qh&\leq&\yh S
\hspace{1.cm} {\mr{if}} \, \, \, \yh \leq y_{h_{d}},
\label{eq41b}
\\  \vph
\Qh_{\mr{II}}&\leq&\Qh&\leq& \qh_{\mr I}
\hspace{1.cm} {\mr{if}} \, \, \, \yh \geq y_{h_{d}},
\label{eq4bb}
\\  \vph
y_h^{\min}(\yl,\ql)
&\leq&\yh&\leq&y_h^{\max}(\epsilon).
\end{array}
\label{ea42}
\ea

When calculating the twofold integrals in~(\ref{irhard}) over
the kinematical domain, which is derived here,
one will need
the substitutions of $\sqrt{C_{\mr 1}}$ and $\sqrt{C_{\mr 2}}$
at the kinematical boundaries.
The square roots at $ ( \Qh )_{\mr{I,II}} $
can be taken exactly:
\ba
\sqrt{C_1(\qh_{\mr I})}
&=&
(\yl-\yh)S
\left[S\frac{\sqrt{\lambda_q}+\yl S}{2M^2}+\ql\right],
\label{c33}
\\ 
\sqrt{C_2(\qh_{\mr I})}
&=&
(\yl-\yh)S
\left[(1-\yl)
S\frac{\sqrt{\lambda_q}+\yl S}{2M^2}-\ql\right],
\\
\sqrt{C_1(\qh_{\mr{II}})}
&=&
(\yl-\yh)S
\left[S\frac{\sqrt{\lambda_q}-\yl S}{2M^2}-\ql\right],
\\
\sqrt{C_2(\qh_{\mr{II}})}
&=&
(\yl-\yh)S
\left[(1-\yl)
S\frac{\sqrt{\lambda_q}-\yl S}{2M^2}+\ql\right].
\label{c12h}
\ea
The corresponding roots at $\qh = \yh S $ may be
taken in the
ultra-relativistic approximation in the electron mass;
see the expressions~(\ref{eq11}).
One arrives at the following absolute values:
\ba
\sqrt{C_1} &\simeq& S^2 | \yh - \xl \yl (1 + \yh - \yl) |,
\label{DD12}
\\
\sqrt{C_2} &\simeq& S^2 | \yh (1 - \yl) - \xl \yl (1 - \yh) |,
\label{D12}
\ea
which vanish inside the region of integration
over \yh, ~(\ref{ea42}), at
\ba
y_{h_{1}}&=&\frac{ \xl \yl (1 - \yl)}{ 1 - \yl \xl  },
\\
y_{h_{2}}&=&\frac{ \xl \yl }{ 1 - \yl (1 - \xl)  },
\ea
correspondingly.
Within the kinematical
limits of figure~\ref{ilep}, the following inequalities hold:
\ba
y_h^{\min} < \yh_1 < \yh_2 < \yh_d < y_h^{\max}(\epsilon).
\ea
 
The integration region
${\cal{I}}_l = (\yh,\Qh)$
is shown in figure~\ref{ilep}.
\subsubsection
{Integration variables
${\cal{I}}_l^{'} = (M_h^2,\Qh)$
\label{b14}
}
The boundaries for $M_h^2$ have been derived in~(\ref{lmh}):
\ba
\nonumber
\left(M+m_{\pi}\right)^2 \leq M_h^2 \leq W^2.
\ea
The infrared singularity is located at
${\cal F}(W^2,\ql)$.
When calculating the hard part~(\ref{irhard}) of the infrared divergent
cross section contribution, one has to take into account the additional
condition for the photon energy
~(\ref{fo6}):
\ba
k^{0,{R}}
&=&
\frac {\displaystyle W^2-M_h^2}{\displaystyle 2\sqrt{W^2}}
 \geq \epsilon > 0.
\ea
This condition may be transformed into the following
modification of the upper bound of $M_h^2$:
\ba
\left(M+m_{\pi}\right)^2 \leq M_h^2 \leq W^2 -2\epsilon\sqrt{W^2}.
\label{lmh2}
\ea
The limits of the
variable $\qh$ for given values of $W^2, M_h^2, \ql$ will be derived now.
For that purpose, one has to replace
in~(\ref{eq39}) the $\yl$ with~(\ref{w2})
by $W^2$ and \ql\, and the $\yh$ with~(\ref{eq28}) by $M_h^2$
and \qh\, and
solve the resulting inequality for $\qh$ in terms of $M_h^2, \ql,
W^2$~\cite{IV}: 
\ba
\qh_{\mr{I,II}} =
\qh^{\max(\min)}(W^2, M_h^2, \ql)
=
\frac{1}{2 W^2} \left[
\left(W^2-M_h^2\right)\left(\yl S \pm \sqrt{\lambda_q}\right)
+ 2 M_h^2 \ql
                \right].
\label{fotqhm}
\ea
The two straight lines meet in
the infrared point  while the two extreme values
of \qh\ at given external variables are reached for $M_h^2=(M+m_{\pi})^2
$.
 
\begin{figure}[thbp]
\begin{center}
\mbox{
\epsfysize=9.cm
\epsffile[0 0 530 530]{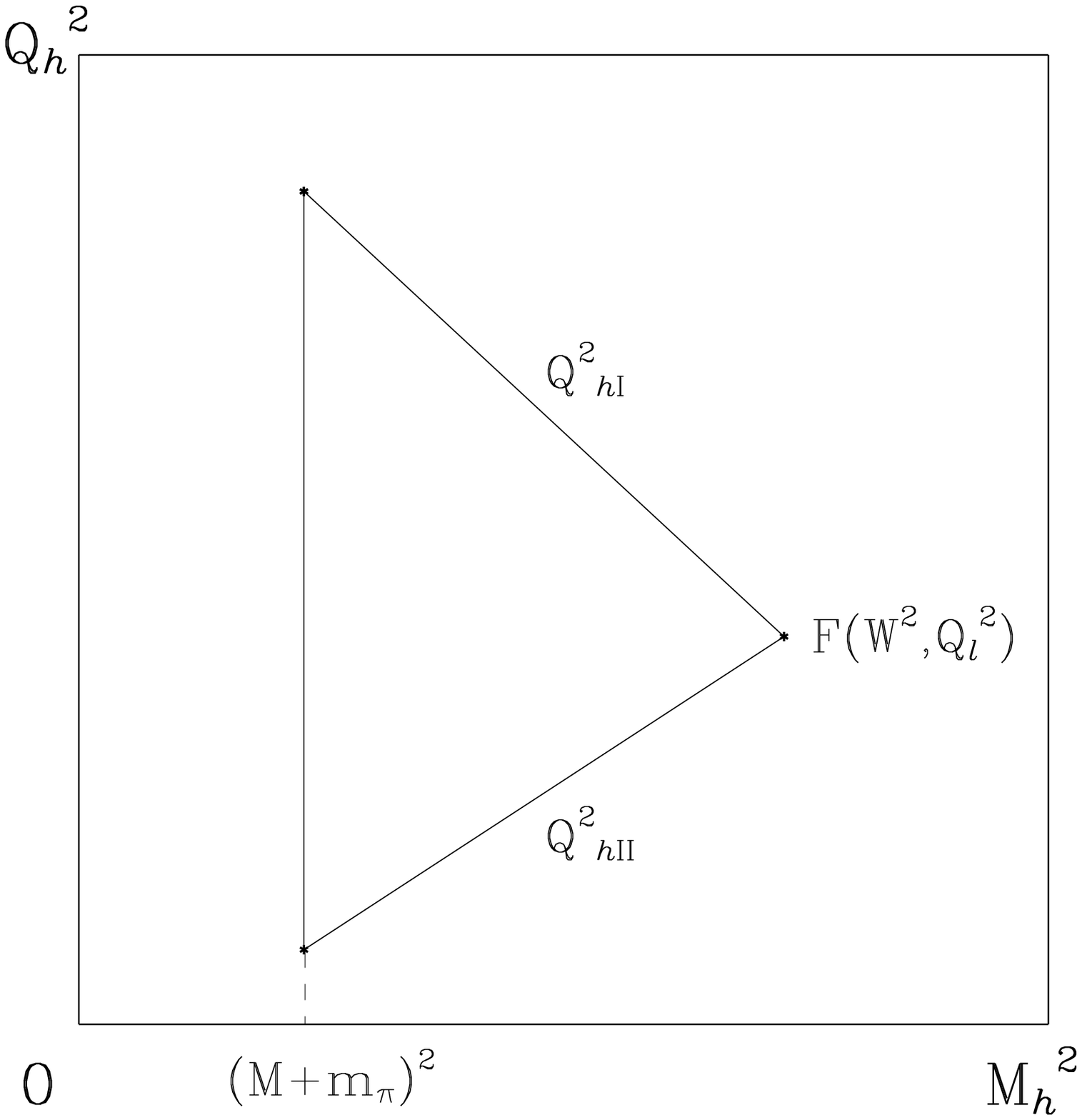}
}
\end{center}
\caption{\it
Integration region ($M_h^2,\qh$)
for the cross section in leptonic variables.
\label{figfot2}
}
\end{figure}

The physical region
${\cal{I}}_l^{'} = (M_h^2,\Qh)$
is shown in figure~\ref{figfot2}.
 
%
\begin{figure}[bhtp]
\begin{center}
\mbox{
\epsfysize=9.cm
\epsffile[0 0 530 530]{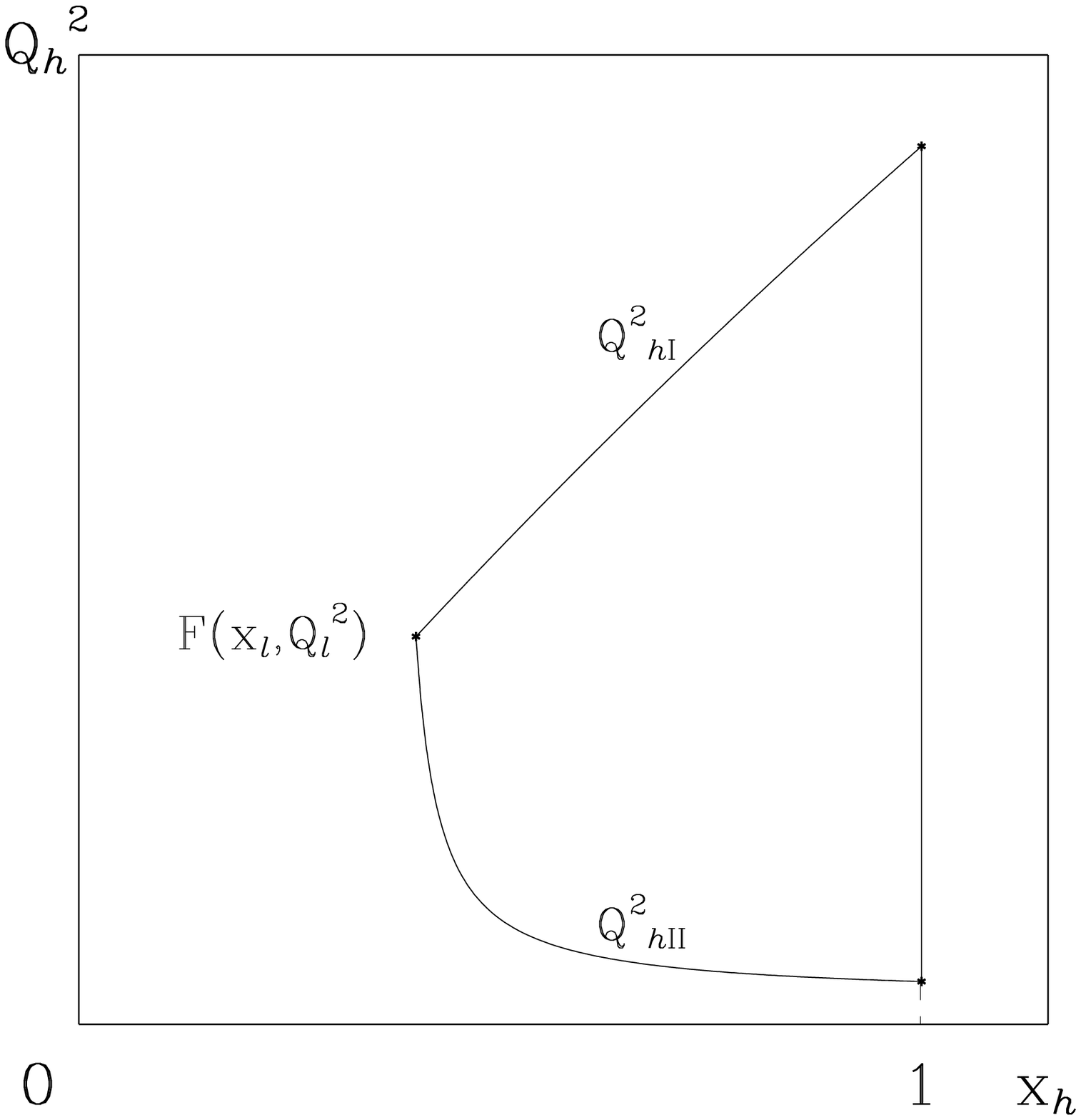}
}
\end{center}
\caption{\it
Integration region ($\xh,\qh$)
for the cross section in leptonic variables.
\label{intxq}
}
\end{figure}

\subsubsection
{Integration variables
${\cal I}_l^{''}
 = (\xh,\Qh)$
\label{b1x}
}
For the discussion of the Compton peak we need the integration
boundaries for the variables \xh\ and \qh\, where the latter has to be
integrated over first.
 
Using in~(\ref{eq39}) the relation $\yh = \qh / (\xh S)$, one
immediately arrives at the boundaries for \qh\ at given values of
\xh, \yl, \ql:
\ba
\qh_{\mr{I,II}}=
\qh^{\max(\min)} (\yl,\ql,\xh)
&=&
\frac{\xh(\xh\yl S- \ql)(\yl S \pm \sqrt{\lambda_q}) + 2 \ql(\xh M)^2}
     {2[\xh\yl S -\ql + (\xh M)^2]}.
\label{e4l}
\ea
At the lower limit of \xh\, the two curves $\qh_{\mr{I,II}}$ meet;
see figure~\ref{intxq}.
{}From~(\ref{e4l}) one derives that this is realized if $\xh=\xl$, which
is at the same time the infrared point.
The upper limit for \xh\ may be got by rewriting~(\ref{eq28}) and
inserting~(\ref{lmh}):
\ba
\qh/\xh - \qh = M_h^2 - M^2 \geq m_{\pi}(2M+m_{\pi}).
\ea
Neglecting here the pion mass, the absolute maximum of \xh\ is evidently
equal to one:
\ba
\xl \leq \xh \leq 1,
\label{eq4x}
\ea
Taking the pion mass into account, the value $\xh=1$ is forbidden;
in fact $\xh=1$ corresponds to the kinematics of the
elastic channel.
 
The integration region
${\cal{I}}_l = (\yh,\Qh)$
is shown in figure~\ref{intxq}.
\subsection
{Boundaries for
mixed variables 
\label{appb2}
}
\subsubsection
{External variables
${\cal E}_m = (\yh,\Ql)$
\label{b21}
}
The boundaries for \ql\ are the same as in the case of leptonic
variables; see~(\ref{eq23}):
\ba
0 \leq \Ql \leq \frac {\lambda_S} {S+M^2+m^2} \equiv
{\bar{\ql}}.
\nonumber
\ea
 
The maximum of \yh\ in the two-dimensional distribution
[$\yh(\ql);\yl(\yh,\qh,\ql)]$  is, independent of the value of \ql,
equal to
\yl. Thus, its absolute maximum is given by the value for
$y_l^{\max}(\ql)$ of~(\ref{eq25}).
The minimal value of $\yh(\ql)$ may be obtained by
studying the functional dependence of
$y_h^{\min}(\yl,\ql)$ in~(\ref{ea42}) on \yl.
The extremum is reached at
$\yl=y_l^{\max}(\ql)$. The explicit expressions are:
\ba
\begin{array}{rcccl}
y_h^{\min}(\ql)&\leq&\yh&\leq&y_h^{\max}(\ql),
\label{eq34}
\end{array}
\ea
where
\ba
y_h^{\min}(\ql) 
&=&
\frac {\Ql \left( \sqrt{(y_l^{\max} S)^2+4M^2 \Ql}- y_l^{\max} S
\right) }
{ S \left[
\sqrt{( y_l^{\max} S)^2+4M^2 \Ql}+ y_l^{\max} S - 2 \Ql
\right]}
\nll
&=&~
\frac{\ql \left(\sqrt{\lambda_m}+\ql\right)\left(S-\sqrt{\lambda_S}
\right)}
{S\left(\sqrt{\lambda_m}-\ql\right)\left(
S+\sqrt{\lambda_S}-\ql-\sqrt{
\lambda_m}\right)} \leq \frac{\ql}{S},
\label{eq35}
\\  
y_h^{\max}(\ql)
&=&\frac{1}{2m^2} \left[ \frac{1}{S}\sqrt{\lambda_S
\lambda_m
} -
\Ql \right] \leq {\bar{\yl}}.
\label{eq2x}
\ea
At the right hand side of~(\ref{eq35}), the
$y_l^{\max}$ stands for $y_l^{\max}(\ql)$ [see~(\ref{eq25})] and
\ba
\lambda_m = \ql\left(\ql+4m^2\right).
\label{laam}
\ea
The second part of~(\ref{eq35}) has been obtained from the identity
\ba
\nonumber
\sqrt{(y_l^{\max}S)^2+4M^2\Ql}  =
\frac{1}{2m^2} \left(
S\sqrt{\lambda_m}-\ql\sqrt{\lambda_S}\right).
\label{ide1}
\ea
The two curves $y_h^{\max}(\ql)$ and $y_h^{\min}(\ql)$ meet each
other at
$\Ql={\bar{Q}_l^2}$ as defined in~(\ref{eq23}).
 
With the variables ${\cal E}_m$ one may define a
corresponding set of scaling variables:
\bq
\qm = \ql, \hspace{1cm}
y_m \equiv y_h, \hspace{1cm}
x_m = \frac{\Ql}{-2 p_1 Q_h} = \frac{\Ql}{\yh S}.
\label{eq36}
\eq
 
\begin{figure}[tbhp]
\begin{center}
\mbox{
\epsfysize=9.cm
\epsffile[0 0 530 530]{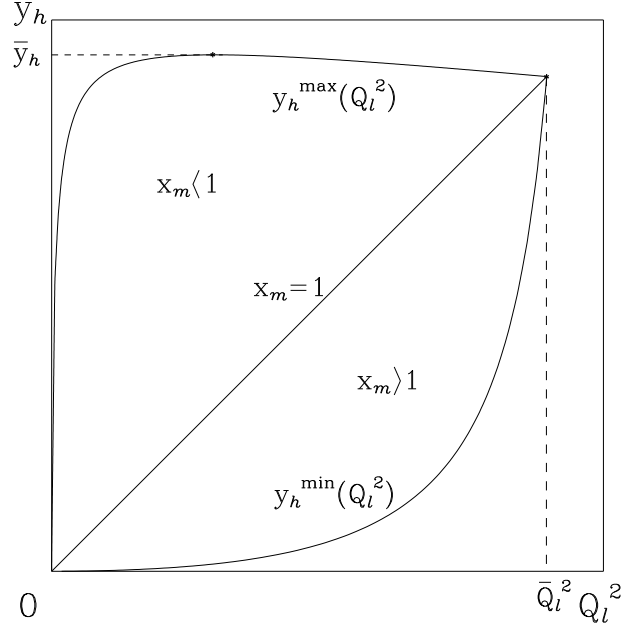}
}
\end{center}
\caption{\it
Physical region ($\yh,\ql$) for the cross section in mixed variables;
lower part for $\xm>1$, upper part for $\xm<1$.
\label{emix}
}
\end{figure}
 
Here the mixed variable $x_m$ is defined from both leptonic
and hadronic variables, while $y_m$ is defined exclusively by the
hadronic kinematics.
For $ \yh S \geq \Ql $, the scaling variable $x_m$ fulfills
$0 \leq x_m \leq 1$,
while at $ \yh S \leq \Ql $ it is $x_m \geq 1$.
 
Values of $x$  which are larger than 1 seem to be exceptional.
Nevertheless they are physical in the case of $\xm$.
This statement may be checked e.g. by reproducing the phase space
volume~(\ref{gami5}) in mixed variables.
 
The physical region
${\cal{E}}_m = (\ql,\yh)$
is shown in figure~\ref{emix}.
 
In the kinematical region with $\xm>1$ it is ensured that
$ \Ql - \Qh > 0 $ since
$ \yh S = \Qh +M_h^2-M^2 < \Ql $. Thus, the infrared
point may not be reached and only the radiative process is possible.
The equality in~(\ref{eq35}) is realized for $y_l^{\max}S= \Ql$;
then it is $y_h^{\min} S = \Ql$.
 
\subsubsection
{Integration variables
${\cal{I}}_m = (\yl,\qh)$
\label{b2z}
}
The {\em complete} analytical
integration of the hard part of the infrared divergent cross
section will be performed first over \qh\, and then over \yl.
We consider only the case $\xm \leq 1$ explicitly.
 
\begin{figure}[tbhp]
\begin{center}
\mbox{
\epsfysize=9.cm
\epsffile[0 0 530 530]{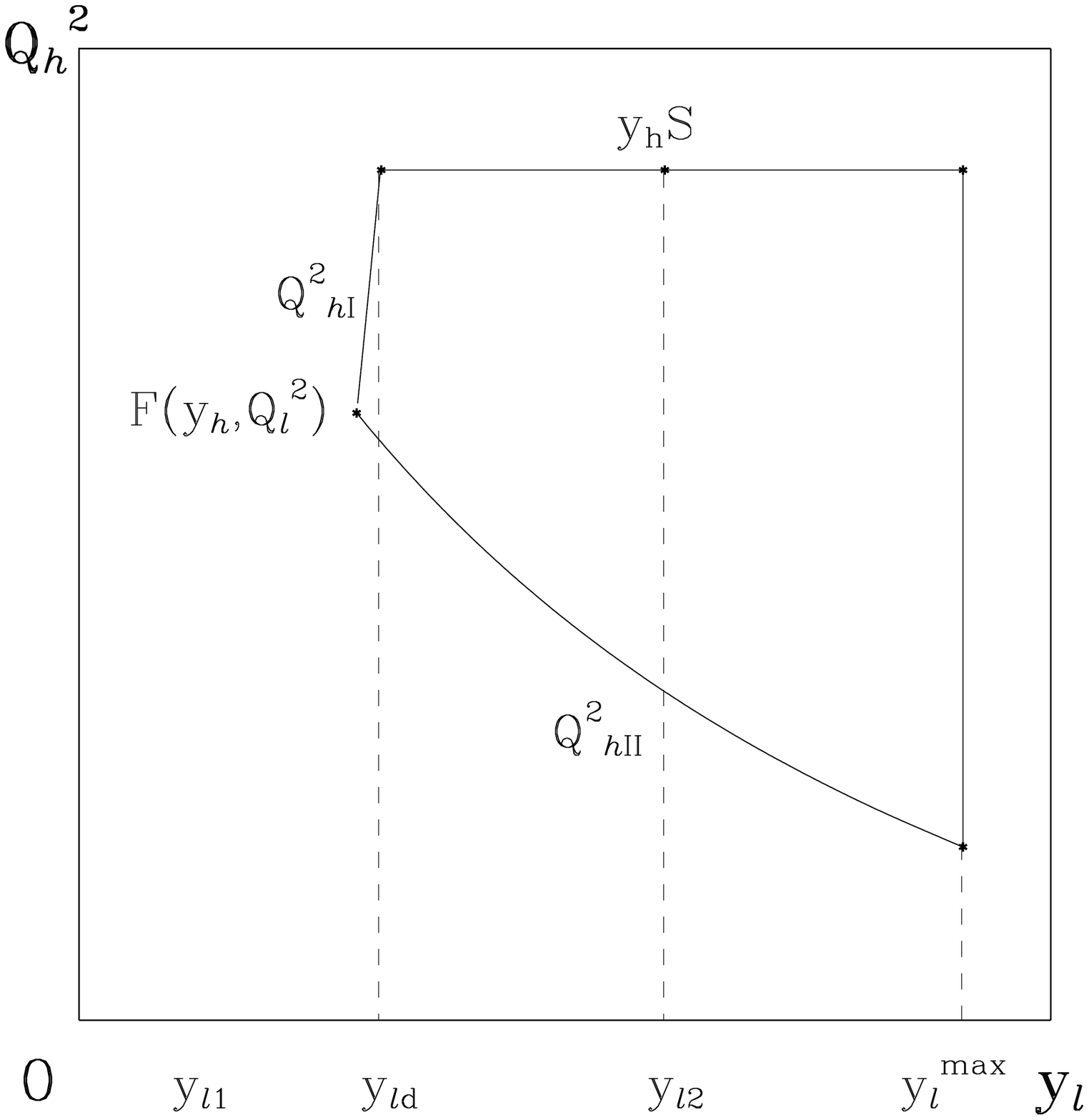}
}
\end{center}
\caption{\it
Integration region ($\yl,\qh$)
for the cross section in mixed variables, $\xm \leq 1$.
\label{ylqha}
}
\end{figure}

As in the case of leptonic variables, the bounds for \qh\ at given
values of $\yl,\yh,\ql$
may be derived from~(\ref{eq27}) and~(\ref{eq30}), and figure~\ref{ylqha}.
The
$\Qh^{\max}(\yl,\yh,\ql)$ is either equal to $\qh_{\mr I}$ or to $\yh S$;
both curves meet at the value $\yl_d$:
\ba
\yl_d &=& \frac{1}{2}  \yh \left[ 1 + \xm + ( 1 - \xm )
                             \sqrt{ 1 + \frac{4r}{\yh}} \right],
\ea
with $r=M^2/S$.
 
The lower bound
$\Qh^{\min}(\yl,\yh,\ql)$ is equal to $\qh_{\mr{II}}$:
\ba
\nonumber
\Qh_{\mr {I,II}}&=&\Ql + \frac{S}{2M^2}( \yl - \yh)
(\yl S \pm \sqrt{\lambda_q}).
\label{ea30}
\ea
 
The upper bound for \yl\ is dependent only on \ql; see~(\ref{eq25}):
\ba
\nonumber
y_l^{\max}(\ql) &=&
\frac{1}{2m^2} \left[ \frac{1}{S}\sqrt{\lambda_S\lambda_m} -
\Ql \right].
\label{eq25a}
\ea
The lower bound for \yl\ is reached at
the infrared singularity
${\cal F}(\yh,\ql).$
In the calculation of the hard part~(\ref{irhard}) of the infrared divergent
cross section contribution one has to take into account an additional
condition:
\ba
k^0 \geq \epsilon > 0.
\ea
The infrared problem is treated in the
proton rest frame. Corresponding to~(\ref{eq06}),
this condition may be transformed into the following
modification of the integration boundaries~(\ref{ea42}):
\ba
y_l^{\min} = \yh   \rightarrow
y_l^{\min} (\epsilon) = \yh + \frac{2 \epsilon M}{S}.
\label{epsh}
\ea
 
The integration boundaries are:
\ba
\begin{array}{rcccl}
\vph
\Qh_{\mr{II}}&\leq&\Qh&\leq&\yh S
\hspace{1.5cm} {\mr{if}} \, \, \, \yl \geq \yl_d,
\label{eq41a}
\\  \vph
\Qh_{\mr{II}}&\leq&\Qh&\leq& \qh_{\mr I}
\hspace{1.5cm} {\mr{if}} \, \, \, \yl \leq \yl_d,
\label{eq41c}
\\  \vph
y_l^{\min}(\epsilon)
&\leq&\yl&\leq&y_l^{\max}(\ql).
\end{array}
\label{ec42}
\ea
 
\begin{figure}[tbhp]
\begin{center}
\mbox{
\epsfysize=9.cm
\epsffile[0 0 530 530]{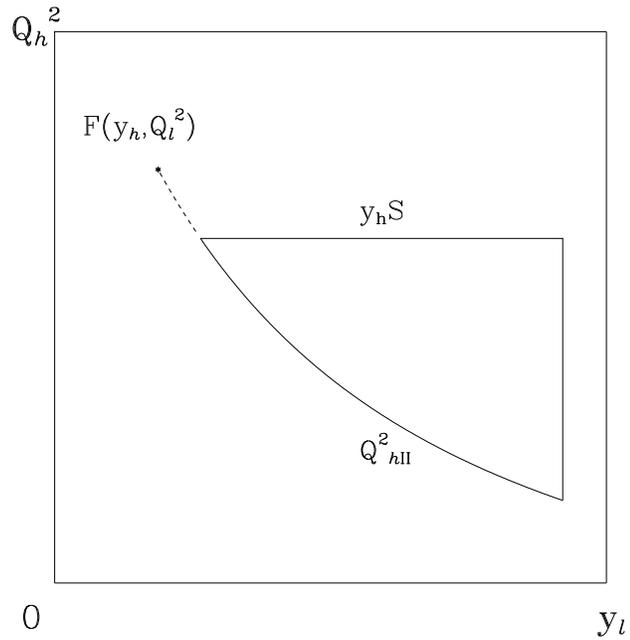}
}
\end{center}
\caption{\it
Integration region ($\yl,\qh$)
for the cross section in mixed variables, $\xm >1$.
\label{ylqhb}
}
\end{figure}

The integration region
${\cal{I}}_m = (\yl,\qh)$ is shown in figures~\ref{ylqha} and~\ref{ylqhb}.
 
In appendix~\ref{e32}, the square roots of the functions $C_1$ and $C_2$
will be needed.
In~appendix~\ref{b12}, we discussed these roots
at the integration boundaries for the case of leptonic variables.
Here, exactly the same discussion applies with inclusion of
the relations~(\ref{c33})--(\ref{D12}).
Although, there is one modification  which
concerns
the location of the zeros of $C_1$ and  $C_2$ at the boundary $\qh=\yh S$.
Differing from the leptonic case  here they are at some values of \yl\ for
fixed external values of \yh\ and \ql:
\ba
\yl_1 &=&  1 + \yh  - \frac{1}{\xm},
\nll
\yl_2 &=&  1 - \xm ( 1 - \yh).
\ea
They satisfy the inequalities
\ba
 \yl_1 < \yh < \yl_d < \yl_2 < y_l^{\max},
\ea
if additionally the condition $\yh < 1/(1+r)$ is fulfilled.
In this case, the $\yl_1$ does not belong to the physical interval
($\yh $, $\yl^{max} $) of variation of $\yl$, see figure ~\ref{ylqha}.
 
We only mention that one needn't know the values of the roots
of $C_1$ and  $C_2$ on the fourth boundary since the integration is performed along its
direction in the phase space.
\subsubsection
{Integration variables
${\cal I}_m^{'} = (\qh,\yl)$
\label{b22}
}
Here the boundaries                 for
the integration of the hard, infrared finite
part of the bremsstrahlung cross section are derived.
In this case,
the integrals over \yl\ are performed analytically in
appendix~\ref{appe4}.
The last one       over \qh\ has to be done numerically.
 
The lower limit of \yl\
at fixed values of $ \Ql $, $ \yh $, and \qh\
may be found from~(\ref{eq39}):
\bq
y_l^{\min}
(\yh,\qh,\ql) =
   \frac{1}{2\Qh} \left[ \yh(\Ql+\Qh) + \frac{1}{S} |\Ql-\Qh|
   \sqrt{\lambda_h} \right ].
\label{eq40}
\eq
There is only one solution if $\xm>1$ since then it is
definitely $\ql>\qh$.
For the more common case $\xm<1$ the solution has two
branches.
Geometrically, they coincide with the curves $\qh_{\mr{I,II}}$ in
figure~\ref{ylqha}.
The two curves~(\ref{eq40}) meet at the infrared point
${\cal F}(\yl,\qh) $.
This point marks the lowest allowed value for \yl\ since $\yl \geq \yh$;
see~(\ref{eq06}).
There is no need to exclude the infrared singularity from the integration
boundary, since the integrand has been regulated there.
The upper limit $y_l^{\max}(\yh,\qh;\ql)$
of \yl\ does not depend on \yh\ or \qh\ and is as derived
earlier [see~(\ref{eq25})]:
\ba
\nonumber
y_l^{\max}(\ql)
&=&\frac{1}{2m^2} \left[ \frac{1}{S}\sqrt{
\lambda_S \lambda_m}-\Ql \right] \leq {\bar{\yl}}.
\ea
 
The upper limit
of \qh\ derives from the inequality
$ M_h^2 \geq M^2 $ together with~(\ref{eq28}). It is independent of
the actual value of \ql:
$\qh^{\max}(\yh,\ql) = \yh S$.
The curves
$y_l^{\min}(\yh,\qh,\ql)$ and $y_l^{\max}(\ql)$ meet at
$\qh = \Qh^{\min}$; this condition may be used for the calculation of
$\qh^{\min}(\yh,\ql)$:
\ba
\Qh^{\min}(\yh,\ql)
&=&
\Ql + \frac{S}{2M^2} \left(y_l^{\max}-\yh\right)
\left[ y_l^{\max} S -\sqrt{(y_l^{\max} S)^2+4M^2 \Ql} \right]
\nll    \vph
&=&~
\ql - \frac{S\left(y_l^{\max}-\yh\right)
\left(\ql+\sqrt{\lambda_m}\right)}{S+\sqrt{\lambda_S}}.
\label{qhmin}
\ea
 
Taking all the above conditions together, the physical region for the
integration variables is found:
\ba
\begin{array}{rcccl}
\vph
\Qh^{\min}(\yh,\ql)&\leq&\Qh&\leq&\yh S,
\label{eq41}
\\  \vph
y_l^{\min}(\yh,\ql,\qh)&\leq&\yl&\leq&y_l^{\max}(\yh,\ql,\qh).
\end{array}
\label{eq42}
\ea
 
For the analytical integration over \yl\ in appendix~\ref{appe4},
we need the expressions for the roots of $C_1$ and $C_2$ at the integration
boundaries.
 
At  $y_l^{\max}$, they are:
\ba
   \sqrt{ C_{1}(y_l^{\max})}
&=&\frac{ {\sqrt{\lambda_S}}
               [Q^4_l  + 2 m^2 (\Ql+\Qh)]  -{\sqrt{\lambda_m}}
               S (\Ql + 2 m^2 \yh )}{2 m^2}    ,
\label{b84}
\\ 
   \sqrt{ C_{2}(y_l^{\max})}
&=&\frac{{\sqrt{\lambda_S}}
                     [ \Ql \Qh + 2 m^2 (\Ql+\Qh)]-{\sqrt{\lambda_m}}
                    S ( \Qh + 2 m^2 \yh )} {2 m^2},
\\ 
   \sqrt{ \lambda_{q}(y_l^{\max})}
&=&\frac{   S{\sqrt{\lambda_m}}
                  - {\sqrt{\lambda_S}} \Ql}{2 m^2}.
\label{b86}
\ea
 
Since the lower boundary of \yl\ has two branches,
the roots have
different analytical expressions there and
the integration region has to be divided into two subregions:
\begin{itemize}
\item
Region I: $ \Ql \geq \Qh $
\item
Region II: $ \Ql \leq \Qh $
\end{itemize}
The following relations hold:
\ba
\sqrt{ C_{1}(y_l^{\min})}
^{\,\mr{I}}
&=& \frac{(\Ql-\Qh)
             \left[ 2 S \Qh-\Ql(\TT+{\sqrt{\lambda_h}})\right]}{2 \Qh} ,
\label{b87}
\\
   \sqrt{ C_{2}(y_l^{\min})}^{\, \mr{I}}
&=& \frac{(\Ql-\Qh)
             \left[2 S \Qh-\Qh (\TT-
                 {\sqrt{\lambda_h}})\right]}{2 \Qh},
\\
   \sqrt{ \lambda_{q}(y_l^{\min})}^{\, \mr{I}}
             &=&
\frac{\TT (\Ql-\Qh)+(\Ql+\Qh)
                 {\sqrt{\lambda_h}}}{2 \Qh},
\\
   \sqrt{ C_{1}(y_l^{\min})}^{\, \mr{II}}
&=& \frac{(\Qh-\Ql)
                   \left[2 S \Qh-\Ql
                    (\TT-{\sqrt{\lambda_h}})\right]}{2 \Qh},
\\
   \sqrt{ C_{1}(y_l^{\min})}^{\, \mr{II}}
&=&\frac{(\Qh-\Ql)
                   \left[2 S \Qh-\Qh
                    (\TT+{\sqrt{\lambda_h}})\right]}{2 \Qh},
\\
   \sqrt{ \lambda_{q}(y_l^{\min})}^{\, \mr{II}}
&=&\frac{\TT
        (\Qh-\Ql) + (\Ql+\Qh) {\sqrt{\lambda_h}}}{2 \Qh}.
\label{b92}
\ea
All these roots are positive definite.
Again, the values of the roots at the fourth boundary are not needed
since the integration is performed along its direction.
\subsection
{Boundaries for hadronic variables
\label{apph}
}
\subsubsection
{External variables
${\cal E}_h = (\yh,\Qh)$
\label{b31}
}
The physical region of ${\cal E}_h = (\yh,\Qh)$ may be derived from the
boundary condition ~(\ref{eqc5})
and from the inequality
\bq
0 \leq \yh S -\Qh = W^2-M^2 \leq {\bar{W}}^2 - M^2,
\label{eqc2}
\eq
where ${\bar{W}}^2$ may be taken from~(\ref{eq21}).
This proceeds exactly as in the case of the leptonic variables,
by a replacement of $(\yl,\ql)$ by $(\yh,\qh)$ everywhere in the
argumentations.
 
One gets
for ${\cal E}_h$ the following physical region:
\ba
\begin{array}{rcccl}
\vph
0&\leq&\Qh&\leq&
\frac{
\displaystyle
\lambda_S} {
\displaystyle
S+M^2+m^2} \equiv
{\bar{Q}_h^2},
\label{hqh1}
\\
y_h^{\min}(\qh) = \frac{
\displaystyle
Q_h^2}{
\displaystyle
S}&\leq&\yh&\leq&y_h^{\max}(\qh),
\label{hyh1}
\end{array}
\ea
where
\bq
y_h^{\max}(\qh)
= \frac{1}{2m^2} \left[ \frac{1}{S}\sqrt{\lambda_S \qh(\qh+2m^2)}
-
\Qh \right] \leq {\bar{y}_h}.
\label{new1}
\eq
Here, the ${\bar{y}_h}$ is the maximal possible value of $\yh$:
\ba
{\bar{y}_h}= \frac{S-2mM}{S}.
\label{new2}
\ea
 
The physical region
${\cal E}_h = (\qh,\yh)$ coincides with that for the case of leptonic
variables, which is shown in figure~\ref{xxxxx}.
\subsubsection
{Integration variables
${\cal I}_h = (\yl,\ql)$
\label{b3g}
}
Similar to the case of the mixed variables, for the determination of the
hard part of the infrared divergent contribution
we first integrate over \ql\ and afterwards over \yl.
This will be done in appendix~\ref{sshv}.
 
Two bounds for \ql\ at given values of $\yl,\yh,\qh$ may be derived
{}from~(\ref{eq30}) by interchanging \yl\ with \yh\ and \ql\ with \qh\ as was
indicated in connection with~(\ref{eq39}),
\ba
\Ql_{\mr{I,II}} &=&
\Qh + \frac{S}{2 M^2}(\yl- \yh)
      \left( \pm \sqrt{\lambda_h } - \yh S\right),
\ea
and two others from~(\ref{eq22}),
\ba
\Ql_{\mr{III,IV}}
&=&
\frac{1}{2M^2}
\left[
(1-\yl)S^2 -4m^2M^2\pm\sqrt{\lambda_S\lambda_l}
\right]
.
\ea
As may be seen in figure~\ref{yhqh}, the absolute maximum
${\bar{y}}_l$ of $\yl$ divides the curved boundary into two parts.
The ${\bar{\yl}}$ is defined in~(\ref{baryl}).
It is the root of $\lambda_l$:  $\lambda_l({\bar y}_l)=0$.
Thus, the two curves $\Ql_{\mr{III,IV}}$ meet there.
Further, the minimum of \yl\ at given \yh\ is equal to \yh\ and independent of
\qh.
 
When calculating the hard part~(\ref{irhard}) of the infrared divergent
cross section contribution, one has to take into account an additional
condition:
\ba
k^0 \geq \epsilon > 0.
\ea
The infrared problem is treated in the
proton rest frame. With the expression for $k^0$
in~(\ref{eq06}),
this condition may be transformed into the following
modification of the lower bound for \yl:
 
\begin{figure}[tbhp]
\begin{center}
\mbox{
\epsfysize=9.cm
\epsffile[0 0 530 530]{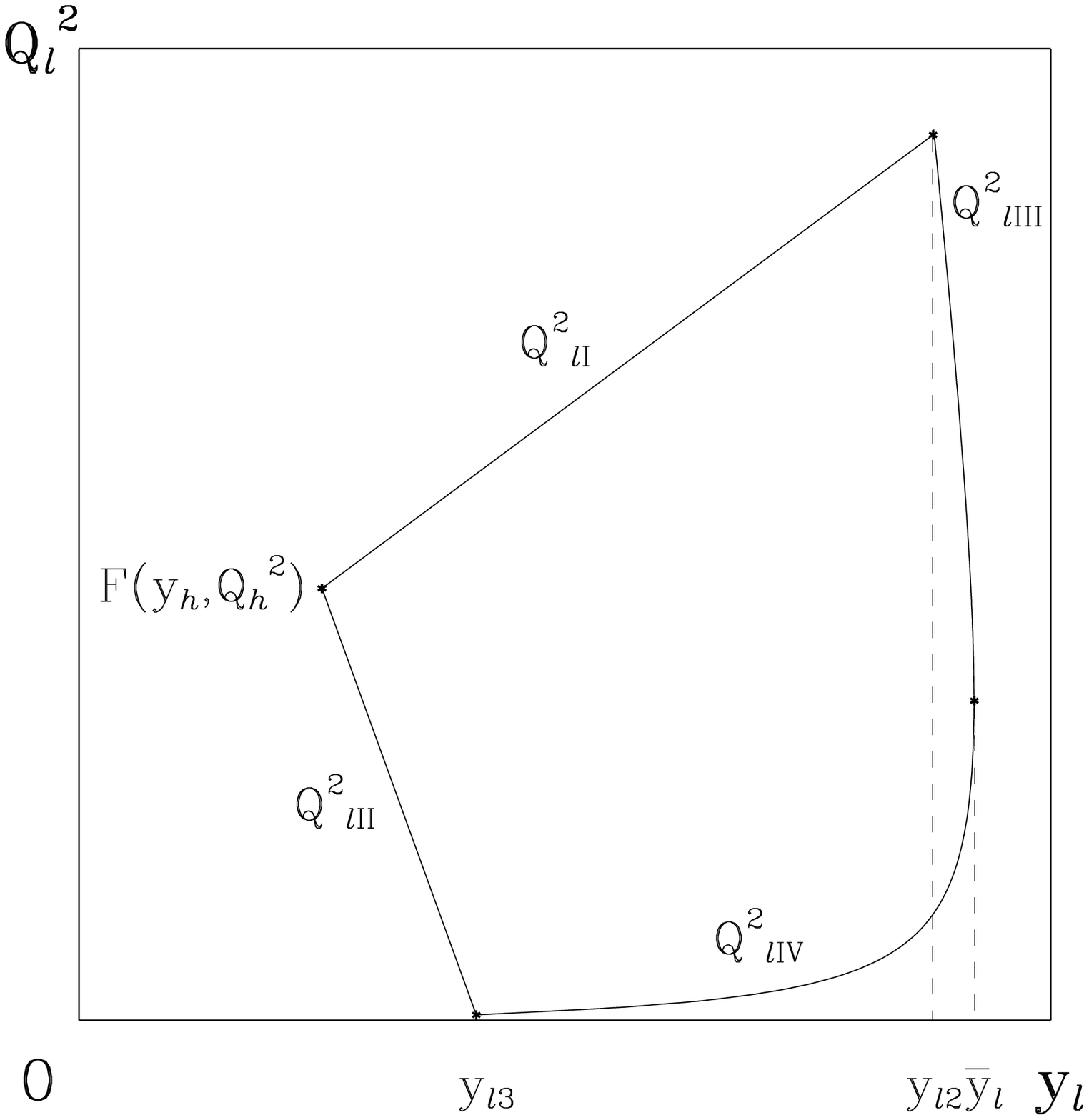}
}
\end{center}
\caption{\it
Integration region ($\yl,\ql$) for the cross section in hadronic
variables.
\label{yhqh}
}
\end{figure}
 
\ba
y_l^{\min} = \yh   \rightarrow
y_l^{\min} (\epsilon) = \yh + \frac{2 \epsilon M}{S}.
\label{e2sh}
\ea
For a complete definition of the integration region one needs yet the points
where each of the two straight boundaries meets one of the two curved.
The points $\yl_3$ and $\yl_2$ are determined from the conditions
$\Ql_{\mr{II}}=\Ql_{\mr{IV}}$ and $\Ql_{\mr{I}} = \Ql_{\mr{III}}$,
respectively:
\ba
\yl_{2(3)}
&=&
\frac{
[(1-\yh)(S\pm\sqrt{\lambda_S}+2\yh S)S-2M^2(\qh+2m^2)]
(\sqrt{\lambda_h}\mp\yh S)
}
{
2S[(1-\yh)(\sqrt{\lambda_h}\mp\yh S)S\pm2M^2(\qh+m^2)]
}
\nll
& &
\mp~\frac
{
2M^2\qh[(1-2\yh)S\mp\sqrt{\lambda_S}]
}
{
2S[(1-\yh)(\sqrt{\lambda_h}\mp\yh S)S\pm2M^2(\qh+m^2)]
}
.
\label{y23}
\ea
In the ultra-relativistic approximation,
\ba
\yl_2
&\approx&
\frac{
2S+\yh(\sqrt{\lambda_h}-\yh S)-2r\qh
}{
2S+\sqrt{\lambda_h}-\yh S
},
\\
\yl_3 &\approx&
\yh + \frac{2r\qh}{\yh S+\sqrt{\lambda_h}}.
\ea
 
The integration limits may be summarized as follows:
\ba
\begin{array}{rcllccl}
\ql^{\min}(\yl,\yh,\qh)&=&
\Ql_{\mr{II}} \hspace{1.cm}&{\mr{if}} \, \, \,  y_l^{\min}(\epsilon)
&\leq&\yl&\leq \yl_3,
\\
\ql^{\min}(\yl,\yh,\qh)&=&
\Ql_{\mr{IV}} \hspace{1.cm}&{\mr{if}} \, \, \,  \yl_3&\leq&\yl&\leq {\bar y}_l,
\\
\ql^{\max}(\yl,\yh,\qh)&=&
\Ql_{\mr{I}} \hspace{1.cm}&{\mr{if}} \, \, \,   y_l^{\min}(\epsilon)
&\leq&\yl&\leq \yl_2,
\\
\ql^{\max}(\yl,\yh,\qh)&=&
\Ql_{\mr{III}} \hspace{1.cm}&{\mr{if}} \, \, \,  \yl_2&\leq&\yl&\leq
{\bar{\yl}}.
\label{hsec}
\end{array}
\ea
 
The integration region ${\cal I}_h = (\yl,\ql)$
is shown in figure~\ref{yhqh}.
The infrared divergence is located at ${\cal F}
(\yh,\qh) $.
 
In the course of the integrations, the values of the roots of $C_1$ and
$C_2$ are needed at the boundaries.
They may be determined exactly:
\ba
\sqrt{C_1(\ql_{\mr{I,II}})}
&=&
(\yl-\yh)S\left[(1-\yl+\yh)S\frac{\sqrt{\lambda_l}\mp\yh S}{2M^2}
\mp\qh\right],
\\
\sqrt{C_2(\ql_{\mr{I,II}})}
&=&
(\yl-\yh)S\left[(1-\yh)S\frac{\sqrt{\lambda_l}\mp\yh S}{2M^2}
\pm\qh\right],
\\
\sqrt{C_1(\ql_{\mr{III,IV}})}
&=&
-\frac{1}{2M^2}
\Biggl\{
\sqrt{\lambda_S}\left[(1-\yl+\yh)(1-\yl)S^2-2M^2(\qh+2m^2)\right]
\nll
& &
\pm~
\sqrt{\lambda_l}\left[(1-\yl+\yh)S^2 -4m^2M^2\right]
\Biggr\},
\\
\sqrt{C_2(\ql_{\mr{III,IV}})}
&=&
 \frac{1}{2M^2}
\Biggl\{
\sqrt{\lambda_S}\left[(1-\yl)(1-\yh)S^2-4m^2M^2\right]
\nll & &
\pm~
\sqrt{\lambda_l}\left[(1-\yh)S^2-2M^2(\qh+2m^2)\right]
\Biggr\}.
\label{c12l}
\ea
\subsubsection
{Integration variables
${\cal{I}}_h^{'} = (\ql,\yl)$
\label{b32}
}
Here, the
            boundaries for the integration over leptonic variables will be
derived  for the case where
the integration over \yl\ is performed before that over \ql.
This is needed in appendix~\ref{appe4}
                           and will be used
for the calculation of  the finite hard part of the cross section.
We remind that in this case
the boundaries for \yl\
are the same as in the case of the mixed variables.
Thus, the minimal value of \yl\ for given values of \ql, \yh, and \qh\
is given by~(\ref{eq40}),
\ba
\nonumber
y_l^{\min}
(\yh,\qh,\ql) =
   \frac{1}{2\Qh} \left[ \yh(\Ql+\Qh) + \frac{1}{S} |\Ql-\Qh|
   \sqrt{\lambda_h} \right ],
\label{eq4a}
\ea
and the maximal value by~(\ref{eq25}):
\ba
\nonumber
y_l^{\max} (\yh,\qh,\ql)
= \frac{1}{2m^2} \left[ \frac{1}{S}\sqrt{\lambda_S\lambda_m} -
\Ql \right] \leq {\bar{\yl}}.
\label{eq2a}
\ea
For fixed $\yh,\Qh$, one may see in figure~\ref{yhqh}  that
$y_l^{\max}$
and $y_l^{\min}$ are equal for the extreme values of \ql.
Solving the resulting equation, one gets:
\ba
\Ql^{\min(\max)}(\yh,\qh) =
\frac {Q_h^4 \left(S \mp \sqrt{\lambda_S} \right)
\left(S \mp \sqrt{\lambda_S} -\yh S \mp \sqrt{\lambda_h} \right)}
{ 2S\left(\yh S \mp \sqrt{\lambda_h} \right) \left( \Qh
+ m^2 \yh  \right) + 4M^2 \Qh(\Qh+m^2)}.
\label{eqc10}
\ea
Concluding, the physical region of ${\cal I}_h^{\prime} = (\Ql,\yl)$ is:
\ba
\begin{array}{rcccl}
\vph
\Ql^{\min}(\yh,\qh) & \leq & \Ql & \leq & \Ql^{\max}(\yh,\qh),
\label{eqc11}
\\ \vph
y_l^{\min}(\ql;\yh,\qh) &\leq&\yl&\leq& y_l^{\max}(\ql;\yh,\qh).
\end{array}
\label{eqc12}
\ea
\subsubsection
{Variables in section~7:
${\cal E}_h = (\yh,\Qh)$,
${\cal I}_h = (\tau)$
\label{b33}
}
In this section, we restrict ourselves to the ultra-relativistic
approximation.
 
The physical region for the external variables is as derived
in section~\ref{b31}:
\ba
\begin{array}{rcccl}
           \vph
0 &\leq&y_h&\leq&  1,
\label{qhyh1}
\\
0 &\leq&Q_h^2 &\leq& y_h S.
\label{qhyh2}
\end{array}
\ea
 
The limits for $\tau=z_2+m^2$ at given values of \yh\ and \qh\
follow from those for $z_2$ which are derived
in appendix~\ref{c14}.
For the case of the hard photon contribution to the infrared singular
cross section part one has also to take into account
the condition
\ba
k^0 \geq \epsilon > 0.
\ea
The infrared problem is treated in the rest frame of the $(\gamma e)$ compound
system.
There, this condition may be fulfilled by the demand
\ba
z_2 \geq {\bar z}_2 = 2m\epsilon.
\label{z2e}
\ea
The boundaries are then:
\ba
m^2 + 2m\epsilon  \leq \tau \leq m^2 +
\xh (1-\yh)S \approx
\xh (1-\yh)S.
\label{tau2}
\ea
At the lower boundary,
one has to retain the influence of the electron mass
in order to ensure the positive definiteness of the phase space measure
and to prevent unphysical singularities during the integrations.
\subsection
{
Boundaries for
Jaquet-Blondel variables:  \, \, \, 
\mbox{
${\cal{E}}_{\mr{JB}}=(\yjb,\qjb)$,}
\mbox{
${\cal{I}}_{\mr{JB}}=(\tau)$}
\label{appjb}}
In this section, we restrict ourselves to the ultra-relativistic
approximation.
The definitions of the Jaquet-Blondel variables are:
\ba
\yjb &=& \yh,
\\
\qjb
&=& {\cal N} ({\vec p}_{2\perp})^2.
\label{yxjb}
\ea
The normalization $\cal N$
has to be adjusted such that in the limiting
case of the Born kinematics it will hold $\qjb = \qh$.
We will immediately see that
\ba
{\cal N} = \frac{1}{1-\yh + \ldots},
\label{njb}
\ea
where the dots stand for some terms proportional to $m^2$ and $M^2$.
An exact expression for $\qjb$
in terms of $\yh,\qh, \tau$ may be obtained in the rest frame of the
proton. In the momentum tetrahedron of figure~\ref{tetra}
it is           $\vec{k_1} - \vec{p_2} = \vec{\Lambda}$.
In this system, these three three-vectors span a triangle whose
area may be expressed in two ways:
\ba
 \frac{1}{2} |\vec{k_1}| |{\vec p}_{2\perp}|
=
\sqrt{-\frac{1}{4}\lambda(\vec{k_1}^2, \vec{p_2}^2, \vec{\Lambda}^2)}.
\label{exqj}
\ea
The three three-vectors squared are expressed by invariants
in~(\ref{eq06}):
\ba
({\vec p}_{2\perp})^2
=
-\frac{1}{4 M^2 \lambda_S} \lambda(\lambda_S, \lambda_h, \lambda_{\tau}).
\label{p2perp}
\ea
With this expression, the \qjb\ is determined after the normalization
has been adjusted such that for $z_2=0$ the \qjb\ becomes \qh:
\ba
\qjb
&=& \qh \frac{[\qh - \yh (\qh+z_2)]S^2 -M^2(\qh+z_2)^2-m^2\lambda_h}
{(1-\yh)\qh S^2-M^2 Q_h^4 - m^2 \lambda_h}.
\label{qjbex}
\ea
If the electron mass is neglected  the \qjb\ becomes
\ba
\qjb = \frac{(1-\yh)\qh - \yh z_2 - (\qh+z_2)^2 r/S}
{(1-\yh) - \qh r/S},
\label{qjbm}
\ea
where it is $r=M^2/S$.
Neglecting also the proton mass, one finally obtains
\ba
\qjb
=
Q_h^2 - \frac{\yh}{1-\yh} z_2.
\label{qjexa}
\ea
The hadronic and Jaquet-Blondel--transferred momentum squares are related by
$\yh=\yjb$ and the
  additional variable $z_2$.
This property may be used for a derivation of the kinematical boundaries.
We start from the hadronic integration limits  which were derived
in appendix~\ref{b33}.
A chain of simple changes of the integration region allows
a derivation of the boundary values for Jaquet-Blondel variables:
\ba
\begin{array}{rcccl}
\Vph
\Gamma_3
&=&
\displaystyle
\int \limits _0^{S} d\qh
\int \limits _{Q_h^2/S}^1 d\yh
 \int \limits _0^{(1-y_h)Q_h^2/y_h} dz_2
&=&
\displaystyle
\int
\limits
_0^1 d\yh  \int
\limits
_0^{y_h S} d\qh  \int\limits
 _0^{(1-y_h)Q_h^2/y_h} dz_2
\nll \Vph
\displaystyle
&=&
\displaystyle
\int\limits
_0^1 d\yh \int\limits
_0^{(1-y_h)S} dz_2 \int\limits
 _{y_h z_2 / (1-y_h)}^
{y_h S} d\qh
&=&
\displaystyle
\int\limits
_0^1 d\yjb \int\limits
_0^{(1-y_h)S} dz_2 \int\limits
_{0}^{y_h S-
 y_h z_{2} / (1-y_h)} d\qjb
\nll \Vph
\displaystyle
&=&
\displaystyle
\int\limits
_0^1 d\yjb  \int\limits
_0^{y_h S} d\qjb \int\limits
_0^{(1-x_{\mr{JB}})(1-y_{\mr{JB}})S} dz_2
&=&
\displaystyle
\int\limits _0^{S} d\qjb
\int\limits ^1_{Q_{\mr{JB}}^2/S}d\yjb
\int\limits _0^{(1-x_{\mr{JB}})(1-y_{\mr{JB}})S} dz_2.
\label{limjd}
\nll
\end{array}
\nll
\ea
Thus, we derived:
\ba \vph
0 \leq & \qjb & \leq S,
\label{bounj1}
\\      \vph
\frac{\qjb}{S} \leq & \yjb & \leq 1,
\label{bounj2}
\\         \vph
m^2 + 2m\epsilon  \leq &\tau &\leq
m^2 + (1-\xjb)(1-\yjb)S
\nll & &
\approx
(1-\xjb)(1-\yjb)S.
\label{boutau}
\ea
Here we took into account that the lower bound for the $\tau$ integration
is modified in case of the hard contribution to the infrared singular
part; see the corresponding remarks in the foregoing appendix.
 
\begin{figure}[tbhp]
\begin{center}
\mbox{
\epsfysize=9.cm
\epsffile[0 0 530 530]{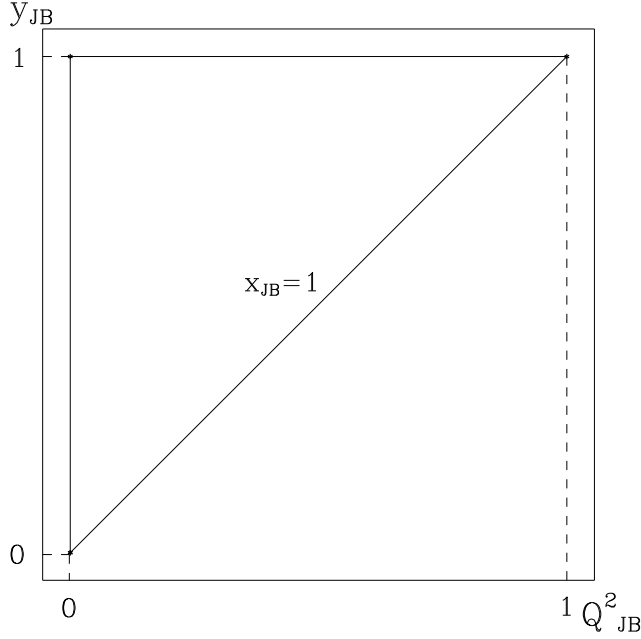}
}
\end{center}
\caption{\it
Physical region ($\qjb,\yjb$) for the cross section in Jaquet-Blondel
variables.
\label{exjb}
}
\end{figure}

The physical region ${\cal E}_{\mr{JB}} = (\qjb,\yjb)$
is shown in figure~\ref{exjb}.
 
\section
{The phase space volume
\label{b15}
}
\ezero
In the above appendices, many boundaries for external variable sets and
for integration variables have been introduced in different combinations
of variables and in different order.
A useful tool for numerical
checks of the correctness of all these boundaries is
the integral over the phase space volume, which must be identical
for all the different presentations.
 
The phase space
volume~(\ref{gamie}) is calculated now
in terms of the variables which were introduced for
the study of the photoproduction process:
\ba
\Vph
\Gamma
&=& \frac{\pi S^2}{4 \sqrt{\lambda_S}}
\int d\yl \, d\ql \,
d\yh \, d\qh \frac{dz_{2(   1)}}{\sqrt{R_z}}
\nll \Vph
&=&  \frac{\pi}{4 \sqrt{\lambda_S}}
\int
_{(M+m_{\pi})^2}^{(\sqrt{s}-m)^2}  dW^2
\int
_{Q_l^{2\min}}^{Q_l^{2\max}} d\ql
\int
_{(M+m_{\pi})^2}^{W^2} dM_h^2
\int
_{Q_h^{2\min}}^{Q_h^{2\max}} d\qh
\int
_{z_2^{\min}}^{z_2^{\max}} \frac{dz_{2}}{\sqrt{R_z}}.
\label{gami2}
\ea
The boundaries for $z_2$ are~(\ref{ax}), those for $\qh$
are~(\ref{fotqhm}), and those for $\ql$ are~(\ref{q2le}).
We first apply~(\ref{eqe2}). Next, the integration over \qh\
yields
\ba
\int d\qh = \qh^{\max} - \qh^{\min},
\label{ga2a}
\ea
which together with the result of the first integration and with
        ~(\ref{eq09}), $A_2=\lambda_q$, leads to
fortunate cancellations:
\ba
\Gamma
&=&  \frac{\pi}{4 \sqrt{\lambda_S}} \int
 dW^2 \, d\ql \, dM_h^2 \frac{W^2-M_h^2}{W^2}.
\label{gami3}
\ea
The two subsequent integrals over $M_h^2$ and \ql\ are also trivial;
after neglecting $m_{\pi}$:
\ba
\Gamma
&=&  \frac{\pi}{8s} \int
 dW^2 \frac{(W^2-M^2)^2}{W^2} \sqrt{(s-W^2+M^2)^2-4m^2W^2}.
\label{gami4}
\ea
The last integration yields
\ba
\Gamma
&=&
\frac{\pi^2}{8S} \Biggl\{
\frac{\sqrt{\lambda_S}}{6} \left[ s^2+10m^2s-5M^2s+m^4-5m^2M^2-2M^4
 \right]
\nll
\vph
& &~-\left[ m^2s(s+m^2-2M^2) +\frac{1}{2}(s+m^2)M^4 \right]
\ln \frac{s+m^2-M^2+\sqrt{\lambda_S}} {s+m^2-M^2-\sqrt{\lambda_S}}
\nll
\vph
& &~+\frac{1}{2}(s-m^2)M^4
\ln
\frac{
\left(s-m^2+M^2+\sqrt{\lambda_S}\right)
\left(s-m^2-M^2+\sqrt{\lambda_S}\right)}
{
\left(s-m^2+M^2-\sqrt{\lambda_S}\right)
\left(s-m^2-M^2-\sqrt{\lambda_S}\right)
}
\Biggr\},
\ea
with $s=S+m^2+M^2$.
Neglecting the electron mass $m$
 one gets
\ba
\vph
\Gamma &\approx&
\frac{\pi^2}{8s}\left[ \frac{1}{6}(s^2-5sM^2-2M^4)(s-M^2)+
sM^4\ln\frac{s}{M^2} \right],
\ea
 and after a subsequent neglection of the proton mass
\ba
\vph
\Gamma &\approx& \frac{\pi^2}{48} s^2.
\label{gami5}
\ea
\section
{\label{appe}
Tables of integrals for the approach of section~3
}
\ezero
\subsection
{First series of $z_{1(2)}$-integrals
\label{appe1}
}
This first series of integrals over $z_1$ or $z_2$ is used for the
common treatment of the cross section in leptonic, mixed, and
hadronic variables.
The first integration is over $z_1$ or $z_2$.
It corresponds
to the angular integration over the angle $\varphi$ of the photon in the
rest system of the proton as discussed in~\ref{appa1}. At
this stage of the integration, there is no         infrared
problem. The necessary integrals are:
\ba
[{\cal A}]_{z} &=& \frac{1}{\pi} \int_{z_{1(2)}^{\min}}^
{z_{1(2)}^{\max}}
\frac{dz_{1(2)}}{\sqrt{R_Z}} {\cal A},
\label{aint}
\\
\left[ 1 \right]_{z}&=& \frac{1}{\sqrt{ A_{1,2}}},
\label{eqe2}
\\
\left[\frac{1}{z_{1(2)}}\right]_{z}&=& \frac{1}{\sqrt{C_{1(2)}}},
\label{eqe3}
\\
\left[ \frac{1} {z_{1(2)}^2} \right]_{z}&=& \frac {B_{1(2)}}
{C_{1(2)}^{3/2}}.
\label{eqe4}
\ea
Using the relation $z_1=z_2+\qh-\ql$, one may derive
\ba
\frac{1}{z_1z_2} =
\frac{1}{\qh-\ql} \left(\frac{1}{z_2} -\frac{1}{z_1} \right),
\ea
and thus reduce the corresponding  integral type to known ones.
The functions $A,B,C$ are defined in~(\ref{eq09})--(\ref{eq11}).
The integration boundaries $z_{1(2)}^{\max(\min)}$
may be found in~(\ref{ax}).
 
In appendix~\ref{appd}, we perform the integration over
$z_{1(2)}$ with dedicated cuts for the photonic energy and
scattering angle.
In this case, the above table of integrals becomes modified:
\ba
\left[ 1 \right]
_{\hat z}
&=&
- \frac{1}{\pi \sqrt{\lambda_q}} {\mr{asin}}
\frac{B_{1(2)}-z_{1(2)}\lambda_q}{\sqrt{D_z}}
\left|_{\hat z},  \right.
\label{eqe2a}
\\
\left[\frac{1}{z_{1(2)}}\right]_{\hat z}&=&
\frac{1}{\pi \sqrt{C_1}}{\mr{asin}}
\frac{z_{1(2)}B_{1(2)}-C_{1(2)}}{z_{1(2)}\sqrt{D_z}}
\left|_{\hat z},         \right.
\\
\left[ \frac{1} {z_{1(2)}^2} \right]_{\hat z}&=&
\frac{1}{\pi C_1}  \frac{\sqrt{R_z(z_{1(2)})}}{z_{1(2)}}
\left|_{\hat z}  \right.
+ \frac{B_{1(2)}}{C_{1(2)}}
\left[\frac{1}{z_{1(2)}}\right]_{\hat z}.
\ea
In the table, the abbreviation
\ba
\left|_{\hat z}  \right.
\equiv \left|_{ {\hat z}_{1(2)}^{\min}} ^{ {\hat z}_{1(2)}^{\max}}
\right.
\label{hatz}
\ea
is used, where the boundaries ${\hat z}_{1(2)}^{\min,\max}$
have been derived in~\ref{appd}.
Further, we use~(\ref{rz}) in the  slightly rewritten form~(\ref{rzm})
and a similar relation holds for $D_z$:
\ba
D_z
&=&
4sW^2
\left[ \ql^{\max}(W^2) - \ql \right]
\left[ \ql - \ql^{\min}(W^2) \right]
\nll & &~
\times~
\left[ \qh^{\max}(W^2,\ql,M_h^2) - \qh \right]
\left[ \qh - \qh^{\min}(W^2,\ql,M_h^2) \right] .
\label{dzm}
\ea
\subsection
{The integrals for the infrared problem
\label{c2}
}
For the angular integration,
\ba
[{\cal A}]_{\xi}
&=&
\frac{1}{2}\int_{-1}^{1} d \xi
{\cal A},
\label{intxi}
\ea
one needs the following integrals which are calculated with
account of the exact kinematics:
\ba
\left[ {1} \right]_{\xi} &=& 1,
\\
\left[ \frac{1}
{1-\beta_i \xi}
 \right]_{\xi} &=&
 \frac{1}{2\beta_i}
\ln \frac{1+\beta_i}{1-\beta_i},
\\
\left[ \frac{1}
{(1-\beta_i \xi)^2}
 \right]_{\xi} &=& \frac{1}{1-\beta_i^2},
\\
\left[\ln (1-\xi^2) \right]_{\xi} &=& 2\ln2 - 2,
\\
\left[ \frac{\ln (1-\xi^2)}
{1-\beta_i \xi}
\right]_{\xi} &=&
\ln2 \frac{1}{\beta_i} \ln \frac{1+\beta_i}{1-\beta_i}
+
\frac{1}{2\beta_i} \left[
\litwo\left(\frac{2\beta_i}{\beta_i-1})\right)
-
\litwo\left(\frac{2\beta_i}{\beta_i+1})\right) \right] ,
\\
\left[ \frac{\ln (1-\xi^2)}
{(1-\beta_i \xi)^2}
\right]_{\xi} &=& \frac{1}{1-\beta_i^2}
\left( 2 \ln2 - \frac{1}{\beta_i}
\ln \frac{1+\beta_i}{1-\beta_i}
\right).
\label{irt1}
\ea
Since the definition of the velocities $\beta_i$ depends on the lorentz
system, some of the above integrals do so.
 
Further,                       integrals over the Feynman parameter
$\alpha$ are used:
\ba
[{\cal A}]_{\alpha}
&=&
\int_{0}^{1} d \alpha {\cal A}.
\label{intx2}
\ea
The first one is
\ba
\left[ \frac{1}{-k_{\alpha}^2} \right] _{\alpha}
&=&
2 \lm.
\label{lm}
\ea
with
\ba
-k_{\alpha}^2
= Q^2\alpha(1-\alpha) + m^2,
\label{mikal}
\ea
and the $\lm$ is defined in~(\ref{lmlamm}).
Due to the invariance of $
-k_{\alpha}^2$, the integral~(\ref{lm}) is relativistic invariant.
 
A second integral of the type~(\ref{intx2}) occurs
in~(\ref{*}). It
depends on the system  in which it is calculated:
\ba
\left[ \frac{1}{
\beta_{\alpha}
 (-k_{\alpha}^2)}
\mbox{ln}{ \frac{1-\beta_{\alpha}}{1+\beta_{\alpha}} }
\right]_{\alpha}
&=&
\frac{2}{Q^2+2m^2}
{\cal S}_{\Phi},
\label{sfi0}
\\ 
{\cal S}_{\Phi}
&=& \frac{1}{2} \left(Q^2+2m^2\right)
 \int_{0}^{1} \frac {d\alpha}{ \beta_{\alpha}(-k_{\alpha}^2)} \;
  \mbox{ln}{ \frac{1-\beta_{\alpha}}{1+\beta_{\alpha}} } .
\label{soft3}
\ea
The exact answer for this type
of integrals may be found following the hints in~\cite{BS}.
For the proton rest system,
the explicit result in the ultra-relativistic limit is~(\ref{soft4}).
For the case of photoproduction, which was calculated in another rest
system, the exact expression has been calculated numerically.
Finally, we mention that for the case of the $(\gamma e)$ rest frame,
which was used e.g. in the case of Jaquet-Blondel variables,
the kinematical simplifications are such that the $S_{\Phi}$-function
need not be introduced; see section~\ref{hjbx}.
\subsection{The second and third integrations for
$\delta_{\mr{ {hard}}}^{\mr{ IR}} $
\label{appe3}
}
In this appendix, integrals over (first) $Q^2$ and (then) $y$ are collected.
They are used for the calculation of the hard part of the infrared
divergent cross section contribution.
\subsubsection
{Leptonic  variables
\label{appe31}
}
The necessity to work out the absolute values in~(\ref{DD12})--(\ref{D12})
explicitly for a subsequent integration over $ y_h$
forces us to perform the integrations
in the following four regions
 separately:
\ba
\begin{array}{lccccclcrcccl}
{\mr{Region~I:}}
&\hspace{.7cm}&
y_h^{\min}
& < &
\yh
& < &
\yh_1,
&\hspace{.7cm}&
\qh_{\mr {II}}
&\leq&
\Qh
&\leq&
\yh S,
\nll
{\mr {Region~II:}}
&\hspace{.7cm}&
\yh_1    &<& \yh &<&   \yh_2,
&\hspace{.7cm}&
\qh_{\mr {II}}  & \leq &  \Qh   & \leq& \yh S,
\nll
{\mr {Region~III:}}
&\hspace{.7cm}&
\yh_2  &<& \yh  &<& \yh_d,
&\hspace{.7cm}&
\qh_{\mr {II}}  & \leq &  \Qh   & \leq& \yh S,
\nll
{\mr {Region~IV:}}
&\hspace{.7cm}&
\yh_d &<& \yh  &<& y_h^{\max}(\epsilon),
&\hspace{.7cm}&
\qh_{\mr {II}}  & \leq  &
               \Qh   & \leq &\qh_{\mr {I}}.
\end{array}
\ea
We introduce now the operation
\ba
\left\{ {\cal A} \right\}_{Qz}
= S \int^{ \min \{Q^2_{h\mr I}, y_h S \} }_{Q^2_{h\mr {II}}}
d\Qh [{\cal A}]_z,
\label{D18}
\ea
where the operation $[{\cal A}]_z$ is defined in~(\ref{aint}).
 
The integrals $\left\{ {\cal A}\right\}_{Qz}$ in the regions
${\mr {I}}$--${\mr {IV}}$ are:
\ba
\Biggl\{ \frac{\Ql}{z_1 z_2}\Biggr\}_{Qz}^{\mr I}
&=&  \frac{ 1 }{ \yl_h } \ln \frac{b}{(-a_1)(-a_2)},
\\
\Biggl\{ \frac{m^2}{z_1^2}  \Biggr\}_{Qz}^{\mr I}
&=&
\Biggl\{ \frac{m^2}{z_2^2}  \Biggr\}_{Qz}^{\mr I}   = 0,
\\
\Biggl\{ \frac{\Ql}{z_1 z_2}\Biggr\}_{Qz}^{\mr {II}}
&=&
\frac{ 1 }{ \yl_h } \Biggl(
           \ln \frac{\Ql}{m^2}+ \ln \frac{a_1}{-a_2} \Biggr),
\\
\Biggl\{ \frac{m^2}{z_1^2} \Biggr\}_{Qz}^{\mr {II}}
&=&
\frac{ 1 }{ \yl_h },
\\
\Biggl\{ \frac{m^2}{z_2^2} \Biggr\}_{Qz}^{\mr {II}}
&=& 0,
\\
\Biggl\{ \frac{\Ql}{z_1 z_2}\Biggr\}_{Qz}^{\mr {III}}
&=&
\frac{ 1 }{ \yl_h } \Biggl(
          2 \ln \frac{\Ql}{m^2}+ \ln \frac{a_1 a_2}{b} \Biggr),
\\
\Biggl\{ \frac{m^2}{z_1^2} \Biggr\}_{Qz}^{\mr {III}}
&=&
\Biggl\{ \frac{m^2}{z_2^2} \Biggr\}_{Qz}^{\mr {III}}
= \frac{ 1 }{\yl_h },
\\
\Biggl\{ \frac{\Ql}{z_1 z_2}\Biggr\}_{Qz}^{\mr {IV}}
&=&
\frac{ 2 }{ \yl_h }
          \ln \frac{\Ql}{m^2},
\\
\Biggl\{ \frac{m^2}{z_1^2} \Biggr\}_{Qz}^{\mr {IV}}
&=&
\Biggl\{ \frac{m^2}{z_2^2} \Biggr\}_{Qz}^{\mr {IV}} = \frac{ 1 }{\yl_h },
\ea
with
\ba
 a_1 &=& ( 1 - \yl \xl )( \yh - \yh_1 ) ,
 \nll
 a_2 &=& [ 1 - \yl (1-\xl )]( \yh - \yh_2 ) ,
 \nll
 b   &=& ( \yh - \xl \yl)^2 (1- \yl - \xl \yl r ),
 \nll
 \yl_h&=&y_l-y_h.
\ea
 
  Now  we define the third integration over $ y_h $:
\ba
\left({\cal A}\right)_{yQz} =
\int^{y_{h1}}_{y_h^{\min}} d\yh
\left\{{\cal A} \right\}_{Qz}^{\mr I}
                 + \int_{y_{h1}}^{y_{h2}} d\yh
                 \left\{{\cal A} \right\}_{Qz}^{\mr {II}}
                 + \int_{y_{h2}}^{y_{hd}} d\yh
                 \left\{{\cal A} \right\}_{Qz}^{\mr {III}}
                 + \int_{y_{hd}}^{y_h^{\max}(\epsilon)} d\yh
                 \left\{{\cal A} \right\}_{Qz}^{\mr {IV}}.
\nll
\ea
The result of this:
\ba
\Biggl( \frac{\Ql}{z_1 z_2}  \Biggl)_{yQz}
&=&
\label{D26}
          \ln \frac{\Ql}{m^2}
          \ln \frac{y_l^2 ( 1 - \xl )^2 ( 1 - \yl )S^2 }
                   {4 M^2 { \epsilon^2}(1-\yl\xl)[1-\yl(1-\xl) ] }
\nll
 & &+~\frac{1}{2}  \ln^2 (1-\yl)
 -  \frac{1}{2}  \ln^2 \left[\frac{ 1-\yl (1-\xl)}{ 1-\yl \xl} \right]
 \nll
 & &-~\litwo \left[  \frac{ 1-\yl }{ (1-\yl\xl)[1-\yl(1-\xl)]} \right]
 -  \litwo(1),
\\
\Biggl( \frac{m^2}{z_1^2}  \Biggr)_{yQz}
&=&
          \ln \frac{\yl ( 1 - \xl )S}
                   {2 M {\epsilon}( 1 - \yl \xl )},
\\
\Biggl( \frac{m^2}{z_2^2}  \Biggr)_{yQz}
&=&
          \ln \frac{\yl ( 1 - \xl )( 1 - \yl )S }
                 {2 M {\epsilon}[ 1 - \yl (1-\xl) ]}.
\label{D28}
\ea
{}From~(\ref{D26})--(\ref{D28}),~(\ref{eqn421}) follows immediately.
\subsubsection
{Leptonic variables. Photoproduction
\label{appfo}
}
The first integration of~(\ref{ir3}) with
$Q^2 = \ql$,  is over $z_2$.
Then, one has to integrate~(\ref{point}), which becomes~(\ref{eqn052a})
for leptonic variables,
 over $\qh$ within
 the integration region as shown in figure~\ref{figfot2}.
We begin with these two integrations:
\ba
\left\{ {\cal A} \right\}_{Qz}
\equiv
\int_{Q_h^{2 \min}}^{Q_h^{2 \max}} d\qh [{\cal A}]_z,
\label{intf2}
\ea
where the boundaries of the integration over $\qh$ are~(\ref{fotqhm}),
and the $[{\cal A}]_z$ is defined and calculated in
appendix~\ref{appe1}.
Two such twofold integrals have to be performed:
\ba
\left\{ \frac{\ql}{z_1 z_2} \right\}_{Qz}
&=&
\frac{1}{W^2-M_h^2}\,  \frac{1}{\beta}\,  {\mr L}_{\beta},
\\
\left\{ \frac{m^2}{z_1^2}  \right\}_{Qz}
=
\left\{ \frac{m^2}{z_2^2}  \right\}_{Qz}
&=&
\frac{1}{W^2-M_h^2}.
\label{ffi3}
\ea
The
L$_{\beta}$ and
 $\beta$
are defined in~(\ref{k120v2}).
There is only one third integration:
\ba
\left\{ \frac{1}{W^2-M_h^2}   \right\}_M
&\equiv&
\int
_{M_h^{2 \min}}
^{M_h^{2 \max}} dM_h^2 \frac{1}{W^2-M_h^2}
\nll
&=&
\ln \frac{W^2-(M+m_{\pi})^2}
{2 {\epsilon}\sqrt{W^2}}.
\label{intf3}
\ea
The boundaries for this integration are
~(\ref{lmh2}).
After three integrations, we finally arrive at~(\ref{eqhard}).
\subsubsection{Mixed variables
\label{e32}
}
W consider the case $\xm \leq 1$.
{}From the discussion
of the roots of $C_{1}$ and $C_{2}$
in appendix~\ref{b2z}
it may be seen that there are three integration regions:
\ba
\begin{array}{lccccclcrcccl}
{\mr{Region~I:}}
&\hspace{.7cm}&
y_l^{\min}(\epsilon)
& < &
\yl
& < &
\yl_d,
&\hspace{.7cm}&
\qh_{\mr {II}}
&\leq&
\Qh
&\leq&
\qh_{\mr I},
\nll
{\mr {Region~II:}}
&\hspace{.7cm}&
\yl_d    &<& \yl &<&   \yl_2,
&\hspace{.7cm}&
\qh_{\mr {II}}  & \leq &  \Qh   & \leq& \yh S,
\nll
{\mr {Region~III:}}
&\hspace{.7cm}&
\yl_2  &<& \yl  &<& y_l^{\max},
&\hspace{.7cm}&
\qh_{\mr {II}}  & \leq &  \Qh   & \leq& \yh S,
\end{array}
\ea
with $\Qh_{\mr {I, II}}$  given by~(\ref{eq30}).
 
Using the definition
~(\ref{D18}) for   $ \{ A \}_{Qz} $, one gets the
following results in the three
 regions ${\mr I}$--${\mr {III}}$:
\ba
\Biggl\{ \frac{\Ql}{z_1 z_2}\Biggr\}_{Qz}^{\mr I}
&=&
                              \frac{2}{\yl_h}  \ln \frac{\Ql}{m^2},
\\
\Biggl\{ \frac{m^2}{z_1^2} \Biggr\}_{Qz}^{\mr I}
&=&
\Biggl\{ \frac{m^2}{z_2^2} \Biggr\}_{Qz}^{\mr I}
=  \frac{ 1 }{ \yl_h },
\\
\Biggl\{ \frac{\Ql}{z_1 z_2} \Biggr\}_{Qz}^{\mr {II}}
&=&
\frac{ 1 }{ \yl_h } \Biggl[
          2 \ln \frac{\Ql}{m^2}+
            \ln \frac{[1-\xm( 1+\yh-\yl)] (\yl_2-\yl)}
                     {(1-\xm)^2 ( 1 - \yl - \xl \yl r)} \Biggr],
\\
\Biggl\{ \frac{m^2}{z_1^2}  \Biggr\}_{Qz}^{\mr{II}}
&=&
\Biggl\{ \frac{m^2}{z_2^2}  \Biggr\}_{Qz}^{\mr{II}}
=
\frac{ 1 }{ \yl_h },
\\
\Biggl\{ \frac{\Ql}{z_1 z_2} \Biggr\}_{Qz}^{\mr{III}}
&=&
\frac{ 1 }{ \yl_h } \Biggl[
           \ln \frac{\Ql}{m^2}
          + \ln \frac{ 1 - \xm (1+\yh-\yl)} {\yl -\yl_2} \Biggr],
\\
\Biggl\{ \frac{m^2}{z_1^2}\Biggr\}_{Qz}^{\mr{III}}
& = & \frac{ 1 }{\yl_h },
\\
\Biggl\{ \frac{m^2}{z_2^2}\Biggr\}_{Qz}^{\mr{III}}
&=& 0.
\ea
The third integration is defined     by
\ba
\left( {\cal A}\right)_{yQz} =
       \int^{ y_{l_d}}_{y_l^{\min}(\epsilon)} d\yl \{ {\cal A}\}_{Qz}^{\mr I   }
     + \int^{y_{l_2}}_{y_{l_d}} d\yl \{ {\cal A}\}_{Qz}^{\mr{II} }
     + \int^{ y_l^{\max} }_{y_{l_2}} d\yl \{ {\cal A}\}_{Qz}^{\mr{III}}.
\ea
The results of the integrations are:
\ba
\Biggl( \frac{\Ql}{z_1 z_2}  \Biggr)_{yQz}
&=&
\label{D36}
          \ln \frac{\Ql}{m^2}
          \ln \frac{  ( 1 - \yh )^2 ( 1 - \xm )S^2 }
                   { 4 {\epsilon^2}M^2  }
          +   \ln  ( 1 - \yh ) \ln ( 1 - \xm )
\nll
 & &~-
\frac{1}{2} \ln^2( 1 - \xm )
  -  \litwo \left[ \frac{ \xm ( 1-\yh ) }
{ \xm -1 } \right] ,
\\
\Biggl( \frac{m^2}{z^2_1}  \Biggr)_{yQz}&=&
          \ln \frac{  ( 1 - \yh     )S}
                   { 2  { \epsilon}M}  ,
\\
\Biggl( \frac{m^2}{z^2_1}  \Biggr)_{yQz}&=&
          \ln \frac{ ( 1 - \yh     )( 1 - \xm     )S}
                   { 2 {\epsilon}M} .
\label{D38}
\ea
 {}From~(\ref{D36})--(\ref{D38}),~(\ref{eqn422}) immediately follows.
%
\subsubsection{Hadronic variables
\label{sshv}
}
 
For the hadronic  variables one has to integrate
$\delta_{\mr{ {hard}}}^{\mr{ IR}} $
over the kinematical region of figure~\ref{yhqh}.
With the aid of the boundary conditions  which are derived in
appendix~\ref{b3g}   the integration region may be
split into the following three subregions:
\ba
\begin{array}{lllllll}
{\mr{Region~I:}}&    y_l^{\min}(\epsilon) \; &\leq \; \yl \; &\leq \; \yl_3,
           \qquad  & \Ql_{\mr{II}} \;   &\leq \;
                        \Ql \; &\leq \; \Ql_{\mr{I}},     \nll
{\mr{Region~II:}}&   \yl_3 \; &\leq \; \yl \;&\leq \; \yl_2,
           \qquad  & \Ql_{\mr{IV}} \; &\leq \;
                        \Ql \; &\leq \; \Ql_{\mr{I}}, \nll
{\mr {Region~III:}}&  \yl_2 \;&\leq \; \yl \;&\leq \; {\bar{y}}_l,
           \qquad  & \Ql_{\mr{IV}} \; &\leq \;
                         \Ql \; &\leq \; \Ql_{\mr{III}}.
\end{array}
\ea
 
The operation  $ \{ {\cal A} \}_{Qz} $ is defined
as follows:
\ba
\{ {\cal A}\}_{Qz} = S\int^{
\min\{Q^2_{l\mr I},Q^2_{l\mr{III}}\}
}_{
\max\{Q^2_{l\mr{II}},Q^2_{l\mr{IV}}\}
}
   d\Ql [{\cal A}]_z.
\ea
The  integrals $ \{ {\cal A}\} $ in the three  regions
${\mr{I}}$--${\mr{III}}$
are:
\ba
\Biggl\{ \frac{\Qh}{z_1 z_2}\Biggr\}_{Qz}^{\mr I}
&=&
                    \frac{2}{ \yl_h} \ln \frac{\Qh}{m^2},
\\
\Biggl\{ \frac{m^2}{z_1^2}  \Biggr\}_{Qz}^{\mr I}
          &=&  \frac{ 1 }{ \yl_h(1-\yl_h)},
\\
\Biggl\{ \frac{m^2}{z_2^2}    \Biggr\}_{Qz}^{\mr I}
          &=&  \frac{ 1-\yl }{ \yl_h( 1-\yh )},
\\
\Biggl\{ \frac{\Qh}{z_1 z_2}\Biggr\}_{Qz}^{\mr {II}}
&=&
\frac{ 1 }{ \yl_h} \Biggl[
          2 \ln \frac{\Qh}{m^2}+
            \ln \frac{ (1 - \yl)(2 +
 \sqrt{\lambda_h}/S
- \yh ) }
                {[ 2 (1 - \yl) +
       \sqrt{\lambda_h}/S
+ \yh] (1- \yh- \xh\yh r)} \Biggr],
\\
  \Biggl\{ \frac{m^2}{z_1^2}\Biggr\}_{Qz}^{\mr {II}}
  &=&
\frac{ 1 }{ \yl_h(1-\yl_h) },
\\
\Biggl\{ \frac{m^2}{z_2^2}  \Biggr\}_{Qz}^{\mr {II}}
&=&
\frac{ 1 - \yl }{ \yl_h( 1 - \yh ) },
\\
\Biggl\{ \frac{\Qh}{z_1 z_2}\Biggr\}_{Qz}^{\mr {III}}
&=&
\frac{ 1 }{ \yl_h } \Biggl[
           \ln \frac{\Qh}{m^2}
          + \ln \frac{ (1 - \yl)^2 }
                     {\xh \yh r - (1-\yl)(1 - \yl_h)} \Biggr],
\\
  \Biggl\{ \frac{m^2}{z_1^2}\Biggr\}_{Qz}^{\mr {III}}
  &=& 0,
  \\
\Biggl\{ \frac{m^2}{z_2^2}\Biggr\}_{Qz}^{\mr {III}}
&=&
         \frac{ 1-\yl }{ \yl_h( 1-\yh ) }.
\ea
 
The third integration is:
\ba
\left( {\cal A}\right)_{yQz} =
       \int^{y_{l_{3}}}_{y_l^{\min}(\epsilon)}d\yl\{ {\cal A}\}_{Qz}^{\mr I}
     + \int^{y_{l_{2}}}_{y_{l_{3}}} d\yl \{ {\cal A}\}_{Qz}^{\mr {II}}
     + \int^{ {\bar{y}}_l }_{y_{l_{2}}} d\yl \{ {\cal A}\}_{Qz}^{\mr {III}}.
\ea
And  the results of the integrations are
\ba
\Biggl( \frac{\Qh}{z_1 z_2} \Biggr)_{yQz} &=&
\label{D46}
            \ln \frac{\Qh}{m^2}
            \ln \frac{(1 - \yh)^2 S^2 }
                     { 4 { \epsilon^2}M^2}
                          +\litwo(1-\yh)- \litwo(1),
\\
\Biggl( \frac{m^2}{z_1^2} \Biggr)_{yQz}  &=&
          \ln \left(
          \frac{S}{2 {\epsilon}M}
 \frac{ 1 - \yh }{\yh}\right),
\\
\Biggl( \frac{m^2}{z_2^2} \Biggr)_{yQz}  &=&
          \ln \frac{(1-\yh)S}{2 {\epsilon}M} - 1.
\label{D48}
\ea
{}From~(\ref{D46})--(\ref{D48}),~(\ref{eqn423}) follows immediately.
 
\subsection{The second integration for $\delta_{\mr{ {R}}}^{\mr{F}} $.
Mixed and hadronic variables
\label{appe4}
}
When performing the integration over $\yl$ at given values of
\yh, \qh, and \ql\
within  the limits $ \yl^{min} $ and $ \yl^{max} $,
one faces eleven types of integrals which should be treated separately in regions
 I and II  of figure~\ref{ylqha}.
 
For details of their definitions
see the appendices~\ref{b22} for the case of mixed and~\ref{b32}
for hadronic variables.
 
The integrals are:
\ba
\left[ {\cal A} \right]_y   &=&
     S \int_{y_l^{\min}}^{y_l^{\max}}
 d\yl \, {\cal A}.
\ea
%
Integrals for region I with $\ql >\qh$:
\ba
\left[  \frac{1}{\sqrt{\lambda_q}} \right]_y^{\mr I} &=&
      \ln\left(\frac{ \Qh}{\yh \Ql}\right), \\
\left[ \frac{1} {\sqrt{C_1}}  \right]_y^{\mr I}    &=&
 \frac{1}{\Ql}
  \ln\left[\frac{\Ql}{m^2}
 \frac{\left(1-\frac{\yh\Ql}{\Qh}\right)}
      {\left(1-\frac{\Qh}{\Ql}\right)}\right], \\
\left[  \frac{1} {\sqrt{C_2}} \right]_y^{\mr I}    &=&
    \frac{1}{\Qh}
   \ln{ \left[ \frac{1-\yh}
  { 1-\frac{\Qh}{\Ql} }\right] },             \\
\left[  \frac{\yl}{\sqrt{C_1}} \right]_y^{\mr I}   &=&
       \frac{\Qh-\yh \Ql}{Q_l^4}
    \left(2-\frac{\Ql}{\Qh}\right)
               +\frac{ \Ql-\Qh+ \yh\Ql}{\Ql}
\left[  \frac{1} {\sqrt{C_1}}   \right]_y^{\mr I}, \\
\left[  \frac{\yl}{\sqrt{C_2}} \right]_y^{\mr I}   &=&
      \frac{ \Qh -\yh \Ql}{Q_h^4}
             + \frac{ \Qh - \Ql + \yh\Ql )}{\Qh}
\left[   \frac{1} {\sqrt{C_2}}  \right]_y^{\mr I}, \\
\left[  \sqrt{C_1}             \right]_y^{\mr I}   &=&
      S^2(  \Qh-\yh \Ql)^2 \left( \frac{\Ql}{2 Q_h^4}
              +\frac{1}{\Ql}-\frac{1}{\Qh}\right),  \\
\left[  \sqrt{C_2}             \right]_y^{\mr I}   &=&
S^2 (\Qh-\yh \Ql)
\left(-\frac{\yh\Ql}{2\Qh}+\frac{\Ql}{\Qh}-\frac{1}{2}\right), \\
\left[ \frac{ m^2}{C_1^{3/2}}  \right]_y^{\mr I}   &=&
     \frac{\Ql}{2 S^2 ( \Qh-\yh \Ql) (\Ql-\Qh)^2},  \\
\left[ \frac{ m^2}{C_2^{3/2}}  \right]_y^{\mr I}   &=&   0,  \\
\left[ \frac{m^2 \yl}{C_1^{3/2}}\right]_y^{\mr I}  &=&
     \frac{  \Ql-\Qh + \yh \Ql  }
          { 2 S^2 ( \Qh-\yh \Ql) (\Ql-\Qh)^2 },
\\
\Biggl[ \frac{m^2 \yl}{C_2^{3/2}} \Biggr]_y^{\mr I}
&=&  0.
\ea
Integrals for region II with $\Ql < \Qh$:
\ba
\Biggl[ \frac{1}{\sqrt{\lambda_q}}\Biggr]_y
^{\mr{II}}
 &=&
      \ln\left(\frac{1}{\yh}\right),               \\
 \left[ \frac{1} {\sqrt{C_1}}     \right]_y^{\mr{II}} &=&
   \frac{1}{\Ql}
        \ln\left[ \frac{ 1-\frac{\yh \Ql}{ \Qh} }
             { 1-\frac{\Ql}{\Qh} } \right],         \\
\left[ \frac{1} {\sqrt{C_2}}     \right]_y^{\mr{II}} &=&
      \frac{1}{\Qh}
      \ln\left[ \frac{\Ql (1-\yh)}
                 {m^2 \left(1-\frac{\Ql}{\Qh}\right)} \right],\\
\left[ \frac{\yl}{\sqrt{C_1}}    \right]_y^{\mr{II}}  &=&
 \frac{1-\yh}{\Ql}
               +\frac{ \Ql-\Qh+ \yh\Ql }{\Ql}
\left[  \frac{1} {\sqrt{C_1}}   \right]_y^{\mr{II}} , \\
\left[ \frac{\yl}{\sqrt{C_2}}    \right]_y^{\mr{II}}  &=&
       \frac{( 1-\yh) (2 \Ql-\Qh)}{Q_h^4}
                +\frac{ \Qh-\Ql+ \yh \Ql }{\Qh}
\left[  \frac{1} {\sqrt{C_2}}   \right]_y^{\mr{II}}, \\
\left[ \sqrt{C_1}                \right]_y^{\mr{II}} &=&
        (1-\yh)S^2\left[\Qh-\frac{(1+\yh)\Ql}{2}\right], \\
\left[ \sqrt{C_2}                \right]_y^{\mr{II}} &=&
       (1-\yh)^2S^2\left[\frac{Q^4_l}{\Qh}-\Ql+\frac{\Qh}{2}\right], \\
\left[ \frac{ m^2}{C_1^{3/2}}    \right]_y^{\mr{II}} &=& 0, \\
\left[ \frac{ m^2}{C_2^{3/2}}    \right]_y^{\mr{II}} &=&
       \frac{1}{2 (1-\yh) (\Ql-\Qh)^2S^2},
\\
\left[ \frac{m^2 \yl}{C_1^{3/2}} \right]_y^{\mr{II}}      &=&   0,\\
\left[ \frac{m^2 \yl}{C_2^{3/2}} \right]_y^{\mr{II}}      &=&
      \frac{ \Qh-\Ql+\yh \Ql }{2 \Qh (1-\yh) (\Ql-\Qh)^2S^2}.
\ea
\section
{
Tables of integrals for section~7
\label{appg}
}
\ezero
 
\subsection
{The $\cos \vartheta_{R}, \varphi_{R}$-integrals
\label{d1}
}
The integrals are defined as follows:
\ba
[{\cal A}]_{\varphi_{R}}
&=&
\frac{1}{2\pi} \int_{0}^{2\pi} d \varphi_{R} {\cal A},
\label{intphi}
  \\
\; [ 1 ]_{\varphi_{R}}
&=& 1,
           \\
\left[ 1- \yl \right]_{\varphi_{R}}
&=& A,
           \\
\left[ (1 - \yl)^2 \right]_{\varphi_{R}}
&=& A^2 + \frac{1}{2} B^2.
\label{ipht}
\ea
The following abbreviations are used:
\ba
A &=& \frac{1}{2 \tau} \left\{ (1-\yh)(\tau+m^2)
+ \frac{z_2}{\sqrt{\lambda_1}}
\left[\qh-z_2-\yh(\qh+\tau+m^2)\right]\cos \vartheta_{R} \right\},
\\
B &=& \frac{z_2}{2\tau S} \sqrt{\lambda_{\tau}} \sin \vartheta_p
\sin \vartheta_{R}.
\label{AB}
\ea
After applying these exact formulae, in a second step
the
following integrals are applied:
\ba
[{\cal A}]_{\vartheta_{R}}
&=&
\frac{1}{2} \int_{-1}^{1} {\cal A},
\label{intthe}
\\
\; [ 1 ]_{\vartheta_{R}} &=& 1,
\\
\left[ \cos \vartheta_{R} \right]_{\vartheta_{R}} &=& 0,
\\
 \left[ \frac{1}{z_1} \right]_{\vartheta_{R}} &=& \frac{\tau}{z_2}
 \frac{1}{\sqrt{\lambda_1}}
       \ln \frac{\qh+\tau+m^2+\sqrt{\lambda_1}}
{\qh+\tau+m^2-\sqrt{\lambda_1}},
\\
\left[ \frac{1}{z_1^2} \right]_{\vartheta_{R}} &=& \frac{\tau}{z_2}
\frac{1}{m^2 z_2}.
\label{itht}
\ea
In ultra-relativistic approximation:
\ba
   \ln \frac{\qh+\tau+m^2+\sqrt{\lambda_1}}
{\qh+\tau+m^2-\sqrt{\lambda_1}} \approx \ltau,
\label{ltz}
\ea
and
\ba
 \left[ \frac{1}{z_1} \right]_{\vartheta_{R}}
                  \approx \frac{1}{Q_{\tau}^2}\ltau.
\label{z1u}
\ea
The $\ltau$ and $Q_{\tau}^2$ are defined in~(\ref{ltau}) and~(\ref{qtau2}),
respectively.
 
The twofold integrals over the two photon angles,
${\cal J}^R[{\cal A}]=(\tau-m^2)/(4\pi\tau) \int d \Omega_R {\cal A}$,
with ${\cal A} = 1/z_1$ and ${\cal A} = 1/z_1^2$ agree exactly with those
which have been calculated for bremsstrahlung problems in
$e^+e^-$ annihilation in~\cite{777}.
\subsection
{The third integration for
$\delta_{\mr{ {hard}}}^{\mr{ IR}} $
\label{d2}
}
In the integration of the infrared divergent part of the cross section
we use the following regularized
integrals for the integration of the hard, finite
contributions in hadronic variables:
\ba
[{\cal A}]_{z}^h &=& \int_{{\bar z}_{2}}^{\tau^{\max}} dz_{2} {\cal A},
\ea
with ${\bar z}_2=2m\epsilon$, and $\tau^{\max}=\qh(1-\yh)/\yh$,
see~(\ref{tau2}).
 
In the case of Jaquet-Blondel variables, the upper boundary differs:
\ba
[{\cal A}]_{z}
^{\mr{JB}}
 &=& \int_{{\bar z}_{2}}^{\tau^{\max}} dz_{2} {\cal A},
\ea
with $\tau^{\max}=(1-\xjb)(1-\yjb)S$, see~(\ref{boutau}).
 
The integrals are:
\ba
\left[\frac{1}{z_{2}}\right]_{z}^h&=&
\ln \left( \frac{\qh}{2 m \epsilon}
\frac{1-\yh}{\yh} \right),
\\
\left[\frac{{\mr L}_{\tau}}{z_2}\right]_{z}^h&=&
2\ln\frac{\Qh}{m^2} \ln \left( \frac{\qh}{2 m \epsilon}\frac{1-\yh}{\yh}
 \right)
-\frac{1}{2}
\ln^2\left( \frac{\qh}{ m^2}\frac{1-\yh}{\yh} \right)
-2 \litwo \left( 1-\frac{1}{\yh} \right) - \litwo(1),
\nll
\label{intz}
\\
\left[\frac{1}{z_{2}}\right]_{z}^{\mr{JB}}
&=& \ln \frac{(1-\xjb)(1-\yjb)S}{2m\epsilon},
\\
\left[\frac{{\mr L}_{\tau}}{z_{2}}\right]_{z}^{\mr{JB}}
&=&
2 \ln\frac{\qjb}{m^2} \ln \frac{(1-\xjb)(1-\yjb)S}{2m\epsilon}
-\frac{1}{2} \ln^2 \frac{(1-\xjb)(1-\yjb)S}{m^2}
\nll
& &-~2 \litwo\left(1-X_{\mr{JB}}\right) - \litwo(1),
\label{izjb}
\ea
where
\ba
X_{\mr{JB}} = \frac{1-\xjb(1-\yjb)}{\xjb\yjb}.
\label{Xjb}
\ea
The regular integrals may be found in appendix~\ref{d3}.
\subsection
{The third integration for $\delta_{\mr{ {R}}}^{\mr{F}} $
\label{d3}
}
A second series of integrals over $z_2$ is used in order to perform
analytically even the last, third integration
in case of hadronic variables.
Here, this integration corresponds to an integral over the invariant
mass $\tau$; see the discussion in section~\ref{irIIh}.
 
The finite integrals over $z_2$ or, equivalently, over $\tau$,
are defined as follows:
\ba
\left[ {\cal A} \right]_{\tau}^h
&=&
\int_{m^2}^{\tau^{\max}} d\tau {\cal A},
\\
\left[ {\cal A} \right]_{\tau}^{\mr{JB}}
&=&
\int_{m^2}^{\tau^{\max}} d\tau {\cal A}.
\ea
Here, at the upper integration limit the ultra-relativist approximation
may be applied:
$\tau^{\max} \approx z_2^{\max}
=\qh(1-\yh)/\yh$ in the case of hadronic variables,
and
$\tau^{\max} \approx z_2^{\max}=(1-\xjb)(1-\yjb)S$
in the case of Jaquet-Blondel variables.
 
The integrals are:
\ba
\left[ 1 \right]_{\tau}^h&=&
\qh \frac{1-\yh}{\yh},
\\
\left[\frac{1}{\tau}\right]_{\tau}^h&=&
\ln \left( \frac{\qh}{m^2} \frac{1-\yh}{\yh}\right),
\\
\left[\frac{1}{\tau^2}\right]_{\tau}^h&=&
\frac{1}{m^2},
\\
\left[\frac{1}{Q_{\tau}^6}\right]_{\tau}^h&=&
\frac{1}{2} (1- y_h^2) \frac{1}{Q_h^4},
\\
\left[\frac{1}{Q_{\tau}^4}\right]_{\tau}^h&=&
            (1-\yh) \frac{1}{Q_h^2},
\\
\left[\frac{1}{Q_{\tau}^2}\right]_{\tau}^h&=&
\ln \frac{1}{\yh},
\\
\left[\ltau \right]_{\tau}^h&=&
\frac{\qh}{\yh} \left\{2
\ln\frac{1}{\yh}
+(1-\yh)\left[2\mbox{ln}\frac{\Qh}{m^2}-\ln\left(\frac{\Qh}{m^2}
\frac{1-\yh}{\yh}
\right) -1 \right] \right\},
\\
\left[\frac{\ltau}{Q_{\tau}^2} \right]_{\tau}^h&=&
\ln^2 \yh
 -\ln\yh\mbox{ln}\frac{\Qh}{m^2}+\litwo\left(\frac{1}{\yh}\right)
 -
 \litwo(1),
\\
\left[ \frac{\ltau}{Q_{\tau}^4} \right]_{\tau}^h&=&
\frac{1}{\qh} \left[ -(1-2\yh)\ln\yh
 +(1-\yh)\left(\mbox{ln}{\left(\frac{\Qh}{m^2}
    \frac{\yh}{1-\yh}\right)}       +2 \right) \right],
\\
\left[ \frac{\ltau}{Q_{\tau}^6} \right]_{\tau}^h&=&
\frac{1}{2 Q_h^4} \left[-(1 - 2 y_h^2)\ln\yh
 +(1- y_h^2)\mbox{ln}{\left(\frac{\Qh}{m^2}
     \frac{\yh}{1-\yh}\right)}        +(1-\yh)(2+\yh)\right].
\nll
\ea
Further,
\ba
\left[\frac{1}{\tau}\right]_{\tau}
^{\mr{JB}}
&=&
\ln \frac{(1-\xjb)(1-\yjb)S}{m^2},
\\
\left[ \frac{\ltau}{Q_{\tau}^2} \right]_{\tau}^{\mr{JB}}
&=&
(1-\yjb)
\left[
\ln^2
X_{\mr{JB}}
+ \ln X_{\mr{JB}} \ln\frac{\qjb}{m^2(1-\yjb)} +\litwo(X_{\mr{JB}})
-\litwo(1) \right],
\ea
where the $X_{\mr{JB}}$ is defined in~(\ref{Xjb}).
\section
{Leading logarithmic approximations
\label{lla}
}
\ezero
In the leading logarithmic approximation, the \oa\ corrections consist of
initial state radiation, final state radiation, and the Compton peak
contribution,
\ba
\frac{d^2 \sigma^{\mr R}}{dxdy}
&=&
\frac{d^2 \sigma^{i}}{dxdy}
+
\frac{d^2 \sigma^{f}}{dxdy}
+
\frac{d^2 \sigma^{C}}{dxdy},
\label{lla1}
\ea
where the latter contribution exists only for
leptonic and mixed variables. For leptonic variables it is numerically
important:
\ba
\frac{d^2 \sigma^{C}}{d\xl d\yl}
&=&
\frac{\alpha^3}{\xl S} \left[1+(1-\yl)^2\right] \ln\frac{\ql}{M^2}
\int \limits_{x_l}^{1} \frac{dz}{z^2}
\frac{z^2+(\xl-z)^2}{\xl(1-\yl)}
\sum_f \left[q_f(z,\ql) + {\bar q}_f(z,\ql)\right].
\nll
\label{compt1}
\ea
The other two cross sections have the following structure:
\ba
\frac{d^2 \sigma^{a}}{dxdy}
&=&
\int\limits_{0}^{1}dz \left[\frac{\alpha}{2\pi}
\left( \ln\frac{Q_a^2}{m^2} \right)
\frac{1+z^2}{1-z} \right]
\left\{
\theta(z-z_0)
{\cal J}(x,y,Q^2)
\left.
\frac{d^2 \sigma^{0}}{dxdy}\right|_{x={\hat x}, y={\hat y}, S={\hat S}}
-
\frac{d^2 \sigma^{0}}{dxdy}
\right\} .
\nll
\label{lla2}
\ea
For the different sets of variables, the definitions
of the ${\hat x},{\hat
y},{\hat S}$, as well as the lower integration boundary $z_0$, differ.
For leptonic variables, they are well-known
since long; see, e.g.~\cite{consoli,kuraev}.
For the cases of mixed and Jaquet-Blondel variables,
they have been derived
in~\cite{johmix} and for further sets of variables
 in~\cite{bluemho} where also the
formulae for higher order LLA corrections with soft photon exponentiation
are collected and numerically discussed.
 
The formulae which were used for our comparisons in section~\ref{ll}
may be found in table~\ref{LLAtab}.
 
In the case of Jaquet-Blondel variables, the hadronic final state is treated
totally inclusive. Thus, in accordance with the Kinoshita-Lee-Nauenberg
theorem ~\cite{KNL} there is no LLA correction from final state radiation.
One should also mention that a gauge invariant
separation of initial and final
state radiation is possible for the leading logarithmic corrections but
not for the complete order \oa.
 
In~(\ref{lla2}), it is
\ba
{\hat x} &=& \frac{{\hat Q}^2}{{\hat y}{\hat S}},
\\
{\cal J}(x,y,Q^2) &=&
\left|\frac{\partial({\hat x},{\hat y})}{\partial(x,y)}\right|,
\\
{\hat x}(z_0) &=& 1.
\label{lla3}
\ea
{}From the last equation, the lower integration boundary derives.

More details on LLA corrections in a large variety of variables may be
found in~\cite{HECTOR}.
 
\begin{table}[thbp]
{\small
\begin{tabular}[]{|c|c|c|c|c|c|c|}
\hline 
     &  \multicolumn{3}{|c|}{} & \multicolumn{3}{|c|}{}
\\   &  \multicolumn{3}{|c|}{Initial state radiation}
     &  \multicolumn{3}{|c|}{Final   state radiation}
\\   &  \multicolumn{3}{|c|}{} & \multicolumn{3}{|c|}{}
\\ \hline 
   &&&&&&
\\ Variables
  & leptonic & Jaquet-Blondel & \hspace{.3cm} mixed \hspace{.3cm}
  & leptonic & Jaquet-Blondel & \hspace{.3cm} mixed \hspace{.3cm}
\\ &&&&&&
\\ \hline 
 &&&&&&
\\ ${\hat y}$ & $\frac{\displaystyle z+y-1}{\displaystyle z}$
              & $\frac{\displaystyle y}{\displaystyle z}$
              & $\frac{\displaystyle y}{\displaystyle z}$
              & $\frac{\displaystyle z+y-1}{\displaystyle z}$
              &    --       & $y$
\\ &&&&&&
\\ \hline 
 &&&&&&
\\ ${\hat S}$     & $zS$              & $zS$          & $zS$
                  & $               S                  $         & -- & $S$
\\ &&&&&&
\\ \hline 
 &&&&&&
\\ ${\hat Q^2}$   & $zQ^2$
                  & $\frac{\displaystyle 1-y}{\displaystyle 1-y/z}Q^2$ & $zQ^2$
                  & $\frac{\displaystyle Q^2}{\displaystyle z}$
                  &  --& $\frac{\displaystyle Q^2}{\displaystyle z}$
\\ &&&&&&
\\ \hline 
\end{tabular}
} 
\caption{The definition of scaling variables for the
different leading logarithmic
cross sections.
\label{LLAtab}
}
\end{table}
 
\newpage

\end{document}